\documentclass[11pt]{article}
\pdfoutput=1
\usepackage{jheppub}
\usepackage{graphicx}
\usepackage{array,tikz-cd}
\usetikzlibrary{positioning, quotes, arrows}
\usepackage{color}
\usepackage{bm, dsfont, braket}
\usepackage{enumitem}

\usepackage{etoolbox, textcomp}

\graphicspath{{./figures/}}

\newcommand*\diff{\mathop{}\!\mathrm{d}}
\newcommand{\pd}{{\partial}}
\newcommand{\ol}{\overline}
\newcommand{\wt}{\widetilde}
\newcommand{\mb}{\mathbf}
\newcommand{\ds}{\displaystyle}
\newcommand{\cp}{{\mathbb{CP}}}
\newcommand{\id}{{\mathds{1}}}
\newcommand{\actson}{\raisebox{-.4ex}{$\includegraphics[width=2.2ex]{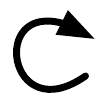}$}}

\newcommand{\catH}{\mathcal{C}^H}
\newcommand{\UH}{\overline{\mathbf{U}}^H}
\renewcommand{\AA}{\mathbb{A}}

\newtheorem{PTh}{Physics Theorem}
\newtheorem{conj}{Conjecture}
\newtheorem{ex}{Example}
\newtheorem{rmk}{Remark}
\newcommand{\FT}{\mathcal{FT}}
\newcommand{\FF}{{\text{FF}}}
\newcommand{\CCab}{\CC^{\rm ab}}
\newcommand{\CCabg}{{\CC^{\rm ab}_g}}
\newcommand{\M}{W}

\newcommand{\g}{\mathfrak{g}}

\newcommand{\Hom}{\text{Hom}}

\newcommand{\Tr}{\text{Tr}}

\newcommand{\Id}{\text{Id}}
\newcommand{\sln}{\mathfrak{sl}_n}
\newcommand{\sltwo}{\mathfrak{sl}_2}
\newcommand{\rep}{\text{Rep}}

\newcommand{\cf}{\emph{cf.}}
\newcommand{\ie}{\emph{i.e.}}
\newcommand{\eg}{\emph{e.g.}}

\newcommand{\be}{\begin{equation}}
	\newcommand{\ee}{\end{equation}}
\newcommand{\bp}{\begin{pmatrix}}
	\newcommand{\ep}{\end{pmatrix}}
\newcommand{\bsp}{\left(\begin{smallmatrix}}
	\newcommand{\esp}{\end{smallmatrix}\right)}

\newcommand{\R}{{\mathbb R}}
\renewcommand{\P}{{\mathbb P}}

\newcommand{\W}{{\mathbb W}}
\newcommand{\C}{{\mathbb C}}
\newcommand{\Z}{{\mathbb Z}}
\newcommand{\X}{{\mathbb X}}

\newcommand{\CA}{{\mathcal A}}
\newcommand{\CB}{{\mathcal B}}
\newcommand{\CC}{{\mathcal C}}
\newcommand{\CD}{{\mathcal D}}
\newcommand{\CE}{{\mathcal E}}
\newcommand{\CF}{{\mathcal F}}
\newcommand{\CG}{{\mathcal G}}
\newcommand{\CH}{{\mathcal H}}
\newcommand{\CI}{{\mathcal I}}

\newcommand{\CM}{{\mathcal M}}
\newcommand{\CN}{{\mathcal N}}
\newcommand{\CO}{{\mathcal O}}

\newcommand{\CR}{{\mathcal R}}
\newcommand{\CS}{{\mathcal S}}
\newcommand{\CT}{{\mathcal T}}

\newcommand{\CV}{{\mathcal V}}  
\newcommand{\CW}{{\mathcal W}}  
\newcommand{\CX}{{\mathcal X}}
\newcommand{\CZ}{{\mathcal Z}}
\newcommand{\CL}{{\mathcal L}}

\newcommand{\bA}{\mathbf{A}}
\newcommand{\bB}{\mathbf{B}}

\newcommand{\bW}{\mathbf{W}}
\newcommand{\bX}{\mathbf{X}}
\newcommand{\bY}{\mathbf{Y}}
\newcommand{\bGamma}{\mathbf{\Gamma}}

\newcommand{\bPsi}{\mathbf{\Psi}}
\newcommand{\bPhi}{\mathbf{\Phi}}

\newcommand{\bLambda}{\mathbf{\Lambda}}

\newcommand{\bOmega}{\mathbf{\Omega}}

\newcommand{\norm}[1]{{{:\!{#1}\!:}}}

\title{A QFT for non-semisimple TQFT}

\author[1]{Thomas Creutzig}
\author[2,3]{Tudor Dimofte}
\author[3,4]{Niklas Garner}
\author[5]{Nathan Geer}

\affiliation[1]{Department of Mathematical \& Statistical Sciences, University of Alberta, 632 CAB, Edmonton, Alberta, Canada T6G 2G1}
\affiliation[2]{School of Mathematics, University of Edinburgh, James Clerk Maxwell Building, Peter Guthrie Tait Road, Edinburgh EH9 3FD, UK}
\affiliation[3]{Department of Mathematics and Center for Quantum Mathematics and Physics (QMAP), University of California, Davis, CA 95616, USA}
\affiliation[4]{Department of Physics, University of Washington, Seattle, WA 98195, USA}
\affiliation[5]{Department of Mathematics \& Statistics, Utah State University, Logan, Utah 84322, USA}

\abstract{We construct a family of 3d quantum field theories $\CT_{n,k}^A$ that conjecturally provide a physical realization --- and derived generalization --- of non-semisimple mathematical TQFT's based on the modules for the quantum group $U_q(\mathfrak{sl}_n)$ at an even root of unity $q=\text{exp}(i\pi/k)$. 
The theories $\CT_{n,k}^A$ are defined as topological twists of certain 3d $\CN=4$ Chern-Simons-matter theories, which also admit string/M-theory realizations. They may be thought of as $SU(n)_{k-n}$ Chern-Simons theories, coupled to a twisted $\CN=4$ matter sector (the source of non-semisimplicity).
We show that $\CT_{n,k}^A$ admits holomorphic boundary conditions supporting two different logarithmic vertex operator algebras, one of which is an $\mathfrak{sl}_n$-type Feigin-Tipunin algebra; and we conjecture that these two vertex operator algebras are related by a novel logarithmic level-rank duality. (We perform detailed computations to support the conjecture.) We thus relate the category of line operators in $\CT_{n,k}^A$ to the derived category of modules for a boundary Feigin-Tipunin algebra, and --- using a logarithmic Kazhdan-Lusztig-like correspondence that has been established for $n=2$ and expected for general $n$ --- to the derived category of $U_q(\mathfrak{sl}_n)$ modules. We analyze many other key features of $\CT_{n,k}^A$ and match them from quantum-group and VOA perspectives, including deformations by flat $PGL(n,\mathbb C)$ connections, one-form symmetries, and indices of (derived) genus-$g$ state spaces.}

\begin{document}
\today
\maketitle


\newpage

\section{Extended introduction: perspectives on non-semismiple and derived TQFT}

\subsection{Brief introduction}
\label{sec:simpleintro}

Quantum invariants of links and three-manifolds rose to prominence three decades ago, incited by the discovery of Jones polynomials \cite{Jones-poly}, their physical realization via 3d Chern-Simons theory and the 2d WZW model due to Witten \cite{Witten-Jones}, and the reformulation of Chern-Simons partition functions via representation theory of quantum groups by Reshetikhin and Turaev~\cite{RT}. The interaction among the three emerging perspectives on quantum invariants
\be \label{persp}  \begin{array}{c} (1) \\ \text{topological QFT} \end{array}  \quad \leftrightarrow\quad 
  \begin{array}{c} (2) \\ \text{vertex operator algebras (VOA's)} \end{array}    \quad \leftrightarrow\quad  \begin{array}{c} (3) \\ \text{quantum groups} \end{array}  \ee
inspired countless surprising developments. Early examples include the equivalent construction of Hilbert spaces associated to surfaces via geometric quantization \cite{Witten-GQ}; WZW conformal blocks \cite{MS-poly, Segal}; and the modular-tensor-category structure of quantum-group representations~\cite{RT}.
More modern examples include an evolving network of approaches to categorification of quantum invariants, beginning with work of Khovanov \cite{Khovanov} in representation theory and a construction of Gukov-Schwarz-Vafa \cite{GSV} in QFT/string theory.

A central algebraic object in each of the above perspectives --- which contains all the necessary data to construct invariants of links and 3-manifolds --- is a braided tensor category~$\CC$.%
\footnote{We are only giving a rough picture here. More precisely, $\CC$ should have the structure of a ``modular'' tensor category, satisfying additional properties that ultimately lead to a definition of invariants of framed, oriented links in framed three-manifolds. See \eg\ the classic lectures \cite{BakalovKirillov} for further details.} %
In 3d topological QFT, $\CC$ is the category of line operators; while from the VOA and quantum-group perspectives, $\CC$ is a category of modules,
\be \CC \;\simeq\; \text{line operators in 3d QFT} \;\simeq\; \text{VOA modules} \;\simeq\; \text{quantum-group modules}\,. \ee
More precisely, in the original constructions of quantum invariants labelled by a compact Lie group $G$ and integer $k$, the braided tensor category could equivalently be described as %
\footnote{We use ``critically shifted'' levels throughout this paper. Thus  $k = k_{UV} + h$, where $k_{UV}$ is the level that appears in the UV Chern-Simons action and $h$ is the dual Coxeter number of $G$. Correspondingly, the OPE of currents in $V^k(\mathfrak g)$ is $J^a(z)J^b(w)\sim k g^{ab} /(z-w)^2+if^{ab}_c/(z-w)$. We also assume that $k\geq h$.}
\begin{align} \CC_{\text{s.s.}} & \;:=\; \text{Wilson lines in Chern-Simons theory with gauge group $G$, at level $k-h$} \notag \\
 & \;\simeq\; \text{modules for the WZW VOA $V^k(\mathfrak g)$ (a simple quotient of $\hat {\mathfrak g}_k$ current algebra)} \label{C-ss} \\
 & \;\simeq\; \text{a semisimplification of modules for $U_q(\mathfrak g)$ at an even root of unity $q =  e^{i\pi/k}$} \notag
\end{align}
A key property of $\CC_{\text{s.s.}}$ is that it is  \emph{semisimple}. We will elaborate momentarily on precisely what this means, but note for now that semisimplicity is a consequence of Chern-Simons theory having no local operators, and of the VOA $V^k(\mathfrak g)$ being rational. Semisimplicity was also built into the category of quantum group $U_q(\mathfrak g)$ modules used by \cite{RT}, which is a substantial reduction of the full category of finite-dimensional $U_q(\mathfrak g)$ modules at a root of unity.

One natural non-semisimple generalization of the original constructions of quantum invariants comes from replacing the compact group $G$ with a supergroup (or $\mathfrak g$ by a Lie superalgebra). The basic example of $G=U(1|1)$ Chern-Simons theory and the corresponding WZW model and quantum supergroup was studied in the early 1990's  \cite{KauffmanSaleur, RozanskySaleur}, in relation to Alexander polynomials and Reidemeister torsion. Many new subtleties were encountered, some of which are still under current development (\cf\ the recent \cite{GaiottoWitten-Janus,Mikhaylov,Mikhaylov:2014aoa,AGPS, Quella:2007hr, Gotz:2006qp, Creutzig:2011cu}). 

In this paper, we explore a different, albeit related generalization. Our main goal is to extend the three interconnected perspectives above to a setting that replaces the semisimiple category \eqref{C-ss} with
\begin{align} \label{C-nss} \CC &\;:=\; U_q(\mathfrak g)\text{-mod} \\ &\;=\; \text{all finite-dimensional modules of $U_q(\mathfrak g)$ at an even root of unity $q=e^{i\pi/k}$} \notag \\[-.1cm]
&\hspace{4.5ex} \text{on which the Frobenius center of $U_q(\mathfrak g)$ acts semisimply}\,. \notag \end{align}
(See Sections \ref{sec:flat-intro} and \ref{sec:U-modules} for more on the Frobenius center.) 
 $\CC$ is a very large category, whose structure was initially described by \cite{DCK,DCKP,Beck}. It contains a particularly interesting non-semisimple subcategory
\be \label{C1-intro} \CC_1 := u_q(\mathfrak g)\text{-mod}\,, \ee
consisting  of modules for the so-called ``restricted'' (or ``small'' or ``baby'') quantum group $u_q(\mathfrak g)$,  \cf\  \cite[Sec. XI.6.3]{Turaev:1994xb}.
The restricted quantum group has the $k$-th powers of Serre generators $E_i,F_i$ are set to zero, and the $2k$-th powers of maximal-torus generators $K_i$ set to~$1$.

The quantum-group and VOA perspectives have already been extensively developed. On the quantum-group side, a series of recent papers beginning with work of Costantino, Patureau-Mirand, and the fourth author (CGP) \cite{CGP} have developed systematic techniques for defining axiomatic TQFT's using non-semisimple categories such as $U_q(\mathfrak g)$ at a root of unity. This work unifies and generalizes earlier constructions, such as those of \cite{ADO, Hennings, Lyubashenko, Kashaev} in the 1990's.
 On the VOA side, we will connect with representation theory of logarithmic VOA's, notably the triplet algebra \cite{Kausch} and its generalizations, the Feigin-Tipunin algebras \cite{Feigin:2010xv}.

Our main contribution is to identify a physical, topological QFT $\CT^A_{G,k}$, labelled by a group $G$ and integer $k$, whose category of line operators is (conjecturally) the derived category $D^b\CC$. We mainly restrict our consideration to $G=SU(n)$ and $\mathfrak g=\mathfrak{sl}_n$, though there are natural guesses for how the correspondence may generalize to other groups/algebras.

The QFT $\CT^A_{G,k}$ is a derived, non-semisimple, and necessarily non-unitary, generalization of Chern-Simons theory. It may be defined by starting with the 3d $\CN=4$ S-duality interface $T[G]$ of~\cite{GaiottoWitten-Sduality}, gauging its Higgs-branch $G$ global symmetry with a Chern-Simons term at level $k$, and then taking an A-type topological twist. Schematically,
\be \label{T-intro} \CT_{G,k}^A :=\big( T[G]/G_k\big)^A \,.\ee
For $G=SU(n)$, we provide an explicit Lagrangian for $\CT_{G,k}^A$ in the BV formalism, and define a boundary condition supporting a new logarithmic VOA $\CN_{G,k}$. Motivated by corner constructions in 4d $\CN=4$ super-Yang-Mills \cite{GaiottoRapcak,CreutzigGaiotto-S}, we argue and partially prove that a slight modification%
\footnote{This minor modification involves taking a $\Z_n$ orbifold of a simple current extension of $\CN_{SU(n),k}$ and is analogous to the extension/orbifold appearing in level-rank duality in type A \cite{NS-duality,NS-braid,FvD}. See Section \ref{sec:levelrank} for more details.} %
$\wt{\CN}_{G,k}$ of $\CN_{G,k}$ is dual to a Feigin-Tipunin algebra, whose category of modules is in turn equivalent to \eqref{C-nss}. With the help of supersymmetric localization techniques,
we also check that characters of (derived!) state spaces and the Grothendieck ring of the category of line operators in $\CT_{G,k}^A$ match expected results from $U_q(\mathfrak g)$-mod. We make some predictions for the state spaces themselves using analogues of geometric quantization. An extended summary of our results appears in Section \ref{sec:results} below.

The origin of the duality between the new logarithmic VOA's $\wt{\CN}_{G,k}$ and Feigin-Tipunin algebras $\FT_k(\mathfrak g)$ 
is the same as the origin of level-rank duality in WZW models of type A \cite{NS-duality,NS-braid,FvD}.  Recall that level-rank duality expresses the WZW models $V^{n+k}(\mathfrak{gl}_k)$ (meaning: affine $\mathfrak{gl}_k$ at non-shifted level $n$) and $V^{k+n}(\mathfrak{sl}_n)$ as mutual cosets inside $nk$ pairs of free fermions $\FF(nk)$.
Equivalently, $\FF(nk)$ is a conformal extension of a simple current extension $\widetilde V^{n+k}(\mathfrak{gl}_k)$  of $V^{n+k}(\mathfrak{gl}_k)$  and $V^{k+n}(\mathfrak{sl}_n)$.
 Since $\FF(nk)$ is a ``holomorphic'' VOA, with a trivial module category, this induces a braid-reversed equivalence of module categories $V^{k+n}(\mathfrak{sl}_n)\text{-mod}\simeq \widetilde V^{n+k}(\mathfrak{gl}_k)\text{-mod}$  \cite{MR3162483}. In Section \ref{sec:VOA}, we will establish an equivalence of two deformable families of cosets, whose large-level limits are related to $\wt{\CN}_{SU(n),k}$ and Feigin-Tipunin algebras $\FT_k(\mathfrak{sl}_n)$.
 We conjecture that $\wt{\CN}_{SU(n),k}$ and $\FT_k(\mathfrak{sl}_n)$ are mutual cosets inside many copies of free fermions, with specific branching rules, \cf\ \eqref{eq:decleft}--\eqref{eq:decright}. We support the conjecture in the case of $\mathfrak{sl}_2$ via the computation of branching rules.
In Section \ref{sec:rect} we also point out a possible close connection of $\CN_{G,k}$ to rectangular $W$-algebras, which may be useful for further studies of $\CN_{G,k}$.

A feature of the 3d QFT's $\CT_{G,k}^A$, common to most theories defined via topological twists, is that its category of line operators is intrinsically a dg (differential graded) category (\cf\ \cite{KRS, Kapustin-ICM,Lurie-HA,Lurie}).
Only the dg category makes sense physically, and behaves well under dualities, such as 3d mirror symmetry.
This is why the equivalence of categories we are proposing involves line operators in $\CT_{G,k}^A$ and \emph{derived} categories of $U_q(\mathfrak g)$ modules and VOA modules.
This strongly motivates the existence of a derived/dg enhancement of structures currently studied in much of the axiomatic TQFT literature based on non-semisimple quantum group and VOA categories. Such an enhancement was also recently advocated and partially constructed in certain cases by \cite{SchweigertWoike, SW-Verlinde}. We will explore many derived/dg structures in the current paper.

The search for a physical QFT that computes invariants based on the full, non-semisimple category $U_q(\mathfrak g)$-mod was already instigated last year  \cite{RW-Coulomb}, motivated in part by recent developments in the 3d-3d correspondence, and in particular the discovery of logarithmic VOA characters \cite{CCFGH} in the ``homological blocks'' of \cite{GPV-spectra, GPPV-spectra}. This line of reasoning was developed in \cite{CostantinoGukovPutrov}, which in particular clarified the role of spin$^{(c)}$ structures in physical QFT's underlying CGP/ADO invariants. (We will say very little about spin structures in this paper, aside from observing that the definition of $\CT_{G,k}^A$ generally requires them.) 
As we were completing our paper, we learned of further work in progress by B. Feigin, S. Gukov, and N. Reshetikhin on similar subjects, \cf\ \cite{FGR}.
We expect that the theories $\CT_{G,k}^A$ studied in the present paper are 3d mirrors of the Rozansky-Witten-twisted (or ``B-twisted'') sigma-models with targets $X_k$ described in \cite{RW-Coulomb}, or some enhancements thereof; we expand further on the relation to \cite{RW-Coulomb} from a 6d perspective in Section \ref{sec:4d6d-intro}.

We also note that an obvious 3d mirror of \eqref{T-intro} for $n=2$ --- obtained by gauging a Coulomb-branch (rather than Higgs-branch) $SU(2)$ symmetry of $T[SU(2)]$ at level $k$ and taking a topological B-twist  (rather than A-twist)--- was  already observed to be related to $\FT_k(\mathfrak{sl}_2)$, \emph{a.k.a.} the triplet VOA, in \cite{CCG}.

We will not extract the full data/structure of an axiomatic TQFT from the theory $\CT_{G,k}^A$ in the present paper. In particular, we do not compute mapping-class-group actions on state spaces, or try to define partition functions on general three-manifolds. The latter often evaluate to zero or infinity without careful regularization. Some of these issues were addressed in \cite{RW-Coulomb}, as well as \cite{Mikhaylov} in the related context of supergroup Chern-Simons; and they were one of the principal difficulties to overcome in mathematical work on non-semisimple invariants, which we come to next.
We hope that a full, cohomological TQFT can be (re)constructed directly from the physical QFT  $\CT_{G,k}^A$ in the future.

\subsection{Organization}
\label{sec:org}

The remainder of this introduction is an extended summary of the concepts and results of the main body of the paper --- beginning with a precise definition of ``semisimple'' and ``non-semisimple,''  both mathematically and in terms of QFT.
We then review some key developments in quantum groups and logarithmic VOA's that motivated this paper.
We introduce a central feature of the category $U_q(\mathfrak g)$-mod that ultimately leads not just to invariants of topological three-manifolds, but to three-manifolds with background (classical) complex flat connections. Such flat connections will play an important role in the VOA and QFT perspectives as well.
Finally, we describe multiple physical constructions of the QFT $\CT_{G,k}^A$, and comment on their relations to analytically continued Chern-Simons theory, the 3d-3d correspondence, and the setup of \cite{RW-Coulomb}. In Section \ref{sec:results}, we give a more precise formulation of our new results.

In Section \ref{sec:toymodel}, we review and develop the structure of topologically twisted QFT's that can be coupled to background flat connections, while illustrating this structure in a self-contained toy model: the B-twist of a free 3d $\CN=4$ hypermultiplet $\CT_{\rm hyper}^B$. The theory $\CT_{\rm hyper}^B$ shares many qualitative features with our theories of interest $\CT_{G,k}^A$, but is much easier to study. It is related to ``superalgebra'' $\mathfrak{psu}(1|1)$ Chern-Simons; it couples to background flat $SL(2,\C)$ connections; and its partition functions are known to compute Alexander polynomials and Reidemeister-Ray-Singer torsion \cite{RozanskySaleur, Mikhaylov}. Despite being a free theory locally, $\CT_{\rm hyper}^B$ has a nontrivial dg category of line operators, equivalent to the derived category of modules for the symplectic fermion VOA \cite{Gaiotto-blocks,CostelloGaiotto}, as well as to a quotient of the derived category of $u_q(\mathfrak{sl}_2)$ modules at $q=i$. Its state spaces are easily computed in multiple ways. 
 
In Section \ref{sec:Uqsl2}, we review the structure of the category of $U_q(\mathfrak g)$ modules at even roots of unity $q=e^{i\pi/k}$, focusing on the simplest nontrivial case $\mathfrak g=\mathfrak{sl}_2$. We describe the precise version of the CGP TQFT --- defined by passing through the unrolled quantum group $U_q^H(\mathfrak{sl}_2)$ --- that we expect to be related to the physical QFT $\CT_{SU(2),k}^A$. We also compute the infinite-dimensional derived state spaces assigned to surfaces of genus 0 and 1, and the characters of these spaces for all genus.

In Section \ref{sec:Atwist}, we give several equivalent definitions of $\CT_{G,k}^A$, including via compactifications of 4d $\CN=4$ Yang-Mills theory and brane constructions in IIB string theory. When $G=SU(n)$, we provide a Lagrangian for $\CT_{G,k}^A$ using the twisted BV formalism of \cite{ACMV,CDG}, verify that the stress tensor is (classically) exact, and define Wilson-line operators. We also use the Lagrangian description to define a holomorphic boundary condition for $\CT_{G,k}^A$, and give the first derivation of the boundary VOA $\CN_{G,k}$.

In Section \ref{sec:Hilbert}, we specialize again to $G=SU(n)$, and present quantitative evidence of the relation between $\CT_{SU(n),k}^A$ and the axiomatic TQFT built from $U_q(\mathfrak{sl}_n)$-mod that \emph{doesn't} rely on boundary VOA's. By applying established methods of supersymmetric localization, we compute characters of state spaces (in all genus for $n=2$, in genus-one for general $n$), the Grothendieck group of the category of line operators, and the 't Hooft anomaly of a discrete one-form symmetry, matching quantum-group results from Section \ref{sec:Uqsl2}. We also speculate on the general algebraic structure of the state spaces themselves, and of the full category of line operators.

Finally, in Section \ref{sec:VOA} we discuss the VOA perspective. In particular we explain how certain corner VOA's times many free fermions decompose as modules for $W$-algebras and affine VOA's. This gives us two realizations of  the same deformable family of VOA's (Section \ref{sec:dual-def}). A large level limit gives us many free fermions times a large center. We then conjecture that the modified VOA $\wt{\CN}_{SU(n),k}$ and $\FT_k(\mathfrak{sl}_n)$ form a commuting pair inside the free fermions in such a way that there has to be a braid-reversed equivalence between their module categories. The remainder of the section explains categorical background on which the conjecture relies as well as explicit computations that support the conjecture.

\subsection{Semisimple and non-semisimple dg categories}
\label{sec:ss}

Since much of this paper revolves around non-semisimple and derived generalizations of more familiar TQFT's, we take a moment to lay some algebraic groundwork for discussing these ideas. A key object of study in this paper is the category of line operators in a TQFT; we review what it means for this to be semisimple (or not), from mathematical and physical perspectives.

All categories $\CC$ in this paper will be additive over $\C$, meaning that $\CC$ has a set of objects $\text{Ob}(\CC)=\{L_i\}_{i\in I}$, $\C$-vector spaces of morphisms $\text{Hom}(L_i,L_j)$, and $\C$-linear composition maps $\text{Hom}(L_j,L_k)\otimes \text{Hom}(L_i,L_j)\to \text{Hom}(L_i,L_k)$. Moreover, in an additive category it makes sense to consider finite direct sums of objects; for all $L_i,L_j\in \text{Ob}(\CC)$, $L_i\oplus L_j\in \text{Ob}(\CC)$.

An additive category is further called \emph{abelian} if kernels and cokernels of morphisms, satisfying certain properties, can be defined; in particular,
 every morphism $\alpha \in \text{Hom}(L_i,L_j)$ has a kernel object $K$ (with a morphism to $L_i$) and a cokernel object $K'$ (with a morphism from $L_j$), such that $0\to K \to L_i\overset{\alpha}\to L_j \to K'\to 0$ is an exact sequence of morphisms.
 
 If $A$ is an associative algebra over $\C$ (such as a quantum group) then its category of finite-dimensional modules, denoted $\CC=A$-mod, is automatically abelian. The objects are $A$-modules (\emph{i.e.} vector spaces with a $\C$-linear action of $A$), morphisms are linear maps that commute with the action of $A$, and kernels and cokernels are defined in the usual way for vector spaces. Similarly, if $\CV$ is a vertex algebra, then its category of vertex-algebra modules, denoted $\CC=\CV$-mod is again abelian. The definition of this category is a little trickier; we refer the reader to \emph{e.g.} \cite{FBZ} for details. Its objects are typically infinite-dimensional vector spaces with an action of $\CV$ (usually described in terms of a generalized OPE) satisfying certain regularity properties; morphisms are linear maps commuting with the $\CV$ action; and kernels and cokernels are again defined in the usual way for vector spaces.
 
Semisimplicity is usually defined for abelian categories. An object $S$ of an abelian category $\CC$ is called \emph{simple} if it has no nontrivial quotients. In a module category, the simple objects are the irreducible representations. In general, one has a categorical analogue of Schur's Lemma: if $\{S_i\}$ denotes the set of non-isomorphic simple objects in $\CC$, then
\be \label{S-morph} \text{Hom}(S_i,S_j) = \delta_{ij} \C \,. \ee
The entire category $\CC$ is called \emph{finite} if
\begin{itemize}
\item[SS1.] $\CC$ contains finitely many simple objects $S_i$.
\end{itemize}
and \emph{semisimple} if
\begin{itemize}
\item[SS2.] Every object $L$ of $\CC$ is a direct sum of finitely many $S_i$, equivalently, every short exact sequence of morphisms $0\to L\to L'\to L''\to 0$ splits.
\end{itemize}

We recall some examples. The category of quantum-group modules at a root of unity typically violates both [SS1] and [SS2]; however, it decomposes into blocks that violate only [SS2] (see Sec. \ref{sec:flat-intro}). The category of modules of a VOA is finite semisimiple if and only if the VOA is rational. 
The category of modules of a $C_2$-cofinite VOA (with self-dual vacuum module) is finite but need not be semisimiple.

In order to connect with topologically twisted QFT's, we must also consider derived categories --- or more generally, dg (differential graded) categories. Loosely speaking, a dg category is an additive category whose morphism spaces are dg vector spaces. Namely, each space $\text{Hom}(L_i,L_j)$ has a ``cohomological'' $\Z$-grading and a nilpotent differential $Q$ of degree 1, which behaves as a derivation on compositions of morphisms. Equivalence relations are imposed on morphisms by taking $Q$-cohomology. More subtly, equivalence relations are also imposed on objects. (We refer the reader to the lectures \cite{Toen-lectures,Keller-lectures} for further mathematical details.)

\begin{figure}[htb]
\centering
\includegraphics[width=3.7in]{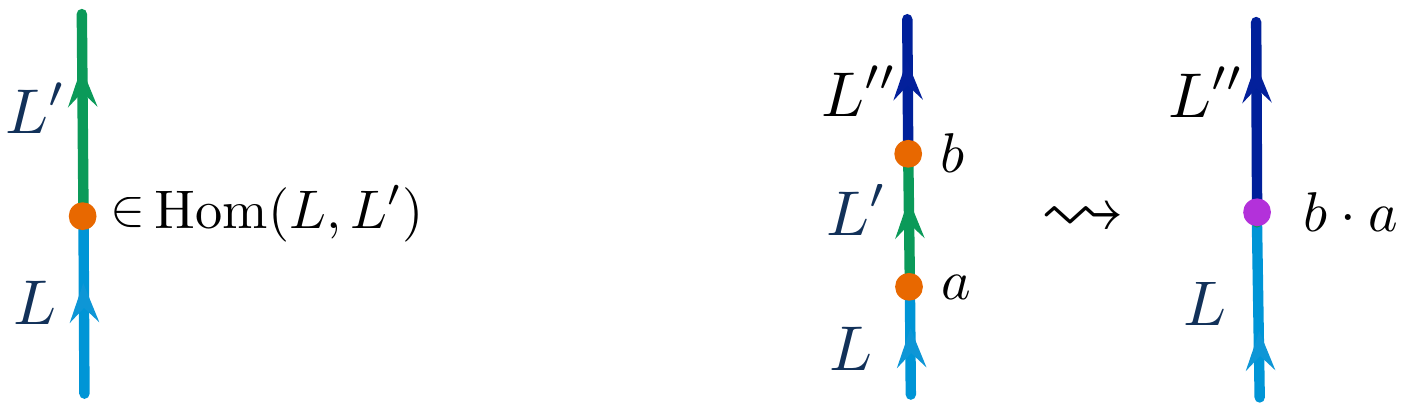}
\caption{Morphisms in the category of line operators, and their composition.}
\label{fig:hom}
\end{figure}

Such a structure arises naturally in topological QFT's of ``cohomological type,'' and in particular in 
the category of line operators of a topologically twisted supersymmetric QFT (\cf\ \cite{KRS} or the recent discussion in \cite{DGGH} for twists of 3d $\CN=4$ theories).%
\footnote{There is much more to say here, largely beyond the scope of this paper. Perhaps the \emph{most} intrinsic description of the category of line operators in a topologically twisted QFT is as an $A_\infty$ category. Mathematically, $A_\infty$ and dg categories are formally equivalent --- in that every dg category is trivially $A_\infty$; and every $A_\infty$ category has a dg model. Physically, the $A_\infty$ structure is natural/intrinsic in the infrared (\cf\ the construction of $A_\infty$ categories in \cite{GMW}); whereas one expects UV Lagrangian descriptions of a QFT to naturally give rise to dg models. Dg categories --- and even more concretely, dg categories constructed as dg enhancements of derived categories of abelian categories --- will be sufficient for us in this paper.} %
In this context, the differential $Q$ generates the ``BRST symmetry'' whose cohomology defines the topological twist. The $\Z$-grading typically comes from a $U(1)$ $R$-symmetry (or ``ghost number symmetry'') for which $Q$ has charge $+1$. The objects of the category of line operators are line operators that preserve $Q$ and the $U(1)$ R-symmetry. Morphisms of a pair of such line operators $L,L'$ are given by the space of local operators at a junction of $L$ and $L'$, as on the left of Figure \ref{fig:hom} which will be a $\Z$-graded vector space with an action of $Q$. Typically one is only interested in $Q$-cohomology of this space. Composition of morphisms is induced by a carefully regularized collision of local operators at consecutive junctions, as on the right of Figure \ref{fig:hom}.
 
As an important special case, we note that any QFT has a trivial (or ``empty'' or ``identity'') line operator. In a topological twist of a supersymmetric theory, it defines an object `$\id$' in the dg category of line operators, whose space of endomorphisms 
\be \label{1bulk-intro} \text{End}(\id) := \text{Hom}(\id,\id) \simeq H^\bullet(\text{bulk local operators},Q) \ee
recovers the $Q$-cohomology of the space of local operators in the bulk.

A dg category $\CD$ can often be represented as (a dg enhancement of) the (bounded) derived category $D^b(\CC)$ of an abelian category~$\CC$. Indeed, in physical contexts, this often happens in several different equally natural ways, say $\CD \simeq D^b(\CC)\simeq D^b(\CC')$.

We recall that the derived category of an abelian category $\CC$ is constructed in two steps. 
Somewhat schematically, one first forms the homotopy category $K(\CC)$, whose objects are chain complexes $L^\bullet$ of objects of $\CC$ and whose morphisms are chain maps (modulo homotopies thereof). 
Then one ``inverts quasi-isomorphisms,'' deeming equivalent any objects $L^\bullet,L'{}^\bullet$ related by a morphisms that induces an isomorphism on their cohomology.
The category $K(\CC)$ acquires a cohomological $\Z$-grading, corresponding to degree in complexes $L^\bullet$; and (perhaps with some extra work, \cf\ \cite{Toen-lectures,Keller-lectures}) $D^b\CC$ acquires the structure of a dg category. When $\CC$ is a category of modules, a dg enhancement of $D^b\CC$ is automatic.

We also recall that if $\CD=D^b\CC$ is represented as (a dg enhancement of) a derived category, the morphisms in $\CD$ correspond to derived morphisms in $\CC$. For example, given objects $L,L'$ of $\CD$ that come from objects in $\CC$, the morphism space
\be \label{Hom-intro} \text{Hom}_\CD(L,L') = \text{Hom}^\bullet_\CC(L,L') \ee
is given by a complex whose cohomology computes extension groups of $L$ and $L'$ in $\CC$, $H^n\big(\text{Hom}^\bullet_\CC(L,L'),Q\big) = \text{Ext}_\CC^n(L,L')$.
In the context of topologically twisted QFT whose category of line operators is $\CD=D^b\CC$, it is the \emph{entire} complex \eqref{Hom-intro} that describes local operators at a junction of $L$ and $L'$. Degree in the complex just corresponds to $U(1)$ R-charge of local operators. The fact that R-charge manifests mathematically in terms of higher extension groups is an artifact of choosing to represent the intrinsic category of line operators $\CD$ as (an enhancement of) the derived category of a particular $\CC$.

We return now to the notion of semisimplicity. Dg categories are typically not abelian, so one cannot directly apply the conditions [SS1]-[SS2] above in the dg setting.
 Instead, we will say that a dg category $\CD$ is finite and/or semisimple if it can be realized as (a dg enhancement of) the derived category $\CD=D^b\CC$ of a finite and/or semisimple abelian category $\CC$. This turns out to be a well defined notion
 due to two standard results in homological algebra:
 
\noindent 1) Finiteness:
If $\CD=D^b\CC$ then K-groups (over $\C$) satisfy $K_0(\CD)=K_0(\CC) \simeq \C^{\text{\# simples in $\CC$}}$

\noindent 2) Semisimplicity:  $\CD=D^b\CC$ is abelian if and only if $\CC$ is semismiple, \cf\ \cite[Sec~III.2.3]{GelfandManin}.

\noindent It follows from these that if $\CD=D^b\CC = D^b\CC'$, then $\CC$ satisfies [SS1] (resp. [SS2]) if and only if $\CC'$ satisfies [SS1] (resp. [SS2]).

It is also useful to observe that if $\CD=D^b\CC$ for semisimple $\CC$, then $\CD$ is just a trivial $\Z$-graded enhancement of $\CC$, and thus essentially equivalent to $\CC$ itself. Concretely, any object $L$ of $\CD$ may be represented as a direct sum of the simples $S_i$ in $\CC$, with different summands possibly shifted in cohomological degree. Moreover, morphisms are simply given by
\be \label{Hom-ss-D} \text{Hom}_\CD(S_i,S_j) = \text{Hom}_\CC(S_i,S_j) = \delta_{ij}\C\,, \ee
with no additional derived structure, since semisimplicity of $\CC$ precludes the existence of higher extensions.

We obtain from this a more intrinsic characterization of semisimplicity in topologically twisted QFT. The category $\CD$ of line operators in a twisted QFT is semisimple if
\begin{itemize}
\item[SS2$'$] There exists a collection $\{S_i\}$ of line operators such that $\text{Hom}(S_i,S_j)=\delta_{ij}\C$ and every line operator is equivalent to a direct sum of $S_i$'s.

In other words, there are no junctions among different $S_i$, and the only local operators bound to a single $S_i$ are multiples of the identity operator; and the insertion of any line operator in a correlation function is equivalent to a sum of $S_i$ insertions.
\end{itemize}
In addition, $\CD$ is finite if
\begin{itemize}
\item[SS1$'$] The collection $\{S_i\}$ has finitely many objects.
\end{itemize}
By applying these properties to the trivial line operator $\id$ and its endomorphisms \eqref{1bulk-intro}, we find that finite semisimplicity requires the space of bulk local operators in the topological QFT to be at most finite-dimensional. The space of local operators will be one-dimensional (generated by the identity operator) if and only if $\id$ itself is simple.

Finally, we remark that unitarity in a topological QFT implies semisimplicity. Unitarity allows one to define orthogonal decompositions of objects in the category of line operators, precluding the existence of non-split short exact sequences. 
 The converse is not true, and many semisimple but non-unitary TQFT's are known, such as the classic Lee-Yang model.

\subsubsection{Braiding, fusion, and state spaces}

\begin{figure}[htb]
\centering
\includegraphics[width=4.8in]{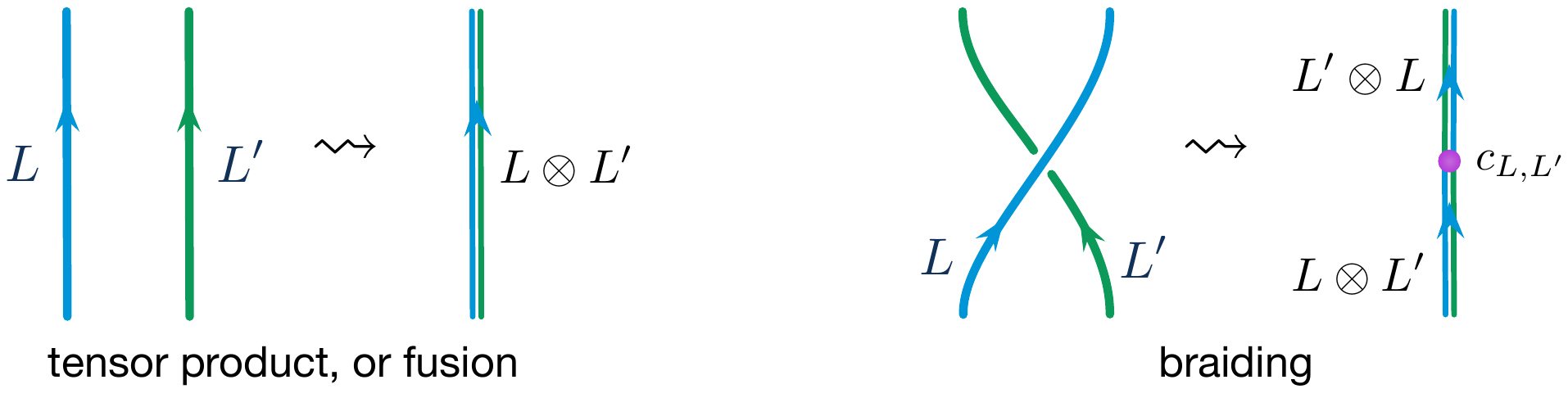} \vspace{-.4cm}
\caption{Tensor product and braiding in the category of line operators.}
\label{fig:fusion-intro}
\end{figure}

The category of line operators in a 3d topological QFT is also expected to be a dg braided tensor category, and optimistically a dg analogue of a modular tensor category. The tensor product and braiding are intrinsically defined by collisions of parallel and crossed line operators, as in Figure \ref{fig:fusion-intro}. The lack of semisimplicity has deep consequences for modular/braided/tensor structure, which have been explored at a non-derived level in \eg\ \cite{Tsuchiya:2012ru, Gainutdinov:2016qhz, Farsad:2017eef, Farsad:2017xgg, Creutzig:2013zza, Auger:2019gts, Creutzig:2013yca} 

Non-semisimplicity also has direct consequences for the structure of state spaces on surfaces $\Sigma_g$, which are closely related to the category of line operators.
  We use the term ``state space'' rather than ``Hilbert space'' throughout the paper, since non-semisimple topological QFT's are generally non-unitary, and we do not assume existence of a positive-definite inner product. In a topological QFT with semisimple category $\CD$ of line operators, the torus Hilbert space is given by its Grothendieck group $K_0(\CD)$, with a basis labelled by simple objects. In higher genus, Hilbert spaces are derived from the fusion algebra of simple objects, and dimensions are given by the Verlinde formula. When $\CD$ is non-semisimple, however, state spaces are much more complicated. The torus state space is given by Hochschild homology of $\CD$
\be \label{HT2-intro} \CH(T^2) \;\simeq\; \begin{cases} K_0(\CD)  & \text{(semisimple)} \\ HH_\bullet(\CD) & \text{(non-semisimple)}\,, \end{cases} \ee
which is typically infinite-dimensional. We will review the physical meaning of Hochschild homology in Section \ref{sec:torus-lines}. Similar generalizations are required in higher genus, \emph{cf.} \cite{SW-Verlinde}.

\subsubsection{Basic examples}
\label{sec:eg-intro}

It is quite special for a 3d topological QFT to have a semisimple category of line operators. 

Chern-Simons theory with compact gauge group $G$ and level $k-h$ is such a special theory. Chern-Simons theory can be expressed as a topological QFT of cohomological type, in the BV-BRST formalism \cite{AKSZ}; thus its category of line operators $\CD$ should in principle be a dg category. However, $\CD$ turns out to be finite semisimple, with trivial dg structure. The simple collection $\{S_i\}$ consists of Wilson lines labelled by irreducible representations of $G$. Only finitely many appear due to an equivalence imposed by large gauge transformations \cite{Witten-Jones, EMSS}. Moreover, there are no local operators available to define gauge-invariant junctions among irreducible Wilson lines; in particular, there are no gauge-invariant bulk local operators besides the identity.

In contrast, topological twists of supersymmetric theories typically have non-finite and non-semisimple categories of line operators. In this paper, we will consider 3d $\CN=4$ gauge theories, which admit two distinct topological twists:
\begin{itemize}
\item[A)] A reduction of Witten's 4d $\CN=2$ Donaldson twist \cite{Witten-Donaldson}, sometimes called the A-twist. (This mixes the spacetime Lorentz group with $SU(2)_{\rm Higgs}$ R-symmetry.)
\item[B)] An intrinsically 3d twist defined by Blau and Thompson \cite{BlauThompson} in gauge theory and extensively explored by Rozansky and Witten for sigma-models \cite{RW}, sometimes called the B-twist. (This mixes the spacetime Lorentz group with $SU(2)_{\rm Coulomb}$ R-symmetry.)
\end{itemize}
The A/B terminology aligns with the fact that 3d A and B twisted theories are naturally related to 2d A and B models upon circle or interval compactification; they also arise from compactifications of 4d $\CN=4$ super-Yang-Mills in the A and B Kapustin-Witten twists \cite{KapustinWitten}.
(For further review of these twists, see the introductory material in \cite{DGGH})

The bulk local operators of a B-twisted gauge theory include the ring of holomorphic functions $\C[\CM_H]$ on its Higgs branch (\emph{a.k.a.} the Higgs-branch chiral ring). As long as there is a noncompact Higgs branch --- the generic situation --- $\C[\CM_H]$ is infinite-dimensional, ruling out finite semisimplicity. Similarly, the bulk local operators of an A-twisted theory include holomorphic functions on the Coulomb branch $\C[\CM_C]$. As long as the theory has a continuous gauge group, $\C[\CM_C]$ is infinite-dimensional, ruling out finite semisimplicity.

Even when moduli spaces are compact, semisimplicity is rare. Consider, for example, the B-twist of a 3d $\CN=4$ sigma model whose target is a smooth complex-symplectic variety $\CX$ (Rozansky-Witten theory).%
\footnote{For general $\CX$, the resulting theory has a $\Z_2$ cohomological grading rather than a $\Z$ grading, which just requires a small modification of the setup outlined above.} %
 Local operators are given by Dolbeault cohomology $H^{0,\bullet}_{\bar\pd}(\CX)$, which is finite-dimensional if $\CX$ is compact \cite{RW}. However, the category of line operators can be represented as the derived category of coherent sheaves $D^b\text{Coh}(\CX)$ \cite{KRS}, which is non-semisimple unless $\CX$ is a collection of (smooth) isolated points --- giving rise to a direct sum of trivial 3d TQFT's.

Semisimple but non-unitary TQFT's have been been associated with point-like but singular moduli spaces.  Examples coming from twists of supersymmetric theories with pointlike Higgs and Coulomb branches appeared recently in \cite{DGNPY, Gang-TQFT, RW-Coulomb}. The examples of \cite{DGNPY, RW-Coulomb} effectively reduce moduli spaces to points using equivariant deformations; while in \cite{Gang-TQFT} moduli spaces are pointlike from the outset.

\subsection{Quantum groups at a root of unity}
\label{sec:Uq-intro}

We next sketch out more of the structure of quantum groups at roots of unity, and the axiomatic 3d TQFT's built from them.

There are several different objects known as the quantum group associated to a simple Lie algebra $\mathfrak g$. In this paper, we focus on what's known as the 
\emph{non-restricted} quantum group or the 
simply connected De Concini-Kac quantum group $U_q(\mathfrak g)$,
see \cite{DCK}.  
The algebra $U_q(\mathfrak g)$ is given by Serre-like generators $\{E_i,F_i,K_i^{\pm1}\}_{i=1}^{\text{rank}\,\mathfrak g}$ and relations.  When $q$ is a generic parameter these relations reduce to the standard generators and relations of $U(\mathfrak g)$ upon setting $K_i=q^{H_i}$ and carefully taking the limit $q\to 1$. 
When $q$ is a root of unity, 
we will also consider the \emph{unrolled} quantum group $U_q^H(\mathfrak g)$, which adjoins the generators $H_i$ themselves to $U_q(\mathfrak g)$, effectively taking a logarithm of the $K_i$. 

For example, $U_q(\mathfrak{sl}_2)$ is generated by $E,F,K^{\pm 1}$, with relations
\be \label{U2-intro} KE=q^2EK\,,\quad KF=q^{-2}FK\,,\quad [E,F] = \frac{K-K^{-1}}{q-q^{-1}}\,, \ee
while $U_q^H(\mathfrak{sl}_2)$ is generated by $E,F,K^{\pm 1},H$, with additional relations
\be [H,K]=0\,,\quad [H,E]=2E\,,\quad [H,F]=-2F\,. \ee

Closely related but \emph{not} studied in this paper is Lusztig's divided-powers version $U_q^L(\mathfrak g)$, which adjoins generators $E_i^\ell/[\ell]_q$ and $F_i^\ell/[\ell]_q$. $U_q^L(\mathfrak g)$ is isomorphic to $U_q(\mathfrak g)$ at generic $q$, but differs upon specializing $q$ to a root of unity. Lusztig's $U_q^L(\mathfrak g)$ has played a major role in representation theory and axiomatic TQFT, and it would be interesting to find a generalization of our QFT construction that includes it.

When $q$ is generic, the category of finite-dimensional modules of $U_q(\mathfrak g)$ is semisimple, and related to modules of a Kac-Moody VOA at generic level by the classic Kazhdan-Lusztig correspondence \cite{KL3}. When $q = e^{\frac{2\pi i}{r}}$ 
is a root of unity, the abelian category of finite-dimensional modules  $\CC = U_q(\mathfrak g)\text{-mod}$ becomes%
\begin{itemize}
\item infinite, due to continuous families of simple modules, violating [SS1]; and
\item non-semisimple, as some simple modules admit nontrivial extensions, violating [SS2].
\end{itemize}
In addition, most modules end up having vanishing quantum dimensions, and the braided tensor structure on 
pieces of $\CC$ (equivalently, the R-matrix on $U_q(\mathfrak g)$) is not defined or becomes extremely subtle to define.
For all these reasons, constructing a full axiomatic TQFT based on $\CC$ has been a difficult mathematical problem.

One way to handle the above problems is to ``semisimplify'' the category $U_q(\mathfrak g)\text{-mod}$. Loosely, this amounts to quotienting out by (or setting to zero) all modules with vanishing quantum dimension. What is left behind is the semisimple category $\CC_{\text{s.s.}}$ used in the original work of Reshetikhin-Turaev \cite{RT}, and related to Chern-Simons theory with compact group.

Well-defined invariants and partial TQFT's based on pieces of the un-semisimplified $\CC = U_q(\mathfrak g)\text{-mod}$ (and the analogous Lusztig divided-powers category) at a root of unity already appeared in the 1990's. Notable examples include
\begin{itemize}
\item Hennings' \cite{Hennings} and Lyubashenko's \cite{Lyubashenko} invariants of 3-manifolds, based on pieces (blocks) of $\CC$ with finitely many simples having nontrivial extensions;
\item Lyubashenko’s invariants were shown to be part of a TQFT in \cite{KerLyu}, however, this TQFT is only defined of connected surfaces and only satisfies a weak monoidal condition (later a monoidal TQFT for non-connected surfaces was defined in \cite{DGGPR},  also see \cite{derenzi2021extended} for De Renzi's general construction for any modular category);
\item The Akutsu-Deguchi-Ohtsuki (ADO) invariants of links \cite{ADO} (see also \cite{Murakami}), based on multiple semisimple pieces of $\CC$ with vanishing quantum dimensions, and related to the Alexander polynomial at $q=i$;
\item Kashaev's invariant \cite{Kashaev} of links in 3-manifolds, shown by \cite{MM-volume} to come from a semisimple piece of $U_q(\mathfrak{sl}_2)$ containing a single, distinguished simple module of vanishing quantum dimension, related to analytic continuation of the Jones polynomial and the Volume Conjecture.  Kashaev's invariant was extended by Baseilhac and Benedetti in \cite{BaseilhacBen2004, BaseilhacBen2007} to  a quantum hyperbolic field theory coming from the Borel subalgebra of quantum $\mathfrak{sl}_2$. 
\end{itemize}
It was also proposed by Kashaev and Reshetikhin \cite{KashaevReshetikhin} that the continuous family of simple modules in the full category $\CC$ should lead to invariants not just of 3-manifolds, but of 3-manifolds with a choice of background flat connection. 

A set of systematic techniques for constructing axiomatic
 link invariants and
 TQFT's using category $\CC = U_q(\mathfrak g)\text{-mod}$ or pieces thereof was then developed in the last decade, in \cite{CGP} and a series of subsequent papers including \cite{CGP2, GP, BCGP, DGGPR,DGP, DGP2, derenzi2019nonsemisimple, GPT,biquandles}. We will refer to the resulting TQFT/invariants as ``CGP TQFT/invariants.'' The rather technical heart of these techniques involves first replacing $\CC$ by the 
  category of modules for the unrolled quantum group $U_q^H(\mathfrak g)$, then taking a suitable equivalence to obtain finite-dimensional state spaces and finite surgery formulas. 
 One motivation for using  $U_q^H(\mathfrak g)$-mod was to obtain a well-defined braiding/R-matrix, though at certain roots of unity braiding also required the introduction of spin structures \cite{BCGP}.  
The problem of vanishing quantum dimensions was dealt with using a regularization procedure, involving ``modified traces'' and ``renormalized quantum dimensions'' \cite{GPT}. All of the previous invariants mentioned above, of Hennings, Lyubashenko, ADO, Kashaev, and 
``abelian'' 
Kashaev-Reshetikhin, were recovered as special cases of CGP invariants in \cite{CGP,CGP2, GP,BCGP, DGGPR,DGP, DGP2, GPT,biquandles}.

\subsubsection{Flat connections}
\label{sec:flat-intro}

The generalization of the Kashaev-Reshetikhin proposal developed in \cite{GP,biquandles}, related to background flat connections, is particularly important for us. Flat connections ultimately originate from the presence of an exceptionally large center in $U_q(\mathfrak g)$ at a root of unity $q=e^{\frac{2\pi i}{r}}$, whose implications for representation theory were originally studied by \cite{DCK,DCKP,Beck}. In particular, the center at $q=e^{\frac{2\pi i}{r}}$ contains a commutative algebra $Z_{\rm Fr}$ generated by (roughly) $r$-th powers of $E_i,F_i,K_i^{\pm 1}$, and known as the Frobenius center. 
 The values of elements in $Z_{\rm Fr}$ parameterize a Zariski-open subset of a $\Gamma$, where $\Gamma$ is the simply connected complex group $G_\C$ when $r$ is odd, and a particular global form of its Langlands dual $G_\C^\vee$ when $r$ is even. (Mathematically, $\overline{\text{Spec} (Z_{\rm Fr})} \simeq \Gamma$.)

We will always impose the additional requirement that $Z_{\rm Fr}$ acts semisimply on modules of $\CC=U_q(\mathfrak g)$-mod.%
\footnote{This is a standard requirement, used in the definition of $U_q(\mathfrak g)$-mod in most of the mathematics literature.
It also seems to be the correct requirement to impose for constructions in this paper relating to 3d QFT. In particular, in the best-understood case $\g=\mathfrak{sl}_2$, the requirement is necessary for identifying $U_q^H(\mathfrak{sl}_2)$-mod with the category of logarithmic modules for the singlet VOA \cite{Creutzig:2016htk}; for $u_q(\mathfrak{sl}_2)$ the semisimple action of $Z_{\rm Fr}$ is automatic and this case  seems to be the correct category to match with line operators in topologically twisted 3d QFT, \cf\ \cite[Section 9]{CCG}.} %
Then central elements of $U_q(\mathfrak g)$ must act by fixed constants on any indecomposable module, and there are no morphisms between modules with different values of the center, so the category $\CC=U_q(\mathfrak g)$-mod decomposes into blocks
\be \label{blocks-intro} \CC = \bigoplus_{g\in \Gamma}  \CC_g\,,\qquad \CC_g:= U_q(\mathfrak g)\big|_g\text{-mod} \ee
where $U_q(\mathfrak g)\big|_g$ is the quantum group at $q=e^{\frac{2\pi i}{r}}$ and elements of the Frobenius center set equal to $g$. Geometrically, $\CC$ becomes a coherent sheaf of categories over the group $\Gamma$,
\be \CC\to \Gamma\,, \label{sheaf-intro} \ee
with `stalk' (or `fiber') categories $\CC_g$.

The case of interest for us is $\mathfrak g$ of ADE type and an even root of unity $r=2k$. Then $\Gamma=G_\C^\vee$, the Langlands-dual of the simply connected group $G_\C$. For example, for $U_q(\mathfrak{sl}_2)$ at $q=e^{\frac{i\pi}{k}}$, the Frobenius center is the commutative algebra freely generated by $E^k,F^k,K^{\pm 2k}$, whose values $(E^k,F^k,K^{2k}) = (e,f,\kappa)\in \C\times \C\times\C^*$ are in 1-1 correspondence with points
\be g =  \bp \kappa & -\kappa e \\ f & \;1-e f \ep\in  PGL(2,\C) \ee
on a Zariski-open subset of $PGL(2,\C)$ (\cf\ \cite[Sec 5.2]{biquandles}).

In general, each block $\CC_g$ contains the same, \emph{finite} number of simple modules, independent of $g$. For generic $g$, $\CC_g$ is semisimple but all its simple modules have vanishing quantum dimension; while non-generic blocks (\emph{e.g.} for $g$ fixed by an element of the Weyl group) are non-semisimple. The ``most'' non-semisimple block $\CC_1$ corresponds to the identity $g=1$, and contains modules for the so-called restricted quantum group $u_q(\mathfrak g) = U_q(\mathfrak g)\big|_1$. 
Generic blocks were used in the original construction of ADO invariants; while parts of $\CC_1$ appeared in Hennings, Lyubashenko, and Kashaev invariants.

The key insight of \cite{KashaevReshetikhin}, translated into QFT terms, was that the various blocks $\CC_g$ of $U_q(\mathfrak g)$-mod at a root of unity behave as if they are line operators in a topological QFT that admits deformations by flat (background) $\Gamma$ connections. In particular, $\CC_g$ should be thought of as the category of line operators in the presence of a vortex defect for a flat background connection, with holonomy $g$. Collision of parallel lines --- as in Figure \ref{fig:hol-fusion} of Section \ref{sec:braiding} --- heuristically suggests that the tensor product of objects in $\CC_g$ and $\CC_{g'}$ should belong to $\CC_{gg'}$, 
\be \otimes: \quad \CC_g\boxtimes \CC_{g'}\to \CC_{gg'} \ee
and that braiding relates $ \CC_g\boxtimes \CC_{g'}$ to $\CC_{gg'g^{-1}}\boxtimes \CC_g$.
These properties were shown to be compatible with the coproduct and R-matrix of $U_q(\mathfrak g)$ in \cite{KashaevReshetikhin,biquandles}.

This structure leads to an axiomatic link invariant with a $\Gamma$ connection $\CA$ on the complement $S^3\backslash K$.  This link invariant conjecturally extends to a 3d TQFT that computes invariants of links $K$ in 3-manifolds $M$ together with the data of a flat $\Gamma$ connection $\CA$ on the complement $M\backslash K$.  
Each strand of $K$ is ``colored'' by an element of $\CC_g$, where $g$ is the basepointed holomomy of $\CA$ around the chosen strand. (For roots of unity divisible by 4, one also requires a choice of spin structure on $M\backslash K$.)
Similarly, the state space on a surface $\Sigma$ depends on a choice of flat $\Gamma$ connection on $\Sigma$. 

We will explore the physical manifestation of these features in topological QFT's with global symmetry in Section \ref{sec:toymodel}.

\subsection{Logarithmic VOA's}
\label{sec:VOA-intro}

Logarithmic conformal field theory dates back to the work of Gurarie \cite{Gurarie:1993xq} and Rozansky-Saleur \cite{Rozansky:1992rx, Rozansky:1992td} almost three decades ago. The term logarithmic refers to the appearance of logarithmic singularities in correlation functions. Such singularities arise if the zero-mode of the Virasoro algebra does not act semisimply and hence logarithmic singularities are tightly connected to non-semisimple modules. By now, one means by a logarithmic conformal field theory a theory that has representations that are reducible but indecomposable, and one calls a module logarithmic if the Virasoro zero-mode does not act semisimply. An introduction to the topic is \cite{Creutzig:2013hma} and a status report on the understanding of conformal blocks and the modular functor in the logarithmic setting is \cite{Fuchs:2019xkv}. The symmetry algebra of a conformal field theory is a vertex operator algebra and so one calls the VOA of a logarithmic theory a logarithmic VOA. 

The best understood logarithmic VOA's are the triplet algebras $W(p)$ (for $p \in \mathbb Z_{\geq 2}$) and close relatives such as symplectic fermions, affine $\mathfrak{gl}(1|1)$, and $\beta\gamma$-ghosts \cite{Kausch:1990vg, Kausch:2000fu, Creutzig:2020zom, Allen:2020kkt}. These and their higher-rank generalizations, the Feigin-Tipunin algebras \cite{Feigin:2010xv}, are also the algebras relevant for the present work. 
The category of ordinary modules of an affine vertex algebra at level not in $\mathbb Q_{> -h^\vee}$ is braided equivalent to a category of modules of the corresponding quantum group at associated root of unity \cite{KL-I, KL-III, KL-IV}. This Kazhdan-Lusztig correspondence was conjectured 15 years ago to have a logarithmic analogue,
involving the triplet algebra $W(p)$ and the restricted quantum group $u_q(\mathfrak{sl}_2)$ at $2p$-th root of unity  \cite{Feigin:2005xs,Feigin:2006iv}. However, proving this conjecture --- and other logarithmic Kazhdan-Lusztig correspondences --- has involved a long and interesting journey.

Following \cite{Feigin:2005xs,Feigin:2006iv}, 
 substantial effort was put into understanding the representation categories of triplet algebras  \cite{Carqueville:2005nu,Adamovic:2007er, Adamovic:2009xn,Adamovic:2007qs,Tsuchiya:2012ru}. 
An equivalence of abelian categories (ignoring braided tensor structure) $W(p)\text{-mod}\simeq u_q(\mathfrak{sl}_2)\text{-mod}$  was formulated in \cite{Nagatomo:2009xp}, though full proofs appeared only recently \cite{mcrae2021structure}. 
 It also came to be understood that $u_q(\mathfrak{sl}_2)$-mod is not braidable with a naive R-matrix \cite{KondoSaito}, and requires a quasi-Hopf modification \cite{Gainutdinov:2015lja,Creutzig:2017khq}. Substantial progress in the theory of vertex tensor categories, in particular \cite{Creutzig:2020zvv, mcrae2021structure, Creutzig:2020smh, Creutzig:2020qvs}, then allowed a Kazhdan-Lusztig correspondence to be established in two very different fashions \cite{Creutzig:2021cpu,GannonNegron}. The approach of \cite{Creutzig:2021cpu} exploits embeddings of triplet algebras in lattice VOA's, and shows that the associator of the former is fixed by the latter.
An equivalence of braided tensor categories was proven in \cite{Creutzig:2021cpu} for $p=2$, and is work in progress for general $p$.
Once one understands enough of the representation theory of the Feigin-Tipunin algebras, it should also be possible to extend the technology of \cite{Creutzig:2021cpu} to higher rank.

\subsubsection{Automorphisms, flat connections, and unrolling}
\label{sec:VOA-flat-intro}

One peculiarity of the triplet algebra and its Feigin-Tipunin analogues is the presence of continuous outer-automorphism groups \cite{sugimoto2021feigintipunin, sugimoto2021simplicities}, certain complex Lie groups. Correspondingly, the OPE algebras --- and module categories --- may be deformed by flat connections for these Lie groups. This is the VOA analogue of the flat connections of Section \ref{sec:flat-intro}. Roughly, each quantum-group stalk category $\CC_g = U_q(\mathfrak g)\big|_g\text{-mod}$ is expected to coincide with modules for a Feigin-Tipunin algebra deformed by a flat connection with holonomy $g$ around the point where modules are inserted.

A useful approach to understanding the outer-automorphism groups and associated deformations --- which we expand on in Section \ref{sec:VOA} --- is to (conjecturally) realize Feigin-Tipunin algebras as large-level limits of deformable families of VOA's, associated to junctions of boundary conditions in $4$d $\mathcal N=4$ super Yang-Mills theory \cite{CreutzigGaiotto-S}. In this context, there are actually multiple ways to take take a large-level limit, which lead either to standard Feigin-Tipunin algebras or to their deformations.
 
The simplest example, developed in the toy model of Section \ref{sec:bdy-VOA}, is symplectic fermions. The module  category of symplectic fermions is a non-semisimple (and thus quite sophisticated) tensor super category. However, symplectic fermions have an $SL(2,\C)$ outer automorphism group, and their OPE can be deformed by a flat $SL(2,\C)$ connection. After a generic deformation, the VOA becomes equivalent to free fermions, whose module category is trivial, \emph{i.e.} equivalent to (graded) vector spaces. In other words, the representation category of the VOA changes drastically if coupled to flat connections. Symplectic fermions arise as a large-level limit of the affine vertex superalgebra of $\mathfrak{osp}(1|2)$, and we illustrate different ways of taking the limit in Section \ref{sec:osp}.  
 
Since the Feigin-Tipunin algebras have large automorphism groups one can also take their orbifolds, \emph{e.g.} orbifolds by a maximal torus of the automorphism group. These have been named \emph{narrow $W$-algebras} and studied in \cite{Creutzig:2016uqu} for higher rank; while in rank one this algebra is the well studied singlet VOA \cite{Ada-singlet, Creutzig:2013zza, Creutzig:2016htk,Creutzig:2020qvs}. Conversely, the Feigin-Tipunin algebras are large simple-current extensions of narrow $W$-algebras. 
These types of extensions are illustrated in the examples of passing from Heisenberg VOA's to lattice VOA's and from the singlet algebra to the triplet algebra in Examples \ref{ex:freeboson} and \ref{ex:singlet} of Section \ref{sec:tensor}.
The quantum groups that supposedly correspond to the narrow $W$-algebras are so-called unrolled quantum groups, see section \ref{sec:Uq-intro}. There is a procedure, called uprolling in \cite{Creutzig:2020jxj}, that recovers quasi-Hopf modifications of the restricted quantum groups \cite{Creutzig:2017khq, Gainutdinov:2018pni}, see also \cite{Negron}. In other words, uprolling is a quantum group version of simple-current extensions and unrolling corresponds to abelian orbifolds on the VOA side.

\subsection{3d topological QFT}
\label{sec:TG-intro}

We are looking for a topological 3d QFT that matches the structure of the CGP TQFT described in Section \ref{sec:Uq-intro}, based on the non-semisimple category $\CC=U_q(\mathfrak g)$-mod at an even root of unity $q=e^{i\pi/k}$.
Assembling the various observations of Sections \ref{sec:ss}-\ref{sec:VOA-intro}, we surmise that:
\begin{itemize}
\item The 3d theory is labeled by a Lie group $G$ and an integer $k$. 

(Note that the quantum group $U_q(\mathfrak g)$ depends on a choice of global form of $G$ with Lie algebra $\mathfrak g$. We have been focusing on the simply connected form of $U_q(\mathfrak g)$.)

\item The theory has $\Gamma$ global symmetry, and may be deformed by $\Gamma$ flat connections, where $\Gamma$  is the complex Lie group over which the category $\CC$ fibers, as in \eqref{sheaf-intro}.
We focus on the simply-connected form $U_q(\mathfrak{sl}_n)$ in type A, with $G=SU(n)$ and $\Gamma=G^\vee_\C = PGL(n,\C)$.

\item Accordingly, for each $g\in \Gamma$, the \emph{derived} category $D^b(\CC_g)$ is equivalent to the category of line operators in the 3d QFT in the presence of a background vortex defect with basepointed holonomy $g$. In the absence of a deformation by a background flat connection, the category of line operators is the non-semisimple $D^b(\CC_1)$.

\item The 3d theory is Chern-Simons-like. In particular, it contains a subset of line operators labelled by the same irreducible representations of $G$ at level $k$ that appear in Chern-Simons theory, matching the modules of $U_q(\mathfrak g)$ that survive semisimplification. However, the fusion and braiding of these line operators is different from $G_k$ Chern-Simons theory.

Another strong hint of a Chern-Simons-like sector comes from recent work proposing \cite{RW-Coulomb} and proving \cite{Willetts,BDGG} (from multiple perspectives) that the $\mathfrak{sl}_2$ ADO invariants of a knot satisfy the same recursion relations as colored Jones polynomials. The recursion relations for colored Jones polynomials were introduced in \cite{GarLe,Gar-AJ,Gukov-A}, and motivated (in \cite{Gukov-A}) by analytic continuation of Chern-Simons theory.

\end{itemize}

\subsubsection{A definition of $\CT_{G,k}^A$}

The theory $\CT_{G,k}^A$ discussed in \eqref{T-intro} has all the properties above. We now supply additional details on how this theory is defined. An expanded discussion appears in Section \ref{sec:Atwist}.

We begin with the 3d $\CN=4$ superconformal theory $T[G]$ originally defined by \cite{GaiottoWitten-Sduality}, in terms of an S-duality interface in 4d $\CN=4$ super-Yang-Mills theory.%
\footnote{One way to define $T[G]$ is by taking 4d Yang-Mills with gauge group $G$ on a half-space with a half-BPS Dirichlet boundary condition, applying S-duality, ``sandwiching'' with a second Dirichlet boundary condition in the new S-dual frame, and flowing to the infrared.} %
The 3d theory $T[G]$ makes sense for any compact simple Lie group $G$, and in fact depends only on the (complexified) Lie algebra $\mathfrak g$. It has $\wt G^\vee\times \wt G$ flavor symmetry, where the factors are the simply connected forms of $G$ and its Langlands dual,
\be \wt G^\vee \; \actson\; T[G]\; \reflectbox{$\actson$}\; \wt G\,. \ee
The respective factors act on the Coulomb and Higgs branches of the moduli space of vacua of $T[G]$, which are Langlands-dual nilpotent cones
\be \CM_{\rm Coul}[T[G]] \simeq \CN^\vee \subset \mathfrak g^\vee\,,\qquad \CM_{\rm Higgs}[T[G]] \simeq \CN \subset \mathfrak g\,. \label{nilp-intro} \ee

We then gauge the simply-connected $\wt G$ symmetry of $T[G]$ by introducing a 3d $\CN=2$ gauge multiplet together with a supersymmetric Chern-Simons term at (UV) level $k$. This defines the theory $\CT_{\wt G,k} := T[G]/\wt G_k$. We require that $k\in \Z\simeq H^4(B\wt G)$ and $k\geq h$ (where $h$ is the dual Coxeter number). The resulting theory retains flavor symmetry given by the adjoint form $G^\vee$ of the Langlands-dual group.
To simplify notation, we will assume that $G=\wt G$ is simply connected to begin with (and drop the tilde). Thus
\be G^\vee \; \actson\; \CT_{ G,k} := T[G]/ G_k\,. \ee

The theory $\CT_{G,k}$ also gains a discrete one-form ``center symmetry'' $Z(G)$ \cite{GKSW}. Indeed, a more refined analysis following \cite{CDI, EKSW, HsinLam-discrete} (closely related to examples in \cite{GHP-global, ABGS, BCH2group}) shows that the full global symmetry of $\CT_{G,k}$ is a 2-group, with one-form part $Z(G)$, zero-form part $\wt G^\vee$, and a nontrivial 2-group structure such that only $Z(G)$ and $\wt G^\vee/Z(\wt G^\vee)=G^\vee$ act as independent 1-form and 0-form symmetries.

Note that in defining $\CT_{G,k}$, we gauge $G$ with a 3d $\CN=2$ --- rather than $\CN=4$ --- vectormultiplet in order to be able to introduce supersymmetric Chern-Simons couplings. (Supersymmetric Chern-Simons theories go back to \eg\ \cite{AKK-CS, Ivanov-CS, ZP-CS}, and their (in)compatibility with higher supersymmetry was discussed in \cite{Schwarz-CS,GaiottoYin}.) Nevertheless, $\CT_{G,k}$ still has 3d $\CN=4$ supersymmetry due to a mechanism found in \cite{GaiottoWitten-Janus}; this relies on the fact that the complex moment-map operators $\mu$ for the $G$ symmetry of $T[G]$ parameterize the Higgs-branch nilpotent cone \eqref{nilp-intro}, and thus satisfy the ``fundamental identity'' $\text{Tr}(\mu^2) = 0$.

The theory $\CT_{G,k}$  has many of the properties we want -- \emph{e.g.} it has Wilson-line operators labelled by representations of $G$, and it has $G^\vee$ global symmetry. However, it is not topological, due to the superconformal ``matter'' from $T[G]$. This is easily remedied, by taking a topological twist.

As reviewed in Section \ref{sec:eg-intro}, there exist two distinct `A' and `B' topological twists of a 3d $\CN=4$ theory. The global symmetry $G^\vee$ behaves differently with respect to the two twists: in our conventions, the B-twist allows deformations by monopole configurations for the $G^\vee$ global symmetry; whereas the A-twist allows deformations by complexified $G^\vee$ flat connections. Thus, we take the A-twist of $\CT_{G,k}$, denoting the resulting theory $\CT_{G,k}^A$. Its B-twisted analogue $\CT_{G,k}^B$ was studied by \cite{KapustinSaulina-CSRW}, and was an important motivation for our work.

It is useful to think of $\CT_{G,k}^A$ as a generalization of ordinary Chern-Simons theory. A direct connection can be established by recalling that a 3d $\CN=2$ Yang-Mills-Chern-Simons theory at level $k$ (with no additional matter) will flow in the infrared to pure, bosonic Chern-Simons $CS[G_{k-h}]$ at level $k-h$ \cite{AHISS}. This is true regardless of twist.%
\footnote{In general, a 3d $\CN=2$ theory only admits a holomorphic-topological twist \cite{ACMV}. However, for 3d $\CN=2$ Chern-Simons theory (with no matter), the holomorphic-topological twist is already topological, and equivalent to what one might call A or B twists; we give some details in Section \ref{sec:Tnk-BV}.} %
Thus, an $\CN=2$ Chern-Simons gauging of $T[G]$ defines $\CT_{G,k}$, whereas an $\CN=2$ Chern-Simons gauging of a trivial theory defines pure Chern-Simons:
\be \label{TCS-intro} (T[G]/G_k)^A = \CT_{G,k}^A \qquad\text{vs.}\qquad   (\,\mathbf{\cdot}\, / G_k)^A \,\simeq\, \text{CS}[G_{k-h}]\,. \ee

In a very rough approximation, one might even think of $\CT_{G,k}^A$ as a product of ordinary Chern-Simons theory and the A-twist of $T[G]$,
\be \CT_{G,k}^A \approx  \text{CS}[G_{k-h}]\otimes T[G]^A\,. \ee
Applying 3d mirror symmetry, the A-twist of $T[G]$ may be further approximated by a B-twisted sigma-model (\emph{a.k.a.} Rozansky-Witten theory) to its Coulomb branch, the nilpotent cone $\CN^\vee$:
\be \label{CS-fact} \CT_{G,k}^A \overset{?}\approx  \text{CS}[G_{k-h}] \otimes \text{RW}(\CN^\vee)\,. \ee
(This of course ignores degrees of freedom at the singular origin of $\CN^\vee$.)
The approximiation \eqref{CS-fact} turns out to give some surprisingly accurate predictions, even if it is not entirely correct! It suggests that the local operators of $\CT_{G,k}^A$ (the main source of non-semisimplicity) correspond to holomorphic functions on the nilpotent cone $\C[\CN^\vee]$, which we will show is indeed true. It also suggests that state spaces factorize
\be \label{Hilb-fact} \CH_{\CT_{G,k}^A}(\Sigma) \approx \CH_{G_{k-h}}(\Sigma)\otimes \CH_{\text{RW}(\CN^\vee)}(\Sigma)\,, \ee
which we find to be approximately true.

\subsubsection{4d constructions and 6d relations}
\label{sec:4d6d-intro}

The purely 3d definition above may be lifted to various ``sandwich'' configurations in 4d $\CN=4$ Yang-Mills theory, employing the BPS boundary conditions and interfaces introduced by \cite{GaiottoWitten-Janus, GaiottoWitten-Sduality, GaiottoWitten-boundary}. 

\begin{figure}[htb]
\centering
\includegraphics[width=5.7in]{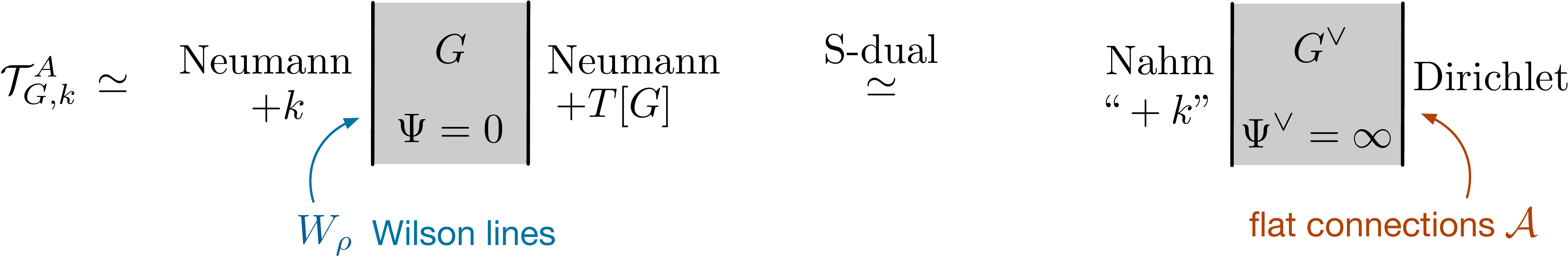}
\caption{Interval compactifications of 4d LG-twisted Yang-Mills theory that reproduce $\CT_{G,k}^A$.}
\label{fig:bc-intro}
\end{figure}

For example, one may consider 4d $G$ gauge theory on an interval $\R^3\times[0,1]$, with a Neumann boundary with a level-$k$ boundary Chern-Simons term at 0, and a Neumann boundary coupled to $T[G]$ at 1, as on the left of Figure \ref{fig:bc-intro}. This flows in the infrared to the 3d theory $\CT_{G,k}$.  Further taking Kapustin-Witten's geometric Langlands (GL) twist \cite{KapustinWitten} of the bulk theory at $\Psi=0$, also known as the 4d A-twist \cite{Witten-Nahm}, induces the 3d A-twist of $\CT_{G,k}$.
 
Dually, one may consider 4d $G^\vee$ gauge theory in the $\Psi=\infty$ twist (the 4d B-twist) sandwiched between a deformed maximal-Nahm-pole boundary condition and a pure Dirichlet boundary condition, as on the right of Figure \ref{fig:bc-intro}.

Each 4d construction makes different features of $\CT_{G,k}^A$ manifest. In the A-twisted sandwich, the Neumann b.c. supports Wilson-line operators, which become the Wilson lines of $\CT_{G,k}^A$. In the B-twisted sandwich, the Dirichlet b.c. has $G^\vee$ global symmetry and may be deformed by flat $G^\vee_\C$ connections, giving rise to the deformations of $\CT_{G,k}^A$.

The setups in Figure \ref{fig:bc-intro} are very similar to those appearing in work on analytic continuation and categorification of Chern-Simons theory \cite{Witten-anal,Witten-path,Witten-fivebranes}, the 3d-3d correspondence \cite{DGG, TerashimaYamazaki, CCV}, and its holomorphic \cite{Pasquetti,BDP} and homological \cite{GPV-spectra,GPPV-spectra} blocks. These various constructions all originate in six dimensions, with the 6d (2,0) theory of ADE type $\mathfrak g$ on a product of a 3-manifold and a twisted cigar (or ``Melvin cigar'') $M\times (D^2\times_q S^1)$.%
\footnote{In 3d-3d correspondences, $D^2\times_qS^1$ is often replaced by other global geometries with transverse holomorphic foliation structures, such as three-spheres or lens spaces. All these geometries have local pieces that  resemble $D^2\times_qS^1$. The local $D^2\times_qS^1$ defines holomorphic and homological blocks, and is closest to our current setup.} %
The 6d theory is topologically twisted along $M$, and given a holomorphic-topological twist (as in \cite{ACMV}) along $D^2\times_q S^1$. At the asymptotic end of the cigar $\pd(D^2)$, one places a boundary condition labelled by a complexified flat connection $\CA$ on $M$ --- irreducible in the original examples of holomorphic blocks, and abelian in the context of homological blocks.

\begin{figure}[htb]
\centering
\includegraphics[width=5in]{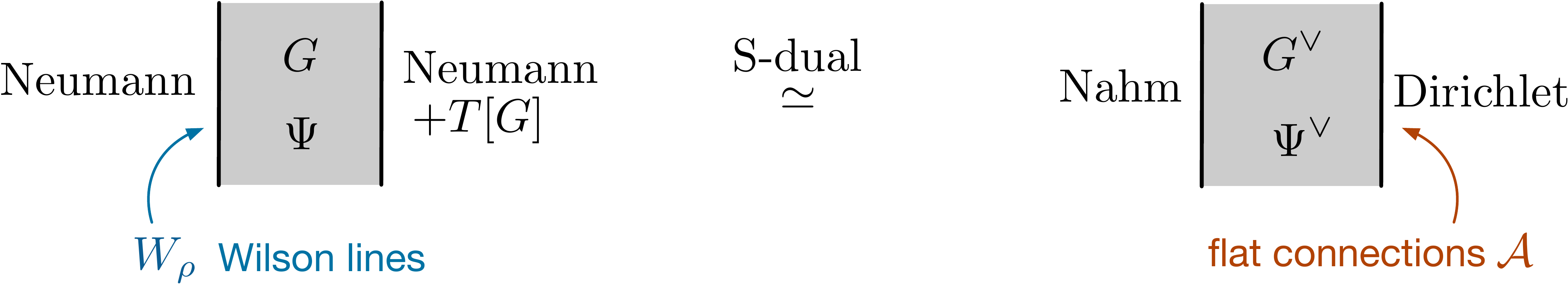}
\caption{Compactifications of the 6d (2,0) of type $\mathfrak g$ on $\R^3\times (D^2\times_qS^1)$, with $q=e^{2\pi i\Psi^\vee}$.}
\label{fig:3d-intro}
\end{figure}

Compactifying on the cigar circle and the noncontractible $S^1$ in various orders (\cf\ \cite{NekrasovWitten}) then leads to GL-twisted 4d Yang-Mills theory on $M\times \R_+$, with various boundary conditions. For example, first compactifying on the cigar and then the noncontractible $S^1$ defines 4d $G^\vee$ Yang-Mills%
\footnote{In this brief discussion, we are not carefully keeping track of discrete data that differentiates different global forms of $G$, $G^\vee$, etc. See \cite{EKSW,GHP-global} for details thereof.} %
with a Nahm-pole b.c. at $0\in \R_+$ and an asymptotic boundary condition at $\infty\in \R_+$ labelled by the flat connection $\CA$. Further replacing the asymptotic boundary condition with a Dirichlet b.c. at finite distance yields the setup on the RHS of Figure \ref{fig:3d-intro}, with GL twist parameter $\Psi^\vee = \frac{1}{2\pi i}\log q$. 
Alternatively, compactifying first on $S^1$ and then on the cigar yields the setup on the LHS, with a Neumann b.c. and the S-dual of a Dirichlet b.c., and GL twist parameter $\Psi=-1/\Psi^\vee$.

These configurations are clearly reminiscent of our $\CT_{G,k}^A$ constructions in Figure \ref{fig:bc-intro}. One might expect them to be closely related upon specializing $q = e^{i\pi/k}$ to a root of unity. Such a relation might connect the appearance of logarithmic VOA's in homological blocks \cite{CCFGH} and in our current work, which we hope to investigate further in the future.

A slightly different compactification from 6d also leads to the 3d $\CN=4$ theory $\CT_{\rm Gr}$ proposed by \cite{RW-Coulomb} to underlie the analytic continuation of ADO invariants. $\CT_{\rm Gr}$ is a 3d sigma-model with target $T^*\text{Gr}_G = T^*(LG/L^+G)$, the cotangent bundle of the affine Grassmannian for $G$. To obtain it, one may start with the 6d theory on a direct product $M\times D^2\times S^1$ (\emph{i.e.} at $q=1$), compactify first on $S^1$, and then on the cigar circle, keeping the latter at finite radius (retaining all KK modes). This produces a 4d $\CN=4$ theory on $M\times \R_+$ with gauge group $LG$ (the loop group), and with a boundary condition at $0\in \R_+$ that breaks $LG$ to the positive loop group $L^+G$. Further replacing the asymptotic boundary condition at $\infty\in \R_+$ with a Dirichlet b.c. at finite distance (that breaks $LG$ completely), one finds a 4d sandwich setup that reduces to the 3d sigma-model $\CT_{\rm Gr}$,
\be \raisebox{-.3in}{$\includegraphics[width=1.5in]{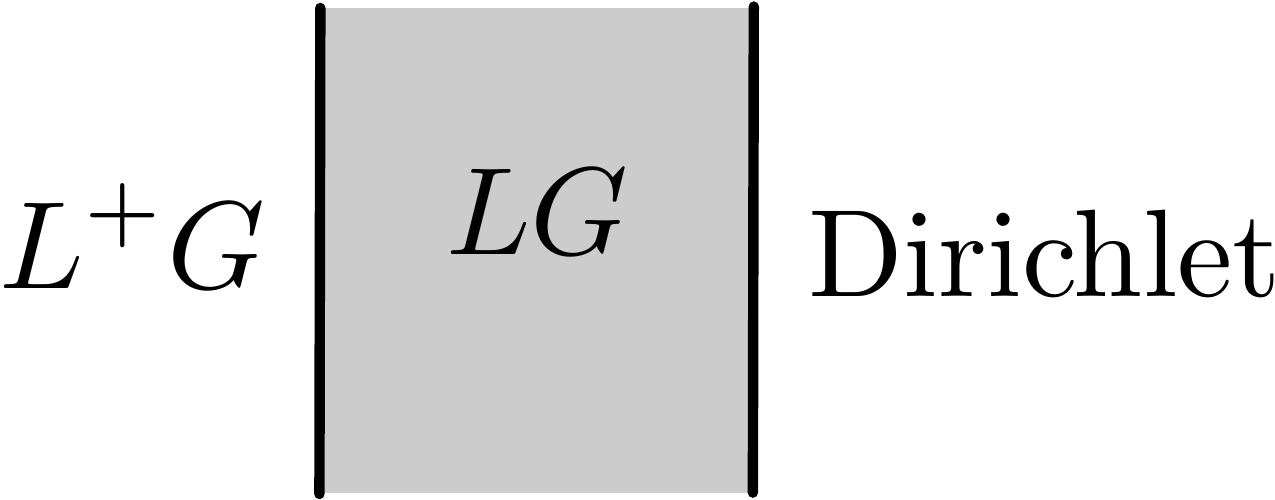}$} \qquad\leadsto\quad T^*(LG/L^+G)\quad\text{sigma-model}\ee
The analysis of \cite{RW-Coulomb} considered the B-twist (Rozansky-Witten twist) $\CT_{\rm Gr}^B$, and the parameter $q$ was re-introduced in the 3d theory as a twisted mass (part of a background flat connection) for $U(1)$ loop rotations of the target $T^*\text{Gr}_G$. 

It was also proposed in \cite{RW-Coulomb} that at roots of unity $q=e^{i\pi/k}$, the theory $\CT_{\rm Gr}^B$ would localize to 3d B-models with finite-dimensional targets `$X_k$' related to cotangent bundles of flag varieties for $G$. This is reminiscent of the B-model factor $\text{RW}(\CN^\vee)$ appearing in the approximation \eqref{CS-fact}, particularly noting that $\CN\simeq \CN^\vee$ for many groups (in particular, in type A) and that the cotangent bundle of the full flag variety is the Springer resolution of the nilpotent cone. This is the most concrete reason for expecting that the construction of \cite{RW-Coulomb} is 3d mirror to our current work. Again, we hope that this relation can be  clarified further in the future.

\subsubsection{BV Lagrangian for $G=SU(n)$}
\label{sec:typeA-intro}

When $G=SU(n)$, the construction of the 3d theory $\CT_{SU(n),k}^A$, which we'll just denote $\CT_{n,k}^A$, can be made even more explicit. The setups of Figure \ref{fig:bc-intro} may be engineered in a familiar way with branes and brane webs in IIB string theory \cite{HananyWitten, AharonyHanany, AharonyHananyKol}, which we'll review in Section \ref{sec:4d}. Correspondingly, $\CT_{n,k}$ has a UV Lagrangian definition as a quiver gauge theory:
\be \label{quiver-intro} \CT_{n,k}^{UV}\quad=\quad \raisebox{-.7in}{$\includegraphics[width=3in]{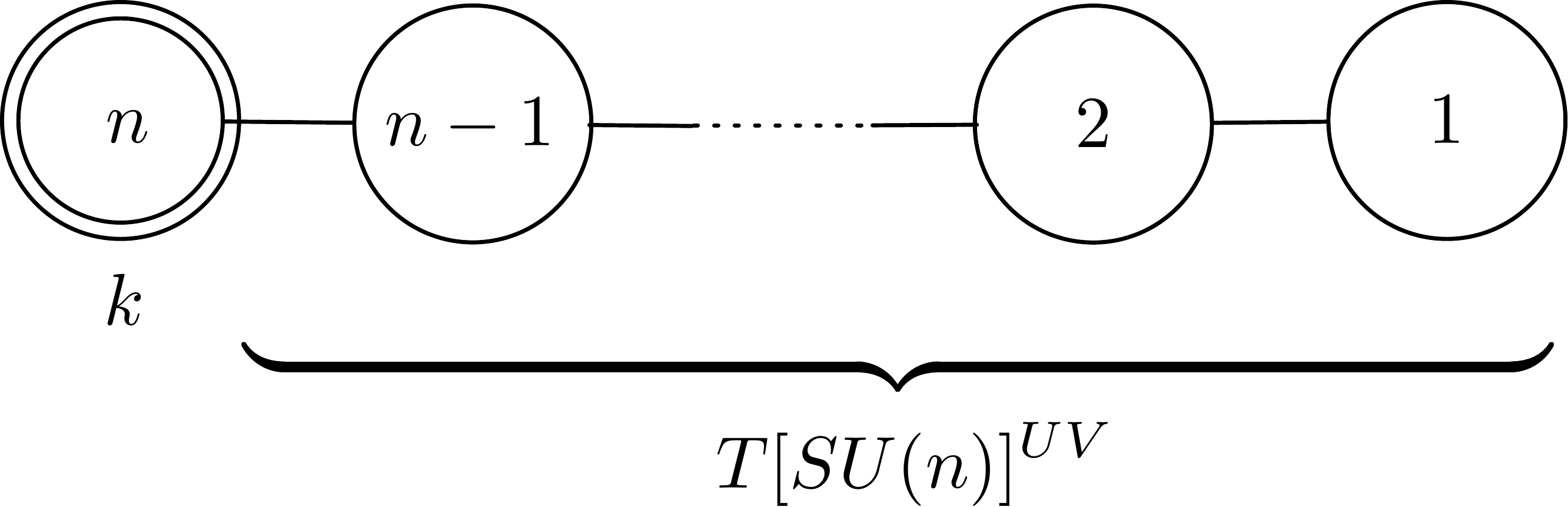}$} \ee
This is the standard 3d $\CN=4$ quiver for $T[SU(n)]$ \cite{GaiottoWitten-Sduality}, with the final $\boxed{n}$ flavor node gauged with an $\CN=2$ vectormultiplet at Chern-Simons level $k$. Altogether, the gauge group is $SU(n)_k\times\prod_{a=1}^{n-1}U(a)$, with hypermultiplet matter in representation $\bigoplus_{a=1}^{n-1}T^*\text{Hom}(\C^a,\C^{a+1})$.

There are two caveats to using the Lagrangian description $\CT_{n,k}^{UV}$: it does not have 3d $\CN=4$ supersymmetry (only 3d $\CN=2$ SUSY acts in the UV); and it does not have full $G^\vee=PSU(n)$ flavor symmetry (only the maximal torus $T^\vee\simeq U(1)^{n-1}$ acts in the UV).  

The first caveat is serious, as not having 3d $\CN=4$ SUSY means there is no BRST operator $Q_A$ with which to define the topological A-twist. We get around this by first passing through a holomorphic-topological (HT) twisted version of $\CT_{n,k}^{UV}$, which only requires $\CN=2$ SUSY \cite{ACMV}. (The 3d HT twist is an analogue of the 4d holomorphic twists developed earlier by \cite{Kapustin-hol, Costello-Yangian}.) Somewhat more precisely, the HT-twisted theory is a different theory that is nonetheless quasi-isomorphic to the holomorphic-topological twist of the original theory. We find that this simplified, HT-twisted version of $\CT_{n,k}^{UV}$, obtained using the twisted BV formalism of \cite{ACMV,CDG}, does admit an additional BRST symmetry $Q_A'$ and we expect that the total cohomology is equivalent to $\CT_{n,k}^A$. Schematically, we conjecture that
\be \label{HT-A-intro} \CT_{n,k}^A \,\simeq\, (\CT_{n,k}^{UV})^{HT+A'}\,, \ee
where the twisted theory on RHS is Lagrangian. Details are given in Section \ref{sec:Tnk-BV}, where we verify that the theory on the RHS is topological at least classically (by showing that the stress tensor is exact).

The twisted Lagrangian theory $(\CT_{n,k}^{UV})^{HT+A'}$ can be defined on any three-manifold $M$ with a transverse-homolorphic-foliation structure. In particular, it makes sense on $M=\Sigma\times \R$, where $\Sigma$ is any Riemann surface, which is sufficient for studying line operators (by taking $\Sigma=\C^*$), state spaces, and boundary VOA's. We will show explicitly in Section \ref{sec:Atwist} that $(\CT_{n,k}^{UV})^{HT+A'}$ admits Wilson-line operators for the Chern-Simons gauge group $SU(n)_k$, as expected. We will also see that $(\CT_{n,k}^{UV})^{HT+A'}$ has manifest global symmetry $T_\C^\vee$, given by the complexified torus of $G^\vee=PSU(n)$, and that it may be deformed by flat $T_\C^\vee$ connections. 

We would expect a similar Lagrangian formulation of $\CT_{G,k}^A$ to exist for any group $G$ such that a UV Lagrangian formulation of $T[G]$ is known. This includes $G=\text{Spin}(2n)$  \cite{GaiottoWitten-Sduality}.

\subsection{Results and conjectures}
\label{sec:results}

In the main part of the paper, we restrict to $G=SU(n)$, and focus on the topologically twisted theories $\CT_{n,k}^A=\CT_{SU(n),k}^A$.

Since $\CT_{n,k}^A$ has global symmetry $PSU(n)$, and couples to complexified $PGL(n,\C)$ background connections, its category of line operators $\CC^{(n,k)}$ forms a coherent sheaf of categories over $PGL(n,\C)$:
\be \CC^{(n,k)} \to PGL(n,\C)\,, \ee 
just as in \eqref{sheaf-intro}. Each stalk $\CC^{(n,k)}_g$ is the dg category of line operators in the presence of a vortex line for the background connection with (basepointed) holonomy $g\in PGL(n,\C)$. We will explain this  structure more carefully in Section \ref{sec:flavor-line}. 

Let us fix integers $n\geq 2$, $k\geq n$, and set $q=e^{i\pi/k}$.
Our main conjecture is
\begin{conj} \label{conj:1} There is an equivalence of coherent sheaves of dg categories
\be \CC^{(n,k)}\,\simeq\, D^b\big(U_q(\mathfrak{sl}_n)\text{-mod}\big) \ee
relating the category of line operators in the topologically twisted theory $\CT_{n,k}^A$ and the derived category of line operators for the simply connected De Concini-Kac quantum group at an even root of unity (with Frobenius center acting semisimply).

More generally, $\CT_{n,k}^A$ defines an extended axiomatic TQFT of cohomological type (a spin TQFT if $k$ is even) whose restriction to cohomological degree zero (to the extent this makes sense) is equivalent to an axiomatic CGP TQFT based on the unrolled quantum group $U_q^H(\mathfrak{sl}_n)$.
\end{conj}
We provide physics proofs and computational evidence for various parts of this conjecture. In particular, we will prove that
\begin{PTh} \label{th:1} There is an equivalence of dg categories
\be \CC^{(2,k)}_{g=1} \,\simeq\, D^b(u_q(\mathfrak{sl}_2)\text{-mod})  \label{thm1} \ee
relating the category of line operators in $\CT_{2,k}^A$ in the absence of background connection to the non-semisimple category of modules of the restricted quantum group. This extends to an equivalence of braided tensor categories, with suitable R-matrix and associator on the RHS.
\end{PTh}

The proof of Theorem \ref{th:1} is where boundary VOA's come in. In Sections \ref{sec:Atwist} and \ref{sec:VOA}, we will define a \emph{pair} of boundary conditions (N, D) for $\CT_{n,k}^A$ that support boundary VOA's (${\CN}_{n,k}, \CD_{n,k}$), respectively. We can identify these VOA's explicitly using 3d-field-theory methods of \cite{CostelloGaiotto, CCG, CDG}, as well as the analysis of corner configurations in 4d $\CN=4$ Yang-Mills theory of \cite{GaiottoRapcak,CreutzigGaiotto-S}. Roughly speaking, ${\CN}_{n,k}$ is a  $(\widehat{\mathfrak{sl}_n})_{k-n}$ coset of the ``S-duality kernel'' VOA of \cite{CostelloGaiotto,CCG,CreutzigGaiotto-S}; while $\CD_{n,k}$ is an extension of the product of a W-algebra and an affine algebra that results of \cite{sugimoto2021feigintipunin} show to be equivalent to a Feigin-Tipunin algebra
\be \CD_{n,k} \simeq \FT_k(\mathfrak{sl}_n)\,. \label{DFT-intro} \ee 
Using \eqref{DFT-intro}, we may then apply the Kazhdan-Lusztig-like correspondence of \cite{Creutzig:2021cpu,GannonNegron}, which established an equivalence of abelian braided tensor categories 
 $\FT_k(\mathfrak{sl}_2)\text{-mod}\simeq u_q(\mathfrak{sl}_2)\text{-mod}$, with monoidal structure on the quantum-group side given by \cite{Gainutdinov:2015lja,Creutzig:2017khq, Creutzig:2020jxj, Gainutdinov:2018pni}.

We further propose in Section \ref{sec:VOA} that
\begin{conj} \label{conj:2}
A slight modification $\wt{\CN}_{n,k}$ of $\CN_{n,k}$ (obtained by a successive extension and orbifold)  and $\CD_{n,k}$ are dual, in the sense that they are mutual commutants inside $nk$ copies of free fermions $\FF(nk)$, 
\be \label{NDFF-intro} \wt{\CN}_{n,k} \simeq \FF(nk)/\CD_{n,k}\,,\qquad \CD_{n,k} \simeq \FF(nk)/\wt{\CN}_{n,k}\,. \ee
\end{conj}
This induces an equivalence between the abelian braided tensor categories  $\wt{\CN}_{n,k}\text{-mod}\simeq \CD_{n,k}\text{-mod} = \FT_k(\mathfrak{sl}_n)\text{-mod}$, which implies an equivalence of corresponding derived categories.

Conjecture \ref{conj:2} proposes a novel logarithmic level-rank duality. 
There is a remarkable property of the quantum-Hamiltonian-reduction functor, namely that it commutes with tensoring with integrable representations \cite{Arakawa:2020oqo}. This allows us  to show that two deformable families of cosets are isomorphic, \emph{cf}. \eqref{voas2ways}. The isomorphism is motivated by a relation between corner configurations in $4$d $\mathcal N=4$ super-Yang-Mills theory \cite{GaiottoRapcak, CreutzigGaiotto-S}. 
If we take a large level limit of one side of this relation, then we get a large center times many pairs of free fermions. The Feigin-Tipunin algebra is by construction a subalgebra of the free fermions and we conjecture that its coset is $\wt{\CN}_{G,k}$. By construction, the coset contains a large subalgebra of this new logarithmic VOA $\wt{\CN}_{n,k}$. In fact, we not only conjecture that these two logarithmic VOA's form a dual pair but also that the decomposition of the free fermions is of a specific form, see \eqref{eq:decleft} and \eqref{eq:decright}. If (and conjecturally also only if) there is indeed a braid-reversed equivalence between the finite tensor categories of two VOA's, then these two VOA's can be extended to a VOA with trivial module category (\eg\ free fermions), and the extension is exactly of the form \eqref{eq:decleft}--\eqref{eq:decright} by \cite{Creutzig:2019psu}. 
In the case of $\mathfrak{sl}(2)$ we are able to perform branching-rule computations that nicely support our conjecture.

Altogether, Conjectures \ref{conj:1} and \ref{conj:2} at $g=1$ may be summarized as
\be \begin{array}{c@{\quad}c@{\quad}c@{\quad}c@{\quad}c} \text{\underline{QFT}} && \text{\underline{VOA}} &&  \text{\underline{qu. group}} \\
\text{line ops $\CC_{g=1}^{(n,k)}$}  & \simeq & D^b\big(\wt{\CN}_{n,k}\text{-mod}\big) \simeq D^b\big(\FT_k(\mathfrak{sl}_n)\text{-mod}\big) & \simeq &  D^b(u_q(\mathfrak{sl}_n)\text{-mod})\,, \end{array}
\label{thm1-exp} \ee
providing a direct analogue to the classic equivalences \eqref{C-ss} in Chern-Simons theory. Even the equivalence of the pair of VOA categories appearing here has a classic analogue, in terms of level-rank duality of WZW algebras. The analogy can be made surprisingly tight, by recalling that $SU(n)_{k-n}$ Chern-Simons theory can be engineered from $SU(n)_{k}$ supersymmetric $\CN=2$ Yang-Mills-Chern-Simons, in the holomorphic-topological twist. The supersymmetric $SU(n)_{k}$ theory admits a pair of holomorphic boundary conditions, Neumann  (N) and Dirichlet (D), described in \cite{DGPdualbdys}. They support the WZW VOA's $V^{k}(\mathfrak{gl}_{k-n})$ and $V^{k}(\mathfrak{sl}_n)$, respectively, which are level-rank dual, and mutual commutants in $\FF(n(k-n))$ \cite{NS-duality,NS-braid,FvD,MR3162483}. Our pair of boundary conditions (N,D) for $\CT_{n,k}^G$ are generalizations of Neumann and Dirichlet b.c. in $\CN=2$ Yang-Mills-Chern-Simons theory, and our pair of VOA's $\big(\wt{\CN}_{n,k},\CD_{n,k}\big)$ are generalizations of the level-rank dual pair $\big(V^{k}(\mathfrak{gl}_{k-n}),V^{k}(\mathfrak{sl}_n)\big)$.

We also describe in Section \ref{sec:VOA} how the VOA's $\big(\wt{\CN}_{n,k},\CD_{n,k}\big)$ and their categories of  modules can be deformed by flat $PGL(n,\C)$ connections. We expect the equivalence of sheaves of categories in Conjecture \ref{conj:1} to be realized via the deformed categories of $\big(\wt{\CN}_{n,k},\CD_{n,k}\big)$ modules.

We of course also expect Conjectures \ref{conj:1}--\ref{conj:2} to have generalizations involving other groups $G$, and various global forms. As mentioned in the preceding quantum-group, VOA, and QFT discussions, we expect multiple subtleties to appear, especially for non-simply-laced $G$. We leave such generalizations to future work.

\subsubsection{Some computations}
\label{sec:comp}

We supplement and support the somewhat abstract equivalences in Conjecture \ref{conj:1} and Theorem~\ref{th:1} with some explicit computations. These are described in Section~\ref{sec:Uqsl2} for $U_q(\mathfrak{sl}_2)$, in Section \ref{sec:Hilbert} for the QFT $\CT_{n,k}^A$ (focusing on $n=2$), and in Section \ref{sec:VOA} for $\FT_k(\mathfrak{sl}_n)$ (focusing on the triplet VOA $\FT_k(\mathfrak{sl}_2)$).

For example, we compute the Grothendieck ring of the category of line operators $\CC_{g=1}^{(2,k)}$ in the QFT $\CT_{2,k}^A$ in terms of the ``Bethe root'' analysis of Nekrasov-Shatashvili \cite{NS-Bethe,NS-int}. We match this with the Grothendieck ring of $u_q(\mathfrak{sl}_2)\text{-mod}\simeq \FT_k(\mathfrak{sl}_2)\text{-mod}$ given \emph{e.g.} in \cite{Feigin:2005xs}.
 We also match the $\Z_n$ one-form symmetry of $\CT_{n,k}^A$ and its 't Hooft anomaly with the $\Z_n$ symmetries generated by invertible modules of $u_q(\mathfrak{sl}_n)$ and $\FT_k(\mathfrak{sl}_n)$.

The category of line operators $\CC_{g=1}^{(n,k)}$ itself should have a direct formulation in the A-twisted QFT $\CT_{n,k}^A$. We make some brief comments/predictions about this in Section \ref{sec:T-cat}. Categories of line operators in topologically twisted 3d $\CN=4$ gauge theories were studied recently by \cite{BFN-lines,DGGH,Webster-tilting,HilburnRaskin}, though unfortunately the results therein do not apply directly to theories with Chern-Simons terms.

We also study the state spaces $\CH(\Sigma_g,\CA)$ associated to genus-$g$ surfaces with a choice of  $G^\vee_\C=PGL(n,\C)$ connection $\CA$ on $\Sigma_g$. Algebraically, `$\CA$' is the data of a local system, $\CA\in \text{Loc}_{PGL(n,\C)}(\Sigma_g)$, and the collection of state spaces for various $\CA$ assembles into a coherent sheaf over the moduli space of local systems, 
\be \CH(\Sigma_g) \to \text{Loc}_{PGL(n,\C)}(\Sigma_g)\,. \ee
(Such sheaves were discussed by \cite{Gaiotto-blocks}, in the general context of 3d $\CN=4$ theories with flavor symmetry.) Each stalk  $\CH(\Sigma_g,\CA)$ is a vector space with a cohomological $\Z$-grading. For generic $\CA$, we expect $\CH(\Sigma_g,\CA)$ to be finite-dimensional and supported entirely in degree zero, while for exceptional $\CA$ (such as $\CA=0$) we expect $\CH(\Sigma_g,\CA)$ to be infinite-dimensional, supported in infinitely many non-negative (say) cohomological degrees, with finite graded dimensions. However, the regularized Euler character (\emph{a.k.a.} Witten index) $\chi\big[\CH(\Sigma_g,\CA)\big]$ should be \emph{independent} of $\CA$.

We compute Euler characters from quantum-group, QFT, and VOA perspectives when $n=2$, finding complete agreement
\be \label{chi-intro} \chi[\CH(\Sigma_g,\CA)] = \begin{cases} 2k & g=1 \\ 2^gk^{3g-3} & g > 1\,. \end{cases} \ee
The quantum-group computation at generic $\CA$ is reviewed in Sections \ref{sec:U-Hilbert} and \ref{sec:decTQFT}.
The QFT computation employs the twisted-index analysis of \cite{BZ-index,BZ-Riemann,CK-comments}, adapted to the topological A-twist. The QFT and VOA perspectives also allow a straightforward refinement of \eqref{chi-intro} by characters of the $PGL(2,\C)$ symmetry, given in \eqref{eqchar} and \eqref{eqchar2}, respectively. For more general $n$, we compute that $\chi[\CH(T^2,\CA)]=nk^{n-1}$ in the QFT $\CT_{n,k}^A$ (Section \ref{sec:SUnBethe}), which again agrees with quantum-group and VOA predictions.

In genus zero, the flat connection $\CA$ is necessarily trivial, and $\CH(\Sigma_0)$ should be isomorphic to the algebra of local operators in our cohomological TQFT. It is infinite-dimensional, and can be computed from the quantum-group perspective to take the form
\be \label{Hg0-intro} \CH(\Sigma_0) \simeq \C[\CN]
 \ee
where $\C[\CN]$ denotes the ring of algebraic functions on the nilpotent cone $\CN$ of $\sln$, with cohomological degree corresponding to weight under the $\C^*$ conical action on $\CN$. 
This quantum-group computation uses a geometric equivalence of \cite{ABG,BL} (see Section \ref{sec:U-Hilbert}). From a QFT perspective, we reproduce the (regularized) Euler character of \eqref{Hg0-intro} by computing the index of the space of local operators of $\CT_{2,k}^A$ (see Section \ref{sec:T-char-g}). The space \eqref{Hg0-intro} is also consistent with the approximation \eqref{Hilb-fact} being \emph{exact} in genus zero: the Chern-Simons state space $\CH_{G_{k-h}}(\Sigma_0)\simeq \C$ is always trivial, while the Rozansky-Witten state space $\CH_{RW}(\CN^\vee)(\Sigma_0)$ is precisely the ring of functions on $\CN^\vee$ (which is isomorphic to $\CN$ when $G=SU(n)$). 

More generally, QFT techniques developed in \cite{Gaiotto-blocks,BF-Hilb,BFK-Hilb,SafronovWilliams} predict that the genus-$g$ state space
\be \label{H-BunG-intro} \CH(\Sigma_g,\CA=0) \simeq H_{\bar\pd}^\bullet\big(\text{Bun}_{SL(n,\C)}(\Sigma_g),\CL^k\otimes \CE_{T[SU(n)]}\big) \ee
will be given by derived sections of a particular sheaf on the moduli space of algebraic $G_\C=SL(n,\C)$ bundles, where the sheaf is a product of a line bundle $\CL^k$ that appears in ordinary Chern-Simons theory and an infinite-rank vector bundle $\CE_{T[SU(n)]}$ determined by the state space of the A-twisted theory $T[SU(n)]^A$. 
The factorization \eqref{Hilb-fact} is equivalent to approximating $H_{\bar\pd}^\bullet\big(\text{Bun}_{SL(n,\C)}(\Sigma_g),\CL^k\otimes \CE_{T[SU(n)]}\big) \approx H_{\bar\pd}^\bullet\big(\text{Bun}_{SL(n,\C)}(\Sigma_g),\CL^k)\otimes \CE_{T[SU(n)]}\big|_{0\in\text{Bun}}$ (see Section \ref{sec:T-Hilb}).

In genus one and $n=2$, the factorization suggests
\be \label{Hg1-approx-intro} \CH(\Sigma_1,\CA=0) \approx \C^{k-1}\otimes H_{\bar \pd}^{\bullet,\bullet}(T^*[2]\mathbb{P}^1)\,, \ee
where the second factor is total (algebraic) Dolbeault cohomology of $T^*[2]\mathbb{P}^1$. (Here $T^*[2]\mathbb{P}^1$ is the Springer resolution of the nilpotent cone $\CN$ for $\mathfrak{sl}_2$, with ``$[2]$'' denoting an appropriate shift in cohomological grading.)  On the other hand, from a quantum-group perspective, the genus-one state space is given by Hochschild homology of $u_q(\mathfrak{sl}_2)$-mod, which the geometric equivalence of \cite{BL,LQ} identifies as
\be \CH(\Sigma_1,\CA=0) \simeq HH_\bullet(u_q(\mathfrak{sl}_2)\text{-mod}) \simeq \C^2 \oplus \big[ \C^{k-1}\otimes H_{\bar \pd}^{\bullet,\bullet}(T^*[2]\mathbb{P}^1)\big]\,. \ee
This is just a small correction to \eqref{Hg1-approx-intro}.

The subspace of $\CH(\Sigma_g,\CA=0)$ in cohomological degree zero should be equivalent to the state space of  the CGP TQFT based on $U_q^H(\mathfrak{sl}_2)$ and to the space of conformal blocks of $\FT_k(\mathfrak{sl}_2)$. It is easy to check that the dimensions $\text{dim}\,\CH(\Sigma_0)\big|_{\text{deg 0}} = 1$ and $\text{dim}\,\CH(\Sigma_1,\CA=0)\big|_{\text{deg 0}} = 3k-1$ agree with known results in the literature; the CGP computation is reviewed in Section \ref{sec:decTQFT}.

From a VOA perspective, the full state space $\CH(\Sigma_g,\CA=0)$ should coincide with \emph{derived} conformal blocks of the triplet algebra. This has not yet been studied. In principal, derived conformal blocks may be defined via Beilinson-Drinfeld's chiral homology \cite{BD}, but effective computational techniques are still being developed, \eg\ in the recent \cite{LiZhou,EkerenHeluani}.

\subsection{Acknowledgements}

We are grateful to many colleagues for discussions and explanations of the many disparate ideas that played a role in this paper; we would in particular like to thank R. Bezrukavnikov, M. Bullimore, M. Dedushenko, D. Jordan, P. Etingof, B. Feigin, S. Gukov, H. Kim, W. Niu, B. Patureau-Mirand,  S. Lentner, P. Safranov, S. Schafer-Nameki, M. Rupert, and B. Willett. We are especially grateful to V. Mikhaylov and N. Reshetikhin for collaboration in the early days of this project.

This project arose from the NSF FRG collaboration \emph{Homotopy Renormalization of Topological Field Theories} (DMS 1664387), on which T. D. and N. Geer are PI's, and which brought all the coauthors together at the event \emph{New Directions in Quantum Topology} (Berkeley, 2019).
 The work of T. D. and N. Garner was also supported by NSF CAREER grant DMS 1753077. The work of T. C.  is supported by NSERC Grant Number RES0048511.

\section{Topologically twisted 3d theories with flavor symmetry}
\label{sec:toymodel}

In this section, we develop some general expectations about the structure of 3d TQFT's defined by topologically twisting a 3d $\CN=4$ supersymmetric theory with flavor symmetry. Much of what we say is review and/or application of existing ideas from the math and physics literature. Some features we seek to emphasize include:
\begin{itemize}
\item The role of flavor symmetry in topological twists of 3d $\CN=4$ theories; in particular, the way that $G$ flavor symmetry can lead to topologically twisted theories coupled to complexified $G_\C$ background flat connections.
\item The dg (differential graded) nature of the braided tensor category of line operators in a topological twist, and the way this category interacts with deformations by flat connections coming from flavor symmetry.
\item How the category of line operators may be represented as a derived category of modules for a boundary VOA.
\item The dg nature of spaces of states on a surface $\Sigma$, and their dependence on a choice of flat $G_\C$ connection on $\Sigma$.
\item How characters of state spaces, which are independent of choices of flat connections, may be computed using established techniques of supersymmetric localization.
\item The relation between state spaces and the category of line operators; in particular, how the genus-one state space may be obtained as Hochschild homology (as opposed to the Grothendieck group/K-theory) of the category of line operators, and what this means physically.
\end{itemize}
Our treatment will be somewhat one-sided, in that we focus on flavor symmetries that give rise to flat connections in a topological twist. There are other flavor symmetries that give rise to deformations by monopole backgrounds, which we do not consider, as they are not ultimately relevant for $\CT_{G,k}^A$ TQFT's.

We will illustrate the above features using a fully explicit and computable toy model: the 3d topological B-twist of a free hypermultiplet. This theory, which we'll denote $\CT_{\rm hyper}^B$,
is known by several other names, including Rozansky-Witten \cite{RW} theory with target $\C^2$, and $\mathfrak{psl}(1|1)$ Chern-Simons theory \cite{Mikhaylov} (related to $U(1|1)$ Chern-Simons at level one \cite{RozanskySaleur}). %
This deceptively simple theory turns out to have many qualitative features in common with the $\CT_{G,k}^A$ TQFT's that we study in the remainder of the paper. In particular, it has a non-semisimple dg category of line operators, has infinite-dimensional state spaces with nontrivial cohomological degree (or ghost number), and admits semisimple deformations by nonabelian flat connections. 
We will eventually propose an even more direct relation between $\CT_{\rm hyper}^B$ and $\CT_{G,k}^A$ theories in Section \ref{sec:SO3-ferm}, namely that there is a duality
\be \CT_{\rm hyper}^B \simeq \CT_{SO(3),k=2}^A\,. \label{hyper-SO(3)} \ee

We will say very little about partition functions on general 3-manifolds, and make no claims about when or whether partition functions (and other correlation functions) can be suitably regularized to give finite results. These are subtle matters. Some recent results on using flavor symmetry/equivariance to regularize partition functions appeared in \cite{RW-Coulomb}.

\subsection{Twisting and the toy model}

We recall that the 3d $\CN=4$ supersymmetry algebra is generated by eight supercharges $Q_\alpha^{a\dot a}$, transforming as a tri-spinor of the Euclidean spin group $SU(2)_E$ (index $\alpha\in\{+,-\}$), a `Higgs' R-symmetry $SU(2)_H$ (index $a\in \{+,-\}$) and a `Coulomb' R-symmetry $SU(2)_C$ (index $\dot a\in \{+,-\}$). In the absence of central charges, the algebra is
\be \{Q_\alpha^{a\dot a},Q_\beta^{b\dot b}\} = \epsilon^{ab}\epsilon^{\dot a\dot b}\sigma_{\alpha\beta}^\mu P_\mu\,. \label{eq:N=4}\ee

Any 3d $\CN=4$ theory that preserves $SU(2)_C$ R-symmetry admits a topological ``B-twist.''
In flat space, the B-twist amounts to working in the cohomology of the nilpotent supercharge%
\footnote{More generally, there is a $\mathbb{CP}^1$ family of B-twists, corresponding to supercharges $Q_B^a := \delta^\alpha{}_{\dot a}Q^{a\dot a}_\alpha$ for any linear combination of indices $a$. Different elements in the family are related by $SU(2)_H$ rotations, and we have fixed this freedom by selecting $a=+$. 
}
\be Q_B := \delta^\alpha{}_{\dot a}Q^{+\dot a}_\alpha = Q_+^{+\dot +} + Q_-^{+\dot -}\,. \label{QB} \ee 
In curved space, the supercharge $Q_B$ may be preserved by introducing an $SU(2)_C$ R-symmetry background equal to the $SU(2)_E$ spin connection. The supercharge $Q_B$ also has charge $+1$ under a maximal torus $U(1)_H\subset SU(2)_H$. 
In any theory that preserves $U(1)_H$, one can then use this symmetry to endow the B-twist with a $\Z$-valued cohomological grading.

A 3d $\CN=4$ sigma-model with hyperk\"ahler target $\CX$ locally parameterized by hypermultiplets preserves $SU(2)_C$, and thus admits a B-twist, known as Rozansky-Witten theory~\cite{RW}. When $\CX$ has an additional $U(1)$ isometry that rotates its $\mathbb{CP}^1$ of hyperk\"ahler structures, the theory preserves $U(1)_H$, and thus has a $\Z$-valued cohomological grading. This was not the case for the compact targets initially studied by Rozansky and Witten (hence only $\Z_2$ fermion-number gradings appeared in \cite{RW}), but it will be the case for us.

We are interested in a single free hypermultiplet, whose two complex, bosonic scalars $X,Y$ parameterize a noncompact target $\C^2$. The 2$\times$2 matrix of scalars and their conjugates
\be SU(2)_H\;\; \actson\;\;  \begin{pmatrix} X & Y \\ \ol Y & -\ol X \end{pmatrix} \;\; \reflectbox{\actson}\;\; SU(2)_m \label{hyperHm} \ee
admits two commuting actions, of $SU(2)_H$ R-symmetry (on the left) and $SU(2)_m$ flavor symmetry (on the right). Since $X,Y$ are invariant under $SU(2)_C$, they remain scalars in the B-twist, even in curved space.
From the action of the (diagonal) maximal torus $U(1)_H\subset SU(2)_H$, we find that both $X$ and $Y$ have cohomological degrees $+1$.

The hypermultiplet fermions transform as tri-spinors of $SU(2)_E\times SU(2)_C\times SU(2)_m$, and may be denoted $\psi^{X\dot a}_\alpha,\psi^{Y\dot a}_\alpha$ (of $U(1)_m\subset SU(2)_m$ flavor charges $+1,-1$, respectively). In the B-twist on curved spacetimes, they reorganize into two scalars $\eta^{X}=\delta^\alpha{}_{\dot a}\psi^{X\dot a}_\alpha,\,\eta^{Y}=\delta^\alpha{}_{\dot a}\psi^{Y\dot a}_\alpha$ and two 1-forms $\chi^X_\mu = (\sigma_\mu)^\alpha{}_{\dot a}\psi^{X\dot a}_\alpha,\, \chi^Y_\mu = (\sigma_\mu)^\alpha{}_{\dot a}\psi^{Y\dot a}_\alpha$. Since the fermions are invariant under $SU(2)_H$, they have cohomological degree $0$.

\subsubsection{Twisted action}

It is enlightening to rewrite the B-twisted hypermultiplet theory $\CT_{\rm hyper}^B$ in the Batalin-Vilkovisky \cite{BV} formalism. Schematically, this involves introducing anti-fields for all physical fields and adding the supercharge $Q_B$ to the BV differential, with a corresponding deformation of the action. (This was derived for general B-twisted sigma models (Rozansky-Witten theories) in \cite{QZ-RW}, and B-twisted gauge theories in \cite{KQZ}.) After further integrating out half the fields and anti-fields, one ends up with the following simplified description of the theory.%
\footnote{This description is directly analogous to the simplified BV action for the holomorphic-topological twist of 3d $\CN=2$ theories developed in \cite{ACMV,CDG} and 4d $\CN=1$ theories in \cite{Costello-Yangian}.}

On a 3d Euclidean spacetime $M$, the fields of $\CT_{\rm hyper}^B$ consist of two mixed-degree differential forms
\be \mb X,\mb Y\in \Omega^\bullet(M)[1]\,. \ee
where `$[1]$' denotes a shift in cohomological degree. The action is simply
\be S = \int_M \mb X \diff \mb Y\,, \label{SXY} \ee
the BV bracket is $\{\mb X(x),\mb Y(x')\}_{BV} = \delta^{(3)}(x-x')\diff\text{Vol}$, and the combined BV/B-twist differential $Q = \{-, S\}_{BV}$ acts as
\be Q \mb X = \diff\mb X\,,\qquad Q\mb Y = \diff\mb Y\,.\ee

To relate this to physical fields, we may expand $\mb X,\mb Y$ in local coordinates as
\be \begin{array}{l}
	\mb X = X + \chi^X_\mu \diff x^\mu + \rho^{X}_{\mu\nu} \diff x^\mu \diff x^\nu + \xi^X  \diff x^1 \diff x^2 \diff x^3 \\[.1cm]
	\mb Y = Y + \chi^Y_\mu \diff x^\mu + \rho^{Y}_{\mu\nu} \diff x^\mu \diff x^\nu + \xi^Y  \diff x^1 \diff x^2 \diff x^3\,.
\end{array}
\ee
The conventions are such that $\mb X$ and $\mb Y$ \emph{and} all differentials $\diff x^\mu$ have cohomological degree ($U(1)_H$ charge) $+1$. Moreover, $\mb X,\mb Y$ are bosonic and the differentials are fermionic. Thus, $X,Y$ are bosons of degree $+1$, coinciding with the physical fields of the same name; and the 1-forms $\chi_\mu$ are fermions of degree $0$, coinciding with the physical fermions of the same name. The two-forms $\rho^X,\rho^Y$ are anti-fields of $\chi^Y,\chi^X$, and are cohomologous to the physical fields $*\diff\ol Y$ and $*\diff(-\ol X)$, respectively. The 3-forms $\xi^{X,Y}$ are anti-fields of the bosons $Y,X$, and are cohomologous to the physical $*\ol\eta^Y,*(-\ol\eta^X)$.

We note that the simplified action \eqref{SXY} naturally generalizes to any B-twisted sigma model with exact holomorphic-symplectic target $\CX$, meaning that the holomorphic-symplectic form $\omega$ on $\CX$ may be written globally as $\omega = \diff \lambda$.
The space of fields of the B-twisted sigma-model may be identified as maps $\bm \Phi: T[1]M\to \CX$ (where $T[1]M$ denotes the shifted tangent bundle of $M$), and `$\mb X \diff \mb Y$' is replaced by the pullback of the holomorphic Liouville 1-form
\be S = \int_M \bm\Phi^*(\lambda)\,. \ee
When $\CX$ has a holomorphic $\C^*$ action that acts on $\omega$ and $\lambda$ with weight $2$, this theory can be endowed with a $\Z$-valued homological grading.

\subsubsection{Flat $SL(2,\C)_m$ connections}
\label{flat-def}

The B-twisted theory $\CT_{\rm hyper}^B$ admits a family of deformations by flat $SL(2,\C)$ connections. Thus $\CT_{\rm hyper}^B$ may be defined on any 3-manifold $M$ with a choice of spin$^c$ structure (which shall be implicit), \emph{and} a choice of flat $SL(2,\C)$ connection $\CA$ on $M$. 
The partition function of $\CT_{\rm hyper}^B$ (if finite/well defined) will compute an invariant of the pair $(M,\CA)$. This is in fact well known, \emph{cf.} \cite{RozanskySaleur,Mikhaylov}: when $\CA$ is generic, the invariant in question is the Reidemeister-Ray-Singer torsion associated to the flat connection $\CA$,
\be Z(M,\CA) = \text{torsion of $(M,\CA)$}\,.\ee

We recall the origin of the deformation by a flat connection.
The 3d $\CN=4$ hypermultiplet has a flavor symmetry $SU(2)_m$ discussed above. In flat space, background connections $A_\mu$ for this symmetry sit in a 3d $\CN=4$ vectormultiplet, which also includes three $\mathfrak{su}(2)$-valued scalars $m_{i=1,2,3}$, usually known as twisted masses. In the B-twist on curved space, the masses $m_i$ become components of an $\mathfrak{su}(2)$-valued 1-form, which combines with a real $SU(2)_m$ connection to form the complexified $SL(2,\C)_m$ connection
\be \CA := A_\mu \diff x^\mu +  i\, m_\mu \diff x^\mu\,. \label{A-mass} \ee
The deformation by a background connection  $\CA$ is compatible with the $Q_B$ supercharge so long as the connection is \emph{flat}, \emph{i.e.} $\CF_\CA:=(\diff+\CA)^2=0$.\,%
\footnote{The flatness condition can be derived by promoting $SU(2)_m$ to a dynamical gauge symmetry, and looking for $Q_B$-fixed points of the associated vectormultiplet fields. One finds that $Q_B$ acts on gauginos to generate $\CF_\CA$ \cite{BlauThompson, DGGH}, whence $Q_B$-preserving backgrounds require $\CF_\CA=0$.}

Alternatively, the deformation by a flat connection may be explicitly incorporated into the simplified BV action \eqref{SXY}. 
Let us group the fields $(\mb X,\mb Y)$ into an $SU(2)_m$ doublet $\bm \Phi^i$, with  $\bm\Phi^1 = \mb X, \bm\Phi^2=\mb Y$. Then the action on a 3-manifold $M$ is $\frac12 \int_M \epsilon_{ij} \bm\Phi^i \diff \bm\Phi^j$. Given a complexified $SL(2,\C)$ connection $\CA$ on $M$, we may deform the action to
\be S = \frac12\int_M \epsilon_{ij} \bm\Phi^i (\diff_\CA \bm\Phi)^j 
\label{SA} \ee
with covariant exterior derivative $\diff_\CA = \diff+\CA$. The BV/B-twist differential becomes
\be  Q\bm\Phi^i = (\diff_\CA\mb \Phi)^i \label{QBV-A} \ee
and squares to zero if and only if $\CA$ is flat.

We note that there was nothing special about the group $SL(2,\C)$ in the above discussion.
More generally, we may consider the B-twist of any 3d $\CN=4$ theory with a flavor symmetry $G$ whose connection  sits in a background vectormultiplet (or, dually, the A-twist of a 3d $\CN=4$ theory with a flavor symmetry whose connection sits in a background twisted vectormultiplet). The resulting TQFT may be deformed by a flat $G_\C$ connection, and should produce invariants of (spin$^c$) 3-manifolds equipped with such flat connections.

\subsubsection{Bulk local operators}
\label{sec:hyper-bulk}

The bulk local operators in the B-twist of a 3d $\CN=4$ sigma-model with complex-symplectic target $\CX$ were identified in \cite{RW} as Dolbeault cohomology classes
\be \text{Ops}_B := H^\bullet(\text{Ops},Q_B) \simeq H_{\bar\pd}^{0,\bullet}(\CX)\,. \ee
For $\CT_{\rm hyper}^B$, we have $\CX=\C^2$ with coordinates $X,Y$ (the hypermultiplet scalars), and expect to find
\be \text{Ops}_B = H_{\bar\pd}^{0,\bullet}(\CX) =\C[X,Y]\,, \label{hyper-ops} \ee
a polynomial algebra in $X,Y$. The commutative product in this algebra comes from collision/OPE, which is well defined in $Q_B$-cohomology.

We note that it is also simple to derive \eqref{hyper-ops} perturbatively from the twisted action above, and to show that the result is independent of the choice of background flat connection.
Starting from \eqref{SA}, we may construct a space of (perturbative) local operators as polynomials in the components of $\mb X, \mb Y$, and their spacetime derivatives. Under the transformation \eqref{QBV-A}, only the zero-form components $X,Y$ are $Q$-closed. Moreover, from
\be Q \bp  \chi^X \\ \chi^Y \ep = (\diff+\CA) \bp X \\ Y \ep\,, \label{hyper-Qchi} \ee
we see that covariant derivatives of $X,Y$ are $Q$-exact. The operator algebra may then be generated by $X,Y$ alone.

The presence of nontrivial bulk local operators guarantees that the category of line operators in $\CT_{\rm hyper}^B$ cannot be semisimple. At the very least, $X,Y$ will show up as endomorphisms of the trivial/identity line operator. We revisit this in Section \ref{sec:C1-hyper}.

\subsection{Line operators and flavor symmetry: generalities}
\label{sec:flavor-line}

The set of line operators of a topologically twisted 3d $\CN=4$ theory that preserve both the topological supercharge and the cohomological $U(1)$ symmetry has the structure of a dg braided tensor category, as outlined in Section \ref{sec:ss}. We saw above that the B-twist of a 3d $\CN=4$ theory with $G$ flavor symmetry may be deformed by a flat $G_\C$ connection. The category of line operators in such a theory correspondingly gains some additional structure.

\subsubsection{Families of categories}
\label{sec:holonomy}

Locally, the neighborhood of a line is $D^*\times I$, where $D^*$ denotes a small punctured disc and $I$ a small interval. Given any $G_\C$ flat connection $\CA$ on $D^*\times I$, we may consider line operators preserving the B-twist in the background $\CA$. These line operators are endowed with the structure of a dg category $\CC_{\CA}$ in the usual way. In particular, morphisms among lines are defined as the vector spaces of local operators at junctions of the lines (Figure \ref{fig:hom}). These spaces are complexes, with an action of the differential $Q_B$, and quasi-isomorphism is imposed as an equivalence relation. Being slightly imprecise, we will usually just take cohomology to represent an equivalence class of local operators:
\be \label{HomLL'} \text{Hom}_{\CC_\CA}^\bullet(L,L') :=  \begin{array}{c} \text{$Q_B$-cohomology of the space of local operators} \\[.1cm] \text{at a junction of $L$ \& $L'$ in background $\CA$} \end{array} \ee

There are equivalences among the categories $\CC_\CA$, induced by flavor gauge transformations. This is not entirely obvious or familiar, since the flavor symmetry is not dynamical. We analyze the situation as follows. If $\CA$ and $\CA^ {h}:= {h}\CA {h}^{-1}- \diff {h}{h}^{-1}$ are related by a complexified gauge transformation $ {h}:D^*\times I\to G_\C$, the corresponding B-twisted theories will be related by a field redefinition, up to $Q_B$-exact terms. For example, in $\CT_{\rm hyper}^B$, the twisted actions $\int \bm\Phi\cdot \diff_\CA\bm\Phi$ and $\int \bm\Phi\cdot \diff_{\CA^ {h}}\bm\Phi$ in the neighborhood of a line are related by the redefinition $\bm \Phi \mapsto  {h}\bm\Phi$. The field redefinition should induce an isomorphism of categories
\be \hspace{.5in} \CC_{\CA} \overset{\sim}\longrightarrow \CC_{\CA^ {h}} \qquad \forall\;  {h}:D^*\times I\to G_\C\,. \label{C-iso} \ee

We may use these isomorphisms to simplify the infinite-dimensional family of categories $\{\CC_\CA\}$. However, we must be careful not to simplify too much, or we will lose important information about morphisms, coming from junctions of lines.

\begin{figure}[htb]
\centering
\includegraphics[width=1.8in]{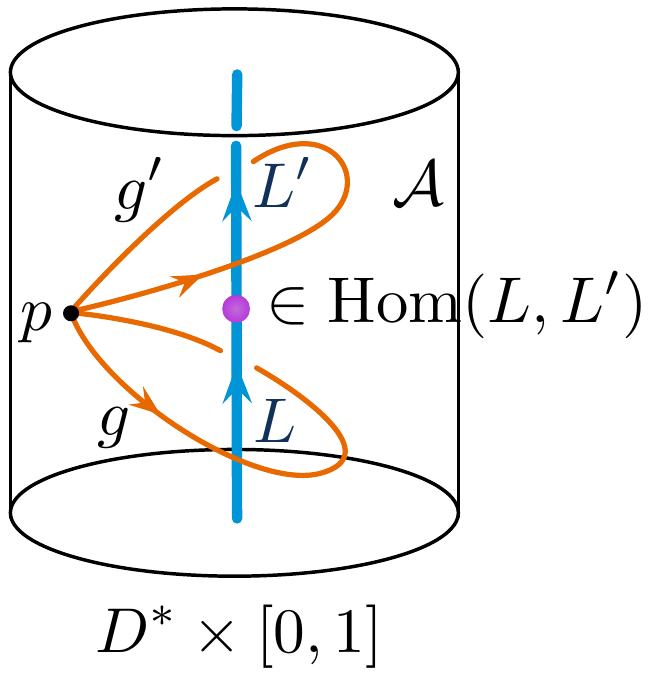}
\caption{Basepointed holonomies above and below a junction must agree.}
\label{fig:A-junction}
\end{figure}

We propose to fix a basepoint $p$ on $D^*\times I$ in the neighborhood of a line operator and/or a junction of line operators, and to explicitly quotient by isomorphisms \eqref{C-iso} corresponding to gauge transformations on $D^*\times I$ that are trivial at $p$. This leads to equivalence classes of categories $\CC_g$ labelled by the basepointed holonomy $g\in G_\C$ measured from $p$. We arrive at a finite-dimensional family of categories. In the neighborhood of a junction, the basepointed holonomy of a flat connection must stay unchanged, as illustrated in Figure \ref{fig:A-junction}; thus we expect 
\be \text{Hom}^\bullet_\CC(L,L')=0\qquad  \text{if $L\in\CC_g$ and $L'\in \CC_{g'}$ with $g\neq g'$}\,. \ee
Correspondingly, we may assemble the category $\CC$ of all possible line operators as a sum of ``blocks''
\be \label{Cg-family} \CC = \bigoplus_{g\in G_\C} \CC_g\,, \ee
with no morphisms among different blocks.

Note that \eqref{C-iso} does imply that categories $\CC_g$, $\CC_{g'}$ with $g,g'$ conjugate in $G_\C$ \emph{are} isomorphic as well. However, had we quotiented/simplified all the way down to conjugacy classes in $G_\C$, and merely labelled categories by conjugacy classes (forgetting information about isomorphisms), we would have lost control over the computation of morphisms at junctions.

\subsubsection{A coherent sheaf of categories}
\label{sec:sheaf}

We may further think of the full category $\CC$ of line operators \eqref{Cg-family} as a sheaf of categories
\be \CC\longrightarrow G_\C \label{C-sheaf} \ee
over a base $G_\C$, with the stalk over each $g$ given by $\CC_g$. 
We expect this to have the structure of a \emph{coherent} sheaf of categories.
The concept of a coherent sheaf of categories was discussed physically by \cite{KRS} in a closely related context; a modern mathematical treatment appears in~\cite{Gaitsgory-sheaves}.

To explain this more concretely, suppose we choose a flat $G_\C$ connection $\CA$ on a small punctured disc $D^*$ in a transverse slice to a putative line operator, representing the element $g\in G_\C$. Using a gauge transformation, we may fix the connection to have the form
\be \CA = \frac{1}{2\pi} a\, \diff \theta\,,  \label{CAa} \ee
where $\theta$ is the angular direction in $D^*$ and $a\in \mathfrak g_\C$ is a constant, chosen so that $e^a=g$. 

Now let us write spacetime close to a line operator as
\be D^*\times I \simeq S^1\times \R_+\times I\,, \ee
where $\R_+$ is the radial direction in the punctured disc $D^*$.
The 3d B-twisted theory on $D^*\times I$ may be rewritten (somewhat abstractly) as a 2d B-model $\CT_{2d}^B$ on a half-space $\R_+\times I$, whose target is the loop space of the 3d target. 
 Such rewritings/reductions were considered in \cite{KSV, BDGH}, and we shall see an explicit example momentarily in the case of $\CT_{\rm hyper}^B$.
The category $\CC_g$ of line operators in the 3d theory is then identified with the category of boundary conditions for $\CT_{2d}^B$.

It was found in \cite{BDGH} that the $\theta$-component of the complexified connection $\CA$ (along the reduction/compactification direction) becomes part of a 2d $\CN=(2,2)$ background \emph{chiral} multiplet in the rewriting. In an effective 2d action, it appears in a superpotential, coupled to a complexified, chiral moment-map operator~$\mu$. 
We thus find that $\CT_{2d}^B$ is a 2d B-model defined over a chiral parameter space $G_\C$, with local coordinate $a$.  Correspondingly, we call such flavor symmetries ``B-type" flavor symmetries, in analogy with chiral deformations of branes in 2d B-model.
 
It was explained in \cite{KRS} (with similar arguments in \cite{KSV}) that the category of boundary conditions in a family of 2d B-models over a chiral parameter space defines a coherent sheaf of (dg) categories. In particular:
\begin{itemize}
\item The dependence of $\CC_g$ on $g=e^a$ is holomorphic. This is because varying the effective 2d $\CN=(2,2)$ theory with respect to $\bar a$ amounts to an insertion of the anti-chiral operator $\bar\mu$ (or more accurately, by a higher component in the $\bar \mu$ multiplet that includes a current), and this insertion is $Q_B$-exact.
\item There is an action on $\CC$ of the (dg) category $\text{Coh}(G_\C)$ of coherent sheaves on $G_\C$, thought of as a monoidal category under tensor product of sheaves.%
\footnote{We are not being careful here about coherent vs. quasi-coherent sheaves, and refer the reader to \cite{Gaitsgory-sheaves} for mathematically precise statements.} %
In particular, for any $\CE\in \text{Coh}(G_\C)$, the action on $L\in \CC_g$ is given by
\be \CE * L := \CE_g \otimes L \ee
where $\CE_g$ denotes the derived stalk of $\CE$ at $g\in G_\C$.

This generalizes (and categorifies) the idea that a coherent sheaf of vector spaces on a variety $X$ is a module for algebraic functions on $X$.

\end{itemize}
In addition, we see that the holomorphic dependence of $\CC_g$ on $g=e^a$ is controlled by integrated insertions of (a component of) the chiral operator $\mu$. Schematically,
\be \pd_a = \oint \mu\,. \ee

\subsubsection{Fusion and braiding with flat connections}
\label{sec:braiding}

The category of line operators $\CC$ in a 3d TQFT should also be equipped with a tensor product
\be \otimes:  \begin{array}{ccc}\CC\boxtimes \CC &\to& \CC \\[.1cm]
 (L,L') &\mapsto & L\otimes L' \end{array} \ee
and with braiding isomorphisms $c_{L,L'}:L\otimes L'\to L'\otimes L$. These operations are defined locally, by collision of parallel line operators (in a chosen 2d plane) and by crossing of line operators (in the neighborhood of a 2d plane):
\be \raisebox{-.6in}{\includegraphics[width=4.7in]{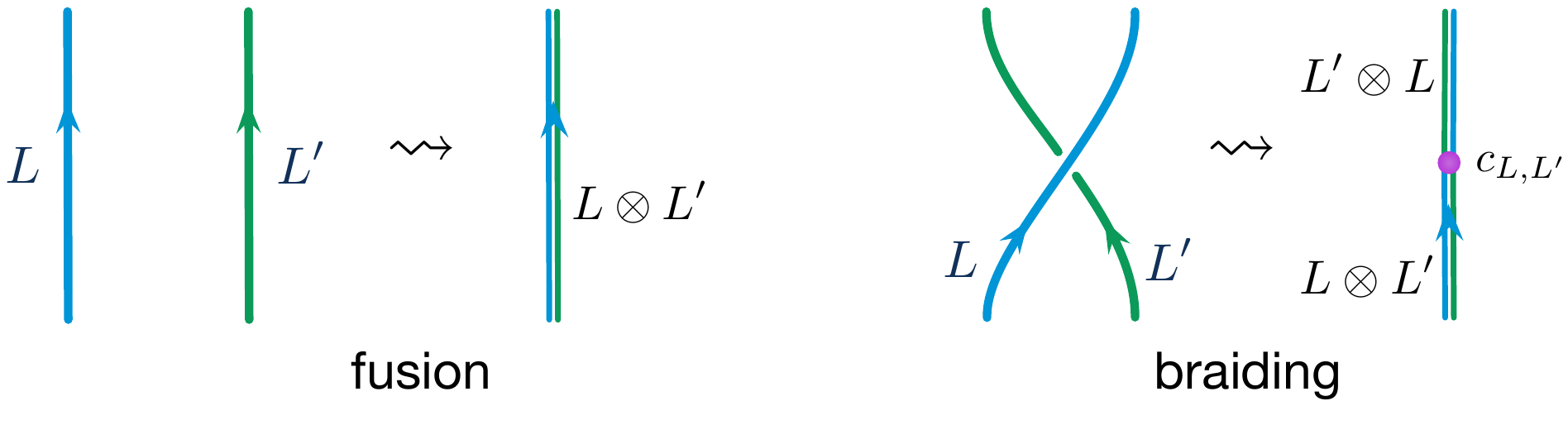}} \ee

When there is flavor symmetry, so that $\CC$ becomes a sheaf of categories over $G_\C$, fusion and braiding may involve line operators in multiple stalks $\CC_g$. This is easy to analyze once we choose a basepoint $p$ from which to measure the holonomy of a background connection $\CA$ around lines.

\begin{figure}[htb]
\centering
\includegraphics[width=4.4in]{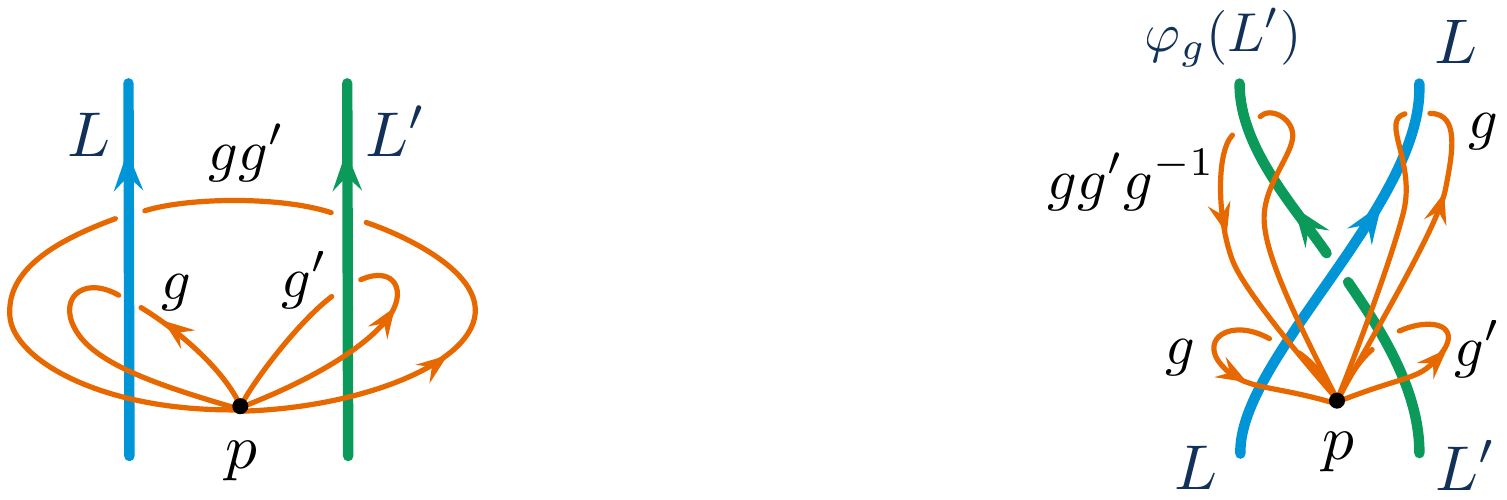}
\caption{Left: during fusion, the basepointed holonomies of a flat connection are multiplied. Right: at a crossing, the holonomy around the under-strand is conjugated.}
\label{fig:hol-fusion}
\end{figure}

Suppose we have two parallel lines $L\in \CC_g$ and $L'\in \CC_{g'}$. Fusion produces a new line operator around which the basepointed holonomy is $gg'$, as in Figure \ref{fig:hol-fusion}. Thus
\be \otimes: \CC_g\boxtimes \CC_{g'} \to \CC_{gg'}\,. \ee
The braiding morphism corresponding to a right-handed (say) crossing of $L$ and $L'$ can no longer be an element of $\text{Hom}^\bullet(L\otimes L',L'\otimes L)$, because $L\otimes L'\in \CC_{gg'}$ and $L'\otimes L\in \CC_{g'g}$ are objects of different stalks (if $g,g'$ do not commute). Instead, keeping track of holonomies as in Figure \ref{fig:hol-fusion}, we find
\be c_{L,L'} : L\otimes L' \to \varphi_g(L')\otimes L \ee
(in other words, $c_{L,L'}\in \text{Hom}^\bullet( L\otimes L' ,\varphi_g(L')\otimes L)$), where $\varphi_g :\CC_{g'} \overset\sim\to \CC_{gg'g^{-1}}$ is the isomorphism of stalks \eqref{C-iso} induced by a flavor gauge transformation. Now $L\otimes L'$ and $\varphi_g(L')\otimes L$ are both objects in $\CC_{gg'}$.

This type of structure is called a ``holonomy braiding''  or a ``$G_\C$-crossed braiding,'' \emph{cf.} \cite[Sec. 8.24]{tensorcategories}. It was used in \cite{Turaev-homotopy, KashaevReshetikhin, biquandles} to formulate invariants of 3-manifolds with background flat connections. The similarity in structure between topological twists of 3d $\CN=4$ theories with flavor symmetry and the work  of  \cite{biquandles} was a large hint that the two should be related.

\subsection{Line operators for $\CT_{\rm hyper}^B$}
\label{sec:hyper-lines}

We now return to our example of the B-twist of a free hypermultiplet, and derive a concrete algebraic description of  its category of line operators $\CC$.  The complexified flavor symmetry is $G_\C=SL(2,\C)_m$, and we thus expect stalks of  $\CC$ to be labelled by basepointed holonomies $g\in  SL(2,\C)_m$.
We will identify the various $\CC_g$ by implementing the reduction to a 2d B-model discussed abstractly in Section \ref{sec:sheaf}. 

At the end of the day, we will rederive the result of \cite{KRS} that at trivial holonomy the category
\be \CC_{g=1} \simeq \text{Coh}(\C^2) \ee
is the derived category of coherent sheaves on $\C^2$, \emph{i.e.} the target space of $\CT_{\rm hyper}$, with coordinates $X,Y$. Equivalently, and perhaps more simply,
\be \CC_{g=1} = \C[X,Y]\text{-mod} \label{XYmod} \ee
is the dg category of modules for a polynomial algebra in the two hypermultiplet scalars $X,Y$, both of  cohomological degree $+1$ due to their $U(1)_H$ charges. This is a non-semisimple category whose basic objects were discussed in \cite{Mikhaylov} from the perspective of $\mathfrak{psl}(1|1)$ Chern-Simons theory.

At generic holonomy, we instead find the trivial (dg) category
\be \CC_{g} \simeq \text{Vect} \qquad \text{$g$ generic} \label{CgVect} \ee
of graded vector spaces. This is a semisimple category, generated by one simple object, the one-dimensional vector space $\C$. This coincides with predictions of \cite{Mikhaylov} and \cite{RW-Coulomb}. Physically, \eqref{CgVect} means that in the presence of generic holonomy defect, there is a unique line operator compatible with the B-twist, up to tensoring (trivially) with auxiliary 1d quantum-mechanics theories, \emph{a.k.a.} adding Chan-Paton bundles.

The generic and non-generic answers are unified in a sheaf of categories depending on a parameter $g$. When $g = \bp e^\alpha & 0 \\ 0 & e^{-\alpha} \ep$ is diagonal and $\alpha$  is small, we find that the category of line operators takes the form of a matrix factorization category
\be \label{MF-alpha} \CC_g \simeq \text{MF}(\C^2,W = \alpha XY)\,,\qquad  \text{for }\;g = \bp e^\alpha & 0 \\ 0 & e^{-\alpha} \ep \approx 1\,. \ee
This indeed reduces  to \eqref{XYmod} when $\alpha=0$, and to the semisimple category \eqref{CgVect} when $\alpha\neq 0$.

\subsubsection{Loop space and matrix factorizations}
\label{sec:hyper-loop}

We analyze the category of line operators by reducing to a 2d B-model.
Consider Euclidean spacetime of the form $D^*\times \R_t$, with polar coordinates $r,\theta$ on the punctured  disc $D^*$. This is the local neighborhood of a line operator. Let $\CA$ be a flat $SL(2,\C)_m$ connection on  $D^*\times \R_t$. The twisted action \eqref{SA} in this background takes the form
\be \label{SAR2} S = \frac12\int_{D^*\times \R_t} \epsilon_{ij}\mb \Phi^i\big((\diff+\CA)\mb \Phi\big)^j\,, \qquad  \mb \Phi := \bp \mb X \\ \mb Y \ep\,.\ee

Next, we deform the metric on $D^*$ to a cylinder $S^1_\theta\times \R_{r>0}$. Notably, the twisted action \eqref{SAR2} does not depend explicitly on the metric, and will be unchanged under this deformation.
We then reinterpret the 3d theory on $(S^1_\theta\times \R_{r>0})\times \R_t$ as a 2d theory on the half-space $\R_{r>0}\times \R_t$, whose fields are configurations of $\mb \Phi$ on $S^1_\theta$ (\ie\ the loop space of the 3d hypermultiplet target space).

We expect from \cite{KRS} to find a 2d B-model. To derive it concretely, let us split the shifted de Rham complex of $M=\R_{r>0}\times S^1_{\theta}\times \R_t$ as
\be \Omega^\bullet(M)[1] \simeq C^\infty(M)[\diff r,\diff t][1] \oplus C^\infty(M)[\diff r,\diff t]\diff \theta\,. \ee
(The first summand has no $\diff \theta$'s, and the second has exactly one $\diff \theta$.) We correspondingly decompose the fields as
\be \label{S1forms} \bm \Phi^i = \hat{\bm \Phi}^i + \epsilon^{ij}\hat{\bm\Psi}_j \diff \theta\,,\ee
with $\hat {\bm \Phi}^1 = X + \chi^X_r\diff r + \chi^X_t \diff t+...\in C^\infty(M)[\diff r,\diff t][1]$, $\bm \Psi_1  = -\chi^Y_\theta +... \in C^\infty(M)[\diff r,\diff t]$, etc. Then the twisted action \eqref{SAR2} becomes
\be \label{SBmodel} S =  \int_{ \R_{r> 0}\times \R_t } \Big[ \int_{S^1_\theta}\diff \theta\, \hat{\bm\Psi}_i \diff_\CA'\hat{\bm \Phi}^i +\frac12\int_{S^1_\theta}\diff \theta\, \epsilon_{ij} \hat{\mb \Phi}^i \big((\pd_\theta+\CA_\theta)\hat{\mb\Phi}\big)^j \Big] \ee
with $\diff'_\CA =(\pd_r+\CA_r)\diff r+(\pd_t+\CA_t) \diff t$ the two-dimensional covariant exterior derivative.

The action \eqref{SBmodel} is a simplified BV action for a 2d B-model on $\R_{r>0}\times \R_t$
with target space $L(\C^2)=\text{Maps}(S^1_\theta,\C^2)$ (\cf\ \cite{Li-B}). The first term $\int_{S^1_\theta}\diff \theta\, \bm\Psi_i \diff_\CA'\hat{\bm \Phi}^i$ is a standard 2d B-model kinetic term (for a loop-space target), while the second term $W_\CA(\hat{\bm \Phi}) := \frac12\int_{S^1_\theta}\diff \theta\, \epsilon_{ij} \hat{\mb \Phi}^i \big((\pd_\theta+\CA_\theta)\hat{\mb\Phi}\big)^j$ is a superpotential on the loop space.

Now, our category of line operators $\CC_\CA$ with background $\CA$ should be equivalent to the category of boundary conditions for the 2d B-model \eqref{SBmodel}. By classic results of Kontsevich and Kapustin-Li \cite{KapustinLi}, we expect the latter to be given by the derived category of matrix factorizations of $W_\CA$,
\be \CC_\CA = \text{MF}(L(\C^2),W_\CA)\,,\qquad W_\CA(\phi) = \frac12\int_{S^1_\theta}\epsilon_{ij} \phi^i(\diff+\CA)\phi^j\,, \label{CAW} \ee
where $\phi^i(\theta)$ are coordinates on the loop space $L^2(\C^2)$.

We expect the objects of $\CC_\CA$ to be pairs $(\CE,d_\CE)$, where $\CE$ is a complex of coherent sheaves on the loop space $L^2(\C)$ that is
\begin{itemize}
\item equivariant with respect to the complexifed $\C^*_H$ R-symmetry acting on $(\phi^1,\phi^2)$ with weights $(1,1)$, and
\item  equivariant with respect to the $SL(2,\C)_m$ flavor symmetry, or a subgroup thereof (\eg\ the $\C_m^*$ subgroup acting on $(\phi^1,\phi^2)$ with weights $(1,-1)$), if we want line operators to preserve some of the flavor symmetry.
\end{itemize}
The cohomological grading is a \emph{sum} of the standard cohomological grading on a complex $\CE$ and $\C^*_H$ weight. 
The operator $d_\CE:\CE\to \CE$ is a chain map of total cohomological degree $1$, satisfying $(d_\CE)^2 = W_\CA(\phi)$, which is compatible with $W_\CA$ having degree $2$. We thus expect to obtain a derived category of matrix factorizations with a $\Z$-valued cohomological grading, which lifts the usual $\Z_2$ grading on matrix factorizations \cite{Orlov-MF}.

Making precise sense of such matrix factorizations in an infinite-dimensional setting takes a bit of work. A mathematical definition of \eqref{CAW} and related categories appears in \cite{HilburnRaskin, BN-betagamma}, based in part on the mathematical methods of \cite{Raskin}. For our current purposes, we will content ourselves with a heuristic analysis of \eqref{CAW}.

To simplify \eqref{CAW} further, let us fix the holonomy $g$ of the connection $\CA$, choose a constant $a\in \mathfrak{sl}_2$ such that $g=e^a$, and gauge-fix the background connection (using isomorphisms \eqref{C-iso}) to have the form
\be \CA = \frac{1}{2\pi} a \,\diff\theta\,. \ee
Let
\be \phi^j(\theta)  =  \sum_{n\in \Z} \phi^j_n e^{in\theta}\qquad (j=1,2) \ee
denote the complex-scalar bosonic coordinates on loop space $L(\C^2)$. Then the superpotential becomes
\be W_a(\phi) = \frac12\sum_{n\in \Z} \epsilon_{jk}  \phi^j_{-n}(2\pi i n +a)\phi^k_n = W_a^{(0)}(\phi_0) + \sum_{n=1}^\infty W_a^{(n)}(\phi_n,\phi_{-n})\,, \ee
\be \label{Wn} W_a^{(0)}(\phi_0) :=  \frac12 a_{jk}\phi_0^j\phi_0^k\,,\qquad W_a^{(n)}(\phi_n,\phi_{-n}) := 2\pi i n \epsilon_{jk} \phi^j_{-n}\phi^k_{n} + a_{jk}\phi_{-n}^j\phi_n^k\,, \ee
where $a_{jk}:=\epsilon_{j\ell}a^{\ell}{}_k$ is symmetric.
The superpotential is thus a sum of terms depending on either the two zero-modes $\phi_0^j$ or the pairs of four modes $\phi_{\pm n}^j$. We have
\be \label{CAW2} \hspace{.5in} \CC_{g} = \text{MF}\Big(\C^2\times \prod_{n=1}^\infty  \C^4, W_a^{(0)} + \sum_{n=1}^\infty W_a^{(n)}\Big) \qquad (\text{with }g=e^a)\,. \ee

\subsubsection{Finite-dimensional model}
\label{sec:hyper-Cfd}

The category \eqref{CAW2} can be greatly simplified to a finite-dimensional model, though the way it simplifies depends on the choice of parameter $g=e^a$.

Physically, we expect to be able to integrate out any sets of modes for which the quadratic forms appearing in the superpotentials $W_a^{(0)}$ or $W_a^{(n)}$ above are non-degenerate. Mathematically, the equivalence of matrix factorization categories induced by integrating out fields in this manner is known as Kn\"orrer periodicity \cite{Knorrer}. 

A brief inspection of \eqref{Wn} suggests that \emph{every} set of modes can be integrated out (and set to the critical value $\phi\equiv 0$) as long as $a$ is sufficiently generic. Thus, generically, we expect $\CA_g$ to be equivalent to the category of boundary conditions in a trivial B-model, whose target is the point $\phi=0$. This is the trivial (dg) category of $\Z$-graded vector spaces,
\be \CC_g \simeq \text{Vect}\qquad (\text{$g$ generic})\,. \ee

To find more interesting behavior, let's look at a small neighborhood of trivial holonomy $g=1$, or $a=0$. As long as $|a|\ll 2\pi$, the $n$-dependent term of $W_a^{(n)}$ will dominate for all $n\geq 1$, allowing us to integrate out all nonzero modes. Then we are left with matrix factorizations on the two-dimensional space with coordinates $\phi_0^1,\phi_0^2$\,,
\be \CC_g = \text{MF}(\C^2, \tfrac12a_{ij}\phi^i_0\phi^j_0)\,. \ee
If the symmetric matrix $a_{ij} =\epsilon_{ik}a^k{}_j$ is non-degenerate, this again reduces to the trivial category of  vector spaces. If the matrix $a_{ij}$ has rank one, then one linear combination of $\phi_0^1,\phi_0^2$ can be integrated out, leaving a category of coherent sheaves on $\C$ (parameterized by the independent linear combination). If $a=0$, then we get $\text{MF}(\C^2,0) =\text{Coh}(\C^2)$:
\be \label{Cg-gen} \CC_g \simeq \begin{cases} \text{Vect} & \text{rank}(a) = 2 \\
 \text{Coh}(\C) & \text{rank}(a)=1 \\
 \text{Coh}(\C^2)  & \text{rank}(a)=0 \end{cases} \qquad \text{for $g=e^a\approx 1$}\,.\ee
Note that these are all dg categories with a $\Z$-valued cohomological grading, compatible with the $\C^*_H$ R-symmetry with weight $1$ on $\C$, or weights $(1,1)$ on $\C^2$.

In the subsequent discussion of state spaces, and the generalization to the 3d TQFT related to $U_q(\mathfrak{sl}_n)$, we will be particularly interested in flat connections with abelian (diagonal) holonomy. If we set $g=e^a$ with $a  = \text{diag}(\alpha,-\alpha)$, then a pair of modes $(\phi^1_{-n},\phi^2_n)$ in \eqref{Wn} becomes massless (the quadratic form on these modes vanishes) precisely when $\alpha = 2\pi i n$.
Thus,
\be \label{Ca-gen} \text{for }g = \bp e^{\alpha} & 0 \\ 0 & e^{-\alpha} \ep\,,\qquad  \CC_g \simeq \begin{cases} \text{Vect} & \alpha \in \C\backslash 2\pi i \Z \\
	\text{Coh}(\C^2) & \alpha \in 2\pi i \Z\,. \end{cases} \ee
In particular, for small $\alpha$, we may integrate out all nonzero modes and recover the description
\be \CC_g \simeq \text{MF}(\C^2,\alpha XY) \ee
from \eqref{MF-alpha}, where $X = \phi^1_0$ and $Y=\phi^2_0$.

\subsubsection{Objects in $\CC_{g=1}$ and representations of $\mathfrak{psl}(1|1)$}
\label{sec:C1-hyper}

To provide some additional intuition into the structure of the non-semisimple category of line operators at $g=1$, we describe some of its basic objects, and relate them to the $\mathfrak{psl}(1|1)$ representations studied by \cite{Mikhaylov}.

\begin{figure}[htb]
\centering
\includegraphics[width=4.5in]{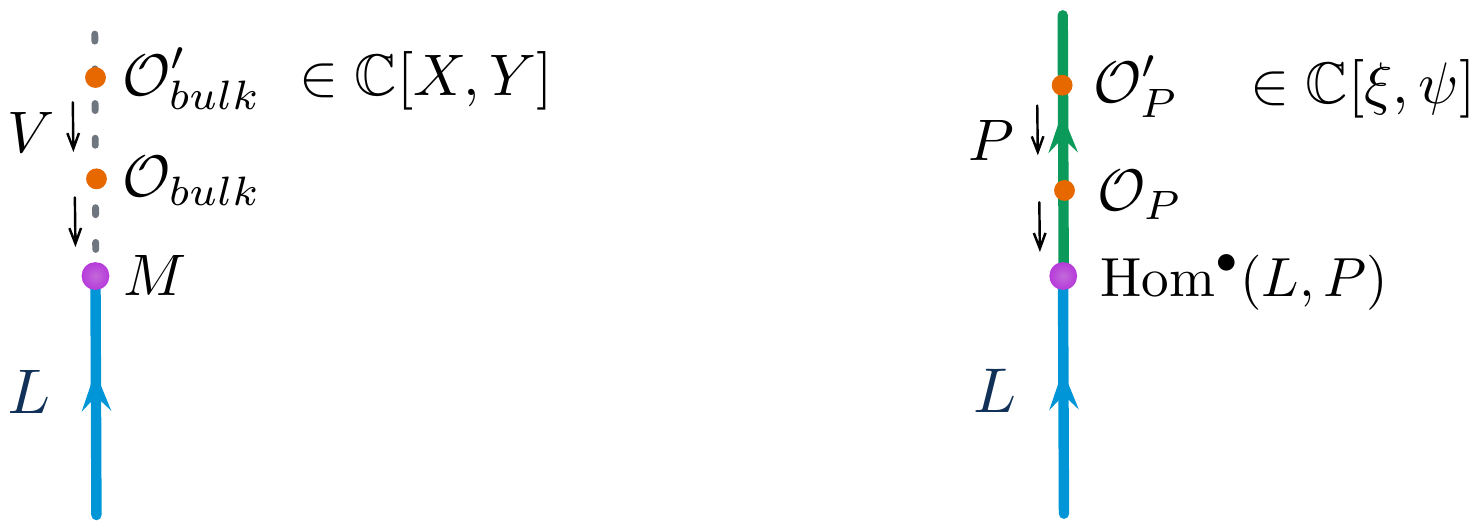}
\caption{Left: any line operator $L$ defines a module $M$ for the algebra of bulk local operators $\C[X,Y]$. (Algebraically, $M = \text{Hom}^\bullet(L,V)$, where $V$ represents the trivial line operator.) Right: any line operator $L$ also defines a module $\text{Hom}^\bullet(L,P)$ for the algebra $\C[\xi,\psi]$ of local operators bound to the line operator $P$. In each case, the action of the algebra on the module comes from collision/OPE.}
\label{fig:XYmod}
\end{figure}

The category at $\CC_{g=1}$ is the derived category of graded modules for the polynomial algebra $\C[X,Y]$ of hypermultiplet scalars.%
\footnote{We note again that we are not being careful about the distinction between coherent and quasi-coherent sheaves. There are some choices to be made about this, both physically and mathematically; a more precise discussion appears in \cite{BN-betagamma}. The category that matches $\C[X,Y]$-mod is actually $\text{QCoh}(\C^2)$.} %
Recall that $\C[X,Y]$ is the algebra of bulk local operators in $\CT_{\rm hyper}^B$.
Physically, the module for $\C[X,Y]$ associated to a particular line operator $L$ is the space of local operators at an endpoint of $L$, as on the left of Figure \ref{fig:XYmod}.
We restrict ourselves to line operators (and thus modules) that preserve the complexified $\C^*_H$ R-symmetry (for which $X,Y$ have charges $1,1$) and a $\C^*_m$ flavor symmetry (for which $X,Y$ have charges $1,-1$).
This amounts to considering only graded modules for $\C[X,Y]$.

Consider the following four line operators:
\begin{itemize}
\item $V=\C[X,Y]$. This is the trivial or ``identity'' line operator. The associated module $\C[X,Y]$ is local operators at an endpoint of $V$, \emph{a.k.a.} bulk local operators.
\item $W_+=\C[X,Y]/(Y)\simeq \C[X]$. This is a line operator on which the hypermultiplet $Y$ is set to zero. The associated module only contains polynomials in $X$.
\item $W_-=\C[X,Y]/(X)\simeq \C[Y]$. This is a line operator on which the hypermultiplet $X$ is set to zero. The associated module only contains polynomials in $Y$.
\item $P = \C[X,Y]/(X,Y) \simeq \C$. This is a line operator on which both $X$ and $Y$ are set to zero. The only local operator at its endpoint is the identity.
\end{itemize}
Any of these line operators can generate the full derived category of graded $\C[X,Y]$-modules.
Physically, this means that we can obtain any other line operator by coupling one of these with some 1d quantum-mechanics system along the line.

All these line operators have nontrivial algebras of local operators bound to them, \emph{a.k.a.} derived endomorphism algebras $\text{End}^\bullet(L)$. Standard computations give:
\begin{itemize}
\item $\text{End}^\bullet(V) = \C[X,Y]$, \emph{i.e.} local operators bound to the identity line are just bulk local operators. Both $X$ and $Y$ are in cohomological degree 1.
\item $\text{End}^\bullet(P) = \C[\xi,\psi]$ is an exterior algebra in two fermionic variables of cohomological  degree zero. 
 In the twisted formalism \eqref{SXY}, these operators are integrals of the 1-form fermions around an infinitesimally small circle linking the line $P$,
\be \xi = \oint_{S^1_\theta} \chi^X\,,\qquad \psi = \oint_{S^1_\theta}\chi^Y\,. \ee
(They are descendants of the bulk $X,Y$, as discussed in \cite{descent}, and are $Q$-closed by Stokes' Theorem, $Q\xi = \oint_{S^1_\theta} Q\chi^X = \oint_{S^1_\theta} \diff X$, etc.)
\item $\text{End}^\bullet(W_+) = \C[X,\psi]$, containing the bulk scalar $X$ and one of the fermions above.
\item $\text{End}^\bullet(W_-) = \C[\xi,Y]$, containing the bulk scalar $Y$ and the other fermion.
\end{itemize}

The free exterior algebra $\C[\xi,\psi]$ appearing in derived endomorphisms of $P$ may be thought of as the enveloping algebra of $\mathfrak{psl}(1|1)$. We recall that $\mathfrak{psl}(1|1)$ is the abelian Lie algebra with two odd generators $\xi,\psi$ and trivial Lie brackets $\{\xi,\xi\}=\{\xi,\psi\}=\{\psi,\psi\}=0$.
By mapping every object $L$ of $\CC_1$ to the space $\text{Hom}^\bullet(L,P)$ of its derived morphisms with $P$ --- or physically, by associating to every line operator $L$ the space of local operators at a junction of $L$ and $P$ as on the right of Figure \ref{fig:XYmod} --- we obtain a functor
\be  \CF:\begin{array}{ccc} \CC_1\simeq \C[X,Y]\text{-mod} & \overset\sim\longrightarrow & \mathfrak{psl}(1|1)\text{-mod} \\
  L & \mapsto & \text{Hom}^\bullet(L,P) \end{array} \label{C1psl} \ee
This functor induces a derived equivalence of $\Z\times \Z$ graded categories. The equivalence is one of the classic examples of Koszul duality \cite{Priddy, BGS} (see \cite{TD-Koszul, CostelloPaquette, PaquetteWilliams} for further physical context).

The images of our four basic line operators under the functor \eqref{C1psl} are
\begin{itemize}
\item $\CF(V) = \C[\xi,\psi]/(\xi,\psi) = \C$ the trivial module of $\mathfrak{psl}(1|1)$, which is the unique simple object in the abelian category of $\mathfrak{psl}(1|1)$ modules.
\item $\CF(W_+) = \C[\xi,\psi]/(\xi)$ 
a two-dimensional module of $\mathfrak{psl}(1|1)$ on which $\xi$ acts as zero.
\item $\CF(W_-) = \C[\xi,\psi]/(\psi)$ 
 a two-dimensional module of $\mathfrak{psl}(1|1)$ on which $\psi$ acts as zero.
\item $\CF(P) = \C[\xi,\psi] = \C\langle 1,\xi,\psi,\xi\psi\rangle$ a four-dimensional module of  $\mathfrak{psl}(1|1)$, which is a projective object in the abelian category of $\mathfrak{psl}(1|1)$ modules, the projective cover of $\CF(V)$.
\end{itemize}
These are all the indecomposable, cyclic, graded modules of $\mathfrak{psl}(1|1)$, and the basic modules studied by \cite{Mikhaylov}. (Other indecomposable noncyclic modules exist, \emph{cf.} \cite{GQS} and references therein, and the derived category additionally contains many nontrivial complexes of modules, which we do not discuss here.)

Braiding and fusion of line operators is quite nontrivial in the category $\CC_1$, despite $\CT_{\rm hyper}^B$ being a free theory. One way to access braiding and fusion is via boundary chiral algebras, which we come to next. For further discussion of these operations in $\CT_{\rm hyper}^B$, see \cite{Mikhaylov} and the upcoming \cite{BN-betagamma}.

\subsection{Boundary VOA}
\label{sec:bdy-VOA}

A new family of boundary conditions for topologically twisted 3d $\CN=4$ theories was introduced in \cite{CostelloGaiotto}. These boundary conditions behave \emph{holomorphically} rather than topologically. In particular, they depend on a choice of complex structure along the boundary, and their spaces of boundary local operators have the structures of vertex operator algebras (VOA's).

The boundary VOA's of \cite{CostelloGaiotto} are closely related to the 4d $\CN=2$ VOA's of \cite{VOA-S} and \cite{OhYagi} (by circle compactification); as well as to the 3d $\CN=4$ corner VOA's of \cite{GaiottoRapcak, CreutzigGaiotto-S} (by interval compactification). They are also directly analogous of the WZW VOA's that appear on holomorphic boundary conditions in Chern-Simons theory \cite{Witten-Jones}.

We will use boundary VOA's to obtain an alternative perspective on categories of line operators and state spaces. We focus on line operators in this section. We begin in Section \ref{sec:VOA-lines} by recalling how bulk line operators are related to modules for a boundary VOA. In Section \ref{sec:hyper-VOA} we explain how boundary VOA's may be constructed in the twisted BV formalism. We introduce the useful technique of first taking a holomorphic twist of both bulk and boundary theories and then deforming the holomorphic twist to a topological twist. We will apply this technique to $\CT_{G,k}^A$ theories in Section \ref{sec:Atwist}; here we illustrate it for our toy model $\CT_{\rm hyper}^B$. Then in Section \ref{sec:VOA-flavor} we consider the effects of deforming by background flavor flat connections. (We will revisit flavor deformations of VOA's from several other perspectives in Section \ref{sec:VOA}.)

\subsubsection{VOA modules and line operators}
\label{sec:VOA-lines}

\begin{figure}[htb]
\centering
\includegraphics[width=3.5in]{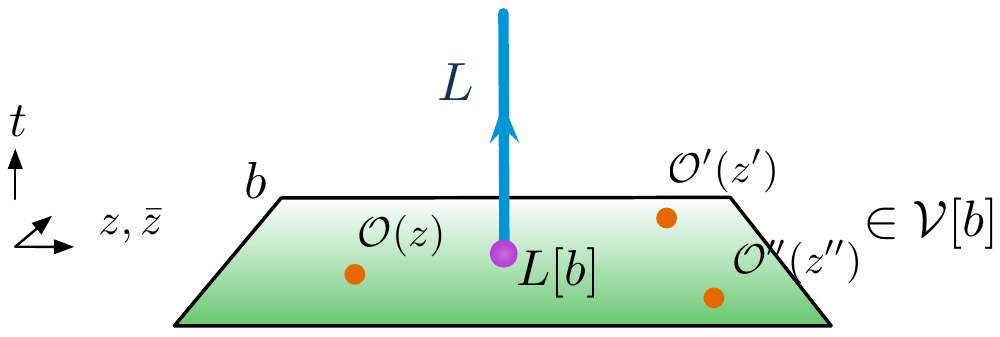}
\caption{Given a holomorphic boundary condition $b$, each line operator $L$ defines a module $L[b]=\CF_b(L)$ for the boundary vertex algebra $\CV[b]$.}
\label{fig:VOA}
\end{figure}

In general, given a topological 3d QFT $\CT$, with a boundary condition $b$ supporting a VOA $\CV[b]$, one expects to have a functor of braided tensor categories
\be \begin{array}{cccc}\CF_b : &\CC[\CT] &\to& \CV[b]\text{-mod} \\
  & L & \mapsto & L[b]\,, \end{array} \label{F-VOA} \ee
relating the category of bulk line operators $\CC[\CT]$ to the category of modules for the boundary VOA. The logic behind \eqref{F-VOA} is illustrated in Figure \ref{fig:VOA}: for any bulk line operator $L$, one can define a vector space $L[b]$ of local operators at the junction of $L$ and the boundary~$b$. The space $L[b]$ has an action of the $\CV[b]$ (by collision/OPE of boundary local operators), and thus defines a $\CV[b]$-module. More so, the map $L\to L[b]$ intertwines bulk morphisms, tensor products, and braiding, with the corresponding operations the the VOA module category. For example, a morphism of line operators $\mu\in\text{Hom}_{\CC[T]}(L,L')$, coming from a bulk junction, may collide with $L[b]$ to define a map $\mu:L[b]\to L'[b]$ that commutes with the action of the boundary VOA $\CV[b]$; thus ``bringing $\mu$ to the boundary'' defines a morphism in the category $\CV[b]$-mod.

The functor \eqref{F-VOA} not in general guaranteed to be an isomorphism. Indeed, it may not be possible for some lines $L$ to end on a given boundary $b$ at all, in which case $\CF_b(L)=0$. However, for a sufficiently rich choice of $b$, one may optimistically assume an equivalence, and then proceed to use VOA modules to study $\CC$.

In the case of Chern-Simons theory with compact gauge group $G$, all bulk line operators (Wilson lines) \emph{can} end on the WZW boundary condition, and it is well known that \eqref{F-VOA} is an isomorphism. More precisely, it is an isomorphism of semisimple, abelian categories.

In the case of topologically twisted 3d $\CN=4$ theories, it is also expected that the boundary conditions of \cite{CostelloGaiotto} are also sufficiently rich for \eqref{F-VOA} to be an equivalence. The relevant VOA's are non-rational, and some care must be taken in defining their module categories in a way that matches bulk physics. This is discussed in \cite{CostelloGaiotto} and especially \cite{CCG}. In particular:
\begin{itemize}
\item One must enlarge the VOA categories to include logarithmic modules (\emph{cf.} Section 9 of \cite{CCG}).

Note, however, that this enlargement is trivial for both the symplectic fermion VOA that's relevant for our $\CT_{\rm hyper}^B$ toy model (see further below) and for the triplet and more general Feigin-Tipunin algebras studied in the remainder of this paper.

\item One must pass to derived categories of the usual abelian VOA module categories, with a suitable dg enhancement. The functor \eqref{F-VOA} should be an equivalence of dg categories.
\end{itemize}

\subsubsection{Holomorphic twists and the toy model}
\label{sec:hyper-VOA}

Any 3d $\CN=2$ theory with $U(1)_R$ R-symmetry admits a holomorphic-topological (HT) twist. This twist, studied recently in \cite{ACMV, CDG}, is a reduction of Kapustin's holomorphic twist of 4d theories~\cite{Kapustin-hol}. 

Any 3d $\CN=4$ theory $\CT$ may be viewed as a 3d $\CN=2$ theory and thus HT-twisted to obtain $\CT^{HT}$. With appropriate choices of 3d $\CN=2$ subalgebra and R-symmetry, the A and B topological supercharges can both be expressed as deformations of a HT supercharge
\be \label{HT-A-B} \begin{array}{ccccc}\CT^A &\quad \;\reflectbox{$\leadsto$}\phantom{xx}& \CT^{HT} &\quad\leadsto\quad& \CT^B \\
 Q_A=Q_{HT}+Q_A' &&Q_{HT}&& Q_B=Q_{HT}+Q_B' 
\end{array}
 \ee
Thus, the B-twisted theory $\CT^B$ (resp. $\CT^A$) may be thought of as a further $Q_B'$ (resp. $Q_A'$) twist of $\CT^{HT}$.

In more concrete terms, given the 3d $\CN=4$ algebra $\{Q_\alpha^{a\dot a},Q_\beta^{b\dot b}\} = \epsilon^{ab}\epsilon^{\dot a\dot b}\sigma_{\alpha\beta}^\mu P_\mu$ in \eqref{eq:N=4}  and a 3d $\CN=2$ subalgebra generated by $(Q_\pm,\bar Q_\pm):=(Q^{-\dot -}_{\pm},Q^{+\dot+}_{\pm})$ satisfying the usual $\CN=2$ relations $\{Q_\alpha,\bar Q_\beta\} = \sigma^\mu_{\alpha\beta}P_\mu$, the HT and topological supercharges may be chosen as
\be \label{QHT-A} \begin{array}{lll}
Q_A &=\;  \delta^\alpha{}_a Q^{a\dot +}_\alpha & \,=\; Q^{+\dot +}_+ + Q^{-\dot+}_- \\
Q_{HT} & = \;\bar Q_+ & \,=\; Q^{+\dot +}_+ \\
Q_B &=\; \delta^\alpha{}_{\dot a} Q^{+\dot a}_\alpha &\,=\; Q^{+\dot +}_{+}+Q^{+\dot -}_-\,,
\end{array} \ee
with the deformations given by $Q_A' = Q^{-\dot +}_-$  and $Q_B' = Q^{+\dot -}_-$.

This provides us with a practical technique for identifying boundary VOA's, which works even for 3d $\CN=4$ theories (like our main Chern-Simons-matter example) that do not admit Lagrangians with manifest 3d $\CN=4$ supersymmetry. We illustrate the approach in our toy model.

Consider the theory $\CT_{\rm hyper}^B$, with its twisted BV action $S = \int_M \mb X \diff \mb Y$ as in \eqref{SXY},
on a spacetime of the form $M=\C_{z,\bar z}\times \R_{t\geq 0}$, which is the local neighborhood of a boundary. 
Let us decompose the shifted de Rham complex on $M$ as%
\footnote{This decomposition is similar but not identical to \eqref{S1forms}, in which we separated out a real (rather than holomorphic) direction.}
\be \Omega^\bullet(M)[1] \simeq C^\infty(M)[\diff \bar z,\diff t][1] \oplus C^\infty(M)[\diff \bar z,\diff t]\diff z\,. \ee
(The first summand has no $\diff z$'s, and the second has exactly one $\diff z$.) We correspondingly decompose the fields as
\be \mb X  = \hat {\mb X} + \hat{\bm \Psi}^Y \diff z\,,\qquad  \mb Y  = \hat {\mb Y} - \hat{\bm \Psi}^X\diff z \,, \ee
with $\hat {\mb X}\in C^\infty(M)[\diff \bar z,\diff t][1]$, $\bm \Psi^X \in C^\infty(M)[\diff \bar z,\diff t] \diff z$, etc. The bulk action becomes
\be S = \int_M \diff z\big[ \underbrace{ \hat{\mb \Psi}^X \diff' \hat{\mb X} + \hat{\mb \Psi}^Y \diff' \hat{\mb Y}}_{S_{HT,kin}} + \underbrace{\hat{\mb X}\pd_z\hat{\mb Y}}_{W_B}\big]\,, \label{S-hyper-HT} \ee
with $\diff' :=\pd_{\bar z}\diff \bar z + \pd_t \diff t$.

The first two terms contain the standard kinetic term of an HT-twisted action for two chiral multiplets (the 3d $\CN=2$ decomposition of a 3d $\CN=4$ hypermultiplet) \cite{CDG}. The third term $W_B := \hat{\mb X}\pd_z\hat{\mb Y}$ is interpreted as a superpotential in the HT action. It is responsible for the deformation from the HT twist to the B twist.

For example, the bulk local operators of two HT-twisted chiral multiplets (with action $S_{HT,kin}$) form a commutative vertex algebra $\CV_{\rm bulk}$, generated by the bottom components $X(z),Y(z),\chi_z^Y(z),-\chi_z^X(z)$ of $\hat{\mb X},\hat{\mb Y},\hat{\mb \Psi}^X,\hat{\mb \Psi}^Y$. The superpotential $W_B$ deforms the bulk algebra by adding a new differential
\be Q_B' \hat{\mb \Psi}^X= \pd_z\hat {\mb Y} \,,\quad  Q_B' \hat{\mb \Psi}^Y= -\pd_z\hat {\mb X}\ee
whose cohomology $H^\bullet(\CV_{\rm bulk},Q_B') \simeq \C[X,Y]$ contains just the zero-modes of $X,Y$ and thus reproduces the topological bulk algebra \eqref{hyper-ops}.

Now consider a ``Dirichlet'' boundary condition $\CD$ that sets 
\be \CD\,:\quad  \hat {\mb X}\big|_{t=0} = \hat {\mb Y}\big|_{t=0} = 0\,, \label{bdyXY} \ee
while leaving $\bm \Psi^X$ and $\bm\Psi^Y$ free at $t=0$. This is a standard boundary condition for the HT twist of two chiral multiplets, and supports a boundary vertex algebra \cite{CDG}
\be \CV[\CD] = \C\big\langle\!\!\big\langle \chi_z^X(z),\chi_z^Y(z)  \big\rangle\!\!\big\rangle\,, \ee
generated by the fermions $\chi_z^X(z)$ and $\chi_z^Y(z)$. In the absence of a superpotential, it is again a commutative vertex algebra.

The Dirichlet boundary condition is compatible with the superpotential $W_B$ because it forces $W_B\big|_{t=0} = 0$. Equivalently, the conditions \eqref{bdyXY} are preserved by $Q_B'$. According to \cite[Sec 5]{CDG}, the addition of $W_B$ then deforms the boundary algebra by introducing a boundary OPE of the form $\chi_z^X(z)\chi_z^Y(0) \sim \frac{1}{z} \pd_X\pd_Y W_B$, \emph{i.e.}
\be \chi_z^X(z) \chi_z^Y(0) \sim \frac{1}{z^2}\,. \label{XYOPE} \ee 
Thus the boundary vertex algebra becomes
\be \label{hyper-VOA}\CV_B[\CD] =SF= \C \big\langle\!\!\big\langle \chi^X_z(z),\chi^Y_z(z)\;\big|\; \chi_z^X(z) \chi_z^Y(0) \sim \tfrac{1}{z^2}  \big\rangle\!\!\big\rangle \,,\ee
otherwise known as the symplectic-fermion VOA.

It was discussed in \cite{CDG} that, on general grounds, a boundary vertex algebra may have a standard 2d stress tensor only if the bulk algebra is topological (\emph{i.e.} the bulk stress tensor is exact). It is interesting to note here that the deformation by $W_B$ implements both of these features simultaneously: it removes non-zero (non-topological) modes from the bulk algebra, and makes the boundary algebra non-commutative, with the standard symplectic-fermion stress tensor $T=:\chi_z^X\chi_z^Y:$.

The boundary condition above for  $\CT_{\rm hyper}$, supporting the symplectic-fermion VOA, was first introduced in  \cite{CostelloGaiotto,Gaiotto-blocks}. The module category $SF\text{-mod}$ was then identified with the category of bulk line operators $\CC_{g=1}=\C[\xi,\psi]\text{-mod}$ in \cite[Sec. 2.1]{CCG}. Key to the identification is the observation that modules for symplectic fermions are equivalent to modules for their zero modes $\chi^X_z(0),\chi^Y_z(0)$, which are precisely the two fermionic variables $\xi,\psi$.

\subsubsection{Flavor background and line operators revisited}
\label{sec:VOA-flavor}

If a 3d $\CN=4$ theory has a $G$ flavor symmetry that allows deformations by flat $G_\C$ connections in a topological twist, and if $b$ is a holomorphic boundary condition that preserves the bulk flavor symmetry, then one expects the boundary VOA $\CV[b]$ to admit deformations by $G_\C$ flat connections as well. Some basic aspects of these deformations were discussed in \cite{Gaiotto-blocks}.

In general, in the presence of bulk $G_\C$ symmetry, a boundary VOA $\CV[b]$ will have an action by a $G_\C$ Kac-Moody algebra. The action need not be internal (the Kac-Moody algebra need not be a subalgebra of $\CV[b]$). The Kac-Moody action can then be used to apply \emph{meromorphic} $G_\C$ gauge transformations to $\CV[b]$, deforming the VOA by flat connections in the neighborhood of fixed points on the boundary, where putative line operators may end. The deformation will modify the OPE.

The setup relevant for relating bulk lines and VOA modules involves flat connections on an infinitesimal punctured disc $D^*\times \R_+$. These can be gauge-transformed to the form $\CA = \CA_z \diff z$, where $\CA_z(z) \in \g(\!(z)\!)$ is a Laurent series. For $a\in \mathfrak g$, let $b_a$ denote the boundary condition deformed by the flat connection $\CA = \frac{a}{2\pi iz}$ near $z=0$, and  $\CV[b_a]$ the corresponding VOA.
Then we expect that each stalk of the category of line operators $\CC=\oplus_{g\in G_\C}\CC_g$ is mapped to the deformed VOA category
\be \label{Fb-A} \CF:\; \CC_{g=\exp(a)} \to \CV[b_a]\text{-mod}\,. \ee
Again, optimistically, this is an isomorphism.

We explore this deformed family of identifications in our toy model $\CT_{\rm hyper}^B$, whose flavor symmetry is $G_\C=SL(2,\C)_m$. 
In the presence of a meromorphic connection $\CA = \CA_z \diff z$, the superpotential in the twisted action \eqref{S-hyper-HT} is deformed to $W_{B}=\tfrac12\epsilon_{ij}\big[\hat{\mb \Phi}^i \pd_z \hat{\mb \Phi}^j + \hat{\mb \Phi}^i (\CA_z \hat{\mb \Phi})^j\big]\diff z$, where $(\hat{\mb \Phi}^1,\hat{\mb \Phi}^2) = (\hat{\mb X},\hat{\mb Y})$. Correspondingly, the boundary algebra generated by $(\chi_1,\chi_2) := (\chi_z^X,\chi_z^Y)$ has OPE deformed to
\be \CV_B[\CD_\CA]= SF_\CA:\qquad  \chi_i(z)\chi_j(w) \sim \frac{\CA_{ij}(w)}{z-w} + \frac{\epsilon_{ij}}{(z-w)^2}\,, \label{SF-A}\ee
where $\CA_{ij} := \epsilon_{ik}\CA^k{}_j$ is symmetric. (This deformed OPE was also derived in \cite[Sec. 3.6]{Gaiotto-blocks}.)

Now let us specialize to the meromorphic connection
\be \CD_a:\qquad  \CA = \frac{1}{2\pi i z}  a\, \diff z\,,\qquad a\in \mathfrak{sl}_2\,, \label{CAaz} \ee
which is a local model for a defect with holonomy $g=e^{a}$, as in \eqref{CAa}, but written in holomorphic gauge. From \eqref{Fb-A}, we expect an isomorphism of derived categories
\be \CF:\CC_{g=e^a} \to \CV_B[\CD_a]\text{-mod}\,. \ee
We know, however, that the bulk category is trivial when $a$ is generic (as in \eqref{Cg-gen}). Let us explain how this comes about, from a VOA perspective.

Denote by $\psi^\pm$ a pair of free fermions ($FF$) with OPE
\be 
FF:\qquad \psi^\pm(z) \psi^\mp(w) \sim \frac{1}{z-w}\,, \qquad  \psi^\pm(z) \psi^\pm(w) \sim 0\,. 
\ee
Note that for any $\bsp \mb a &\mb b \\ \mb c & \mb d \esp$ in SL$(2, \mathbb C)(\!(z^\pm)\!)$ and any $\bm \alpha, \bm \beta$ in $\mathbb C(\!(z)\!)$ one can define the map 
\be
\chi_1 \mapsto  \mb a\psi^+ + \mb b\pd \psi^- + \bm \alpha\psi^-, \qquad 
\chi_2 \mapsto  \mb c\psi^+ + \mb d\pd \psi^- + \bm \beta\psi^-\,,
\ee
which is an injective vertex-algebra homomorphism $SF_\CA\to FF$ from the deformed symplectic fermions \eqref{SF-A} to free fermions so long as
\be \CA_{11} = 2\mb a \bm \alpha\,,\qquad \CA_{22}=2 \mb c \bm \beta\,,\qquad \CA_{12}=\CA_{21} = \mb a \bm \beta+\mb c \bm \alpha\,.\ee
If $\CA$ is sufficiently generic, \emph{e.g.} of the form \eqref{CAaz} for nondegenerate $a$, and we enlarge the vacuum modules to a completion that allows for infinite sums, then the homomorphism is also surjective. We thus learn that $\CV_B[\CD_\CA]=SF_\CA\simeq FF$. The category of modules for $FF$, however, is well known to be trivial, $FF\text{-mod}\simeq \text{Vect}$.

To make this more concrete, consider the case of abelian holonomy $\CA = \frac{1}{2\pi iz}\text{diag}(\alpha,-\alpha)\,\diff z$\,. Note that $\CA_{ij}=\epsilon_{ik}\CA^k{}_j=\frac{1}{2\pi iz}\bsp 0 & \alpha \\ \alpha & 0 \esp \diff z$. Then we can embed the deformed symplectic fermions
\be \label{SFtoFF1} \chi_1(z)\chi_2(w) = \frac{\alpha}{2\pi i w(z-w)} + \frac{1}{(z-w)^2} \ee
into free fermions via
\be  \label{SFtoFF2} \chi_1(z)\mapsto \psi^+(z)\,,\qquad \chi_2(z) \mapsto \big(\pd + \frac{\alpha}{2\pi iz}\big)\psi^-(z)\,.\ee
This map is invertible when $\alpha\neq 0$, provided we use a formal series to invert the covariant derivative $(\pd + \frac{\alpha}{2\pi iz})^{-1} = \frac{2\pi iz}{\alpha}(1+\frac{2\pi iz}{\alpha}\pd - (\frac{2\pi iz}{\alpha})^2\pd^2+...)$.

\subsection{State spaces and indices}
\label{sec:Hilb-gen}

State spaces of topologically twisted 3d $\CN=4$ theories were initially discussed in \cite{BlauThompson, RW,KRS}. They have found renewed interest in many recent works, including \cite{Gaiotto-blocks, BFK-quasimaps, BFK-Hilb, SafronovWilliams, RW-Coulomb,Gang-TQFT}. We wish here to review the general structure of state spaces on closed surfaces, particularly their dependence on background flavor connections, and to relate counts of states to the ``twisted indices'' introduced by \cite{BZ-index} (and further developed in many works, including \cite{GP-Verlinde, BZ-Riemann, CK-comments,BFK-quasimaps}). We will use $\CT_{\rm hyper}^B$ as an explicit, illustrative example.

Throughout this paper, by ``state space'' of a theory $\CT$ twisted by a nilpotent supercharge $Q$, we implicitly mean the $Q$-cohomology of the full/physical Hilbert space of $\CT$,
\be \CH_{\CT^Q}(\Sigma) := H^\bullet(\CH_\CT(\Sigma),Q)\,.  \label{H-HQ}\ee

\subsubsection{Flavor symmetry}
\label{sec:Hilb-flavor}

In a topologically twisted 3d $\CN=4$ theory that can be deformed by flat connections for the complexified flavor group $G_\C$, the state space on a surface $\Sigma$ should depend on a choice of flat connection $\CA$ on $\Sigma$.

A careful analysis of this dependence proceeds just as in Section \ref{sec:holonomy}, where our effective surface was the punctured disc $D^*$. In general, we may quantize a theory on $M=\Sigma\times \R_t$ in the presence of a flat background connection $\CA$. Gauge transformations of $\CA$ are equivalent to field redefinitions, and thus induce isomorphisms of state spaces, as in \eqref{C-iso}. We will pass to temporal gauge, so that $\CA$ is a flat connection on $\Sigma$ alone, and write
\be \CH_Q(\Sigma,\CA)\,,\qquad \CA\in \text{Flat}_{G_\C}(\Sigma)\,.  \ee
for (the $Q$-cohomology of) the state space on $\Sigma$ with background connection $\CA$. 
Further flavor gauge transformations along $\Sigma$ should induce isomorphisms
\be \CH_Q(\Sigma,\CA) \simeq \CH_Q(\Sigma,\CA^g)\,. \ee

The state spaces for various $\CA$ may be assembled into a bundle over the moduli space of flat connections on $\Sigma$,
\be \begin{array}{c} \CH_Q(\Sigma)  \\ \downarrow \\ \text{Flat}_{G_\C}(\Sigma) \end{array} \ee 
with stalks $\CH_Q(\Sigma,\CA)$. It was argued in \cite{Gaiotto-blocks} that this bundle has the structure of a \emph{coherent sheaf}. In particular, the local dependence of $\CH_Q(\Sigma,\CA)$ on $\CA$ is holomorphic.

The holomorphic dependence on $\CA$ can be seen in explicit models for the state space of (say) a B-twisted gauge theory $\CT$ with (vectormultiplet) flavor symmetry $G$ \cite{Gaiotto-blocks, BF-Hilb, BFK-quasimaps}. The twisted theory $\CT^B$ on $\Sigma\times \R_t$ is equivalent to a B-twisted quantum mechanics on $\R_t$ whose fields are valued in maps from $\Sigma$ to the original 3d target. The connection $\CA$ appears in an effective superpotential, implying (by standard techniques in B-twisted quantum mechanics) that dependence of the state space on $\CA$ is holomorphic. More so, one expects variations of $\CA$ to simply modify the differential acting on the full physical Hilbert space, as in \eqref{H-HQ}.

Holomorphic dependence on $\CA$ suggests that the state space may jump along complex loci in $\text{Flat}_{G_\C}(\Sigma)$, much as we saw categories of line operators jumping in Section \ref{sec:hyper-lines}. This is an interesting feature, also observed in many of the non-semisimple TQFT's studied in recent years, see for example \cite{BCGP}.  Nevertheless, the index, or Euler character, of the state space should be \emph{independent} of the choice of $\CA$, due to the usual arguments governing invariance of a Witten index under continuous deformations \cite{Witten-index}.

In a B-twisted 3d $\CN=4$ theory, the state space $\CH_{Q_B}(\Sigma,\CA)$ is a representation of $U(1)_H\subset SU(2)_H$ R-symmetry, as well as a subgroup $\text{Stab}_{G_\C}(\CA)\subseteq G_\C$ of constant flavor transformations that leave $\CA$ invariant. Correspondingly, $\CH_{Q_B}(\Sigma,\CA)$ is graded by
\begin{itemize}
\item charge $H\in \Z$ for the $U(1)_H$, which is a cohomological grading
\item charge $e$ for a maximal torus $T_\C$ of $\text{Stab}_{G_\C}(\CA)$, which is non-cohomological
\end{itemize}
One may thus construct a graded Poincar\'e series and a graded index (\cf\ \cite{GP-Verlinde, BFK-quasimaps,RW-Coulomb})
\be \label{P-chi} \begin{array}{rcl}
P\big[\CH_{Q_B}(\Sigma,\CA)\big](t,y) &:=&  \text{Tr}_{\CH_{Q_B}(\Sigma,\CA)} t^H y^e \\[.2cm]
\chi\big[\CH_{Q_B}(\Sigma,\CA)\big](y) &:=& \text{Tr}_{\CH_{Q_B}(\Sigma,\CA)} (-1)^H y^e = P(-1,y) \end{array} \ee
depending on formal variables $t\in \C^*$, $y\in T_\C$.

In general, $\CH_{Q_B}(\Sigma,\CA)$ will be infinite-dimensional. This introduces some subtleties that we discuss further in Section \ref{sec:grading}. For sufficiently well-behaved theories, $\CH_{Q_B}(\Sigma,\CA)$ will still have finite graded dimensions, allowing the Poincar\'e series to be defined as a formal series, though defining the index requires regularization. One optimistically expects that the Poincar\'e series will only be a piecewise-constant function of $\CA$,
 but that the index --- suitably regularized --- will be constant. 
 (Note that to compare the index for flat connections $\CA,\CA'$ with different stabilizers, it may be necessary to set some $y$ fugacities to $1$, so that only dependence on a common maximal torus $T_\C\subseteq \text{Stab}_{G_\C}(\CA)\cap \text{Stab}_{G_\C}(\CA')$ appears.)

\subsubsection{The twisted index}
\label{sec:twistedindex}

By viewing a 3d $\CN=4$ theory $\CT$ as a 3d $\CN=2$ theory, one may also associate to it a \emph{twisted index} $\CI_\CT(\Sigma)$ on any Riemann surface $\Sigma$ \cite{BZ-index}. The relation between twisted indices and 3d $\CN=4$ topological indices (such as $\chi$ in \eqref{P-chi} above) has been discussed in \cite{CK-comments, BFK-quasimaps}. We elaborate on it here.

The twisted index of a 3d $\CN=2$ theory $\CT$ is constructed as a partition function on $\Sigma\times S^1$, introducing a background along $\Sigma$ that preserves two supercharges of the 3d $\CN=2$ algebra \cite[Sec 2.1.2]{BZ-index}, \cite[App A]{CK-comments}. These two supercharges coincide with the holomorphic-topological supercharge $Q_{HT}$ of \cite{ACMV, CDG} (also discussed in Section \ref{sec:hyper-VOA} above) and a conjugate $\wt Q_{HT}^\dagger$, such that the commutator $\{Q_{HT},Q_{HT}^\dagger\}\sim \pd_t$ generates a translation along $S^1$.

This $\Sigma\times S^1$ partition function is usually computed by localizing with respect to the sum $\wt Q:=Q_{HT}+Q_{HT}^\dagger$. The supercharge $\wt Q$, which satisfies $\wt Q^2\sim \pd_t$, is often referred to as the ``3d A-twist'' supercharge, because it coincides with a familiar A-twist of a 2d $\CN=(2,2)$ theory upon compactifying along $S^1$. (It is \emph{not} the same as the nilpotent supercharge used to define the topological A-twist of a 3d $\CN=4$ theory.)
The state space $\CH(\Sigma)$ of the theory on $\Sigma\times \R_t$ is graded by $U(1)_R$ R-symmetry, under which $Q_{HT}$ and $Q_{HT}^\dagger$ have charges $1$ and $-1$. The subspace of supersymmetric ground states may equivalently be described as the kernel of $\wt Q$, or the cohomology of either $Q_{HT}$ or $Q_{HT}^\dagger$,
\be \CH_0(\Sigma) = \text{ker}\, \wt Q\big|_{\CH(\Sigma)} \simeq H^\bullet\big(\CH(\Sigma),Q_{HT})\,. \ee
(This follows from the standard structure of state spaces in supersymmetric quantum mechanics \cite{Witten-Morse}, under an additional assumption that graded components of $\CH$ are finite-dimensional.) The twisted index can then be expressed as the character of $Q_{HT}$-cohomology, using R-charge as the cohomological grading:
\be \CI_{\CT}(\Sigma) = \chi\big[H^\bullet\big(\CH(\Sigma),Q_{HT})\big] = \text{Tr}_\CH (-1)^R\,. \ee
In other words, the twisted index of a 3d $\CN=2$ theory is \emph{the same} as the index of its state space in the holomorphic-topological twist.

Now suppose a 3d $\CN=2$ theory $\CT$ actually has $\CN=4$ supersymmetry.  Then it gains a canonical flavor symmetry $U(1)_\epsilon$ whose charge is a difference of charges for the $\CN=4$ $U(1)_H\subset SU(2)_H$ and $U(1)_C\subset SU(2)_C$ R-symmetries,
\be \epsilon = H-C\,. \label{eHC} \ee
Moreover, the 3d $\CN=2$ R-symmetry $U(1)_R$ may be chosen as $U(1)_C$, or $U(1)_H$, or any other ad-mixture with $U(1)_\epsilon$ and other flavor symmetries. Each choice corresponds to a different background along $\Sigma$ and thus a different state space $\CH$ and a different twisted index. Once a choice is made, the twisted index takes the form
\be \CI_\CT(\Sigma)(t) = \text{Tr}_\CH (-1)^R t^\epsilon\,, \ee
adding a fugacity for the canonical flavor symmetry.

If one chooses $U(1)_R=U(1)_C$  (or $R=C$) to define the twisted index of a 3d $\CN=4$ theory, written as 3d $\CN=2$, then the background on $\Sigma$ is the \emph{same} as the background  that would be used to define the 3d topological B-twist on $\Sigma\times \R_t$. This background preserves both $Q_{HT},Q_{HT}^\dagger$ supercharges and the supercharge $Q_B'$ (and its conjugate) that can deform the HT twist to a 3d topological B twist, as in \eqref{HT-A-B}. 
The state space in the topological B-twist is
\be \CH_{Q_B}(\Sigma) = H^\bullet(\CH,Q_B) = H^\bullet(\CH,Q_{HT}+Q_B')\,, \label{HTB'} \ee
and uses $H = R+\epsilon$ as a cohomological grading. Thus, its index is
\be \label{I-HT-B} \boxed{ \chi\big[\CH_{Q_B}(\Sigma)\big] = \text{Tr}_\CH (-1)^{R+\epsilon} = \CI^{(R=C)}_\CT(\Sigma)(t=-1) }\,, \ee
and may be obtained by specializing $t=-1$ in the twisted index.

Optimistically, the topological state space itself might be computed iteratively as $H^\bullet(\CH,Q_{HT}+Q_B') \simeq H^\bullet(H^\bullet(\CH,Q_{HT}),Q_B')$. In other words, if we denote the state space in the HT twist as $\CH_{HT}:=H^\bullet(\CH,Q_{HT})$, then we might obtain the topological state space as $\CH_{Q_B} = H^\bullet(\CH_{HT},Q_B')$. (This will be true if an appropriate spectral sequence degenerates.) In this representation, $U(1)_\epsilon$ plays the role of a cohomological grading on the complex $\big(\CH_{HT},Q_B'\big)$, and the twisted index $\CI_\CT(\Sigma)(t) = \text{Tr}_{\CH_{HT}} t^\epsilon$ is the Poincar\'e polynomial for this complex. 

\emph{We emphasize, however, that that the twisted index $\CI_\CT(\Sigma)(t)$ is not the same as the Poincar\'e polynomial of the topological twisted state space, as we have defined it in \eqref{P-chi}}.
Their respective cohomological gradings are different: $U(1)_\epsilon$ vs. $U(1)_{R+\epsilon}=U(1)_H$.
We will see the difference explicitly in examples.

Similar to \eqref{I-HT-B}, the index of a state space in the topological \emph{A-twist} of a 3d $\CN=4$ theory  may be obtained by first computing the twisted index with $R=H$, then~setting~$t=-1$:
\be \label{I-HT-A}   \boxed{ \chi\big[\CH_{Q_A}(\Sigma)\big] = \text{Tr}_\CH (-1)^{R-\epsilon} = \CI^{(R=H)}_\CT(\Sigma)(t=-1) }\,. \ee
The cohomological grading in the topological A-twist is $C = R-\epsilon$.

Finally, we note that if additional 3d $\CN=4$ (B-type, say) flavor symmetry is present, both the twisted index and the (B-twisted) topological index may include a fugacity for the flavor symmetry. Relation \eqref{I-HT-B} will hold with this fugacity inserted on both sides, \eg\
\be \chi\big[\CH_{Q_B}(\Sigma)\big](y) =  \text{Tr}_\CH (-1)^R (-1)^\epsilon y^e  = \CI_\CT^{(R=C)}(\Sigma)(t=-1,y)\,.\ee
Neither side depends on deformations by a flat flavor connection along $\Sigma$. The twisted index could in principle be further generalized to include flavor flux through $\Sigma$, corresponding to a deformation by a \emph{non}-flat connection on $\Sigma$; but such deformations are not compatible with the topological twist.

\subsubsection{State spaces and indices for $\CT_{\rm hyper}^B$}
\label{sec:hyper-states}

In order to illustrate the role of flavor symmetry, and the relation between twisted and topological indices, we'll explicitly compute some state spaces and indices in $\CT_{\rm hyper}^B$.

We begin with trivial background connection $\CA=0$. The state space on a genus-$g$ surface $\Sigma_g$ may be computed by reducing the B-twisted theory on $\Sigma_g\times \R_t$ to a 1d B-twisted quantum mechanics on $\R_t$. This can be done very explicitly starting from the twisted BV action \eqref{SXY}, and was also described some time ago in \cite{RW,KRS} (or see the recent \cite{BF-Hilb,BFK-Hilb}).

After integrating out massive degrees of freedom, one finds an effective B-type (or $(0,2)$ type) quantum mechanics with
\begin{itemize}
\item Two chiral multiplets with lowest components $X,Y$, coming from the zero-modes of the 3d complex scalars along $\Sigma_g$.
Here $(X,Y)$ is a doublet for the $SU(2)_m$ flavor symmetry, and has cohomological $U(1)_H$ charge $+1$.

\item $2g$ fermi multiplets with lowest components $\{\chi^X_i,\chi^Y_i\}_{i=1}^g$. These come from the holomorphic modes of the 1-form fermions $\chi^X,\chi^Y$ on $\Sigma_g$. Explicitly, if we choose a basis $\omega^i\in H^{(1,0)}(\Sigma_g)$ of holomorphic 1-forms on $\Sigma_g$, we have $\chi^X = \sum_i \chi^X_i \omega^i$, $\chi^Y = \sum_i \chi^Y_i \omega^i$.
Each $(\chi_i^X,\chi_i^Y)$ is an $SU(2)_m$ doublet, with $U(1)_H$ charge zero.

This decomposition depends on a choice of complex structure on $\Sigma_g$, though the state space itself does not.

\end{itemize}

\noindent Quantization requires a further choice of polarization. In the fermionic sector, two natural $SU(2)_m$-invariant choices lead to a Hilbert space represented either as an exterior algebra in the $2g$ fermions  $\chi^X_i,\chi^Y_i$, or their conjugates. We will take the fermionic Hilbert space to be the exterior algebra $\C[\chi^X_1,\chi^Y_1,...,\chi^X_g,\chi^Y_g] \simeq \C^{2^{2g}}$ for now.

In the bosonic sector, things are more subtle.%
\footnote{We thank M. Bullimore for extended discussions on this point.} %
 We will continue to require $SU(2)_m$ invariance. Then there are again two choices, leading either to a state space $\C[X,Y]$ represented as a symmetric algebra in the zero-modes $X,Y$, or a symmetric algebra $\C[\dot{\ol X},\dot{\ol Y}]$ in (time derivatives of) their conjugates. These spaces, while infinite-dimensional, have many desirable features:

\noindent\hspace{.2cm} - they have an action of $SU(2)_m$ (extending to an action of $SL(2,\C)_m$);

\noindent\hspace{.2cm} - they have semi-bounded cohomological degrees: $\C[X,Y]$ has non-negative cohomological grading, while $\C[\dot{\ol X},\dot{\ol Y}]$ has non-positive cohomological grading;

\noindent\hspace{.2cm} - each subspace of fixed cohomological degree is finite dimensional;

\noindent\hspace{.2cm} - they are directly related to Hochschild homology computations (Section \ref{sec:Hoch})

\noindent However, \emph{neither} $\C[X,Y]$ nor $\C[\dot{\ol X},\dot{\ol Y}]$ can be given the structure of a Hilbert space. In particular, it is clear that cohomological degrees are not symmetric about zero, as would be required for the existence of a nondegenerate bilinear form of degree zero. In a finite-dimensional setting (\emph{e.g.} in quantization of compact bosons), one usually solves this problem by shifting cohomological degree, but no suitable shift is available here.

An alternative quantization is pursued in  \cite{BFK-Hilb}. There, a real mass `$m_\R$' for a maximal torus $U(1)_m\subset SU(2)_m$ is introduced, related to deforming by a complexified connection along the `time' direction. This effectively determines a choice of polarization. Moreover, it has the effect of regularizing bosonic wavefunctions, making them square-normalizable, and thus does lead to an honest Hilbert space of states. For positive $m_\R$, one finds a Hilbert space represented as a symmetric algebra $\C[X,\dot{\ol Y}]$; whereas for negative $m_\R$, one finds a Hilbert space represented as $\C[\dot{\ol X},Y]$. Unfortunately:

\noindent\hspace{.2cm} - the spaces $\C[X,\dot{\ol Y}]$ and $\C[X,\dot{\ol Y}]$  do not have an action of  $SU(2)_m$ (only $U(1)_m$ acts)

\noindent\hspace{.2cm} - they have unbounded cohomological degree

\noindent\hspace{.2cm} - their subspaces of fixed cohomological degree are inifinite-dimensional.

\noindent In this paper, for both $\CT_{\rm hyper}^B$ and the later theories $\CT_{n,k}^A$, we will choose the former sorts of polarizations, which give rise to state spaces that preserve the full flavor symmetry, and have bounded cohomological degrees with finite graded dimensions, at the cost of failing to be honest Hilbert spaces. Having bounded cohomological degrees with finite graded dimensions will in particular let us restrict to cohomological degree zero, and match CGP TQFT's and VOA conformal blocks.

\medskip

Suppose, then, that for $\CT_{\rm hyper}^B$ we choose the polarization leading to the non-negatively graded space
\be \label{HB-hyper} \CH_{Q_B}(\Sigma_g,\CA=0) \simeq \C[X,Y]\otimes \C[\chi^X_1,\chi^Y_1,...,\chi^X_g,\chi^Y_g]\,. \ee
Then the Poincar\'e polynomial and graded character are
\be \label{P-hyper} \begin{array}{lcl} P\big[\CH_{Q_B}(\Sigma,\CA=0)\big](t,y) &= & \ds \frac{(1+y)^g(1+y^{-1})^g}{(1-ty)(1-ty^{-1})}\,, \\[.4cm] \chi\big[\CH_{Q_B}(\Sigma,\CA=0)\big](y) &= & (1+y)^{g-1}(1+y^{-1})^{g-1}\,. \end{array} \ee

Next, let us introduce a nontrivial $SL(2,\C)_m$ background flat connection $\CA$ on $\Sigma$, in a holomorphic flavor gauge, such that $\CA = \CA_z(z)\diff z$. We may expand this in a basis of holomorphic 1-forms as $\CA = \CA_{z,i}\omega^i$.
The flat connection adds an $E$-type superpotential in the effective quantum mechanics, which induces a differential $Q_\CA$ on the state space \eqref{HB-hyper}, acting as
\be Q_\CA \bp \chi^X_i \\   \chi^Y_i \ep = \CA_{z,i} \bp X \\ Y \ep\,.
 \ee
Note that this preserves $U(1)_m$ flavor charges only if $\CA$ is diagonal, as anticipated above \eqref{P-chi}.
We expect that $\CH_{Q_B}(\Sigma_g,\CA)$ can be computed as the $Q_\CA$-cohomology of $\CH_{Q_B}(\Sigma_g,0)$. (This will be true if an appropriate spectral sequence degenerates.)

If $\CA$ is sufficiently generic, then in $Q_\CA$-cohomology of $\CH_{Q_B}(\Sigma_g,0)$ we will find that two linear combinations of the fermions `cancel' with the bosons $X,Y$. For example, to obtain such a cancellation, it is sufficient for $\CA$ to be diagonal, with nontrivial holonomy along at least one cycle of $\Sigma_g$.

In genus zero, no cancellation is possible. Indeed, any flat connection on $S^2$ is gauge-equivalent to the trivial flat connection, so the state space will remain unchanged,
\be \CH_{Q_B}(S^2,\CA) = \C[X,Y] \qquad (\text{any $\CA$})\,. \ee
This of course reproduces the bulk local operators \eqref{hyper-ops}.
 In genus one, a generic flat connection will completely trivialize the state space, while in genus $g\geq 2$, a generic flat connection will leave behind $2(g-1)$ fermionic ``directions'',
 \be \CH_{Q_B}(\Sigma_g,\CA) \simeq (\C^2)^{\otimes 2(g-1)} \qquad \text{(generic $\CA$)} \label{H-hyper-A} \ee
all in cohomological degree zero. The Poincar\'e polynomials are
\be \label{PA-hyper} \begin{array}{lcl} P\big[\CH_{Q_B}(S^2,\CA)\big](t,y) &=& \ds \frac{1}{(1-ty)(1-ty^{-1})} \qquad (\text{any $\CA$}) \\ [.4cm]
 P\big[\CH_{Q_B}(\Sigma_{g\geq 1},\CA)\big](t,y) &=& (1+y)^{g-1}(1+y^{-1})^{g-1}  \qquad (\text{generic $\CA$})\,, \end{array}
\ee
where the extra $y$ grading makes sense if $\CA$ is diagonal. The graded characters are unchanged, as required: $\chi\big[\CH_{Q_B}(\Sigma,\CA)\big](y) =  (1+y)^{g-1}(1+y^{-1})^{g-1}$ for any $\CA$.

Finally, let us consider the twisted index. Choosing the 3d $\CN=2$ R-symmetry to be $U(1)_R = U(1)_C$ so that it is compatible with the topological B-twist, we find from \cite{CK-comments, CK-review} an all-genus formula for the twisted index
\be \CI(t,y) = \big[-t(1-ty)(1-ty^{-1})\big]^{g-1}\,, \ee
which agrees with graded character $\chi\big[\CH_{Q_B}(\Sigma,\CA)\big](y)$ upon setting $t=-1$. Note, however, that $\CI(t,y)$ is \emph{not} the Poincar\'e polynomial of the topologically-twisted state space when $g\geq 2$.

\subsection{Genus 1 and the category of lines}
\label{sec:torus-lines}

In a 3d TQFT, the state space on a genus-one surface  $\Sigma=T^2$ has a special relationship with the category of line operators $\CC$.

The general statement one expects in a cohomological TQFT is that the torus state space is equivalent to the \emph{Hochschild homology} of the category of line operators, \emph{cf.} \cite{KRS,Kapustin-ICM,Lurie}
\be \label{H-HH} \CH(T^2)\simeq HH_\bullet(\CC)\,. \ee
We would like to sketch a physical description of Hochschild homology, and explain why \eqref{H-HH} is a natural manifestation of a state-operator correspondence.

The relation \eqref{H-HH}, in particular its full cohomological version (containing all degrees on the RHS), is still somewhat unfamiliar in some of the quantum topology and physics literature.%
\footnote{Though see \cite{Beliakova-trace, GHW-traces} for recent uses of Hochschild homology in the context of knot homology/categorification, and \cite{LQ-derived} for $SL(2,\Z)$ actions on Hochschild homology of the small quantum group. Recent work of Schweigert and Woike \cite{SchweigertWoike,SW-Verlinde} also explains why derived concepts, including Hochschild homology, should be introduced to the study of non-semisimple TQFT.} 
This is in part for a good reason: in the most studied case of Chern-Simons theory with compact gauge group, the category of line operators is semisimple, and its Hochschild homology is equivalent to its Grothendieck group (or K-theory),
\be HH_\bullet(\CC) \simeq HH_0(\CC) = K_0(\CC) \qquad \text{for semisimple $\CC$}\,. \ee
Thus, in Chern-Simons theory, one finds the more familiar statement that the state space $\CH(T^2)$ is spanned by states associated to simple line operators \cite{Witten-Jones, EMSS}, which also generate $K_0(\CC)$.

In contrast, for the sorts of cohomological TQFT's arising as topological twists of 3d $\CN=4$ theories, the dg category of line operators $\CC$ is generally far from semisimple. There are natural maps
\be K_0(\CC) \overset{ch}\longrightarrow HH_0(\CC) \hookrightarrow HH_\bullet(\CC) \simeq \CH(T^2)\,, \ee
though in general the Grothendieck group only spans a small part of the torus state space. We will review the physical origin of these maps, and describe them explicitly in the example of $\CT_{\rm hyper}^B$. 

We saw in Section \ref{sec:Hilb-gen} that background flat connections for a flavor symmetry can deform state spaces in nontrivial ways. Background connections induce corresponding deformations in Hochschild homology, which we explore in Section \ref{sec:Hoch-flavor}.

\subsubsection{Hochschild homology}
\label{sec:Hoch}

Consider a 3d TQFT quantized on $T^2\times \R_t$. 
A state-operator correspondence relates states in the state space $\CH(T^2)$ and configurations of local and line operators supported along the core of a solid torus $S^1\times \{0\}\subset S^1\times D^2$.

To get from a state to an operator, one may view spacetime as $T^2\times \R_t\simeq S^1\times (S^1\times \R_t)$, deform the second factor to a punctured plane $(S^1\times \R_t)\simeq  \C^*$ (in radial quantization), and then evolve any state $v\in \CH(T^2)$ backwards in time until it defines a configuration of operators on $S^1\times \{0\}\subset S^1\times \C^*$. Conversely, the path integral on a solid torus will define a state on the $T^2$ boundary given any configuration of operators along the core of the solid torus. Altogether, there are complex-linear maps
\be \CH(T^2)\;\underset{\tau}{\overset{\sigma}{\includegraphics[width=4ex]{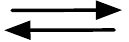}}}\; \{\text{operators in the core of $S^1\times D^2$}\}\,, \ee
guaranteed to satisfy $\tau\circ\sigma = \text{id}$, which implies that $\tau$ is surjective.

\begin{figure}[htb]
\centering
\includegraphics[width=5.5in]{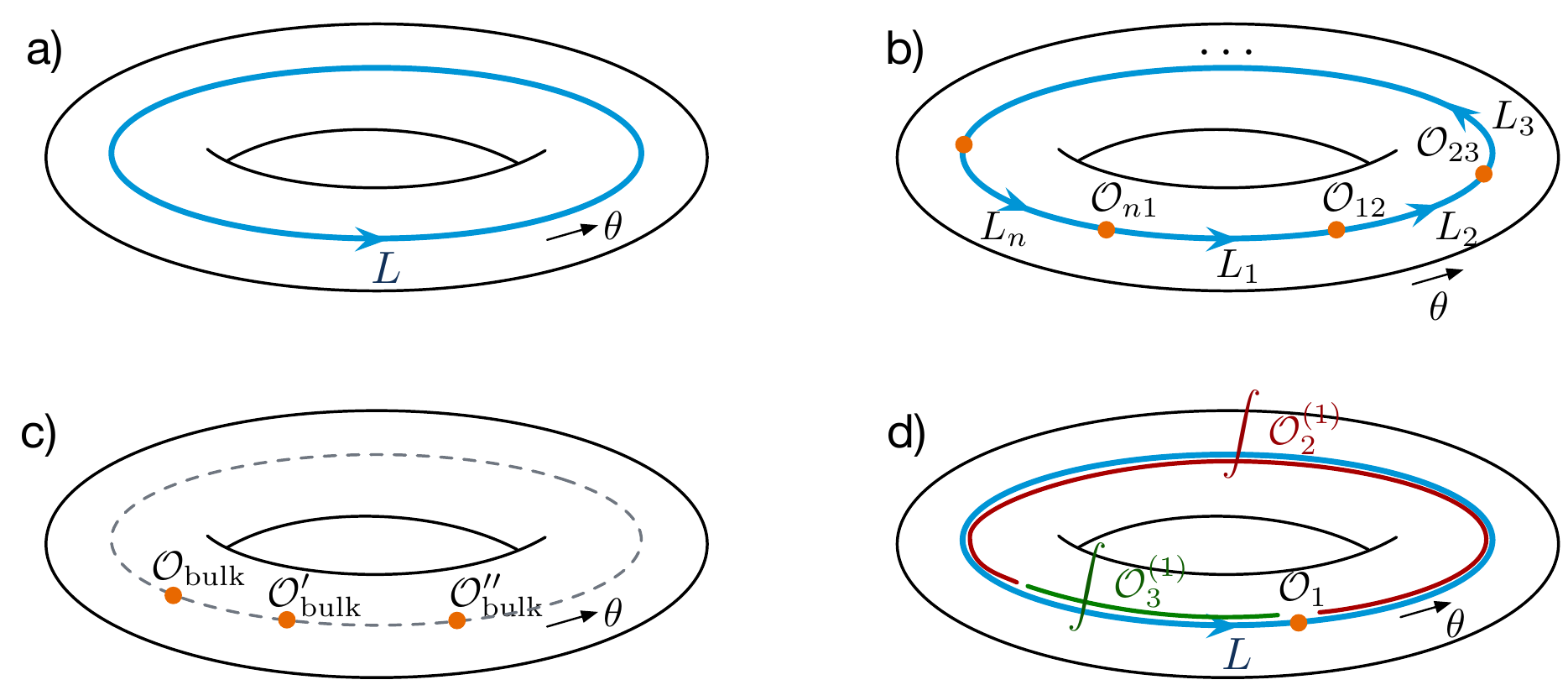}
\caption{Contributions to the torus state space: a) line operators wrapping the core; b) line operators joined by local interfaces; c) bulk local operators (self-interfaces of the identity line); d) integrated descendants of local operators on lines.}
\label{fig:T2Hilb}
\end{figure}

In a cohomological theory with differential/supercharge $Q$, such as a topological twist of a 3d $\CN=4$ theory, the sorts of operators that may appear along the core of a solid torus include (see Figure \ref{fig:T2Hilb}):
\begin{itemize}
\item Line operators $L\in \text{Ob}(\CC)$ wrapping the entire $S^1$ core.
\item Cyclic configurations of line operators $(L_1,L_2,...,L_n)$ placed around the $S^1$, with a choice of $Q$-closed local operators $\CO_{12},...,\CO_{n1}$ at each junction.
\item $Q$-closed bulk local operators placed at some points around $S^1$, which may equivalently be thought of as junctions between the trivial line operator $\id\in \text{Ob}(\CC)$ and itself.
\item Topological 1-form descendants of local operators at junctions of lines, integrated along segments (or along all of $S^1$) to produce a $Q$-closed configuration.
\end{itemize}
The mathematical operation of taking Hochschild homology of the category of line operators, denoted $HH_\bullet(\CC)$, automatically includes all such configurations, \emph{and} is meant to impose appropriate equivalence relations on them such that the induced map
\be \tau:HH_\bullet(\CC) \overset\sim\to \CH(T^2) \ee
(now implicitly taking $Q$-cohomology of the state space on the RHS) becomes an equivalence.

When the category $\CC$ is generated by a particular line operator $L$, in the sense that the category is equivalent to dg modules for the derived endomorphism algebra of $L$,
\be \CC\simeq A\text{-mod}\,,\qquad  A:= \text{Hom}^\bullet(L,L) \quad (\text{as a dg algebra})\,, \ee
Hochschild homology may be defined as follows. One forms the semi-infinite complex
\be \label{HH-complex} \begin{array}{cccccccccc}
C_\bullet(A):= \; \cdots & \overset{d_H}\longrightarrow & A^{\otimes 4} & \overset{d_H}\longrightarrow & A^{\otimes3}  & \overset{d_H}\longrightarrow & A^{\otimes 2} & \overset{d_H}\longrightarrow & A &\to 0\,, \\
\text{degrees:} && -3 && -2 && -1 && 0 \end{array}
 \ee
with differential
\begin{align} d_H(a_1\otimes a_2\otimes \cdots\otimes a_n\otimes a_{n+1}) &= (a_1a_2)\otimes a_3\otimes\cdots \otimes a_{n+1} - a_1\otimes (a_2a_3)\otimes a_4\otimes\cdots \otimes a_{n+1}  \notag \\
& \quad + \ldots + (-1)^{n+1}a_1\otimes \cdots \otimes (a_na_{n+1})  \label{dH} \\
&\quad + (-1)^{|a_{n+1}|(|a_1|+...+|a_n|)+n} (a_{n+1}a_1) \otimes a_2\otimes \cdots \otimes a_n\,.  \notag
\end{align}
In other words, consecutive pairs of elements in $n+1$ copies of the algebra are multiplied together, with alternating signs, and finally $a_{n+1}$ and $a_1$ are multiplied with a sign that accounts for their fermion number in $A$.
 
The degree conventions in the Hochschild complex are a little subtle (see Section \ref{sec:grading} below). Here we use a cohomological convention, so that each term $A^{\otimes n}$ in the complex has its degree shifted by $1-n$, and $d$ has degree $+1$.
The ``Hochschild homology'' is a vector space given by the cohomology
\be HH_\bullet(\CC) := H^\bullet(C_\bullet(A),d+d_H)\,, \ee
where $d$ denotes the internal differential in the dg algebra $A$.

The Hochschild complex is related to physics as follows.%
\footnote{A similar description of the Hochschild complex in the context of twisted 3d supersymmetric theories also appeared recently in \cite[Sec. 4.3.1]{DedushenkoII}.} %
 Consider the generating line operator $L$ placed along the $S^1$ core of a solid torus. The algebra $A$ is the algebra of local operators bound to $L$. 
Each local operator $a\in A$ comes with a 1-form descendent $a^{(1)}$, satisfying
\be Q\big( a^{(1)} \big) = \pd_\theta a\, \diff \theta - (Qa)^{(1)} \,, \ee
where $\theta\in [0,2\pi]$ is a coordinate along the $S^1$ core.%
\footnote{The descendant can be canonically defined by choosing a supercharge $Q^\dagger$ of the 3d $\CN=4$ algebra that satisfies $\{Q,Q^\dagger\} = \pd_\theta$, a translation along the solid torus, and setting $a^{(1)}:=(Q^\dagger a)\,\diff \theta$. See \cite{descent} and the classic \cite{Witten-Donaldson, MooreWitten} for further details.} %
In particular, for $Q$-closed $a$, we simply have $Q\,a^{(1)} = \pd_\theta a\,\diff \theta$, which further implies that the difference of $a$ inserted at different points is $Q$-exact
\be a(\theta_2)-a(\theta_1) = Q\Big[\int_{\theta_1}^{\theta_2} a^{(1)}\Big]\,, \label{atheta} \ee
whence correlation functions are independent of insertion point.

In the Hochschild complex, an element $a_1\otimes a_2\otimes\cdots\otimes a_n\in A^{\otimes n}$ represents the physical configuration of operators given by
\be \label{a-integral}  a_1\otimes\cdots \otimes a_n \quad \leftrightarrow\quad     a_1(0) \int_{0<\theta_2< \theta_3 <\cdots< \theta_n<2\pi} a_2^{(1)}(\theta_2)\, a_3^{(1)}(\theta_3)\cdots a_n^{(1)}(\theta_n)\,. \ee
This ordered integral of descendants, all bound to the line operator $L$, is depicted in Figure \ref{fig:T2Hilb}(d) for $n=3$.%
\footnote{Some additional care is needed when considering the ``boundaries'' of the integral \eqref{a-integral} at which local operators along $L$ come close to each other. These limits might be singular, and need to be regularized, \emph{e.g.} by always keeping local operators a distance $\epsilon$ apart.} %
Note that taking $n-1$ descendants of operators in $A^{\otimes n}$ decreases the cohomological degree by $n-1$, in agreement with the cohomological convention above.
The action of $Q$ on the RHS of \eqref{a-integral} then coincides with the combined action of the internal and Hochschild differentials $d+d_H$ on the abstract element $a_1\otimes\cdots \otimes a_n$ on the LHS!

For example, if $a_1,a_2$ are $Q$-closed (so that we may ignore the internal differential), we find
\be \label{a1a2} d_H(a_1\otimes a_2) = a_1a_2 - (-1)^{|a_1||a_2|} a_2a_1\,, \ee
in agreement with
\begin{align}  Q\Big[ a_1(0)\int_{0<\theta<2\pi} a_2^{(1)}(\theta) \Big] &= a_1(0)\int _{0<\theta<2\pi} Q\big[a_2^{(1)}(\theta)\big] \notag \\
 &  =  a_1(0)(a_2(\epsilon)-a_2(-\epsilon))  \qquad \text{(for small $\epsilon$)} \\
  & = (a_1*a_2)(0) - (-1)^{|a_1||a_2|}(a_2*a_1)(0)\,,  \notag
\end{align}
where $|a|$ denotes fermion number and `$*$' denotes the product of $Q$-closed local operators induced by collision --- which is the mathematical product in the algebra $A$.

Indeed, the zeroth Hochschild homology $HH_0(\CC)$ is the co-center, or ``algebraic trace'' of the algebra $A$. It is generated by $Q$-closed local operators $a$ bound to $L$, modulo the relation that $a_1 a_2 = (-1)^{|a_1||a_2|}a_2a_1$. This equivalence is precisely what one would expect physically for local operators on a circle.

\subsubsection{Computations for $\CT_{\rm hyper}^B$}
\label{sec:Hoch-hyper}

We'll now illustrate how Hochschild homology indeed recovers the correct torus state space for $\CT_{\rm hyper}^B$, at trivial flat flavor connection. The category of line operators $\CC_1$ at trivial connection may be represented in many different ways, as we saw in Section \ref{sec:hyper-lines}. A simple option is to ``generate'' it from the trivial line operator $V$, whose endomorphism algebra $A=\C[X,Y]$ just contains the bulk local operators.  (It will suffice here to simply pass directly to $Q$-cohomology of the algebra $A$).

The algebra $A$ is \emph{already} commutative. Thus the differential $d_H:A^{\otimes 2}\to A$ is zero, and the zeroth Hochschild homology $HH_0(\CC_1)\simeq A$ simply contains the algebra itself. An element $p(X,Y)\in A$ represents an insertion of bulk local operators $p(X(0),Y(0))$ at  $0\in S^1$.

There are higher Hochschild homology groups as well. One can compute them easily from the complex $C_\bullet(A)$ above, but it is more instructive to describe them in terms of integrated descendants. Note that $X^{(1)} = \chi_\theta^X\,\diff \theta$ and $Y^{(1)} = \chi_\theta^Y\,\diff \theta$.  Then $HH_{-1}(\CC_1)$ contains the integrated descendants $\chi_1^X:=\oint \chi_\theta^X\,\diff \theta$ and $\chi_1^Y:=\oint \chi_\theta^Y\,\diff \theta$, as well as their products with arbitrary polynomials $p(X,Y)$ in $X(0)$ and $Y(0)$. Their abstract equivalence classes would be represented as
\be p(X,Y)\otimes X\qquad\text{and}\qquad p(X,Y)\otimes Y \quad \in A^{\otimes 2}\,.\ee
The group $HH_{-2}(\CC_1)$ contains the doubly integrated descendant $\chi_1^X\chi_1^Y = \oint \oint\chi_\theta^X(\theta)\chi_\theta^Y(\theta')\,\diff \theta \diff \theta'$, and its product with arbitrary polynomials $p(X,Y)$. These operators are represented by
\be p(X,Y)\otimes (X\otimes Y-Y\otimes X)\in A^{\otimes 3}\,.\ee
Higher Hochschild homology groups vanish, and we are left with
\be HH_0(\CC_1)\simeq \C[X,Y]\,,\quad HH_{-1}(\CC_1)\simeq \C[X,Y]\chi_1^X\oplus \C[X,Y]\chi_1^Y\,,\quad
  HH_{-2}(\CC_1)\simeq \C[X,Y]\chi_1^X\chi_1^Y\,. \label{HH-hyper-0}\ee

The groups \eqref{HH-hyper-0} precisely recover the torus state space \eqref{HB-hyper} (at $g=1$), aside from one final important detail: the algebra $A$ already had an internal cohomological grading given by $U(1)_H$ charge. To reproduce the correct graded state space, we must add the $U(1)_H$ cohomological grading to that in the Hochschild complex. This simply renames the various Hochschild cohomology groups; their direct sum is still given by $\C[X,Y]\langle 1,\chi_1^X,\chi_1^Y,\chi_1^X\chi_1^Y\rangle$ as in \eqref{HH-hyper-0} (and computed exactly the same way!), but the regraded groups are
\be \label{HH-hyper} HH_n(\CC_1) \simeq \C[X,Y]\big|_{\text{degree $n$}} \langle 1,\chi_1^X,\chi_1^Y,\chi_1^X\chi_1^Y\rangle\,,\qquad n\geq 0\,, \ee 
with $HH_n$ containing homogeneous polynomials of degree $n$ in $X$ and $Y$. More succinctly, the total Hochschild homology is isomorphic to the algebra $HH_\bullet(\CC_1)\simeq \C[X,Y,\chi_1^X,\chi_1^Y]$, with two bosonic/commuting generators $X,Y$ in degree 1 and two fermionic/anticommuting generators $\chi_1^X,\chi_1^Y$ in degree zero.

An alternative option is to generate the category $\CC_1$ from the nontrivial line operator $P$, whose endomorphism algebra was described in Section \ref{sec:hyper-lines} as a graded-commutative $A' = \text{End}^\bullet(P) = \C[\xi,\psi]$ generated by two fermions. Now $A'$ lies entirely in cohomological degree zero, so there is no need to re-grade Hochschild homology groups. However, a different subtlety arises.

The Hochschild homology computed from $A'$ is non-vanishing in all non-negative degrees, and is isomorphic to
\be HH_{-n}(A') \simeq A'\otimes \text{Sym}^{n-1}(A')\qquad n\geq 0\,. \ee
For example,
\begin{align}
HH_0&\simeq A' \notag\,, \\
 HH_{-1} &\simeq A'\otimes \C\langle \xi,\psi\rangle\,, \\
 HH_{-2} &\simeq A'\otimes \C\langle \xi\otimes \xi,\psi\otimes \psi,\xi\otimes \psi+\psi\otimes \xi\rangle\,. \notag
\end{align}
Physically, the descendants of $\xi,\psi$ that appear in the $p_{n-1}$'s and are being integrated around $S^1$ are proportional to the \emph{conjugate} bosonic fields $\dot{\ol X}, \dot{\ol Y}$. They have $U(1)_H$ charge $-1$, hence they are showing up in cohomological degree $-1$. We would obtain from this description a torus state space that takes the form of a symmetric algebra in $\dot{\ol X}, \dot{\ol Y}$ (of degree -1) and an exterior algebra in $\xi,\psi$ (of degree 0),
\be \label{H-hyper-inverted} \CH(T^2) \simeq \C[\bar X,\bar Y,\xi,\psi]\,. \ee
This is the representation of the bosonic Hilbert space corresponding to the other $SU(2)_m$-invariant choice of polarization discussed in Section \ref{sec:hyper-states}. It has non-positive cohomological degree. Its Poincar\'e series is
\be \frac{(1+y)(1+y^{-1})}{(1-t^{-1}y)(1-t^{-1}y^{-1})} = t^2 \frac{(1+y)(1+y^{-1})}{(1-ty)(1-ty^{-1})}\,, \ee
which we observe is related to \eqref{P-hyper} by analytic continuation and an overall shift in cohomological degree.
The Euler character at $t=-1$ remains unchanged.

\subsubsection{A comment on grading and dualization}
\label{sec:grading}

The example of Section \ref{sec:Hoch-hyper} highlights a key subtlety in the definition of state spaces, which it is important to be aware of.

Both $\CT_{\rm hyper}^B$ and theories $\CT_{n,k}^A$ studied later in the paper have
\begin{enumerate}
\item a noncompact moduli space of vacua $\CM$ (the $\C^2$ Higgs branch of $\CT_{\rm hyper}^B$, and the Coulomb branches of  $\CT_{n,k}^A$), which necessarily makes state spaces $\CH(\Sigma)$ infinite-dimensional; and
\item an R-symmetry identified with cohomological degree ($U(1)_H$ for $\CT_{\rm hyper}^B$ and $U(1)_C$ for the A-twisted $\CT_{n,k}^A$) that extends to a contracting $\C^*$ action on the moduli space with compact fixed locus.
\end{enumerate}
Noncompactness of the moduli space $\CM$ means that there will be an infinite-dimensional algebra of functions $\C[\CM]$, corresponding physically to infinitely many bosonic local operators. It is the presence of these local operators that force state spaces  to be infinite-dimensional. The R-symmetry can be used to control the structure of state spaces, to an extent. The fact that the R-symmetry extends to a contracting action with compact fixed locus implies that the local operators $\C[\CM]$ will have non-negative cohomological degree, and moreover that each graded subspace of $\C[\CM]$ will be finite dimensional.

One might expect this to imply that all state spaces $\CH(\Sigma)$ are also non-negatively graded, with finite-dimensional graded components --- since, very roughly speaking, $\CH(\Sigma)$ is obtained by quantizing spaces of maps from $\Sigma$ to $\CM$. This cannot universally be true, for two related reasons. First, defining $\CH(\Sigma)$ requires choosing an orientation of the transverse time direction, and changing this choice (\emph{e.g.} passing from ``outgoing'' to ``incoming'' states) should dualize $\CH(\Sigma)$ as a vector space, which will \emph{invert} the cohomological degree of all states. 
 In a finite-dimensional setting, one usually shifts the cohomological grading on $\CH(\Sigma)$ to make it symmetric, ensuring that $\CH(\Sigma)\simeq \CH(\Sigma)^*$; but when $\CH(\Sigma)$ is infinite dimensional, this is not possible.
Second, when carefully quantizing the bosonic part of $\CH(\Sigma)$, one must make a choice of polarization, which roughly amounts to either including functions $\C[\CM]$ \emph{or} their conjugates as states. We saw examples of this choice in Section \ref{sec:hyper-states}. Changing the choice of polarization will again invert cohomological degrees. 

Altogether, the best we can (and will) expect for the structure of state spaces is that there is a \emph{choice} of orientation and polarization such that a given space $\CH(\Sigma)$ has non-negative cohomological degree, and has finite-dimensional graded components,
\be \label{non-neg} \CH(\Sigma) = \bigoplus_{n\geq 0} \CH(\Sigma)^{(n)}\,,\qquad \text{dim}\, \CH(\Sigma)^{(n)} < \infty\,. \ee
Swapping (say) incoming to outgoing orientations should have the effect of separately dualizing each finite-dimensional graded component, and inverting the overall cohomological degree, 
\be \CH(\Sigma)^* = \bigoplus_{n\leq 0} \big[\CH(\Sigma)^*\big]^{(n)}\quad\text{with}\quad  \big[\CH(\Sigma)^*\big]^{(n)} := \big(\CH(\Sigma)^{(-n)}\big)^*\,. \ee

There is a similar ambiguity that appears mathematically in computing Hochschild homology of a dg category. In the case of $\CT_{\rm hyper}^B$, we saw that the category of line operators had two derived-equivalent descriptions, $\CC_1\simeq D^b(A\text{-mod})\simeq D^b(A'\text{-mod})$, as modules for Koszul-dual algebras $A=\C[X,Y]$ and $A'=\C[\xi,\psi]$. It was shown in classic work \cite{FeiginTsygan} that the Hochschild homologies $HH_\bullet(A)$ and $HH_\bullet(A')$ of Koszul-dual algebras in general are only isomorphic up to an inversion of cohomological degree. This is exactly what we saw in \eqref{H-hyper-inverted}. 

For practical purposes, in this paper, we will take the liberty of inverting the cohomological grading on state spaces and Hochschild homology, where necessary, to always place them in non-negative degree as in \eqref{non-neg}.

\subsubsection{Grothendieck group}
\label{sec:K}

In semisimple TQFT's, such as Chern-Simons theory with compact gauge group, a more familiar statement is that the torus state space is isomorphic to the Grothendieck group (\emph{a.k.a.} K-theory) $K_0(\CC)$ of the category of line operators. This is simply not true in the general setting of topological twists and non-semisimple TQFT. However, there does always exist a ``Chern character'' map $K_0(\CC) \overset{ch}\longrightarrow HH_0(\CC)$ relating the Grothendieck group and the part of the torus state space in cohomological degree zero. We review this here in order to make contact with the constructions of semisimple TQFT.%
\footnote{For further discussion of the Grothendieck group vs. Hochschild homology, in the context of quantum knot invariants, we refer readers to \cite{Beliakova-trace}.} %

Let us begin by recalling the definition of the Grothendieck group. If $\CC$ is a dg category that arises as the derived category of an abelian category $\CC_{\rm ab}$, the (complexified) Grothendieck group of $\CC$ may be constructed as the free abelian group generated by objects $A\in \text{Ob}(\CC_{\rm ab})$, modulo relations $[B]=[A]+[C]$ whenever there is an exact sequence $0\to A\to B\to C\to 0$,
\be \label{K-def} K_0(\CC) \simeq K_0(\CC_{\rm ab}) = \C\big\langle [ A]\,|\,A\in \text{Ob}(\CC_{\rm ab})\big\rangle \Big/ \big([A]-[B]+[C]\big)_{0\to A\to B\to C\to 0}\,. \ee
If $\CC$ is furthermore a monoidal category, the Grothendieck group becomes a ring, with product%
\footnote{This definition assumes that the tensor product $A\otimes-$ is an exact functor in $\CC_{\rm ab}$, which is true for all the categories we will encounter. More generally, the derived tensor product (the Tor functor) must be used to define a product structure on the Grothendieck group.}
\be \label{K-ring}  [A]\cdot [B] := [A\otimes B]\,. \ee
Finally, if $\CC_{\rm ab}$ happens to be semisimple, then $\CC\simeq \CC_{\rm ab}$, and the Grothendieck group has a basis given by the equivalence classes of simple objects
\be \label{K-SS} K_0(\CC) \simeq \C\langle [S_i]\,|\, S_i\in\text{Ob}(\CC_{\rm ab}),\;\text{$S_i$ simple}\rangle\,. \ee

Physically, given any object $L$ in $\CC_{\rm ab}$ (and more generally, any object in $\CC$), one may perform the path integral on a solid torus $D^2\times S^1$ with the line operator $L$ inserted at its core to produce a state in the torus state space. This defines a map
\be \text{Ob}(\CC_{\rm ab})  \overset\kappa\longrightarrow \CH(T^2)\,. \ee
The map has some very nice properties:
\begin{itemize}
\item It is linear, in the sense that direct sums, \ie\, superpositions, of line operators $A\oplus B$ map to sums of states, $\kappa(A\oplus B)=\kappa(A)+\kappa(B)$.
\item It respects the tensor product, in the sense that $\kappa(A\otimes B)=\kappa(A)\cdot\kappa(B)$, where the product of states in $\CH(T^2)$ on the RHS is defined by the path integral on $S^1$ times a pair-of-pants, with incoming boundary $T^2\sqcup T^2$ and outgoing boundary $T^2$.
\item It factors through the Grothendieck ring $K_0(\CC_{\rm ab})$, roughly because the states on the RHS cannot detect the difference between direct sums of line operators $A\oplus B$ and nontrivial extensions (bound states) $0\to A\to C\to B\to 0$.
\item Assuming that $\CC_{\rm ab}$ has a trivial internal cohomological grading, the image of the map $\kappa$ is contained in the cohomological-degree-zero subspace of the state space $\CH(T^2)$. 
\end{itemize}
Putting this together with the general isomorphism $\CH(T^2)\simeq HH_\bullet(\CC)$, we find that $\kappa$ induces  
\be  K_0(\CC) \overset{ch}\longrightarrow HH_0(\CC) \subseteq \CH(T^2)\,. \ee

If the category of line operators is semisimple, then this map is an equivalence. It can be seen rather explicitly. Let $L = \bigoplus_{i=1}^d S_i$ be a direct sum of simple objects in $\CC=\CC_{\rm ab}$. Then $L$ generates the category $\CC$, and its endomorphism algebra is just generated by the projections $\pi_i:L\to S_i$ to each simple summand,
\be A = \text{End}(L) = \C\langle \pi_i\,|\, \pi_i\pi_j = \delta_{ij} \pi_j\rangle \ee
The Hochschild homology of $\CC$, computed from the algebra $A$, is concentrated in degree zero, and has a basis given by the local operators $\pi_i$,
\be HH_\bullet(\CC) = HH_0(\CC) \simeq \C\langle \pi_i\rangle_{i=1}^d \simeq \C^d\,.\ee
More so, placing the simple line operator $S_i$ around the core of a solid torus is equivalent to placing $L$ with a single insertion of $\pi_i$. (This is because we can write $\pi_i = \pi_i \cdot  \pi_i$, and then ``slide'' the second operator $\pi_i$ all the way around the circle to project $L$ to $S_i$.) Therefore, we obtain an isomorphism
\be  ch:K_0(\CC) \to HH_0(\CC)\,,\qquad ch([S_i]) = \pi_i\,. \label{ch-iso} \ee

In contrast, in the theory $\CT_{\rm hyper}^B$, the Grothendieck ring of the non-semisimple category $\CC_1\simeq \C[X,Y]\text{-mod}$, is one-dimensional, spanned by equivalence class of the trivial line operator~$[V]$\,,
\be K_0(\CC_1) \simeq \C\langle [V]\rangle\,, \qquad\text{with}\quad [V]\cdot[V]=[V]\,.\ee
However, the degree-zero part of the state space is four-dimensional, $HH_0(\CC_1)\simeq \C\langle 1,\chi_1^X,\chi_1^Y,\chi_1^X\chi_1^Y\rangle$.
We described several other line operators in Section \ref{sec:C1-hyper}. $W_+$ and $W_-$ are quasi-isomorphic to complexes $V\overset{Y}\to V$ and $V\overset{X}\to V$, respectively; thus there are exact sequences $0\to V\to V\to W_{\pm}\to 0$, forcing the classes of these object to be zero, $[W_\pm]=0$. Similarly, $P$ is quasi-isomorphic to a complex $W_+\to W_-$, forcing $[P]=0$. Altogether, we find
\be ch:\begin{array}{ccc} V & \mapsto & 1 \\ W_\pm,P &\mapsto & 0 \end{array} \in HH_0(\CC_1)\,. \ee
The Grothendieck ring misses all the other fermionic states, and all the states generated by bulk local operators $X,Y$, in the torus state space.

\subsubsection{Flavor symmetry and deformations}
\label{sec:Hoch-flavor}

We have seen that, in the presence of flavor symmetry $G$, both categories of line operators and state spaces may be deformed by flat $G_\C$ connections. In the case of a torus state space, the deformations are compatible with the isomorphism $HH_\bullet(\CC) \overset\sim\to \CH(T^2)$, in the following way.

Let $\CA$ be a flat connection on $T^2$, let us choose basepointed cycles $\gamma_a,\gamma_b$ generating $\pi_1(T^2)\simeq \Z\times \Z$, and let $g_a = P\exp\oint_{\gamma_a}\CA$, $,g_b=P\exp\oint_{\gamma_b}\CA$ be the corresponding holonomies. Note that the holonomies must commute, $g_ag_b=g_bg_a$.

To relate configurations of line operators to states in the torus state space, we must choose a way to fill in $T^2$ to a solid torus, placing line operators along its core. This choice breaks the symmetry of the torus. Suppose we choose a filling such that $\gamma_a$ is contractible in the solid torus, and $\gamma_b$ runs parallel to the core of the solid torus. Then
\begin{itemize}
\item The line operators we insert along the core of the torus must belong to the deformed category $\CC_{g_a}$ described in Section \ref{sec:flavor-line}. 
\item $Q$-closed local operators placed on a line operator $L$ wrapping the core will no longer obey \eqref{atheta}, \emph{i.e.} their translations along the core will no longer be locally constant. Rather, translations will be covariantly constant with respect to the flat connection $\CA$ (or rather, an extension of $\CA$ into the solid torus). In particular, a single local operator $a$ will satisfy
\be a(2\pi) = g_b\cdot a(0) + Q\text{-exact}\,, \ee
and a pair $a_1,a_2$ will obey $a_1(0)a_2(\epsilon) = a_1(0)(g_b\cdot a_2)(2\pi -\epsilon) = (-1)^{|a_1||a_2|} (g_b\cdot a_2)(-\epsilon) a_1(0)$ up to $Q$-exact terms, or simply
\be a_1a_2 = (-1)^{|a_1||a_2|}(g_b\cdot a_2) a_1\,, \ee 
deforming the RHS of \eqref{a1a2}. In general, covariance may be encoded in a deformed Hochschild differential
\begin{align}
d_H^{g_b}(a_1\otimes\cdots\otimes a_{n+1})\; :=\; &  (a_1a_2)\otimes\cdots \otimes a_{n+1} + \ldots +(-1)^{n+1}a_1\otimes \cdots\otimes (a_na_{n+1}) \notag \\
& + (-1)^{|a_{n+1}|(|a_1|+...+|a_n|)+n} ((g_b\cdot a_{n+1})a_1) \otimes\cdots\otimes a_n \label{dHA}
\end{align} 
\end{itemize}
Altogether, if $L_{g_a}$ is a generator of the category $\CC_{g_a}$, with (dg) endomorphism algebra $A_{g_a} := \text{End}^\bullet_{\CC_{g_a}}(L_{g_a})$, then we can define twisted Hochschild homology
\be HH_\bullet^{g_b}(\CC_{g_a}) := H^\bullet(C_\bullet(A_{g_a},d+d_H^{g_b})\,, \label{HHA} \ee
and we expect an equivalence
\be HH_\bullet^{g_b}(\CC_{g_a}) \overset\sim\to  \CH(T^2,\CA)\,.  \label{HHT2A} \ee

Note that having commuting holonomies $g_ag_b=g_bg_a$ is necessary for the deformed Hochschild homology \eqref{HHA} to be defined.
The deformed differential $d_H^{g_b}$ only makes sense if the category $\CC_{g_a}$ actually preserved the one-parameter subgroup of the flavor symmetry generated by $g_b$, so that endomorphism algebras of objects will have an action of $g_b$. This in turn is true precisely when $g_a$ and $g_b$ commute.

We also remark that the symmetry of the two holonomies $g_a,g_b$ that is manifest in the torus state space on the RHS of \eqref{HHT2A} should also hold (nontrivially) in the twisted homology on the LHS. In particular, we expect an isomorphism 
\be \label{HH-iso} HH_\bullet^{g_a^{-1}}(\CC_{g_b}) \simeq HH_\bullet^{g_b}(\CC_{g_a}) \,. \ee
More generally, for any $\varphi = \bsp p&q\\r&s\esp\in SL(2,\Z)$, with $\varphi(g_a,g_b) := (g_a^pg_b^q,g_a^r g_b^s)$, we expect  $HH_\bullet^{\varphi(g_b)}(\CC_{\varphi(g_a)}) \simeq  HH_\bullet^{g_b}(\CC_{g_a})$.

Let's verify a particular instance of \eqref{HHT2A} and \eqref{HH-iso} for our toy model $\CT_{\rm hyper}^B$. Consider a flat connection on $T^2$ with generic diagonal holonomy $g_a=\text{diag}(e^\alpha,e^{-\alpha})$  around one cycle and trivial holonomy $g_b=1$ around the other. We expect that the torus state space \eqref{H-hyper-A} becomes one-dimensional, supported in cohomological degree zero,
\be \CH_{Q_B}(T^2,\CA) \simeq \C\qquad (\text{in degree 0})\,. \label{HT2-hyper-A} \ee
Comparing this with Hochschild homology, we know on one hand from \eqref{Ca-gen} that the category $\CC_{g_a}$ is isomorphic to the ``trivial'' category  $\text{Vect}$. This is a semisimple category with one simple object, so its Hochschild homology is
\be HH_0(\CC_{g_a}) = \C\,,\qquad HH_n(\CC_{g_a}) = 0\quad (n\neq 0)\,, \ee
in agreement with \eqref{HT2-hyper-A}.

On the other hand, if we swap the cycles, we should get the same result by computing $g_a^{-1}$-twisted Hochschild homology of $\CC_1 = \C[X,Y]${-mod}. Let's choose the trivial line operator as a generator, with endomorphism algebra $A = \C[X,Y]$. At the tail end of the Hochschild complex $A^{\otimes 2}\to A$, the twisted differential sends
\be \label{d-twisted-eg} d_H^{g_a^{-1}}:\begin{array}{ccl}p(X,Y)\otimes X &\mapsto& p(X,Y)X-e^{-\alpha} X p(X,Y) = (1-e^{-\alpha})Xp(X,Y) \\[.2cm]
 p(X,Y)\otimes Y &\mapsto & p(X,Y)Y-e^{\alpha} Y p(X,Y) = (1-e^{\alpha})Yp(X,Y)  \end{array} \ee
 for any polynomial $p(X,Y)$. Since $1-e^{\pm \alpha}$ are some nonzero constants, any elements of $A$ that are divisible by $X$ or $Y$ become cohomologically trivial, whence
 \be \label{HH0-twist} HH_0^{g_a^{-1}}(\CC_1) \simeq \frac{\C[X,Y]}{\C[X,Y](X,Y)} \simeq \C\langle 1\rangle\,. \ee
All higher cohomology groups vanish. (For example, the elements $p(X,Y)\otimes X$ and $p(X,Y)\otimes Y$ that used to be nontrivial in $HH_{-1}$ are no longer closed.) After shifting cohomological degree by for $U(1)_H$ charge, which does nothing to the identity operator in \eqref{HH0-twist}, we again find agreement with \eqref{HT2-hyper-A}. Thus $HH_\bullet(\CC_{g_a}) \simeq HH_\bullet^{g_a^{-1}}(\CC_1) \simeq \CH_{Q_B}(T^2,\CA)$.

\subsubsection{Hochschild cohomology, centers, and Drinfeld-Reshetikhin map}
\label{sec:DR}

Finally, we make some remarks on a dual construction of the torus state space in a cohomological TQFT, which connects to discussions in the literature involving centers of quantum groups, as well as the ``Drinfeld-Reshetikhin map'' in the theory of tensor categories \cite{tensorcategories}.

In an unframed 3d TQFT --- meaning physically that a theory preserves the full $SU(2)$ R-symmetry required to twist on arbitrary backgrounds --- the Hochschild homology and cohomology of the category of line operators should be isomorphic, and both should equally well compute the torus state space,
\be \label{HH-isos} HH_\bullet(\CC) \simeq HH^\bullet(\CC) \simeq \CH(T^2)\,. \ee

Algebraically, Hochschild cohomology may be computed by choosing a generator $L$ of $\CC$, with dg endomorphism algebra $A$, constructing the complex
\be \label{HH-dual} \begin{array}{cccccccc} C^\bullet(A):=  & \quad 0\to & \text{Hom}_\C(\C,A) & \overset{d_H^*}\longrightarrow & \text{Hom}_\C(A,A) & \overset{d_H^*}\longrightarrow & \text{Hom}_\C(A^{\otimes 2},A) & \overset{d_H^*}\longrightarrow \cdots \\[.1cm]
\text{degrees:} && 0 && 1 && 2 \end{array} \ee
with differential acting on an element $\varphi:A^{\otimes n}\to A$ as
\begin{align} (d_H^*\varphi)(a_1\otimes \cdots a_{n+1}) =\; &a_1\varphi(a_2\otimes \cdots a_{n+1}) - \varphi(a_1a_2\otimes a_3\otimes\cdots a_{n+1})   \\
 & +\ldots + (-1)^n \varphi(a_1\otimes\cdots\otimes a_na_{n+1}) +(-1)^{n+1}\varphi(a_1\otimes\cdots\otimes a_n) a_{n+1}\,, \notag \end{align}
and setting $HH^\bullet(\CC) = H^\bullet(C^\bullet(A),d^*+d_H^*)$, where $d^*$ is induced from the internal differential on~$A$. The complex $C^\bullet(A)$ is dual to $C_\bullet(A)$ in \eqref{HH-complex}, though not in an entirely obvious way. 

Some care must be taken in interpreting the isomorphism \eqref{HH-isos}. We chose a negative degree convention for the complex \eqref{HH-complex} and a positive one for \eqref{HH-dual} for physical reasons (in order to correlate with the action of supercharges on operators and descendants); the conventions do match in twists of 3d $\CN=4$ theories, but only after adding R-charges (such as $U(1)_H$ charge in $\CT_{\rm hyper}^B$) to the Hochschild cohomological degrees.
There is also some intrinsic ambiguity in defining degrees of states in a torus state space, stemming from a choice of fermionic vacuum as well as from a choice of bosonic polarization, related to hidden choices of real parameters, as in \eqref{H-hyper-inverted}. 

Whereas the zeroth Hochschild homology group computes the co-center of  $A$, the zeroth Hochschild cohomology group computes the center. This follows from noting that $\text{Hom}_\C(\C,A)\simeq A$, and the first differential acts as $(d_H^*\,a)(b) = ba-ab$, whence $\text{ker}\,d_H^*= HH_0(\CC)=Z(A)$.

If the algebra $A$ has trivial  cohomological grading, then its center
\be Z(A) = HH_0(\CC)   \subseteq \CH(T^2) \ee
is isomorphic to the degree-zero part of the torus state space. In the mathematics literature, actions of the modular group $SL(2,\Z)$ on the centers of quantum groups at roots of unity have been defined, initially by \cite{Kerler,Lachowska-center}.  One expects them to correspond to actions of $SL(2,\Z)$ on the degree-zero part of the torus state space in TQFT's whose categories of lines are equivalent to modules for various quantum groups ($\CC\simeq A$-mod with $A$ a quantum group). An $SL(2,\Z)$ action on higher Hochschold cohomology of small quantum groups $u_q(\mathfrak{g})$ at odd roots of unity was constructed in \cite{LQ-derived}, extending \cite{Kerler,Lachowska-center}.

There is an intuitive physical description of the relation between Hochschild homology and cohomology. Let us fix a generator $L$ of the category $\CC$, with endomorphism algebra $A$ (\ie\ $A$ is the algebra of local operators bound to $L$). Rather than wrapping $L$ on a circle, we'll now place $L$ on an infinite straight line.

\begin{figure}[htb]
\centering
\includegraphics[width=5in]{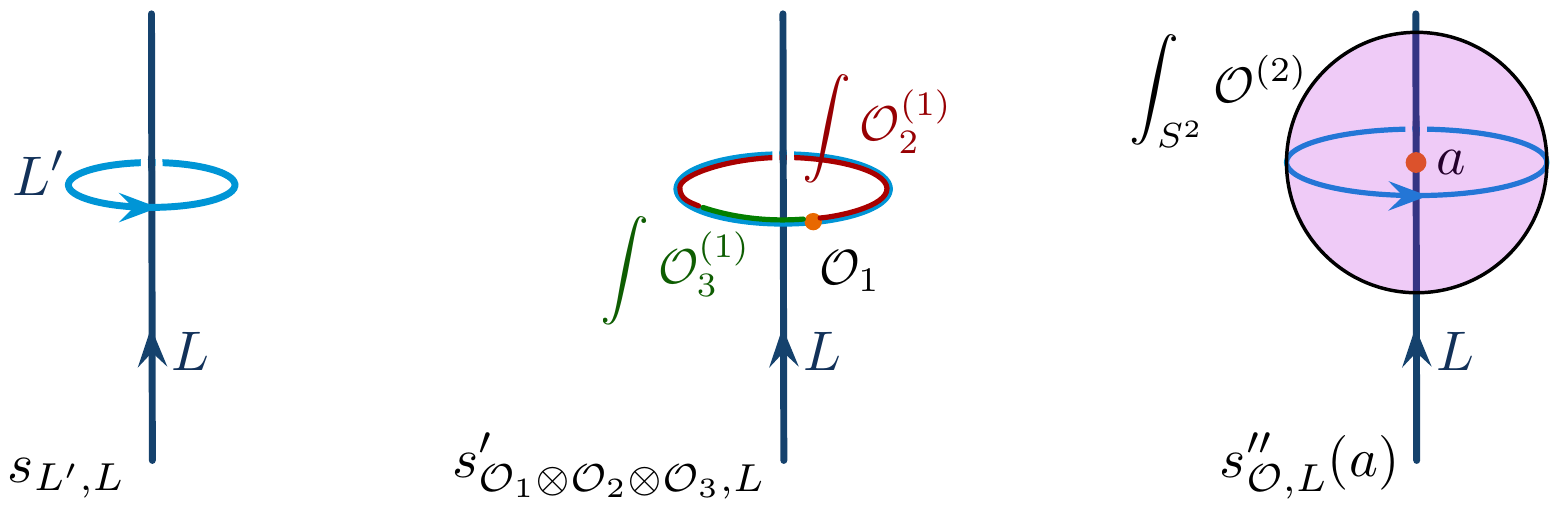}
\caption{Enhancing the Drinfeld-Reshetikhin map to include higher operations.}
\label{fig:DR}
\end{figure}

Given a \emph{second} line operator $L'$, we may wrap $L'$ on a small circle linking $L$ as on the left of Figure \ref{fig:DR}. Shrinking the circle defines a local operator bound to $L$, and thus a map
\be s_{-,L}:\begin{array}{ccc} \CC &\to & A \\ L' &\mapsto & s_{L',L}\,. \end{array} \ee
This is sometimes called the Drinfeld-Reshetikhin map. It factors through the Grothendieck group of $\CC$, and its image necessarily lands in the center of $A$, since loops as in Figure \ref{fig:DR} can be freely moved around any other local operator on $L$; thus,
\be \label{sKZ} s_{-,L} : K_0(\CC)\to Z(A)\,. \ee 
(More so, taking into account the braided-monoidal structure of $\CC$, $s_{-,L}$ becomes a map of commutative algebras.) 
If furthermore the category is semisimple, then $Z(A)\simeq \CH(T^2)$ and \eqref{sKZ} is an isomorphism, giving a dual perspective on the Chern character \eqref{ch-iso}.

If the category of line operators is not semisimple, one needs a suitably derived version of the Drinfeld-Reshetikhin map to obtain an isomorphism with the torus state space. Two steps are required. First, as depicted in the middle of Figure \ref{fig:DR}, arbitrary configurations of line operators and local operators wrapped along $S^1$ should be added to the domain. This induces a map $s'_{-,L}:HH_\bullet(\CC)\to Z(A)$ from Hochschild homology to the center of $A$. Second, as depicted on the right of Figure \ref{fig:DR}, higher descendants of the line-operators-on-loops must be considered, integrated along higher-dimensional cycles around collections of local operators on the straight line $L$. With some care, this should define a chain map
\be s''_{-,L}:C_\bullet(A) \to C^\bullet(A) \label{DR-derived} \ee 
from Hochschild chains to the Hochschild cochain complex \eqref{HH-dual} of $A$, which should induce an isomorphism on (co)homology.

In the case of $\CT_{\rm hyper}^B$, Hochschild cohomology and the derived Drinfeld-Reshetikhin map work as follows. Let us take $L$ to be the trivial line operator, generating $\CC_1=\C[X,Y]$-mod, with $A=\text{End}^\bullet(L)\simeq \C[X,Y]$. Abstractly, the Hochschild cohomology computed from the polynomial algebra $A$ is well known to contain

0) the algebra itself in degree zero (since $A$ is commutative, so $A=Z(A)$)

1) derivations $\C[X,Y]\langle \pd_X,\pd_Y\rangle$ in degree 1  (acting as maps $A\to A$)

2) the Poisson bivectors $\C[X,Y]\langle \pd_X\wedge \pd_Y\rangle$ in degree 2 (acting as maps $A^{\otimes 2}\to A$)

\noindent Accounting for the fact that $X,Y$ have $U(1)_H$ charge +1, the elements $\pd_X,\pd_Y,$ and $\pd_X\wedge \pd_Y$ are all shifted back into cohomological degree zero, giving regraded cohomology groups
\be HH^n(\CC_1) \simeq \C[X,Y]\big|_{\text{degree $n$}} \langle 1,\pd_X,\pd_Y,\pd_X\wedge \pd_Y\rangle\,,\qquad n\geq 0\,, \ee
perfectly matching the structure of the torus state space.

To analyze the Drinfeld-Reshetikhin map, let us place the trivial line operator $L$ along a straight line, and consider wrapping configurations of local operators and integrated descendants on a small loop linking $L$. We know from Section \ref{sec:Hoch-hyper} that the independent wrapped configurations consist of bulk local operators $X,Y$, the integrated fermionic descendants $\chi_1^X=\oint_{S^1} X^{(1)}$ and $\chi_1^Y=\oint_{S^1} Y^{(1)}$, and products thereof. Since $L$ is the trivial line, bringing local operators $X,Y$ to it just produces the corresponding elements of its endomorphism algebra $A$; while wrapping integrated descendants about it is trivial, since their loops are contractible. We obtain a partially derived map
\be s_{-,L}':C_\bullet(A) \to Z(A)\,,\qquad s_{-,L}': \begin{array}{ccc} X,Y &\mapsto& X,Y \\ \chi_1^X,\chi_1^Y &\mapsto& 0\,. \end{array} \ee
To fully derive the map, we must consider higher descendants of the wrapped operators. In particular, we may take a descendant $(\chi_1^X)^{(1)}$ and integrate it on an arc starting and ending on the line $L$, which is equivalent to integrating $\oint_{S^2}X^{(2)}$ on a sphere centered at a point $p$ on $L$. The operator $\oint_{S^2}X^{(2)}$ ``acts'' on any other local operators $\CO\in A$ bound to $L$ by inserting $\CO$ inside the sphere, at the point $p$. Thus $\oint_{S^2}X^{(2)}\in \text{Hom}_\C(A,A) = C^1(A)$. The relevant action was computed in \cite{descent}:
\be \oint_{S^2}X^{(2)}\, X(p) = 0\,,\qquad \oint_{S^2}X^{(2)}\, Y(p) = 1\,, \ee
whence $\oint_{S^2}X^{(2)}$ acts as $\pd_Y$. Similarly, $\oint_{S^2}Y^{(2)}$ acts as $-\pd_X$. The complete map sends
\be s''_{-,L} : C_\bullet(A) \to C^\bullet(A)\,,\qquad s_{-,L}': \begin{array}{cccl} X,Y &\mapsto& X,Y & \in C^0(A)  \\ \chi_1^X,\chi_1^Y &\mapsto& \pd_Y,-\pd_X & \in C^1(A) \\
\chi_1^X\chi_1^Y &\mapsto& \pd_X\wedge \pd_Y & \in C^2(A)\,.   \end{array}
\ee
and induces an isomorphism on (co)homology.

\section{Quantum groups and non-semisimple TQFTs}
\label{sec:Uqsl2}

In this section, we begin by reviewing some of the structure of finite-dimensional representations of the De Concini-Kac quantum group $U_q(\mathfrak{sl}_2)$ at an even root of unity $q=e^{i\pi/k}$, $k\geq 2$. We let
\be \CCab:= U_q(\mathfrak{sl}_2)\text{-mod} \ee
be the non-semisimple \emph{abelian} category of finite-dimensional representations, on which the Frobenius center (defined in \eqref{Frob-alg} below) acts semisimply. We let  $\CC = D^b\CCab$ be its derived category. We explain why $\CC$ has some of the right features to be a category of line operators in a 3d topological QFT that couples to flat $PGL(2,\C)$ connections, and explicitly compute the sphere and torus state spaces that would appear in such a QFT. After deforming by a generic abelian flat connection, we compute the putative state space in any genus.

In the second part of this section, we outline how the CGP construction and its later developments define an axiomatic 3d TQFT coupled to abelian flat connections, for a maximal torus of $PGL(2,\C)$, using an unrolled analogue of $\CCab$. As noted in the Introduction, unrolling provides at least one route to defining a braiding --- which is necessary in order to obtain a full TQFT. We comment on the physical significance of unrolling, comparing with the abstract description of line operators from Section \ref{sec:flavor-line}. We also describe a precise set of choices that define a CGP TQFT whose state spaces correspond to the \emph{cohomological-degree-zero} parts of the state spaces obtained from the derived category $\CC$.

For concreteness, we only consider the Lie algebra $\mathfrak g=\mathfrak{sl}_2$ in this section.  There is a parallel development for $\mathfrak g=\mathfrak{sl}_n$ and algebras of other types, as discussed and briefly reviewed in the Introduction.

\subsection{Modules and flat connections}
\label{sec:U-modules}

Given $k\geq 2$ and a corresponding even root of unity $q=e^{i\pi/k}$, we let $\mb U:= U_q(\mathfrak{sl}_2)$ be the algebra over $\C$ with generators $E,F,K^{\pm1}$ and relations
\be K^{-1}K=KK^{-1}=1\,,\qquad \begin{array}{c} KE=q^2EK \\[.1cm] KF=q^{-2}FK\,, \end{array}\qquad [E,F] = \frac{K-K^{-1}}{q-q^{-1}}\,.\ee
It may further be given the structure of a Hopf algera, with coproduct, counit, and antipode
\begin{align}
	\Delta(E)&= 1\otimes E + E\otimes K\,, 
	&\varepsilon(E)&= 0\,, 
	&S(E)&=-EK^{-1}\,, 
	\notag \\
	\Delta(F)&=K^{-1} \otimes F + F\otimes 1\,,  
	&\varepsilon(F)&=0,& S(F)&=-KF\,,
	\\
	\Delta(K)&=K\otimes K\,,
	&\varepsilon(K)&=1\,,
	& S(K)&=K^{-1}\,.
	\notag 
\end{align}

The center $Z(\mb U)$ is large and rather intricate \cite{DCK,DCKP,Beck}. It includes quantum analogues of Casimir operators, generating what is known as the Harish-Chandra center. It also includes the powers $E^k,F^k,K^{2k}$, generating what is known as the Frobenius center
\be \label{Frob-alg} Z_{\rm Fr}(\mb U) = \C\langle E^k,F^k,K^{\pm 2k}\rangle\,. \ee
As discussed in Section \ref{sec:flat-intro} of the Introduction, $Z_{\rm Fr}(\mb U)$ is isomorphic to the algebra of functions on a Zariski-open subset $PGL(2,\C)'$ of the Langlands-dual group $PGL(2,\C)$ (viewed as a complex variety),
\be \label{Frob} \text{Spec}(Z_{\rm Fr}(\mb U)) \;\simeq\; \C\times \C\times \C^* \simeq PGL(2,\C)' \underset{\text{open}}\subset PGL(2,\C)\,.\ee
One way to specify this subset is by associating points $(E^k=e,F^k=f,K^{2k}=\kappa)\in \text{Spec}(Z_{\rm Fr}(\mb U))$ with \cite{KashaevReshetikhin,biquandles}
\be g =  \bp \kappa & -\kappa e \\ f & \;1-e f \ep\in  PGL(2,\C)\,. \label{efkg} \ee
We will use points $(e,f,\kappa)\in \C\times\C\times \C^*$ and their images $g\in PGL(2,\C)$ given by \eqref{efkg} interchangeably.

(Note that $K^{\pm k}$ are \emph{also} central in $\mb U$. We make the precise choice of center \eqref{Frob} over which to ``fiber'' in order to match the structure that eventually appears in the QFT $\CT_{2,k}^A$. Other choices/modifications are possible on both sides.)

For each $g\in PGL(2,\C)'$, define the central quotient $\mb U_g:= \mb U/(E^k-e,F^k-f,K^{2k}-\kappa)$; and let  $\CCabg := \mb U_g\text{-mod}$  be the corresponding categories of finite-dimensional representations. For each $g$, $\CCabg$ is the subcategory of $\CCab$ containing the modules on which $E^k,F^k,K^{2k}$ take fixed constant values $(e,f,\kappa)$. Then
\be \CCab \to PGL(2,\C)' \label{U-sheaf} \ee
has the structure of a coherent sheaf of categories, with stalk (or `fiber') categories $\CCabg$. In particular, $\CCab$ decomposes as a direct sum of its subcategories $\CCabg$
\be \label{U-sheaf-sum} \CCab\, \simeq \bigoplus_{g\in PGL(2,\C)'} \CCabg\,, \ee
which simply says that every module is a direct sum of modules with fixed values of $E^k,F^k,K^{2k}$, and that there are no morphisms (no linear maps commuting with the action of $\mb U$) between modules with different central values. These are standard results in representation theory.

Each algebra $\mb U_g$ is finite dimensional (of dimension $4k$), which makes its module category $\CCab_g$ particularly nice. It implies, for example, that $\CCab_g$ must have the same number of simple modules and indecomposable projective modules. Relating the categories $\CCab_g$ for \emph{different} $g$ is harder; this was studied in \cite{DCK,DCKP,Beck}, and is part of the overall structure of the coherent sheaf \eqref{U-sheaf}. One finds that
\begin{itemize}
\item Each $\CCabg$ has exactly $2k$ simple modules.
\item As a category (not a braided tensor category), each $\CCabg$ depends only on the conjugacy class of $g$; \emph{i.e.} for each $g\in PGL(2,\C)$ and each $h\in PGL(2,\C)$ such that $hgh^{-1}\in PGL(2,\C)'$, there are isomorphisms
\be \CCabg \simeq \CCab_{hgh^{-1}}\,. \label{U-iso} \ee
\item If $g$ has distinct eigenvalues, $\CCabg$ is a semisimple category. Its simple objects are automatically projective as well.
\item Otherwise, $\CCabg$ may not be semisimple. The most non-semisimple case is $g=1$, meaning $E^k=F^k=0$ and $K^{2k}=1$, giving
\be \CCab_{1}  = \mb u\text{-mod}\,,\qquad \mb u := \mb U_1 =  \mb U/(E^k,F^k,K^{2k}-1)\,, \ee
where $\mb u$ is known as the restricted quantum group. (In part of the literature, $\mb u$ is also just called the small quantum group.) The category $\CCab_{1}$ has $2k$ simple modules that may be extended in interesting ways to produce the $2k$ indecomposable projectives.
 
\end{itemize}

We denote the corresponding derived categories as $\CC=D^b \CCab$ and $\CC_g=D^b \CCabg$. The derived category similarly forms a sheaf
\be \CC \to PGL(2,\C)'\,,\ee
with stalks/fibers $\CC_g$. In particular, the direct sum decomposition \eqref{U-sheaf-sum} continues to hold at the derived level. This is essentially due to the absence of derived morphisms (higher Ext groups) among $\CCabg$ for different $g$.

We would of course like to identify $\CC$ with the category of line operators in a 3d topological QFT that can be deformed by flat $PGL(2,\C)$ connections. Heuristically, each stalk $\CC_g$ should contain the line operators that exist in the presence of a flat connection $\CA$ with holonomy $g$, as in Section \ref{sec:flavor-line}. There are some notable similarities with the abstract structure of Section \ref{sec:flavor-line}. Categories labelled by conjugate holonomies are indeed expected to be isomorphic, by \eqref{C-iso}. Moreover, as one varies the holonomy $g$, the various $\CC_g$ are indeed expected to form a coherent sheaf \eqref{C-sheaf}.

\subsubsection{Tensor products and geometry}

The tensor product of modules in $\CCab$ is defined by using the coproduct in $\mb U$. Namely, given modules with underlying vector spaces $M,N$, their tensor product is just the tensor product of vector spaces $M\otimes N$, with the action of $\mb U$ given by
\be a \cdot (m\otimes n) := \Delta(a)(m\otimes n) \ee
for $a\in \mb U$ and  $m\in M$, $n\in N$. The tensor product in the derived category $\CC$ is defined the same way, upon replacing $M$ and $N$ with complexes (or dg vector spaces).%
\footnote{It is not necessary to further derive the abelian tensor product. Since it is just a tensor product of vector spaces over $\C$, higher Tor groups automatically vanish.}

If each stalk $\CC_g$ indeed corresponds to the category of line operators in a QFT with $PGL(2,\C)$ connections, we would expect that the tensor product multiplies holonomies,
\be \label{expect-tensor} \otimes:\CC_g\boxtimes \CC_{g'} \to \CC_{gg'}\,, \ee
matching the left of Figure \ref{fig:hol-fusion}. In other words, if $M\in \CC_g$ and $N\in \CC_{g'}$ then $M\otimes N\in \CC_{gg'}$.

Part of the key to realizing \eqref{expect-tensor} is a special identity of the coproduct
\be \Delta(K^{2k}) = K^{2k}\otimes K^{2k}\,,\quad \Delta(E^{k}) = 1\otimes E^{k}+E^k\otimes K^k\,,\quad \Delta(F^k) = K^{-k}\otimes F^k + F^k\otimes 1\,, \ee
which holds at a $2k$-th root of unity due to cancellations in the ``cross terms'' of $\Delta(E^k)$ and $\Delta(F^k)$.
It is then easy to check that \eqref{expect-tensor} holds when $g$ is restricted to be diagonal. The matrix multiplication
\be g=\bp \kappa & 0 \\ 0 & 1 \ep\,,\quad  g=\bp \kappa' & 0 \\ 0 & 1 \ep \quad\Rightarrow\quad
 gg' = \bp \kappa\kappa' & 0 \\ 0 & 1 \ep \ee
precisely agrees the coproduct of central elements: if $(E^k,F^k,K^{2k})=(0,0,\kappa)$ on $M$ and $(E^k,F^k,K^{2k})=(0,0,\kappa')$ on $N$, then $(\Delta(E^k),\Delta(F^k),\Delta(K^{2k})) = (0,0,\kappa\kappa')$ on $M\times N$.

For general  $g$, with $e,f\neq 0$, the expected relation  \eqref{expect-tensor} does not hold on the nose. This is ok. The precise relation \eqref{expect-tensor} assumed that we were measuring holonomies from a common basepoint, but there are other combinatorial prescriptions for keeping track of them.  A combinatorial scheme was developed by \cite{KashaevReshetikhin} that \emph{does} correctly relate flat connections in the complement of multiple lines to the tensor-product structure of $\mb U\text{-mod}$; 
it was considered in more detail in \cite{biquandles}. Roughly speaking, it is necessary to keep track of multiple basepoints and partial holonomies among them. We will not require further details of the general construction in this paper.

\subsubsection{Generic stalks}
\label{sec:V-generic}

We elaborate a bit on what the modules in $\CCabg$ actually look like. 

If $g$ is generic, we may assume without loss of generality (thanks to the conjugation isomorphisms \eqref{U-iso}) that $g$ is diagonal, of the form%
\footnote{Since the group is $PGL(2,\C)$, we have the freedom to multiply by any multiple of the identity, and could choose a square root $\kappa^{1/2}$ and equivalently write $g = \text{diag}\big( \kappa^{1/2},1/\kappa^{1/2}\big)$.
 We use the representative \eqref{gkappa} because it has the advantage of being manifestly algebraic.}
\be \label{gkappa} g = \bp \kappa & 0 \\ 0 & 1 \ep\,,\qquad \text{with} \quad \kappa =e^\alpha \in  \C^*\backslash 1\,. \ee
We want to describe the stalk $\CCabg = \mb U_g\text{-mod}$, which contains modules on which $E^k,F^k$ act as zero and $K^{2k}$ acts as multiplication by $\kappa$.

It is useful to choose a logarithm $\alpha\in \C\backslash \Z$ such that $\kappa=e^{2\pi i\alpha}$.  It should be clear that all results below only depend on the class of $\alpha$ in $\C/\Z$.

Let us consider the Verma modules of $\mb U_g$, \emph{i.e.} modules generated by a vector $v$ that satisfies $Ev=0$ and $Kv=b\,v$ for some $b\in \C^*$. Since $K^{2k}=\kappa$, the ``highest weight'' $b$ must satisfy $b^{2k}=\kappa$, whence there are exactly $2k$ choices: $b=q^{\alpha+n}$ for $n=0,...,2k-1$. Let $V_{\alpha,n}$ denote the Verma module with highest weight $q^{\alpha+n}$\,.

Due to the central constraint $F^k=0$, each of the Verma modules $V_{\alpha,n}$ is exactly $k$-dimensional, with basis given by $\{v,Fv,...,F^{k-1}v\}$. Pictorially,
\be  \label{V-generic} V_{\alpha,n} \;=\quad \raisebox{-.7in}{$\includegraphics[width=2in]{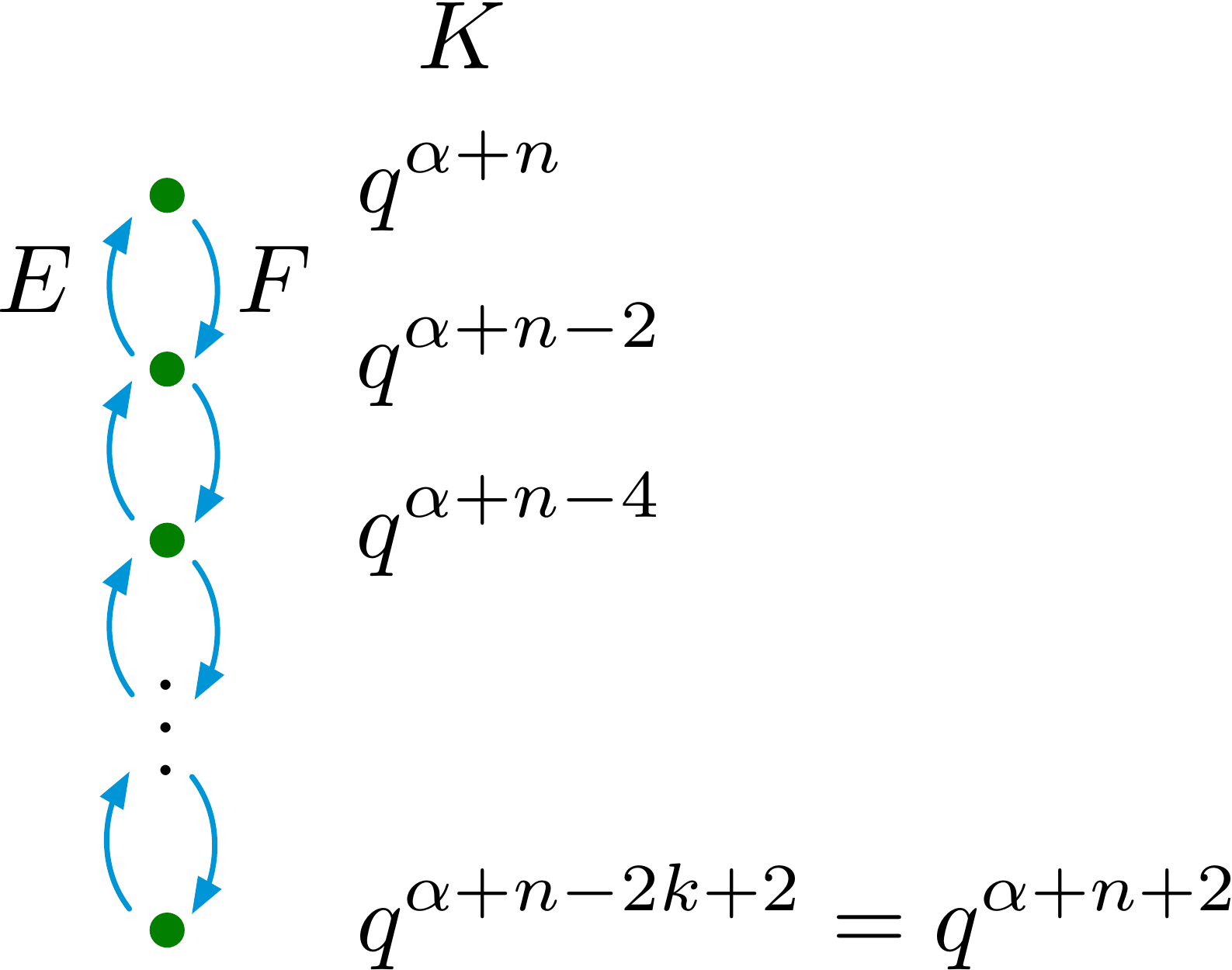}$} \simeq \mb U/(E,K-q^{\alpha+n},F^k)\,. \ee
It is straightforward to see that $\alpha\notin \Z$ implies that each $V_{\alpha,n}$ is irreducible (\emph{i.e.} simple). Each $V_{\alpha,n}$ is also projective, and altogether $\CCabg$ is a semisimple category generated by the $2k$ modules $V_{\alpha,n}$. The derived category $\CC_g$ is thus essentially equivalent to $\CCabg$, as discussed in Section \ref{sec:ss}.

Given two diagonal holonomies $g=\text{diag}(e^{2\pi i \alpha},1)$, $g'=\text{diag}(e^{2\pi i \alpha'},1)$, we know from \eqref{expect-tensor} that the tensor product will map $M\in \CCabg$, $N\in \CCab_{g'}$ to $M\otimes N\in \CCab_{gg'}$. As long as $\alpha,\alpha',\alpha+\alpha'\notin \Z$, all three categories $\CCab_{g}$, $\CCab_{g'}$, $\CCab_{gg'}$ are semisimple, and the tensor product of modules is given by the simple formula
\be V_{\alpha,n}\otimes V_{\alpha',n'} \;=\!\! \bigoplus_{\footnotesize\begin{array}{c} 0\leq m\leq 2k-1\\ n+n'+m\equiv 0\,\text{mod}\,2 \end{array}} \!\! V_{\alpha+\alpha',m} \label{V-ss-fusion} \ee 

The modules $V_{\alpha,n}$ for generic $\alpha$ are the ones used in the construction of the ADO invariant \cite{ADO,Murakami} of links in $S^3$. We recall that two technical challenges to overcome were the vanishing of quantum dimensions $\text{dim}_q V_{\alpha,n} = \text{Tr}_{V_{\alpha,n}}K=0$ and the ill-definedness of the universal R-matrix. These are dealt with systematically in the CGP TQFT.

\subsubsection{Identity stalk, $\mathbb Z_2$ symmetry, and Grothendieck ring}
\label{sec:V-1}

The category $\CCab_1=\mb u\text{-mod}$ of modules for the restricted quantum group has a very different structure.%
\footnote{The second author (T.D.) is very grateful to P. Etingof for first introducing him to the structure of $u_q(\mathfrak{sl}_2)$ modules and their Grothendieck ring some years ago, in the course of joint discussions with S. Gukov and C.~Vafa on the quantum Hall effect. That introduction ultimately inspired some of the current presentation.} %
We now have $E^k=F^k=0$ and $K^{2k}=1$.

The category $\CCab_1$ still contains $2k$ Verma modules, of the same form \eqref{V-generic}. However, most of the Verma modules are reducible. Their maximal simple quotients are $2k$ modules that we denote $S_n^\pm$ for $n=1,2,...,k$, which have the form
\be \label{reps-small} \raisebox{-.7in}{$\includegraphics[width=5.5in]{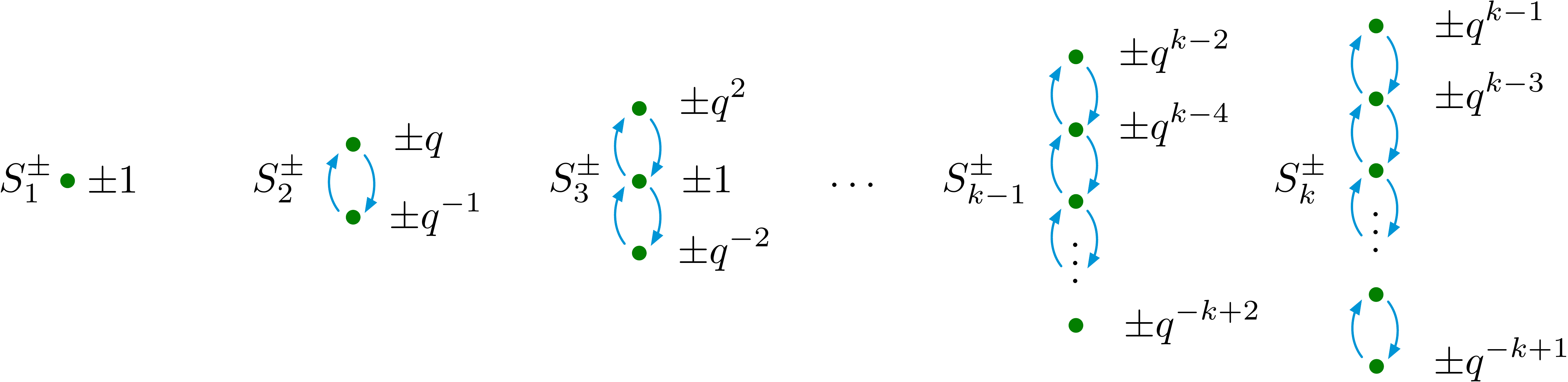}$}\,, \ee
of dimensions $\text{dim}\,S_n^\pm=n$.
The two modules $S_k^\pm$ are Vermas themselves. Each remaining Verma is an extension of a simple $S_n^\pm$ (for $n=1,...,k-1$) by its ``complement'' $S_{k-n}^\mp$. The $2k$ modules in \eqref{reps-small} are all the simple modules of  $\CCab_1$.

There are also $2k$ indecomposable projective modules, which we denote $P_n^\pm$ for $n=1,...,k$.
The two modules $S_k^\pm = P_k^\pm$ are already projective. The remaining projectives are successive extensions involving four simple composition factors, encoded by the Loewy diagrams (see Appendix A.4 of \cite{Creutzig:2013hma} for background on Loewy diagrams)
\be \label{reps-proj} \raisebox{-.7in}{$\includegraphics[width=5.5in]{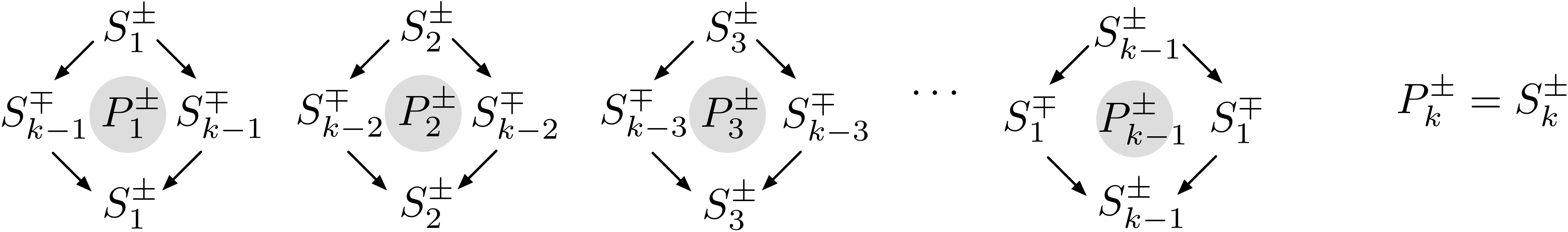}$} \ee
As an example of how to read these diagrams, we note that $P_1^+$ has a copy of $S_1^+$ (appearing on the top) as a submodule. The quotient $P_1^+/S_1^+$ has $S_{k-1}^-\oplus S_{k-1}^-$ as a submodule. And the successive quotient $(P_1^+/S_1^+)/(S_{k-1}^-\oplus S_{k-1}^-)$ is isomorphic to another copy of $S_1^+$ (appearing on the bottom). 

Since the identity matrix $g=1$ satisfies $gg=g$, the tensor product \eqref{expect-tensor} preserves $\CCab_1$, making it a tensor category in its own right. The tensor-identity is given by the trivial representation $\mb 1=S_1^+$. Tensor products of other modules are described \emph{e.g.} in \cite{Feigin:2005xs} and summarized later (from an equivalent VOA perspective) in \eqref{tensorproducts}.

Tensor products involving the two one-dimensional modules $S_1^\pm$ are particularly simple. These modules are invertible objects of $\CCab_1$, and generate a categorical representation of $\mathbb Z_2$: $S_1^+\otimes S_1^+\simeq S_1^+$,\; $S_1^+\otimes S_1^- \simeq S_1^-\otimes S_1^+\simeq S_1^-$,\;  $S_1^-\otimes S_1^-\simeq S_1^+$. By tensoring $S_1^\pm$ with other objects, one obtains a $\Z_2$ action on the entire category $\CC^{\rm ab}$, fixing every stalk $\CC^{\rm ab}_g$.
Such a $\Z_2$ symmetry of a category is usually referred to as a $\Z_2$ one-form symmetry in physics \cite{GKSW}.

The Grothendieck ring $K_0(\CCab_1) = K_0(\CC_1)$ is also fairly simple to describe; it was computed (\eg) in \cite{Feigin:2005xs}, and takes the following form. Let $X=[S_2^+]$ denote the classes of the two-dimensional module. Then $K_0(\CCab_1)$ is isomorphic to the ring of polynomials in $X$, with a single relation $f_k(X)=0$,
\be \label{U-ring}  K_0(\CC_1) \simeq \C[X]/(f_k(X))\,,\qquad f_k(X) := (X^2-4)\prod_{j-1}^{k-1}(X-q^j-q^{-j})^2\,. \ee
For example, $f_2 = (X^2-4)X^2$, $f_3 = (X^2-4)(X^2-1)^2$, $f_4 = (X^2-4) X^2(X^2-2)^2$. We have
\be \text{dim}\,K_0(\CC_1) = \text{deg}\,f_k = \text{\# simples} = \text{\# indecomp. projectives} = 2k \ee
This is true for the module category of any finite-dimensional algebra.

We noted in the Introduction that $\CCab_1$ is related to many classic quantum invariants. We can now be more precise about this.

The semisimplified category $\CC_{s.s.}$ used by Reshetikhin-Turaev \cite{RT} in their classic work on quantum invariants --- and specifically used in the case of $\mathfrak g=\mathfrak{sl}_2$ to construct the colored Jones polynomials of links in the sphere  \cite{KirbyMelvin} --- is obtained from $\CCab_1$ in two steps. First, one quotients out by every module with vanishing quantum dimension. This ``sets to zero'' $S_k^\pm$, and all nontrivial extensions of simples, including all the Verma modules and all the projectives. It leaves behind a semisimple category containing just the simples $S_1^\pm,...,S_{k-1}^\pm$. Then one passes to a subcategory containing just the $S_1^+,...,S_{k-1}^\pm$. These are the ``colors'' that decorate link strands in the RT TQFT; they correspond to the irreducible representations of $SU(2)$ that label Wilson lines in $SU(2)_{k-2}$ Chern-Simons theory.

The simple projective module $S_k^+$ by itself generates a braided tensor subcategory of $\CCab_1$, and it was shown by Murakami-Murakami \cite{MM-volume} that it can be used to construct the Kashaev invariant \cite{Kashaev}. This ultimately related the Kashaev invariant of a link to an analytic continuation of the colored Jones polynomial, and extended Kashaev's Volume Conjecture to a statement about Jones polynomials.

\subsubsection{Identity stalk and the flag manifold}
\label{sec:V-flag}

Since $\CCab_1$ is not semisimple, the derived category $\CC_1=D^b(\CCab_1)$ is quite rich. We can give it geometric characterization using a beautiful result of \cite{ABG,BL}: there is a derived equivalence
\be \label{U-flag} \CC_1 \simeq \text{Vect}^{\oplus 2} \oplus \big[\text{Coh}(T^*[2]\mathbb{P}^1)\big]^{\oplus k-1}\,. \ee
Here `Vect' denotes the semisimple category with a single simple object, and $\text{Coh}(T^*[2]\mathbb{P}^1)$ denotes a dg enhancement of the \emph{derived} category of coherent sheaves on the cotangent bundle $T^*\mathbb{P}^1$, with cotangent fibers shifted in cohomological degree, so that linear functions on the fibers are in degree 2.%
\footnote{The same sorts of cohomological shifts appeared in Section \ref{sec:hyper-lines}, when describing the B-twist of a hypermultiplet. The category of line operators there was $\C[X,Y]\text{-mod}$, with $X,Y$ in degree 1; it could also have been written as $\text{Coh}(T^*[2](\C[1]))$, in a manner analogous to \eqref{U-flag}.} %
The direct sum in \eqref{U-flag} indicates as usual that there are no morphisms between object in different summands.

We emphasize that the equivalence \eqref{U-flag} just applies to categories, and does not respect any braided tensor structure. Giving a natural braided tensor structure to the RHS of \eqref{U-flag} is an interesting open question, which the QFT in this paper might help address.

We briefly explain a bit of the structure behind the decomposition \eqref{U-flag}, and elaborate further in Appendix \ref{app:U}. We saw in Section \ref{sec:V-1} that only certain pairs of simple modules have extensions that build Vermas and projectives. This is more generally true:\,\footnote{We remind the reader that, throughout the paper, Hom$^\bullet$ denotes derived Hom, \ie\ the morphism space in the derived/dg category. In particular, $\text{Hom}^0=\text{Hom}$ and $\text{Hom}^{i>0}=\text{Ext}^i$.}
\be \text{Hom}^\bullet(S_j^{\epsilon},S_{j'}^{\epsilon'}) = 0\quad \text{unless} \quad j=j'\,,\;\epsilon=\epsilon' \quad\text{or}\quad j+j'=k-1\,,\;\epsilon\neq \epsilon'\,.  \ee
The abelian and derived categories thus decompose as direct sums of $k+1$ ``blocks''
\be \label{block-C1}  \CCab_1 \simeq \CB_k^{\text{ab}\,+}\oplus \CB_k^{\text{ab}\,-} \oplus \bigoplus_{j=1}^{k-1} \CB_j^{\rm ab}\,,\qquad\quad  \CC_1 \simeq \CB_k^+\oplus \CB_k^- \oplus \bigoplus_{j=1}^{k-1} \CB_j\,, \qquad \ee
where $\CB_k^{\text{ab}\,\pm}$ are the semisimple subcategories generated by the simple projectives $S_k^\pm$, and each $\CB_j^{\rm ab}$ is the subcategory generated by the pair $(S_j^+,S_{k-j}^-)$. Equivalently, each $\CB_j^{\rm ab}$ contains and may be generated by the pair of projectives $(P_j^+,P_{k-j}^-)$.  Passing to derived categories respects the block decomposition, with $\CB_k^\pm = D^b\CB_k^{\text{ab}\,\pm}$, $\CB_j=D^b\CB_j^{\rm ab}$.

The two semisimple blocks $\CB_k$ are each clearly isomorphic to Vect. The remaining blocks $\CB_j$ are all isomorphic to each other (as are the $\CB_j^{ab}$).
For example, each $\CB_j$ is generated by the pair of simples $(a,b)=(S_j^+,S_{k-j}^-)$, whose derived endomorphism algebra (\ie\ Ext algebra) is the path algebra of the quiver
\be \hspace{.5in} \raisebox{-.4in}{$\includegraphics[width=1.3in]{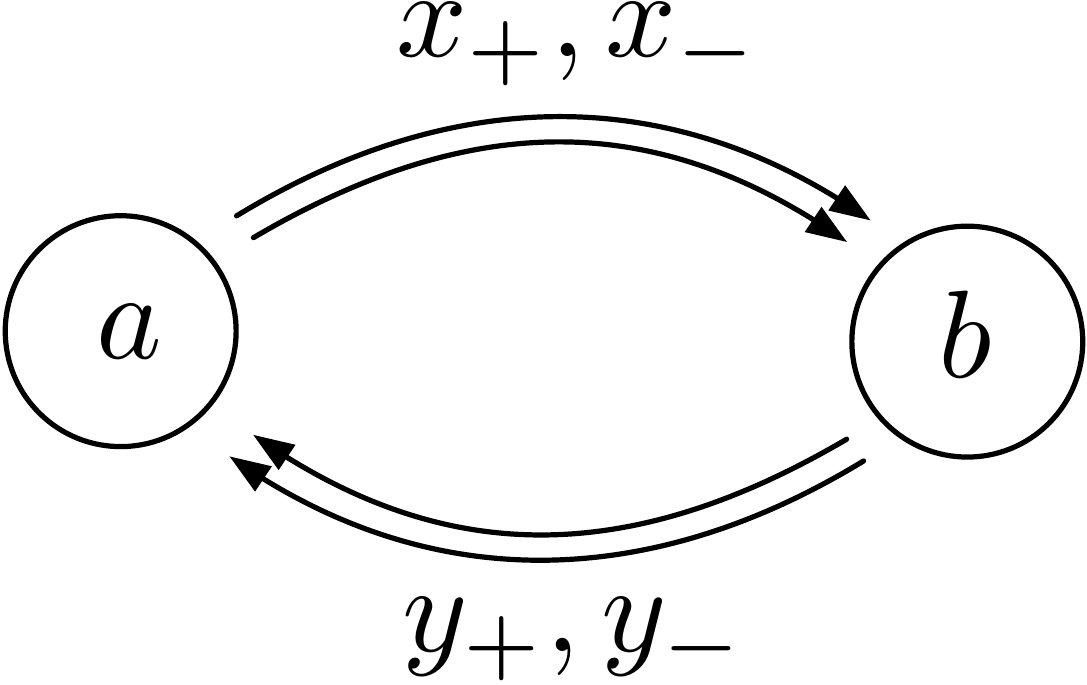}$}  \qquad \text{w/ relations $\begin{array}{c}x_+y_--x_-y_+=0 \\ y_+x_--y_-x_+=0\end{array}$} \label{simple-quiver} \ee
independent of the choice of $1\leq j\leq k-1$. In a similar way, the derived category $\text{Coh}(T^*[2]\mathbb{P}^1)$ can be generated from the structure sheaf $\CO$ and the line bundle $\CO(1)$, which have exactly the same derived endomorphism algebra as \eqref{simple-quiver}, upon identifying $(a,b)\leftrightarrow(\CO,\CO(1))$. Thus
\be \CB_j \simeq \text{Coh}(T^*[2]\mathbb{P}^1) \qquad \forall\;\; 1\leq j \leq k-1\,, \label{Bj-flag} \ee
which implies the decomposition \eqref{U-flag}.

A more general result appears in \cite{ABG,BL}. The authors there focus on the case of odd roots of unity ($q=e^{2\pi i/r}$ with $r$ odd), and consider the ``principal block''  $\CB_1$ of the derived category $u_q(\mathfrak g)\text{-mod}$, for any simple $\mathfrak g$. The principle block is defined to be the block containing the trivial representation. A derived equivalence is established between $\CB_1$ and the $\text{Coh}(\wt \CN_{\mathfrak g})$, the derived category of coherent sheaves on the Springer resolution $\wt \CN_{\mathfrak g}$ of the nilpotent cone of $\mathfrak g$. (Recall that $\wt \CN_{\mathfrak g}$ is isomorphic to the cotangent bundle of the flag variety $T^*(G_\C/B)$; with a cohomological shift, the correct space for the equivalence is $T^*[2](G_\C/B)$. When $\mathfrak g =\mathfrak{sl}_2$, one recovers the space $\wt\CN_{\mathfrak g}=T^*[2]\mathbb{P}^1$ appearing above.)

There should be several modifications to \cite{ABG,BL} at even roots of unity. In general, we would expect the Langlands-dual Springer resolution to play a role. For $\mathfrak g=\mathfrak{sl}_n$, however, the Springer resolution and its Langlands dual are equivalent. Moreover, for $\mathfrak g=\mathfrak{sl}_n$, even though the decomposition of $u_q(\mathfrak g)\text{-mod}$ into blocks differs for even vs. odd roots of unity, it appears that the principal block $\CB_1$ remains unchanged. We expect for any $\mathfrak g=\mathfrak{sl}_n$ and any root of unity that
\be \CB_1 \simeq \text{Coh}(T^*[2](SL(n,\C)/B))\,. \ee
This should be involved in a generalization of \eqref{U-flag}.

\subsubsection{A sheaf of derived categories}
\label{sec:V-sheaf}

We can use a bit of physical intuition and borrowing of future results to give a natural guess for a unified description of the stalks $\CC_g$ of the full category $\CC = D^b(\mb U\text{-mod})$, for all $g$ in a neighborhood of the identity. 

In Section \ref{sec:Atwist}, we will define a theory $\CT_{2,k}^A$ whose category of lines (in the presence of various background $PGL(2,\C)$ flat connections) should be isomorphic to $\CC$. The Coulomb branch of this theory is the nilpotent cone $\CN_{\mathfrak{sl}_2}$, and, by 3d mirror symmetry, the A-twisted theory $\CT_{2,k}^A$ should \emph{very roughly} look the same as the B-twist of a sigma-model to $\CN_{\mathfrak{sl}_2}$.  (This description is not exact, because $\CN_{\mathfrak{sl}_2}$ is singular, and there are extra infinitesimal degrees of freedom related to the level-$k$ Chern-Simons gauging.)
This is the same sort of B-twisted sigma-model that was discussed in \cite{RW-Coulomb}. Flat connections come from the $PGL(2,\C)$ complex-Hamiltonian isometry group of $\CN_{\mathfrak{sl}_2}$. 

In Section \ref{sec:hyper-Cfd}, we considered a toy model, the B-twist of a sigma-model to $T^*\C$. It coupled to flat connections for the isometry group $SL(2,\C)$. For a flat connection with infinitesimal holonomy $g=e^a\approx 1+a$, with $a\in \mathfrak{sl}_2$, we found that each category $\CC_g$ of line operators could be described as the derived category of matrix factorizations
\be \CT_{\rm hyper}^B:\qquad  \CC_{1+a}\simeq   \text{MF}(T^*\C,W_a)\,,\qquad W_a = \langle a,\mu\rangle\,,\ee
with superpotential defined by contracting the complex moment map $\mu:T^*\C\to \mathfrak{sl}_2^*$ with $a\in \mathfrak{sl}_2$.
For a B-twisted sigma-model with \emph{smooth} target $\wt \CN_{\mathfrak{sl}_2} =T^*[2]\mathbb{P}^1$, the category of line operators in the presence of infinitesimal holonomy $g=1+a$, $a\in \mathfrak{pgl}_2$, should similarly be given by the derived category of matrix factorizations $\text{MF}(T^*[2]\mathbb{P}^1,\langle a,\mu\rangle)$, with $\mu:T^*[2]\mathbb{P}^1\to \mathfrak{pgl}_2^*$ the complex moment map.

Combining these observations with the true decomposition of our category \eqref{U-flag} at $g=1$, we conjecture that for any infinitesimal holonomy $g=1+a$, the category $\CC_g$ is
\be \CC_{1+a} \simeq \text{Vect}^{\oplus 2}\oplus \big[\text{MF}(T^*[2]\mathbb{P}^1, \langle a,\mu\rangle)\big]^{\oplus k-1} \ee
Note that the function $\mu$ has cohomological degree 2, as required for the matrix-factorization categories to be $\Z$-graded, just as in Section \ref{sec:hyper-Cfd}.

This conjecture is automatically compatible with the known form of $\CC_1$. It is also pleasantly compatible with the generic stalks $\CC_g$ discussed in Section \ref{sec:V-generic}. If $g=1+a$ is generic and (WLOG) diagonal, the moment map $\langle a,\mu\rangle : T^*[2]\mathbb{P}^1\to \C$ has exactly two, non-degenerate critical points, at the north and south poles of $\mathbb{P}^1$. (The moment map is quadratic in the neighborhood of each critical point.) The category $\text{MF}(T^*[2]\mathbb{P}^1, \langle a,\mu\rangle)$ localizes to the category of coherent sheaves at the critical points, which is the semisimple category $\text{Vect}^{\oplus 2}$. Altogether,
\be \CC_{1+a} \simeq \text{Vect}^{\oplus 2k} \qquad \text{($a$ generic)}\,, \ee
exactly as found in Section \ref{sec:V-generic}.

\subsection{Derived state spaces}
\label{sec:U-Hilbert}

If the sheaf of categories $\CC = D^b(\mb U\text{-mod})$ above indeed corresponds to line operators in a 3d topological QFT, one should be able compute from it state spaces on various surfaces, using the methods reviewed in Section \ref{sec:toymodel}.

Recall from Section \ref{sec:Hilb-gen} that in a theory that couples to $G_\C$ flat connections, state spaces $\CH(\Sigma,\CA)$ depend on a choice of flat connection $\CA$ on $\Sigma$ (up to gauge equivalence). Globally, the state spaces assemble into a coherent sheaf
\be \CH(\Sigma)\to \text{Flat}_{G_\C}(\Sigma) \ee
over the moduli stack of flat connections, modulo complex gauge equivalence. Here we have $G_\C=PGL(2,\C)$, and we want to describe various stalks $\CH(\Sigma,\CA)$.

Note that the relation between the category $\CC$ and state spaces does \emph{not} invoke a braiding on $\CC$. In genus 0 and 1, it does not even use the tensor product, and it is enough to know $\CC$ as an ordinary category. Thus the full power of CGP TQFT and unrolling is not necessary for this analysis. Braiding (and more) \emph{is} required if one wants to describe mapping-class-group actions on state spaces, which we don't address here.

\subsubsection{Genus 0: local operators}
\label{sec:U-0}

We start with $\Sigma=S^2$ a sphere. Any flat connection $\CA$ on $S^2$ is gauge-equivalent to the trivial flat connection, so there is only one state space $\CH(S^2):=\CH(S^2,\CA=0)$ to consider. It is the space of local operators in a putative 3d topological QFT. It may be computed from the category of line operators as the endomorphism algebra of the tensor-identity object $\mb 1$,
\be \CH(S^2) = \text{End}^\bullet_{\CC}(\mb 1)\,. \ee

In our case, $\mb 1=S_1^+$ is the trivial representation of $\mb U$, and belongs to the principal block $\CB_1$ of the stalk category $\CC_1$. Thus $\text{End}^\bullet_{\CC}(\mb 1) = \text{End}^\bullet_{\CB_1}(S_1^+)$. Using the geometric description \eqref{Bj-flag} of $\CB_1$ as coherent sheaves on $T^*[2]\mathbb{P}^1$, with $S_1^+$ identified as the structure sheaf, we then find that
\be \CH(S^2) \simeq \text{End}^\bullet_{\text{Coh}(T^*[2]\mathbb{P}^1)}(\CO) = \C[T^*[2]\mathbb{P}^1] \ee
is just the space of algebraic functions on $T^*[2]\mathbb{P}^1$.

Algebraic functions on $T^*[2]\mathbb{P}^1$ can be described more explicitly by organizing them into representations of Hamiltonian isometry group $PGL(2,\C)$ acting on $T^*[2]\mathbb{P}^1$. (Physically: we expect the state space $\CH(S^2)$ form a representation of the global symmetry group $PGL(2,\C)$.) For every odd $j\in \mathbb N$, let $\rho_j$ denote the $j$-dimensional representation of $PGL(2,\C) \simeq SO(3,\C)$, and let $\rho_j[d]$ denote its shift in cohomological degree by $d$ units. Then
\be \CH(S^2) \simeq \C[T^*[2]\mathbb{P}^1] = \bigoplus_{\text{odd $j\in\mathbb N$}} \rho_j[j-1] = \rho_1+\rho_3[2]+\rho_5[4]+\ldots \ee
is just the regular representation of $PGL(2,\C)$. Its graded Poincar\'e series is
\begin{align} P[\CH(S^2)](y,t) &= 1+(y^2+1+y^{-2})t^2+(y^4+y^2+1+y^{-2}+y^{-4})t^4+\ldots \notag \\
 &= \frac{y-y^{-1}}{y-y^{-1}}+\frac{y^3-y^{-3}}{y-y^{-1}}t^2 + \frac{y^5-y^{-5}}{y-y^{-1}}t^4 + \ldots \notag \\
 &= \frac{1+t^2}{(1-t^2y^2)(1-t^2y^{-2})}\,, \label{P-gen0}
\end{align}
with $y$ a fugacity for (a character of) the maximal torus of $PGL(2,\C)$.

\subsubsection{Genus 1: Hochschild homology}
\label{sec:U-1}

Now consider a torus $\Sigma=T^2$, together with an abelian (\emph{i.e.} diagonal) flat connection $\CA$ having holonomies
\be \text{Hol}_{\gamma_a}(\CA) = g_a = \text{diag}(e^{2\pi i \alpha},1)\,,\qquad  \text{Hol}_{\gamma_b}(\CA) = g_b = \text{diag}(e^{2\pi i \beta},1) \ee
around A and B cycles. As discussed in Section \ref{sec:Hoch-flavor}, the state space will be given by twisted Hochschild homology of an appropriate stalk of the category $\CC$,
\be \CH(T^2,\CA) = HH_\bullet^{g_b}(\CC_{g_a}) \simeq HH_\bullet^{g_a^{-1}}(\CC_{g_b}) \ee

Suppose that at least one of the holonomies is nontrivial, say (WLOG) $g_a\neq 1$. Then the category $\CC_{g_a}\simeq \text{Vect}^{\oplus 2k}$ is semisimple, generated by the $2k$ simple modules $V_{\alpha,n}$, and the computation of the state space reduces to the Grothendieck ring
\be \CH(T^2,\CA\;\text{generic}) = HH_\bullet^{g_b}(\CC_{g_a}) \simeq K_0(\CC_{g_a}) \simeq \C^{2k}\,. \label{Hilb-T2A} \ee

On the other hand, if both holonomies are trivial, the state space is given by untwisted Hochschild homology of the non-semisimple stalk $\CC_1$, 
\be \CH(T^2,\CA=0) = HH_\bullet(\CC_1)\,. \ee
We illustrate how to compute these homology groups by hand in Appendix \ref{app:U}. A complete answer is more readily obtained by using the geometric decomposition \eqref{U-flag} and the fact that Hochschild homology of the category of coherent sheaves on a smooth variety $\CX$ is the total (algebraic) Dolbeault cohomology of $\CX$.

In our case, we need $HH_\bullet(\CB_1)=HH_\bullet(\text{Coh}(T^*[2]\mathbb P^1)) \simeq H_{\bar\pd}^{\bullet,\bullet}(T^*[2]\mathbb P^1)$, where cohomological degree in $HH_\bullet$ is the sum of the $(p,q)$ degrees of Dolbeault cohomology and the internal degree shift in the cotangent fibers. The computation is done in \cite[Prop. 5.8]{LQ-derived}, with a beautiful result given in terms of $PGL(2,\C)$ representations as
\be HH_{i}(\CB_1) = \begin{cases} \rho_1^{\oplus 3} & i=0 \\ \rho_i\oplus \rho_{i+2} & \text{odd}\;i\geq 1\\
    \rho_{i+1}^{\oplus 2} & \text{even}\;i\geq 2\,, \end{cases} \qquad\text{\emph{i.e.}}\qquad
     \begin{array}{rcl} HH_0(\CB_1) &=& \rho_1^{\oplus  3} \\ HH_{1}(\CB_1) &=& \rho_1\oplus \rho_3 \\ HH_2(\CB_1) &=& \rho_3\oplus \rho_3 \\ HH_3(\CB_1) &=& \rho_3\oplus\rho_5 \\ HH_4(\CB_1) &=& \rho_5\oplus \rho_5\;\; \ldots \end{array} \label{dol-P1}  \ee
with graded Poincar\'e series
\begin{align} P[HH_{-\bullet}(\CB_1)](y,t) &= 3+t+\sum_{\text{odd}\;j\geq 3} \frac{y^j-y^{-j}}{y-y^{-1}}t^{j-1}(t^{-1}+2+t) \notag \\
& = 1-t^{-1}+\frac{(1+t)^2(t+t^{-1})}{(1-t^{2}y^2)(1-t^{2}y^{-2})}\,,
\end{align}
and graded Euler character $\chi[HH_\bullet(\CB_1)](y) = P[HH_\bullet(\CB_1)](y,-1) = 2$, equal to the Euler character of $T^*\mathbb P^1$.

Combining this with the block decomposition \eqref{U-flag} finally gives
\begin{align} \CH(T^2,0) &\simeq \rho_1^{\oplus 2}\oplus HH_{\bullet}(\CB_1)^{\oplus k-1}\;,
\qquad \text{\emph{i.e.}}\quad \begin{array}{rcl} \CH_0 &\simeq& \rho_1^{\oplus 3k-1} \\ \CH_1 &\simeq& \rho_1^{\oplus k-1}\oplus\rho_3^{\oplus k-1} \\ \CH_2 &\simeq& \rho_3^{\oplus k-1} \oplus \rho_3^{\oplus k-1} \\ &&\cdots \end{array},
  \label{Hilb-T2} \\
 P[\CH(T^2,0)](y,t^{-1}) &= 2+(k-1)\Big[1-t^{-1}+\frac{(1+t)^2(t+t^{-1})}{(1-t^{2}y^2)(1-t^{2}y^{-2})}\Big]\,, \notag\\
 \chi[\CH(T^2,0)](y) &= 2+2(k-1)=2k\,. \notag
\end{align}
As expected, the Euler character is independent of the choice of flat connection $\CA$. The state space is clearly not, since it jumps from a $2k$-dimensional space (in cohomological degree zero) at generic $\CA$ to an infinite-dimensional space (with unbounded cohomological degree) at $\CA=0$.

Finally, if a flat connection $\CA$ has generic holonomy along a single cycle, say $g_b\neq 1$, we can further think of $\CH(T^2,\CA)$ as a deformation of $\CH(T^2,0)$ induced by twisting the Hochschild differential, analogous to the toy model in \eqref{d-twisted-eg}. We then expect that $\CH(T^2,\CA) =\rho_1^{\oplus 2k}$ can be obtained by adding a differential to the infinite-dimensional $\CH(T^2,0)$. There is an obvious differential that does the job, cancelling pairs of representations within each block:   \vspace{-.1in} 
\be \raisebox{-.5in}{$\includegraphics[width=1.7in]{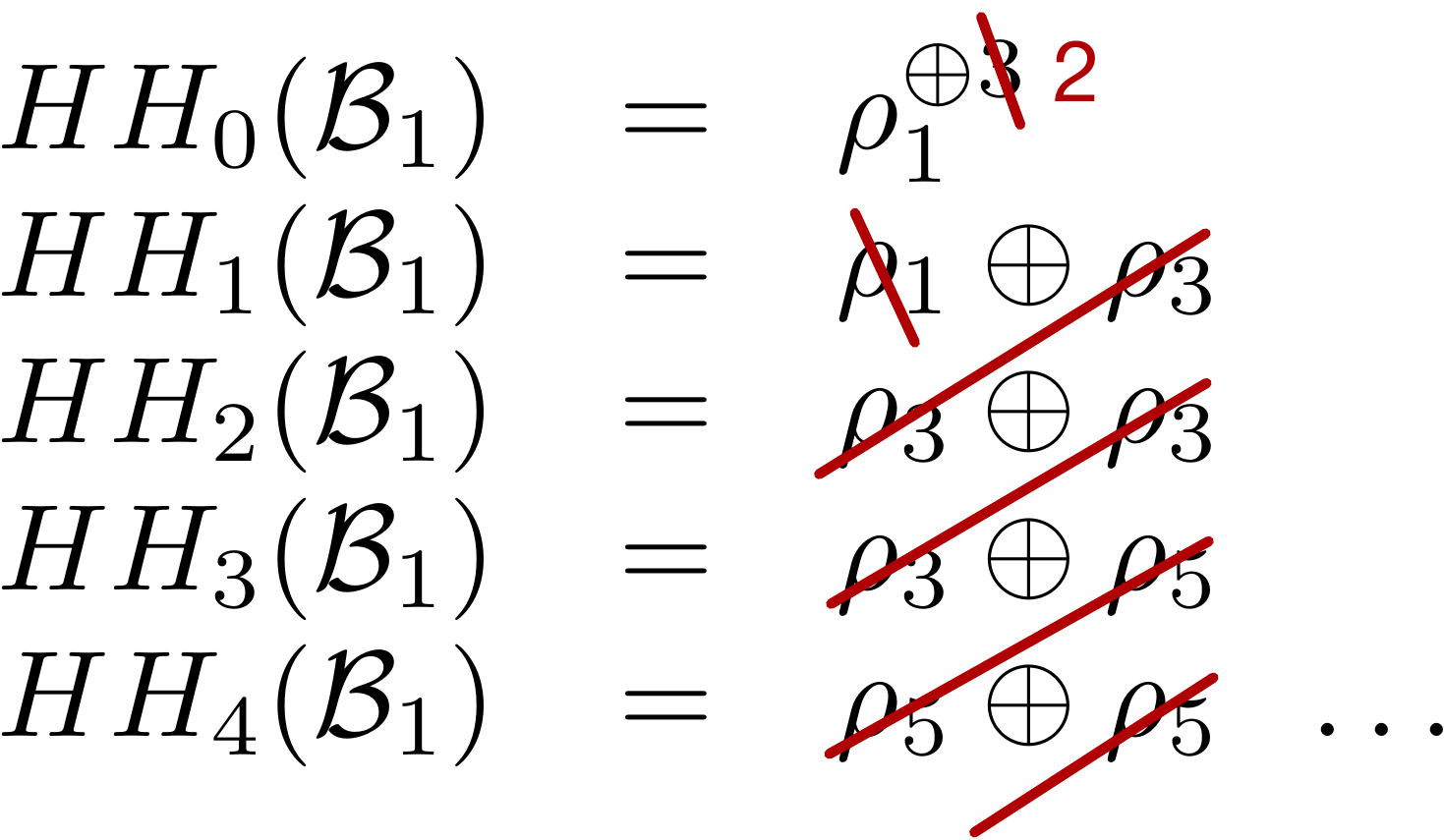}$}  \ee

\subsubsection{Generic flat connection: any genus}
\label{sec:U-g}

In higher genus $g\geq 2$, it is straightforward to compute the state space in the presence of a generic abelian flat connection $\CA$, and much less straightforward at $\CA=0$. We will just do the former here.

\begin{figure}[htb]
\centering
\includegraphics[width=5in]{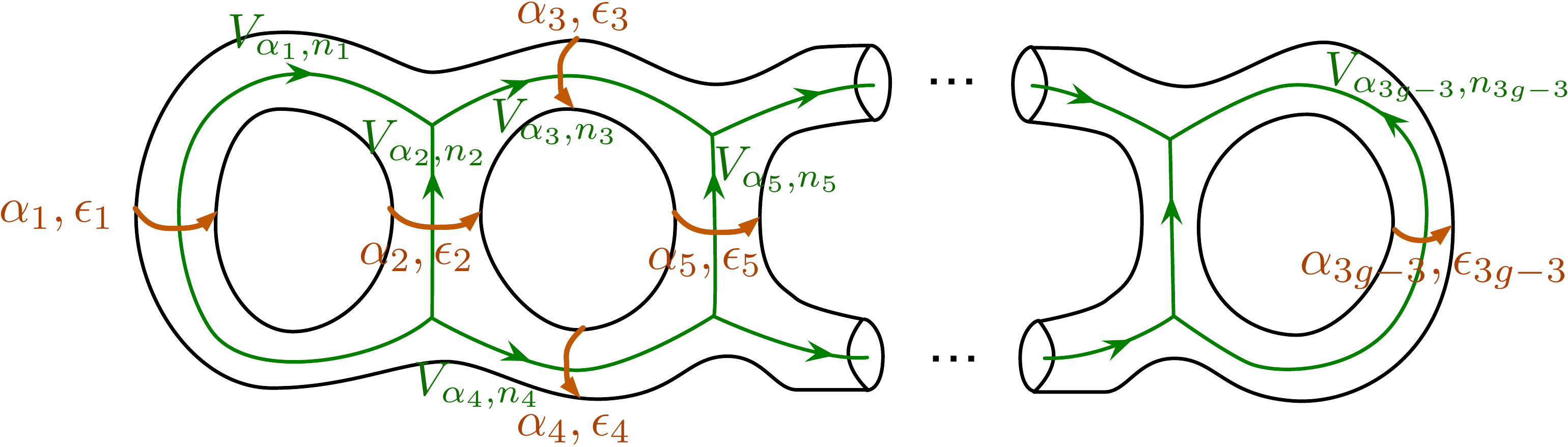}
\caption{Choosing networks of line operators inside a handlebody to produce states in the state space for a genus-$g$ surface with generic abelian flat connection.}
\label{fig:genus-g}
\end{figure}

Consider a genus-$g$ surface $\Sigma_g$ with a ``pants decomposition'' as in Figure \ref{fig:genus-g}, and an abelian flat connection that has generic holonomies $g_i=\text{diag}(e^{2\pi i\alpha_i},1)$ around the $3g-3$ ``cuffs.'' The holonomies must satisfy $\alpha_i\pm \alpha_j\pm \alpha_k=0$ mod $\Z$ at each junction, with appropriate signs corresponding to the orientations. We assume that all $\alpha_i\notin \Z$.%
\footnote{Proposition 6.5 of \cite{BCGP} implies that if a flat connection on $\Sigma_g$ has holonomy around at least \emph{one} cycle not equal to $1$, one can choose a pants decomposition such \emph{all} cuff holonomies obey  $\alpha_i\notin \Z$.}

We expect that states in the state space $\CH(\Sigma_g,\CA)$ are in 1-1 correspondence with trivalent networks of irreducible line operators threaded through the core of a handlebody with boundary $\Sigma_g$. The line passing through a core component linked by holonomy $g_i$ on the outside must be a simple object of the semisimple category $\CC_{g_i}$, and at each trivalent junction the fusion rules \eqref{V-ss-fusion} must be obeyed.

In order to enumerate all possible networks, we can first choose a logarithm $\alpha_i$ for the holonomy around each cuff.  Then we choose a sign $\epsilon_i\in \{\pm\}$, such that $\epsilon_i\epsilon_j\epsilon_k=+$ at every junction. There are $2^g$ possible choices of signs (corresponding to a choice of $\Z_2$-valued homology class on a complementary handlebody). Finally, in each core component, we choose a simple object $V_{\alpha_i,n}\in \CC_{g_i}$, with $n$ even (resp. odd) if $\epsilon_i$ is even (resp. odd). The fusion rules \eqref{V-ss-fusion} imply that all trivalent junctions will exist, and are uniquely determined, independently of how these $V_{\alpha_i,n}$'s are chosen. This gives us an additional $k^{\#\,\text{cuffs}} = k^{3g-3}$ choices of objects, so that finally
\be \text{dim}\, \CH(\Sigma_g,\CA) =2^gk^{3g-3}\,.  \label{Hilb-g}\ee

When all objects are chosen from semisimple stalks of $\CC$ like this, $\CH(\Sigma_g,\CA)$ should lie entirely in cohomological degree zero. Invariance of the Euler character under deformations of $\CA$ then implies
\be \chi[\CH(\Sigma_g,0)](y=1) = \chi[\CH(\Sigma_g,\CA)](y=1) = 2^gk^{3g-3}\,. \label{U-chi-g} \ee
as well. It is not obvious from the above construction how $\CH(\Sigma_g,\CA)$ at generic abelian $\CA$ decomposes into representations of (the maximal torus of) the symmetry group $PGL(2,\C)$; hence we set $y=1$ in \eqref{U-chi-g}. A computation of QFT indices in Section \ref{sec:Hilbert} will give a prediction for this missing grading.

\subsection{The CGP TQFT}
\label{sec:CGP}

If one is to define an axiomatic 3d TQFT that not only assigns vector spaces to surfaces, but also assigns linear maps to 3d cobordisms and numbers to closed three-manifolds, more structure is needed than what we described in Sections \ref{sec:U-modules}--\ref{sec:U-Hilbert}. This is where the techniques of \cite{CGP} come in. Our goal in this section is to outline a particular version `$\CZ_{\rm CGP}$' of a CGP TQFT, based on the unrolled quantum group $U_q^H(\mathfrak{sl}_2)$ at an even root of unity $q=e^{i\pi/k}$, with the following properties:
\begin{itemize}
\item $\CZ_{\rm CGP}$ is a `3-2-1' extended TQFT (just like the Reshetikhin-Turaev TQFT \cite{RT}), defined on oriented manifolds of dimension $d=1,2,3$ enriched by $T$-bundles with flat connections, where $T\subset PGL(2,\C)$ is the maximal torus.

If $k= 0$ (mod 4), a choice of generalized spin structure is also required to define the TQFT \cite{BCGP-spin}; we will assume here that $k\neq 0$ (mod 4).

We will only consider ``abelian'' $T$ flat connections in this section.
A closely related TQFT enriched by general $PGL(2,\C)$ flat connections was partially developed in \cite{biquandles}, combining earlier ideas of \cite{KashaevReshetikhin} with the modified traces of \cite{CGP}. It is not (currently) based on the unrolled quantum group. 

\item To the circle, the TQFT associates an \emph{abelian} category $\catH=\CZ_{\rm CGP}(S^1) \approx U_q^H(\mathfrak{sl}_2)\text{-mod}$ (see Section \ref{sec:defCH} for precise definitions).
It is a sheaf of categories over the torus $T$, thought of algebraically as the additive group $\AA := \C/\Z$. The stalk at $\alpha\in \AA$ (or $\text{diag}(e^{2\pi i\alpha},1)\in T$) is given by
\begin{align} \label{ZCGPS1} \catH_\alpha &= \CZ_{\rm CGP}(S^1, \CA = \big(\begin{smallmatrix} i\alpha&0\\0&0\end{smallmatrix}\big)d\theta) \\ & = \{\text{$U_q^H(\mathfrak{sl}_2)$ moduled with $E^k=F^k=0$, $K^{2k}=e^{2\pi i \alpha}$}\}\,.\notag \end{align}
Roughly speaking, each $\catH_\alpha$ contains infinitely many copies of the objects of the objects of $\CCab_g$ (for $g =  \text{diag}(e^{2\pi i\alpha},1)$) considered in Section \ref{sec:U-modules}.
The category $\catH$ is a \emph{relative modular category} in the terminology of \cite{derenzi2019nonsemisimple}; in  particular, it is a braided tensor category. 

\item The spaces assigned to surfaces with fixed flat connections $\CZ_{\rm CGP}(\Sigma,\CA)$ are finite-dimensional. They coincide with the spaces $\CH(\Sigma,\CA)$ computed more naively from $U_q(\mathfrak{sl}_2)$ in Section \ref{sec:U-Hilbert}, in \emph{cohomological degree zero}
\be \CZ_{\rm CGP}(\Sigma,\CA) = \CH(\Sigma,\CA)\big|_{\text{deg}\; 0}\,. \ee
We will demonstrate this explicitly for genus $g=0,1$ when $\CA=0$, and for any $g\geq 1$ when $\CA$ is generic abelian.

There is an action of the mapping-class group on spaces  $\CZ_{\rm CGP}(\Sigma)$. For $\CA=0$ the action is non-unitary and after projectivization corresponds to the Lyubashenko projective mapping-class group representation given in \cite{Lyubashenko:1994tm}, see \cite{DGP2, derenzi2021mapping}.

\end{itemize}
As stated in the Introduction, we expect that there exists a derived enhancement of the CGP TQFT that fully matches the structure of the topologically twisted QFT $\CT_{2,k}^A$ coupled to background abelian flat connections. We leave this to future work!

\subsubsection{Relative modular categories and flat connections}
\label{sec:rel-flat}

In \cite{derenzi2019nonsemisimple}, Marco De Renzi introduced the notion of a \emph{relative modular category}, as the fundamental algebraic structure underlying extended CGP TQFT's `enriched' by abelian flat connections.
Relative modular categories can have vanishing quantum dimensions and contain an infinite number of non-isomorphic simple modules, a setting in which the usual Reshetikhin-Turaev 3-manifold invariant does not apply.

A relative modular category is a ``generically semisimple graded category'' with a ``free realization'' and an ``m-trace,'' satisfying certain compatibility and non-degeneracy conditions. We briefly and informally explain what this means, connecting the terminology of \cite{derenzi2019nonsemisimple} (as well as \cite{CGP} and many related papers) with the perspective in the rest of this paper.

A ``graded'' category $\CC$ is (in the language of our paper) just a tensor category that forms a sheaf of categories over an abelian group $\AA$,
\be \CC\to \AA\,, \ee
such that the stalks $\CC_\alpha$ $(\alpha\in \AA)$ are full subcategories that satisfy $\text{Hom}_\CC(M,N)=0$ if $M\in \CC_\alpha$, $N\in \CC_\beta$ for $\alpha\neq \beta$, and the tensor product is compatible with addition in $\AA$\,,
\be \otimes: \CC_\alpha\boxtimes \CC_\beta\to \CC_{\alpha+\beta}\,. \ee
The category is generically semisimple if generic stalks $\CC_\alpha$ (say, for $\alpha$ in a Zariski-open subset of $\AA$) are semisimiple.
This roughly matches the structure we expect for a category of line operators in a topological QFT that couples to flat connections for an abelian Lie group $T$, where $T$ is related to the additive group $\AA$ via the exponential map.  In particular, each union of stalks
\be
	\bigoplus_{\alpha\,\text{s.t.}\,q^\alpha=g} \CC_\alpha
\ee
 is an infinite cover of $\CC_g$.

In the formalism of relative modular categories, the precise way in which the infinite covering/exponential map is encoded is in the data of a \emph{free realization}. In physical terms, a free realization implements large gauge transformations.
Mathematically, a free realization partitions objects of $\catH_\alpha$ into equivalence classes. It also keeps track of equivalence relations that should be implemented when defining spaces of states $\CZ_{\rm CGP}(\Sigma,\CA)$ on surfaces, to ensure that these spaces are finite dimensional.

Finally, the \emph{m-trace}, or modified trace, is a systematic tool for regularizing vanishing quantum dimensions. Such regularization is paramount if one is to construct a TQFT with consistent cutting-and-gluing rules, and well-defined invariants of closed 3-manifolds. Partition functions in the sorts of supersymmetric QFT's that we consider also contain new zeroes and infinities, which must be regularized, and a physical analogue of the m-trace will almost certainly be required. We do not explore this in the current paper, however.

\subsubsection{The unrolled quantum group}
\label{sec:defUH}

We recall the unrolled quantum group and its category of weight modules. As usual, we fix an integer $k\geq 2$ and a corresponding even root of unity $q=e^{i\pi/k}$. We also assume that $k\neq 0$ (mod 4), to avoid introducing spin structures.

The unrolled quantum group $\mb U^H:= U_q^{H}(\mathfrak{sl}_2)$ is the $\C$-algebra given by generators $E, F, H, K, K^{-1}$ 
and relations:
\be\label{E:RelDCUqsl}
\begin{array}{c}
	KK^{-1}=K^{-1}K=1\,,\quad KEK^{-1}=q^2E\,,\quad  KFK^{-1}=q^{-2}F\,,\quad
	\ds [E,F]=\frac{K-K^{-1}}{q-q^{-1}}, \\[.2cm]
	HK=KH\,,\quad 
	 [H,E]=2E\,,\quad   [H,F]=-2F\,. \end{array}
\ee
It thus extends $\mb U$ by the Cartan generator $H$.
The partially restricted unrolled quantum group $\UH$ has generators  $E, F, H, K, K^{-1}$ and relations  \eqref{E:RelDCUqsl} together with
\be E^k=0\,,\quad F^k=0\,. \ee
(The relations $E^k=F^k=0$ corresponds to considering only abelian flat connections.)

The algebras $\mb U^H$ and $\UH$ are Hopf algebras, with the coproduct, counit, and
antipode defined by
\begin{align}
	\Delta(E)&= 1\otimes E + E\otimes K\,, 
	&\varepsilon(E)&= 0\,, 
	&S(E)&=-EK^{-1}\,, 
	\notag \\
	\Delta(F)&=K^{-1} \otimes F + F\otimes 1\,,  
	&\varepsilon(F)&=0\,,& S(F)&=-KF\,,
	\\
	\Delta(K)&=K\otimes K\,,
	&\varepsilon(K)&=1\,,
	& S(K)&=K^{-1}\,,
	\notag \\
	\Delta(H)&=H\otimes 1 + 1 \otimes H\,,
	& \varepsilon(H)&=0\,, 
	&S(H)&=-H\,. \notag
\end{align}

\subsubsection{The relative modular category $\catH$}
\label{sec:defCH}

Let $V$ be a finite-dimensional $\UH$-module.  An eigenvalue
$\lambda\in \C$ of the operator $H:V\to V$ is called a \emph{weight}
of $V$ and the associated eigenspace is called a \emph{weight space}.
A vector $v$ in the $\lambda$-eigenspace  
of $H$ is a \emph{weight vector} of \emph{weight} $\lambda$, \ie\ $Hv=\lambda
v$.  We call $V$ a \emph{weight module} if $V$ splits as a direct sum
of weight spaces and $q^H=K$ as operators on $V$, \ie\ $Kv=q^\lambda
v$ for any vector $v$ of weight $\lambda$.  Let $\catH$ be the category
of finite-dimensional $\UH$ weight modules.

Since $\UH$ is a Hopf algebra, $\catH$ is a tensor category, whose unit object $\id$ is the 1-dimensional trivial module $\C$ (on which $E,F,H$ act as zero, and $K$ acts as $1$).  Moreover,  $\catH$ is a linear ribbon category, see \cite{Murakami,Ohtsuki, GPT}.  We use the formulas and notation for this ribbon structure given in 
\cite{CGP2}. We next explain why the category $\catH$ is a relative modular category, so that from \cite{derenzi2019nonsemisimple} it leads to a TQFT.  In Section~2.3 of \cite{AGP} the properties of a relative modular category are briefly reviewed; here we use the notation of \cite{AGP}. 

To this end, we must introduce a grading.
Consider the additive group
\be  \AA = \C/\Z\,, \label{ACZ} \ee
which we identify with the maximal torus $T\subset PGL(2,\C)$ via the exponential map. For each $\bar \alpha \in\C/\Z$, define
$\catH_{\bar \alpha}$ as the full sub-category of weight modules whose weights 
are all in the class $\bar \alpha$ (mod $\Z$).  
Then $\catH=\{\catH_{\bar \alpha}\}_{\bar \alpha\in \AA}$ is an $\AA$-grading. Equivalently, $\catH$ forms a sheaf of categories over $\AA$, with stalks $\catH_{\bar\alpha}$.

It is possible to introduce a more refined grading on $\catH$, given by the group $\hat \AA = \C/2\Z$, identified with the maxmial torus $\hat T\subset SL(2,\C)$. Each $\AA$-graded component of $\catH$ is a direct sum of two $\hat \AA$-graded components. The $\hat \AA$-grading is the only one considered in \cite{BCGP} and previous related papers. Here we use the $\AA$-grading in order to match the structure of physical QFT's $\CT_{2,k}^A$, with $PGL(2,\C)$ symmetry.

For $m\in \Z$, let $\C^H_{mk}$ be the one-dimensional module in $\catH_{\bar 0}$ where both $E$ and $F$ act by zero and $H$ acts by $mk$, and let
\be \sigma(m):=\C^H_{2mk}\,,\qquad m\in \Z\,.\ee
Then $\{\sigma(m)\}_{m\in \Z}$ is a free realization, satisfying Definition 2.10 of \cite{AGP}.
A free realization has two important properties, used in constructing the TQFT:
\begin{itemize}
\item the modules $\sigma(m)$ act freely, via tensoring, on objects of $\catH$, such that within each $\catH_{\bar \alpha}$ the isomorphism classes of simple modules form finitely many orbits; and
\item the double braiding of $\sigma(m)$ with a module in $\catH_{\bar\alpha}$ is a scalar $q^{2\bar\alpha mk}$, \emph{cf.} \eqref{E:Braiding-sigma} below.
\end{itemize}
Therefore, the ``unbraiding'' of a closed component of a link labelled with $\sigma(m)$ is controlled by a computable scalar. With an appropriate normalization, the 3-manifold invariant does not see a closed component labeled with $\sigma(m)$, a notion called \emph{$\sigma$-equivalence}. These properties allow one to define finite Kirby colors and a corresponding 3-manifold invariant.  They also make the state spaces associated to a surface finite dimensional.

The category $\catH$ is closely related to abelian stalks of the category $\CCab=\mb U\text{-mod}$ explored in Section \ref{sec:U-modules}. For each $\bar\alpha\in \AA$, let
\be g(\bar \alpha) := \begin{pmatrix} e^{2\pi i \bar\alpha} & 0 \\ 0 & 1 \end{pmatrix}\in PGL(2,\C)\,. \ee
Then there is a 1--1 correspondence between (isomorphism classes of) objects of the stalk $\CCab_{g(\bar\alpha)}$ and orbits of the free realization $\{\sigma(m)\}_{m\in \Z}$ acting on objects of $\catH_{\bar\alpha}$. Roughly speaking, $\CCab_{g(\bar\alpha)}$ is the quotient of $\catH_{\bar\alpha}$ by the free realization.

For example, every simple module of $\catH$ is isomorphic to exactly one of the modules in the list:
\begin{itemize}
\item $S_n\otimes \sigma(m)$ and $S_n\otimes \C_k^H\otimes \sigma(m)$, for $n=1,...,k-1$ and $m\in \Z$\,, in $\catH_{\bar 0}$, \\
where $S_n$ is the simple $n$-dimensional $\UH$-module of highest weight $n-1$ (that becomes the usual $n$-dimensional module for the $\mathfrak{sl}_2$ subalgebra generated by $E,F,H$).

The images of $S_n\otimes \sigma(m)$ and $S_n\otimes \C_k^H\otimes \sigma(m)$ in $\CCab_{1}$ are $S_n^+$ and $S_n^-$ (respectively), as in \eqref{reps-small}.

\item $V_\alpha$, for all $(\alpha \in \C\backslash \Z)\cup k\Z$, \\
where $V_\alpha$ (for any $\alpha\in \C$) denotes the $k$-dimensional Verma module of highest weight $\alpha+k-1$.

Note that $V_\alpha \in \catH_{\bar \alpha}$ (where $\bar\alpha$ denotes the class of $\alpha$ in $\C/\Z$). If $\alpha\in \C\backslash \Z$, then the images of $V_{\alpha+n+k+1}$ for $n\in \Z/2k\Z$ in $\CCab_{g(\bar\alpha)}$ are the modules $V_{\alpha,n}$ depicted in \eqref{V-generic}. If $\alpha\in 2k\Z$  (resp. $\alpha\in k+2k\Z$), the quotient of $V_\alpha$ in $\CCab_1$ is $V_k^-$ (resp. $V_k^+$), as in \eqref{reps-small}.
\end{itemize}

Let $\X=\Z/\Z=\{\bar 0\} \subset \C/\Z$.  Then $\X$ is \emph{symmetric} (\ie\ $\X=-\X$)  and \emph{small} (\ie\ for any $\bar \alpha_1,\ldots ,\bar \alpha_n\in \AA$ we have 
$\bigcup_{i=1}^n ({\bar \alpha}_i+\X) \neq \AA$).  Let $\bar \alpha\in \AA \setminus \X$.  From Theorem 5.2 of \cite{CGP2} we have that $\catH_{\bar \alpha}$ is semi-simple.     Fix any complex number $\alpha$ whose image in $\AA$ is $\bar \alpha$ and set $I_{\alpha}=\{\alpha, \alpha+1,..., \alpha+2k-1\}$.  The list of simple modules above and the fact that $V_\alpha \otimes \sigma(m) \cong V_{\alpha +2mk}$ imply that $\{V_{z}\otimes  \sigma(m) | z\in I_{\alpha} \text{ and } m\in \Z \}$ is the set of all simple modules (up to isomorphism) in $\catH_{\bar \alpha}$ (note that this set does not depend on the choice of $\alpha$).   In the language of \cite{AGP}  this set is a completely reduced dominating set of $\catH_{\bar \alpha}$ and $\catH$ is generically semisimple. 

Let $\psi: \AA \times \Z \rightarrow \C^*$ be the bilinear map defined by $\psi(\bar \alpha, m)=q^{2\alpha mk}$ where $\alpha$ is any complex number whose image in $\C/\Z$ is $\bar \alpha$.  
Then a direct calculation shows $\psi$ satisfies the compatibility condition:
\begin{equation}\label{E:Braiding-sigma}
	c_{\sigma(m),V}\circ c_{V,\sigma(m)}= \psi(\bar \alpha,m) \cdot  \Id_{V \otimes \sigma(m)}.
\end{equation}
for any $\bar \alpha \in \AA$, $V \in \catH_{\bar \alpha}$ and $m \in \Z$.
There exists a unique m-trace on the ideal of projective objects of $\catH$ which is unique up to multiplication by an element of $\C$, see for example Theorem 5.4 of \cite{CGP2}.   
The results of \cite{GPR} (also see Lemma A.4 of \cite{BCGP}) imply that there exists a modularity parameter satisfying the modular condition given in Definition 1.3 of \cite{derenzi2019nonsemisimple}. Summarizing the above, we have that $\catH$ is a modular $\AA$-category relative to $(\Z, \X)$.

\subsection{State spaces in the CGP TQFT}
\label{sec:decTQFT}

The TQFT associated to the category $\catH$ is defined on suitably decorated surfaces and cobordisms.   A decorated surface $\mathcal S$ is a surface $\Sigma$ with framed colored points $\{p_i\}$ and a compatible 1-cohomology class $\omega \in H^1(\Sigma\backslash\{p_i\} ;\AA)$.   
 A decorated cobordism $\mathcal M$ is a triple (with certain admissibility conditions, see Section 3.2 of \cite{CGP}):
$$
\big(\text{3-manifold $M$, $\catH$-colored ribbon graph $\tau$ in $M$, cohomology class $\omega \in H^1(M\setminus \tau; \AA)$}\big)\,,
$$
where the coloring of $\tau$ is compatible with $
\omega$, \ie\ each oriented edge $e$ of $\tau$ is colored by
an object in $\catH_{\omega(m_e)}$ where $m_e$ is the oriented meridian of $e$.

Note that on any smooth manifold $W$, there is a 1--1 correspondence between classes $\omega \in H^1(W,\AA)$ and gauge-equivalence classes of flat $T$-connections $\CA$ (modulo $T$-valued gauge transformations),  where as usual $T\simeq \exp(\AA)$ is the maximal torus of $PGL(2,\C)$. Thus one could alternatively say that this is a TQFT enriched by flat $T$-connections.

\subsubsection{Decorated space of states: general construction}
\label{sec:CGP-Hilbert}

In Section 6 of \cite{BCGP}, the space of states on a decorated surface was constructed for a closely related TQFT with slightly different grading $\hat \AA$ (rather than $\AA$, as discussed below \eqref{ACZ}). The computation is easy to adapt to the current setting, and we discuss it now.

We restrict for the moment to decorated surfaces $\CS=(\Sigma,\omega)$ such that $\Sigma$ is a genus-$g$ surface without any framed points, and $\omega  \in H^1(\Sigma;\AA)$ takes at least one value not in $\X=\Z/\Z$. We will return to the case $\omega=0$ in Section \ref{sec:CGP-0} below.

Let $\CZ_{\rm CGP}(\CM)$ be the invariant of closed oriented 3-manifolds $M$ with $\catH$-colored ribbon graph $\tau$ and cohomology class $ \omega \in H^1(M\setminus T; \AA)$ defined in \cite{CGP} (denoted `$\mathrm N$' therein). In  \cite{BCGP}, this invariant was extend to a TQFT as follows. For the decorated surface $\CS$, consider the infinite dimensional vector space $\mathcal{V}(\Sigma)$ which is the $\C$-span of all the decorated cobordisms whose underlying manifold bounds $\Sigma$.  Also define $\mathcal{V}'(\Sigma)$ as the $\C$-span of all decorated cobordisms bounding $\overline{\Sigma}$ (the surface with opposite orientation). One can define a pairing $\langle\,,\,\rangle: \mathcal{V}' (\Sigma) \otimes \mathcal{V} (\Sigma) \to \C$ by extending linearly  the  assignment $\langle \CM, \mathcal \CM'\rangle = \CZ_{\rm CGP} (\CM \circ \CM') \in \C$.   Finally, define the CGP state space as $\mathcal{V} (\mathcal  S)$ modulo the right kernel of this pairing:
\be \mathsf{V} (\mathcal S) = \CZ_{\rm CGP}(\CS) :=\mathcal{V}(\mathcal S) / \text{kerR}\langle\,,\,\rangle\,. \ee
The state space has the following nice properties (see \cite{BCGP}).
Let $M$ and $M'$ be 3-manifolds with boundaries  $\Sigma$ and $\overline{\Sigma}$, respectively.  Then 
\begin{enumerate}
\item The vector space $\mathsf{V}(\mathcal S)$ is finite dimensional.
\item The vector space $\mathsf{V}(\mathcal S)$ is generated by cobordisms whose underlying manifold is $M$, \ie\ decorated cobordisms $(M,\tau, \omega)$ such that $\partial{M}=\Sigma$.
\item A linear combination $\sum a_i  \mathcal{M}_i$ of cobordisms is zero in $\mathsf{V} (\mathcal S)$ if and only if for any decorated cobordism $\mathcal {M}'$ with underlying manifold $M'$, one has $\sum a_i \CZ_{\rm CGP} ( \mathcal{M}_i\circ \mathcal{M}')=0$.  
\end{enumerate}
Even though $\mathcal{V}(\mathcal S)$ is infinite dimensional, the above properties allow one to make a finite number of computations to determine $\mathsf{V} (\mathcal S)$.  The main tools used to do such computations are the $\sigma$-equivalence mentioned above and \emph{skein equivalence}, described in Section 4.1 of \cite{BCGP}.  Loosely speaking, a skein relation is a linear combination of graphs whose value by the Reshetikhin-Turaev functor is zero; 
such relations impose equivalences on cobordisms.  

We now use these equivalences and the properties above to compute the state spaces in the several special cases. 
 
Given a closed genus-$g$ surface $\Sigma$, fix a handlebody $\eta$ bounding $\Sigma$ and let $\Gamma$ be an oriented spine of $\eta$ (such that $\eta$ collapses onto $\Gamma$).  Using meridians of the edges of $\Gamma$ the cohomology class $\omega  \in H^1(\Sigma;\AA)$ induces an $\AA$-coloring of the edges of $\Gamma$.  We say a $\catH$-coloring is compatible with this $\AA$-coloring if for each edge $e$ of $\Gamma$ the object in $\catH$ assigned to $e$ is in the graded portion of the category determined by the $\AA$-coloring. A closely related setup was depicted in Figure \ref{fig:genus-g}.

If $g=1$ then let $\Gamma_{V,f}$ be the graph $\Gamma$ which is the circle with a single coupon colored with a morphism $f:V\to V$ and whose edge is colored with an object $V$ of $\catH$ which is a compatible coloring. 

If $g>1$ then let $\Gamma_{\{V_e,f_v\}}$ be the graph $\Gamma$ where each edge $e$ is colored with an object $V_e$ of $\catH$ which is a compatible coloring and each trivalent vertex $v$ is extended to a coupon colored with a morphism $f_v$.  For any cobordism whose underlying manifold is $\eta$ there exists a skein-equivalent cobordism of the form $\Gamma_{V,f}$ or $\Gamma_{\{V_e,f_v\}}$.  We use these cobordisms to find a basis for the state spaces.  

 \subsubsection{The CGP state space for the torus}
 \label{sec:CGP-torus}
 
Consider the torus $\Sigma=S^1\times S^1$.  Let $\eta=B^2\times S^1$ be the solid torus and $\Gamma=\{0\}\times S^1$ be its core.   As above the cohomology class colors $\Gamma$ with an element $\bar\alpha$ of $\AA$. 
 The proof of Proposition 6.5 of \cite{BCGP} applies here, and shows that since  $\omega  \in H^1(\Sigma;\AA)$ takes at least one value not in $\X=\Z/\Z$, we can assume that the edge coloring $\Gamma$ is an element  $\bar\alpha$ of $\AA\setminus\X$. 
Choose a lift $\alpha\in \C$ of $\bar\alpha$.  Let $\Gamma_{z}=\Gamma_{V_z, \Id}$ be the graph $\Gamma=\{0\}\times S^1\subset B^2\times S^1$ colored with $V_z$ for $z \in I_{\alpha}$.  The basis of the CGP state space associated to $\Sigma=S^1\times S^1$ is given by the cobordisms $(\eta,\Gamma_z,\omega)$ for  $z \in I_{\alpha}$.  
In particular, the dimension of this state space is $2k$, in agreement with \eqref{Hilb-T2A}.

To prove these vectors are a basis one needs to show that they generate the CGP state space and that they are linearly independent.  To see they generate, recall that any cobordism whose underlying manifold is $\eta$ is skein-equivalent a cobordism of the form $(\eta,\Gamma_{V,f},\omega)$.  Since  $\bar\alpha\in \AA\setminus \X$,  the object $V\in \catH_{\bar\alpha}$ must be a direct sum of simple modules of the form $V_{z}\otimes \sigma(m) \cong V_{z +2mk}$ for $z\in I_{\alpha}$ and $m\in \Z$.  
It follows that $(\eta,\Gamma_{V,f},\omega)$ is skein-equivalent to a linear combination of  
cobordisms of the form $(\eta,\Gamma_{z+2mk},\omega)$.   
Since $V_{z+ 2mk}\cong V_{z}\otimes \C^H_{2mk}$, then by definition the cobordism whose core is $\Gamma_{z+ 2mk}$ has the same value as the cobordism whose core is $\Gamma_{z}$ with a parallel component colored with $\sigma(m)=\C^H_{2mk}$.  Therefore,  the cobordisms with cores $\Gamma_z$ and $\Gamma_{z+ 2mk}$ are $\sigma$-equivalent, and so from Proposition 4.2 of \cite{BCGP} they give equal vectors in $\mathsf{V}(\mathcal S)$ for all $m\in \Z$.  The $\sigma$-equivalent cobordisms are equal because the compatibility condition given in Equation \eqref{E:Braiding-sigma} implies that the parallel component labeled with $\sigma(m)$ can be unlinked and removed while not changing the value of the 3-manifold invariant $\CZ_{\rm CGP}$.  To see that the vectors  $\mathsf{V}(\eta,\Gamma_z,\omega)$ for  $z \in I_{\alpha}$ are linearly independent one can use the third property above, pairing these vectors with appropriate vectors in $\CV'(\mathcal S)$ and showing that the pairing is non-degenerate.  This is done in Proposition 6.3 of \cite{BCGP} by gluing two solid tori together and reversing the orientation of the graph.  This can also be done by embedding the solid torus $\eta$ into the sphere $S^3$, taking the exterior of $\eta$ as a second solid torus, and using the Hopf pairing induced from $\CZ_{\rm CGP}$.  

 \subsubsection{The CGP state space for higher genus}
 \label{sec:CGP-g}
 
For $g=2$, fix a handlebody bounding the genus 2 decorated surface $\Sigma$ without any points.  The core of this handlebody is a theta graph $\Gamma$ with edges $e_1, e_2, e_3$.   As above, the cohomology class colors each $e_i$ with element $\bar\alpha_i$ of $\AA$.  As above, Proposition 6.5 of \cite{BCGP} implies that since  $\omega  \in H^1(\Sigma;\AA)$ takes at least one value not in $\X$, we can assume that each $\bar\alpha_i$ is not in $ \X$.  
Choose lifts $\alpha_1\in \C$ and $\alpha_2\in \C$ of $\bar\alpha_1$ and $\bar\alpha_2$, respectively.  For $i=1,2$ let  $\beta_i \in I_{\alpha_i}$ and label $e_i$ with $V_{\beta_i}$.  Then the trivalent vertices of $\Gamma$ label elements in the direct-sum decomposition 
\be V_{\beta_1}\otimes V_{\beta_2}\cong \bigoplus_{j\in H_k} V_{\beta_1+ \beta_2+j}\,, \ee
where $H_k =\{-(k-1),-(k-3),...,(k-1)\}$.
(This decomposition assumes that $\beta_1, \beta_2, \beta_1+\beta_2$ are not in $\X$).  These decompositions and corresponding labelings give non-zero vectors and a basis  in the CGP state space.  There are $2k$ choices for the colorings of both $e_1$ and $e_2$; where for each such choice there are $k$ summands.  Thus, the state space for $g=2$ has dimension $k(2k)^2=2^2k^3$.

In general, a similar argument shows that the dimensions of the state space associated to a genus $g>1$ surface $\Sigma_g$  without any framed points and whose cohomology class takes at least one value not in $\X$ is
\be  \text{dim}\, \mathsf V(\Sigma_g,\omega)= 2^g k^{3g-3}\,, \ee
in agreement with \eqref{Hilb-g}.
The combinatorics are closely analogous to those of Section \ref{sec:U-g}.

 \subsubsection{The CGP state space with zero cohomology class}
 \label{sec:CGP-0}

Finally, we consider the decorated surface $\mathcal S$ whose underlying manifold is the torus  $\Sigma=S^1\times S^1$ with no marked points and \emph{zero} cohomology class (trivial flat connection).   A basis for the state space $\mathsf{V}(\mathcal S)$ was conjectured in Proposition 7.3 of \cite{CGP2}. We summarize this result here;
it is closely related to the computation of Hochschild homology of $\CCab_1$ in Appendix \ref{app:U}.

Let $\eta=B^2\times S^1$ be the solid torus and let $\Gamma=\{0\}\times S^1$ be its core.   
Analogous to the generic setting above, the state space $\mathcal{V} (\mathcal S)$ is generated by cobordisms whose underlying manifold is $\eta$, and its core has colorings in  $\catH_{\bar 0}$. Now, however, there may be additional coupons along the core, \ie\ nontrivial morphisms among objects $V\in \catH_{\bar 0}$ coloring the core.

The category $\catH_{\bar 0}$ is generated by projective indecomposable modules $P_n\otimes \sigma(m)$ and $P_n\otimes \C_k^H\otimes \sigma(m)$, for $n=1,...,k$ and $m\in \Z$, whose images in $\CCab_1$ are the modules $P_n^+$ and $P_n^-$ (respectively) shown in \eqref{reps-proj} (also see Proposition 6.2 of \cite{CGP2} for a list of the projective indecomposable modules). The space $\mathcal{V} (\mathcal S)$ is then generated by all cobordisms whose underlying manifold is $\eta$ with the graph $\Gamma_{P,f}$ where $P$ is any direct sum of projective indecomposable modules in the list above and $f:P\to P$ is any morphism.

The projective modules $P_k,\,P_k\otimes \C_k^H$ (and their $\sigma$-images) are simple, and admit no nontrivial morphisms.

The remaining projectives may be divided into $k-1$ blocks
\be \mathsf B_n = \{P_n\otimes \sigma(m),\, P_{k-n}\otimes \C_k^H \otimes\sigma(m)\}_{n\in\{1,...,k-1\},m\in \Z}\,, \ee
with the property that there are no morphisms between modules in different blocks. The endomorphism algebra $\mathsf A$ of the (direct sum of) projective modules within each block is independent of $n\in\{1,...,k-1\}$; moreover, this algebra may be represented as a path algebra of an infinite periodic quiver, depicted in \cite[Sec. 7.1]{CGP2}, which is an infinite cover of the two-vertex quiver appearing in \eqref{Pquiver} of Appendix \ref{app:U}. After quotienting out by all relations of the form $fg-gf$, corresponding to sliding coupons around the core of the torus, one finds that the co-center
\be \mathsf A/[\mathsf A,\mathsf A] = HH_0(\mathsf A) \ee
is generated by identity endomorphisms $1_{n,m}^+$ and $1_{n,m}^-$ of each $P_n\otimes \sigma(m)$ and $P_{k-n}\otimes \C_k^H \otimes\sigma(m)$ (respectively), and by a single nilpotent endomorphism $\tau_{n,0}^+$ of $P_n\otimes \sigma(0)$.

After imposing $\sigma$-equivalence, the CGP state space $\mathsf{V}(\mathcal S)$ is then generated by the following vectors:
 \begin{enumerate}
\item cobordisms $\eta$ whose core is labelled by $P_n$ or $P_n \otimes \C^H_{k}$ ($n=1,...,k$) with no coupon
\item cobordisms $\eta$ whose core is labelled by $P_n$ ($n=1,...,k-1$) with the single coupon $\tau_{n,0}^+$
\end{enumerate}
Conjecturally this generating set is a basis, and if true the dimension of the state space would be $2k+(k-1)=3k-1$, in agreement with the degree-zero part of \eqref{Hilb-T2}. To prove this one needs to find a non-degenerate pairing. Analogous to above, such a pairing should come from embedding the solid torus $\eta$ into the sphere $S^3$,  taking the exterior of $\eta$ as a second solid torus, and using the Hopf pairing induced from $\CZ_{\rm CGP}$.  Interestingly, here the second torus must have cores colored with the non-projective simple modules $S_n$.  We have checked that this pairing proves the conjecture for $k=2$ and $k=3$.

\section{3d topological A-twist of Chern-Simons-matter theory}

\label{sec:Atwist}

In this section we introduce the main family of 3d quantum field theories expected to realize and generalize the mathematical TQFT's of Section \ref{sec:Uqsl2}. 

Gaiotto and Witten \cite{GaiottoWitten-Sduality} introduced a family of 3d $\CN=4$ superconformal field theories $T[G]$ labelled by compact simple groups $G$, defined as decoupling limits of S-duality interfaces in 4d $\CN=4$ super-Yang-Mills theory. The theory $T[G]$ has $G\times G^\vee$ flavor symmetry, where $G^\vee$ denotes the Langlands-dual group. The theories we want to consider, denoted
\be \CT_{G,k}:= T[G]/G_k\,, \ee
gauge the $G$-symmetry with Chern-Simons kinetic term at non-zero level `$k$.' The gauging is done by introducing a 3d $\CN=2$ vectormultiplet, which admits a supersymmetric Chern-Simons term \cite{GaiottoYin}.

Quite nontrivially, the resulting theory $\CT_{G,k}$ has enhanced 3d $\CN=4$ superconformal symmetry in the infrared \cite{GaiottoWitten-Janus}.%
\footnote{The theory is closely related to ABJM theory \cite{ABJM} and the generalized Chern-Simons-matter quivers of \cite{HLLLP1,HLLLP2}, as well as the recent \cite{GangYamazaki} in which a diagonal flavor symmetry of $T[SU(2)]$ is gauged. The superconformal representation theory underlying enhancements to $\CN\geq 4$ SUSY was further developed in~\cite{dMFOME}.} %
 In particular, the 3d $\CN=4$ SUSY algebra and full $SU(2)_C\times SU(2)_H$ R-symmetry are symmetries of $\CT_{G,k}$.
This allows one, in principle, to further define 3d topological A and B twists of $\CT_{G,k}$. As discussed in previous sections, this amounts in flat space to restricting to local and extended operators that are preserved by a nilpotent supercharge $Q_A$ (resp. $Q_B$) in the 3d $\CN=4$ algebra, whose anticommutators $[Q_A,-]$ contain all translations. In curved space, one would also introduce an $SU(2)_H$ (resp. $SU(2)_C$) background matching the spin connection, in order to preserve $Q_A$. We denote the topological $A$ and $B$ twists of $\CT_{G,k}$ as $\CT_{G,k}^A$, $\CT_{G,k}^B$.

Many aspects of the 3d cohomological TQFT $\CT_{G,k}^B$ were developed by Kapustin and Saulina in \cite{KapustinSaulina-CSRW}. More accurately, \cite{KapustinSaulina-CSRW} studied a deformation of $\CT_{G,k}^B$ that came from resolving the Higgs branch of $T[G]$. Their TQFT was termed ``Chern-Simons-Rozansky-Witten'' theory, since it combined aspects of Chern-Simons theory and Rozansky-Witten theory \cite{RW} (\emph{a.k.a.} the B-twist of a 3d $\CN=4$ sigma-model to the Higgs branch). In this paper, we are instead interested in the A-twist $\CT_{G,k}^A$.

The theory $\CT_{G,k}^A$ has several qualitative features that make it a potential candidate for matching the mathematical TQFT of Section \ref{sec:Uqsl2} (when $G=SU(n)$)\,:
\begin{itemize}
\item $\CT_{G,k}^A$ is Chern-Simons-like. In particular, its line operators include a finite set of Wilson lines, labelled by representations of $G$, which we will construct explicitly in Section \ref{sec:line1}.
The Wilson lines match the simple representations of $U_q(\mathfrak g)$ at $q=e^{i\pi/k}$.
\item $\CT_{G,k}^A$ has a nontrivial algebra of bulk local operators, and local operators at junctions of line operators. This makes the category of line operators non-semisimple, one of the key features of  $U_q(\mathfrak g)$-mod.
\item $\CT_{G,k}^A$ has $G^\vee$ flavor symmetry, a ``B-type'' symmetry in the language of Section \ref{sec:sheaf}, which allows the theory to be deformed by flat, complexified $G_\C^\vee$ background connections.
\item $\CT_{G,k}^A$ also has a $\mathbb Z_n$ 1-form symmetry (for $G=SU(n)$), which grades the category of line operators. (The corresponding $\Z_2$ symmetry of $U_q(\mathfrak{sl}_2)$-mod was discussed in Section \ref{sec:V-1}. It generalizes to a $\Z_n$ symmetry of $U_q(\sln)$-mod.)
\end{itemize}

In order to access more quantitative features of $\CT_{G,k}^A$ --- such as its category of line operators, boundary VOA's, and state spaces on various surfaces --- we need a more concrete definition than that given above, in terms of the abstract SCFT $T[G]$. Our main goal in this section is to outline two such definitions, for the case $G=SU(n)$. 

One definition, based on collisions of boundaries and interfaces in 4d $\CN=4$ SYM, is fairly standard; we review it in Section \ref{sec:4d}. The other definition is new; it is based on passing through a holomorphic-topological twist \cite{ACMV, CDG}, which is valid when 3d spacetime has a transverse holomorphic foliation (for example, $M=\Sigma\times \R$ for any Riemann surface $\Sigma$). It will lead to a fully Lagrangian formulation of  $\CT_{G,k}^A$ in such geometries, outlined in Section \ref{sec:Tnk-BV}.

We recall that when $G=SU(n)$, the theory $T[SU(n)]$ admits a UV Lagrangian formulation as a linear quiver gauge theory:
\be \label{TSUn-UV}
 T[SU(n)]^{UV}:\quad \raisebox{-.2in}{$ \begin{tikzpicture}[scale=1]
		\begin{scope}[auto, every node/.style={inner sep=1}, node distance=1cm]
			\node (v3) {};
			\node[right=of v3] (v4) {};
		\end{scope}
		\begin{scope}[auto, every node/.style={minimum size=3em,inner sep=1},node distance=0.5cm]
			\node[draw, circle, left=of v3] (v2) {$n-1$};
			\node[draw, left=of v2] (v1) {$n$};
			\node[draw, circle, right=of v4] (v5) {2};
			\node[draw, circle, right=of v5] (v6) {1};
		\end{scope}
		\draw(v1)--(v2);
		\draw(v2)--(v3);
		\draw[loosely dotted, line width = 1pt] (v3)--(v4);
		\draw(v4)--(v5);
		\draw(v5)--(v6);
	\end{tikzpicture} $}
\ee
One may wonder why this is not good enough for computations. It has manifest $SU(n)$ ``Higgs branch'' flavor symmetry acting on the `$n$' node, which can gauged with a level-$k$ Chern-Simons coupling, producing a theory $T[SU(n)]^{UV}/SU(n)_k$ that will flow to $\CT_{SU(n),k}$ in the IR. The problem is that $T[SU(n)]^{UV}/SU(n)_k$
 has at most 3d $\CN=3$ rather than 3d $\CN=4$ SUSY, the latter only appearing in the IR. In particular, the A-twist supercharge $Q_A$ is \emph{not} a symmetry of the action of $T[SU(n)]^{UV}/SU(n)_k$, so it is not possible to define its topological A-twist.
 
On the other hand, the Lagrangian theory $T[SU(n)]^{UV}/SU(n)_k$ does admit a holomorphic-topological (HT) twist, defined using a supercharge $Q_{HT}$ in the 3d $\CN=2$ algebra.  After passing to a HT-twisted version of $T[SU(n)]^{UV}/SU(n)_k$ --- integrating out some fields and removing $Q_{HT}$-exact terms --- we will find an action that has an additional nilpotent symmetry $Q_A$ with properties matching the desired A-twist supercharge. In particular, $Q_A$ has the correct R-symmetry charges, and the stress tensor is $Q_A$-exact.%
\footnote{We only verify exactness of the stress tensor classically in this paper.} %
One may interpret the existence of the extra $Q_A$ symmetry in the HT-twisted theory as saying that \emph{the failure of the orignal action of $T[SU(n)]^{UV}/SU(n)_k$ to be $Q_A$-invariant is $Q_{HT}$-exact.} 
This is not too surprising, as one expects the HT twist to be nearly invariant under RG flow (\emph{cf.} \cite{CDG,Williams-oneloop}), and $T[SU(n)]^{UV}/SU(n)_k$ recovers $\CN=4$ SUSY (and thus the A-twist) in the IR.

\subsection{Global considerations and higher symmetries}
\label{sec:global}

We remark that there are some subtle choices to be made when gauging a flavor symmetry of $T[G]$. Four-dimensional aspects of this were explained in \cite{AST}, and implications for the 3d theory $T[G]$ were explained in \cite{EKSW}.

The full flavor-symmetry group of $T[G]$ is $\wt G\times \wt G^\vee$, where $\wt G, \wt G^\vee$ denote the \emph{simply connected} forms (the universal covers) of $G$ and $G^\vee$. There is a mixed 't Hooft anomaly among the centers of $\wt G$ and $\wt G^\vee$. 
Any global form $G'$ of $G$ can be gauged with appropriately quantized Chern-Simons levels, leading to a theory
\be \CT_{G',k} \,=\, T[G]/G'_k\,, \ee
and an $A$-type topological twist $\CT_{G',k}^A$.  Roughly speaking, the mixed 't Hooft anomaly will then dictate which global form of $G^\vee$ survives as a flavor symmetry of $\CT_{G',k}$, and thus what sort of background flat connections this TQFT can be deformed by.

A more refined analysis indicates that when $G'$ has nontrivial center, the gauged theory $T_{G',k}$ will have a 2-group symmetry \cite{CDI,HsinLam-discrete,ABGS, BCH2group}. The 2-group consists of a discrete 1-form symmetry \cite{GKSW, HLS1form} dual to the center of $G'$, and a 0-form symmetry $\wt G^\vee$, part of whose center is ``entwined'' with the 1-form symmetry. Inside the 2-group, one finds 1) a standard 0-form symmetry that's a particular quotient of $\wt G^\vee$; and 2) an independent 1-form symmetry dual to the center of $G'$.

We will mainly consider $G=SU(n)$, and gauge the simply connected form $G'=SU(n)$. Then $\CT_{n,k}=T[SU(n)]/SU(n)_k$ has
\begin{itemize}
\item A 1-form $\Z_n$ global symmetry
\item A 0-form $SU(n)$ global symmetry, whose center is entwined with the 1-form symmetry, as part of a 2-group structure. The standard 0-form symmetry (the part independent of the 1-form symmetry) is $G^\vee=SU(n)/\Z_n=PSU(n)$. 
\end{itemize}
In turn, the A-twisted theory $\CT_{n,k}^A$ will have a category of line operators that is $\Z_n$-graded, and can be deformed by flat $PGL(n,\C)=PSU(n)_\C$ connections. It is worth noting that the non-zero Chern-Simons level induces an 't Hooft anomaly in the above 1-form symmetry. This does not mean the category is no longer graded by $\Z_n$, only that the 1-form symmetry generators themselves can have nontrivial $\Z_n$ charges. We will return to this in Section \ref{sec:Hilbert}.

Our main conjecture is that this $\CT_{n,k}^A$ reproduces a (derived version of) the TQFT based on representations of $U_q(\mathfrak{sl}_n)$ at $q=e^{i\pi/k}$ that was discussed in Section \ref{sec:Uqsl2}. There are other choices of $G'$, and other closely-related versions of the TQFT from Section \ref{sec:Uqsl2} (differing in how various quotients are taken in the representation category), which would be interesting  to match.

\subsection{Definition from 4d and branes}
\label{sec:4d}

We briefly review a construction of $\CT_{G,k}$ using BPS boundary conditions and interfaces in 4d super-Yang-Mills, following \cite{GaiottoWitten-Janus, GaiottoWitten-boundary, GaiottoWitten-Sduality}, and their lifts to brane webs in IIB string theory \cite{AharonyHanany, AharonyHananyKol}. The A-type topological twist of $\CT_{G,k}$ is induced by a particular geometric-Langlands twist \cite{KapustinWitten} of the 4d configuration.
We refer the reader to these works as well as discussions in the more recent \cite{GaiottoRapcak, CreutzigGaiotto-S, FrenkelGaiotto} for further details and subtleties. 

4d $\CN=4$ super-Yang-Mills theory admits an array of half-BPS boundary conditions. In the case of $G=(P)SU(n)$,  many may be engineered by junctions of a stack of $n$ D3 branes with $(p,q)$ 5-branes. These include:
\begin{itemize}
\item Neumann $B_{1,0}$, which preserves the bulk gauge symmetry. It is engineered by D3's ending on a single NS5 brane, \emph{a.k.a.} (1,0) 5-brane.%
\footnote{Our notation `$B_{p,q}$' is borrowed 
\cite{GaiottoRapcak, CreutzigGaiotto-S}.} %
\item Neumann with an additional $k$ units of Chern-Simons coupling $B_{1,k}$. It is engineered by D3's ending on a single $(1,k)$ brane.
\item Nahm pole $B_{0,1}$, the S-dual of Neumann, which breaks the bulk gauge symmetry with no residual boundary flavor symmetry. It is engineered by D3's ending on a single D5 brane.
\item Dirichlet $\wt B_{0,1}$, which breaks the bulk gauge symmetry to constant gauge transformations at the boundary, and thus has boundary flavor symmetry $G$. It is engineered by $n$ D3's each ending on an individual D5 brane.
\item $\wt B_{1,0}$, the S-dual of Dirichlet, equivalent in the IR to Neumann coupled to the S-duality interface $T[G]$. It is engineered by $n$ D3's each ending on an individual NS5 brane.
\end{itemize}

Now let's construct $\CT_{G,k}$.
The 3d $\CN=4$ theory $T[G]$, an S-duality interface decoupled from the 4d bulk, may be engineered by ``sandwiching''  the S-duality interface between Dirichlet boundary conditions:
\be  \raisebox{-.3in}{$\includegraphics[width=4.4in]{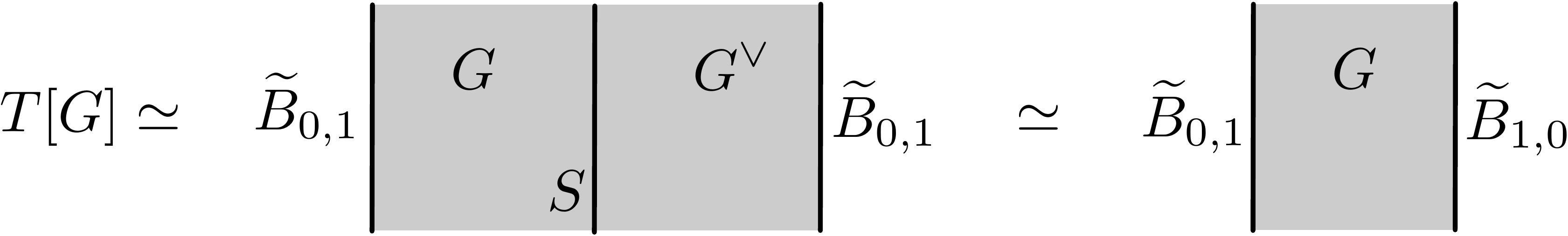}$}\ee
In order to gauge the $G$ flavor symmetry of $T[G]$, at Chern-Simons level $k$, we replace one of the Dirichlet b.c. with a deformed Neumann b.c. $B_{1,k}$,
\be   \raisebox{-.5in}{$\includegraphics[width=5.5in]{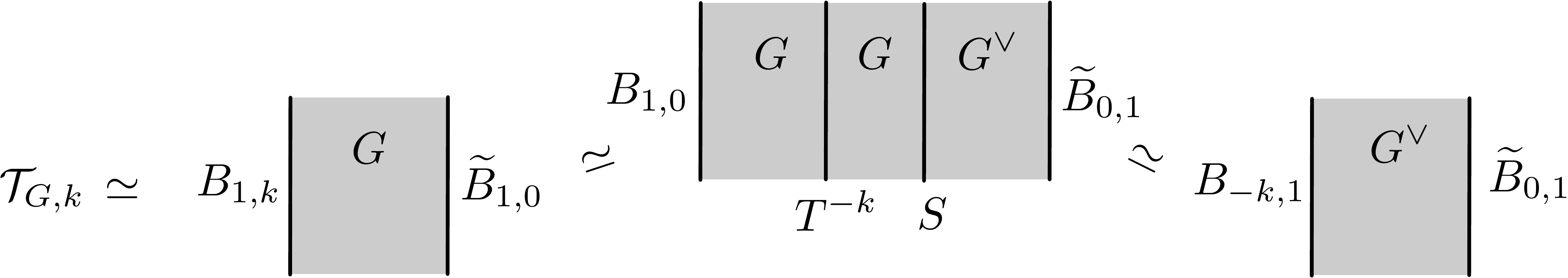}$} \label{sand} \ee
Equivalently, as shown in the middle of \eqref{sand}, we may sandwich a $T^{-k}S$ interface between a Neumann and a Dirichlet b.c.  We may also collide all the interfaces with the Neumann side, obtaining a sandwich between a modified Nahm b.c. $B_{-k,1}$ and pure Dirichlet, as shown on the right. All these configurations become equivalent after a flow to the infrared.

We would further like to engineer the 3d topological A-twist of $\CT_{G,k}$. This can be done by considering the geometric-Langlands twist \cite{KapustinWitten}  of the bulk theory, with canonical parameter $\Psi=0$ by the deformed Neumann b.c. and $\Psi=\infty$ by the Dirichlet b.c. Adding canonical parameters to the previous diagrams, we obtain
\be \hspace{-.2in}   \raisebox{-.5in}{$\includegraphics[width=5.7in]{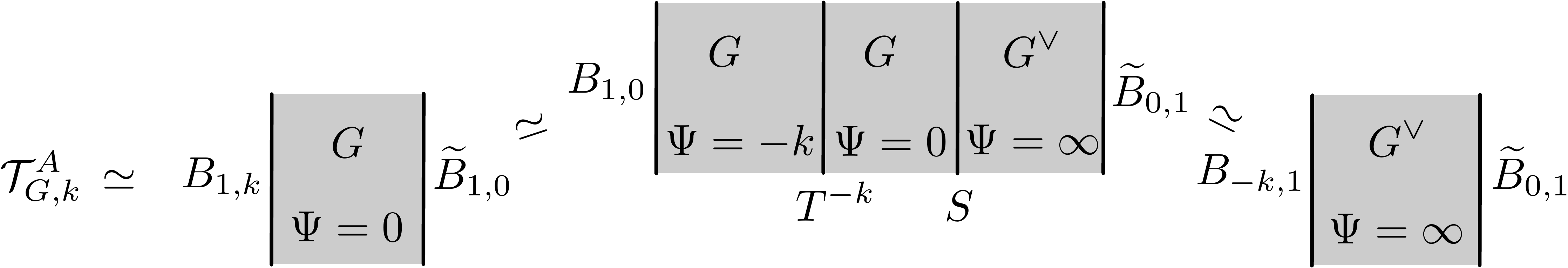}$} \label{sandA} \ee

The $\Psi=0$ twist of the bulk theory, also called the 4d $A$-twist  (\emph{e.g.} in \cite{Witten-Nahm}), induces a 3d topological A-twist on boundary conditions that preserve 3d $\CN=4$ supersymmetry. Thus, the $(B_{1,k},\wt B_{1,0})$ configuration on the left of \eqref{sandA} implements a 3d A-twist of $T[G]$ (on the $\wt B_{1,0}$ boundary), further gauging it at Chern-Simons level $k$.

In the configuration on the right of \eqref{sandA}, we find the $\Psi=\infty$ twist of the bulk theory, also called the 4d B-twist. It induces a 3d topological B-twist on 3d $\CN=4$ boundary conditions.
 The Dirichlet boundary condition $\wt B_{0,1}$ breaks the bulk $G^\vee$ gauge symmetry to constant gauge transformations along the boundary, \emph{i.e.} a $G^\vee$ flavor symmetry. In the B-twist, this allows the boundary to be deformed by complexified $G^\vee_\C$ flat connections --- by essentially the same mechanism as in Section \ref{flat-def}. This deformation on a Dirichlet boundary condition for the 4d B-twist has been discussed in \cite{KapustinWitten,CreutzigGaiotto-S,FrenkelGaiotto}.

When $G=(P)SU(n)$, the sandwiched configurations in \eqref{sand} can be further lifted to ``brane webs'' in type IIB string theory, as first discussed by \cite{AharonyHanany,AharonyHananyKol}. For example, to produce the $(B_{-k,1},\wt B_{0,1})$ sandwich, one considers a stack of $n$ D3 branes ending on single $(-k,1)$ 5-brane on one side, and a stack of $n$ $(0,1)$ branes (\emph{a.k.a.} D5 branes) on the other. In order to preserve four supercharges, the slope of a $(p,q)$ brane in a particular plane along the D3's must equal $q/p$; thus the configuration schematically looks like:
\be \label{brane-D}
\raisebox{-.4in}{$\includegraphics[width=4.5in]{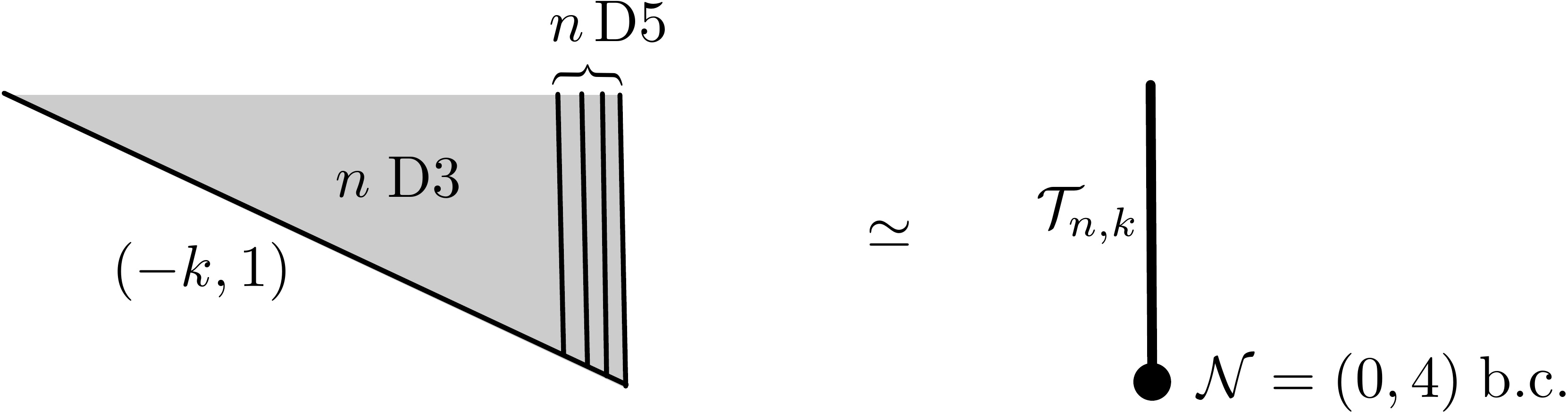}$} 
\ee
Due to the relative angles of the branes, this effectively produces not just the 3d $\CN=4$ theory $\CT_{n,k}$, but also a 2d $\CN=(0,4)$ boundary condition for $\CT_{n,k}$ at the corner where the branes meet. In the A-twist, the corner supports a vertex operator algebra that was constructed in \cite{GaiottoRapcak, CreutzigGaiotto-S}, which we will revisit in Section \ref{sec:VOA}.

\subsection{3d gauge theory}
\label{sec:3d}

We now return to the 3d field-theory construction of $\CT_{G,k}$, in the UV, as a linear Chern-Simons-matter gauge theory. We restrict to the theory for $G=SU(n)$, also called $\CT_{n,k}$.

The construction follows directly from the ``sandwich'' on the left side of \eqref{sand}, which defines the 3d quiver gauge theory
\be \label{Tnk-quiver} \CT_{n,k}^{UV}:\quad \raisebox{-.2in}{$ \begin{tikzpicture}[scale=1]
		\begin{scope}[auto, every node/.style={inner sep=1}, node distance=1cm]
			\node (v3) {};
			\node[right=of v3] (v4) {};
		\end{scope}
		\begin{scope}[auto, every node/.style={minimum size=3em,inner sep=1},node distance=0.5cm]
			\node[draw, circle, left=of v3] (v2) {$n-1$};
			\node[draw, circle, double, left=of v2] (v1) {$n_k$};
			\node[draw, circle, right=of v4] (v5) {2};
			\node[draw, circle, right=of v5] (v6) {1};
		\end{scope}
		\draw(v1)--(v2);
		\draw(v2)--(v3);
		\draw[loosely dotted, line width = 1pt] (v3)--(v4);
		\draw(v4)--(v5);
		\draw(v5)--(v6);
	\end{tikzpicture} $} = T[SU(n)]^{UV}\big/SU(n)_k
\ee
Most of this is a 3d $\CN=4$ quiver gauge theory: there are bifundamental hypermultiplets for every edge, and the symmetries $U(n-1)\times\cdots\times U(1)$ are gauged with 3d $\CN=4$ vectormultiplets. The final $SU(n)$ symmetry corresponding to the doubled node is gauged with a 3d $\CN=2$ vectormultiplet, with a Chern-Simons kinetic term at level $k$.

One explanation for 3d $\CN=4$ SUSY enhancement in the IR (and only in the IR!) goes as follows. Consider the UV gauge theory realizing $T[SU(n)]$ in \eqref{TSUn-UV}. The $SU(n)$ flavor symmetry of $T[SU(n)]$ has a $\mathbb{CP}^1$ of associated complex moment-map operators $\mu_\zeta \in (\mathfrak{sl}_n)^*$, labelled by choices of complex structure `$\zeta$' on the Higgs branch.
It was shown in \cite{GaiottoWitten-Janus} (see also the discussion in \cite{KapustinSaulina-CSRW}) that gauging $SU(n)$ with a 3d $\CN=2$ vectormultiplet at nonzero Chern-Simons level $k$ actually preserves 3d $\CN=4$ supersymmetry so long as the operator identity
\be \text{Tr} (\mu_\zeta{}^2) = \text{constant} \qquad \forall \;\zeta\,\in \cp^1 \label{fund-id} \ee
holds.

In the infrared of $T[SU(n)]$, the LHS of \eqref{fund-id} is thus set equal to a quadratic function of the complex FI parameters on the $U(n-1),...,U(1)$ nodes, in complex structure $\zeta$. Explicitly,
\be \text{Tr} (\mu_\zeta{}^2) \;\overset{IR} \sim \; \sum_{i=1}^n (\hat t^{(i)}_\zeta)^2\,, \label{mom-FI} \ee
where $\sum_{i=1}^n \hat t^{(i)}_\zeta=0$ and the FI parameters are $t_\zeta^{(i)} = \hat t_\zeta^{(i)}-\hat t_\zeta^{(i+1)}$, $i=1,...,n-1$.
Therefore, \eqref{fund-id} holds. More so, at the CFT point where FI parameters are set to zero, we have $  \text{Tr} (\mu_\zeta{}^2) \;\overset{IR} =0$. In the UV, however, \eqref{mom-FI} is not an operator identity, and the full $\CN=4$ SUSY algebra does not act via symmetries of $\CT_{n,k}^{UV}$.

\subsection{The $A$-Twist of $\CT_{n,k}$}
\label{sec:Tnk-BV}

We would have liked $\CT_{n,k}^{UV}$ to have $\CN=4$ SUSY, in order to identify a nilpotent supercharge $Q_A$  whose cohomology defines the topological A-twist. It is not possible to do this directly. We will circumvent the problem by first passing to a holomorphic-topological twisted version of $\CT_{n,k}^{UV}$. The holomorphic-topological twisted theory is presented in the BV-formalism and has a particularly simple action in terms of ``twisted superfields." We identify a nilpotent symmetry $Q_A'$ of the holomorphic-topological twisted theory that we expect to implement its deformation to the $A$-twisted theory $\CT_{n,k}^A$. The resulting theory is a (chiral deformation of) mixed BF theory and thus admits a finite, 1-loop exact, perturbative quantization \cite[Theorem 5.1]{GWoneloop}.

\subsubsection{Holomorphic twist and its deformations}

As reviewed in Section \ref{sec:hyper-VOA}, any 3d $\CN=2$ theory admits a holomorphic-topological (HT) twist \cite{Kapustin-hol, ACMV}. This requires working on a 3d spacetime $M$ with a transverse-holomorphic-foliation structure. Locally, we will assume $M$ is flat Euclidean space, split as $M = \C_{z,\bar z}\times \R_t$.

The $HT$-twist of general $\CN=2$ Chern-Simons-matter theories admits a very clean description in the BV-BRST formalism using the ``twisted superfields" of \cite{ACMV, CDG}. These twisted superfields can be immediately applied to the $HT$-twisted $\CN=4$ theories of interest by choosing an $\CN=2$ subalgebra of the full $\CN=4$ algebra. We will start with this $HT$-twisted theory and then deform it to the $A$-twist. For more details about this deformation, as well as the deformations to the B-twist, see \cite{twistedN=4}. (The HT$\to$B deformation of a hypermultiplet was also discussed in Section \ref{sec:hyper-VOA}.)

One utility of this formalism is to dramatically simplify the field content of the theory without losing any of the derived structures, \eg\ higher operations obtained by descent. In particular, the bulk local operators in the HT twist have the structure of a commutative vertex algebra $\CV$, which is $\Z\times \Z$ -graded by $U(1)_R$ charge (a cohomological grading) and spin in the $\C_{z,\bar z}$ plane (a non-cohomological grading). This algebra is endowed with a Poisson bracket $\{\!\{-,-\}\!\}$  of cohomological degree -1 and spin 0, defined using topological descent \cite{descent, OhYagi}.

Moreover, the bulk algebra $\CV$ contains a secondary stress tensor $G$ such that
\be \{\!\{G,\CO\}\!\} = \pd_z \CO \ee
for any other local operator $\CO$ in $Q_{HT}$-cohomology. This replaces the more familiar OPE $T(z)\CO(w) \sim ...+ \frac{1}{z-w}\pd\CO(w)$ that one would expect in a non-commutative VOA.

A sufficient condition for an HT-twisted theory to be fully topological --- at least in the sense that its operator algebra in flat space becomes fully independent of insertion points --- is that the secondary stress tensor is exact, $G = Q_{HT}(...)$. This will rarely happen (it does not happen for generic gauge theories).
We are interested, however, in a further deformation of the HT-twist $Q_{HT}\leadsto Q_A= Q_{HT}+Q_A'$, where $Q_A'$ is a second symmetry that satisfies 
$Q_A'{}^2 = \{Q_{HT},Q_A'\} = 0$. A sufficient condition for the deformed theory to be topological is that
\be G = Q_A(P)  \label{G-exact} \ee
for some local operator $P$.%
\footnote{There is further interesting structure present in topological deformations of the HT twist, which is discussed in \cite{twistedN=4}. For example, we expect that the Poisson bracket $\{\!\{-,-\}\!\}$ will vanish on (deformed) cohomology, and will be superseded by the even higher products of \cite{descent} obtained from purely topological descent.}
We will verify this in constructions below, at a classical level.

We will ultimately be interested in coupling Chern-Simons fields to the 3d $\CN=4$ linear gauge theory $T[SU(n)]^{UV}$ as in \eqref{TSUn-UV}. We'll set up the building blocks required to construct the HT-twist and A-twist of this theory, and assemble them together in Section \ref{sec:CSGLSM}.
We work throughout on local Euclidean spacetime $M=\C_{z,\bar z}\times \R_t$.

\subsubsection{Yang-Mills gauging of hypermultiplets}
\label{sec:YM-HT}

We begin with a 3d $\CN=4$ (Yang-Mills) gauge theory, with gauge group $G$ and hypermultiplet matter in a quaternionic representation $T^*V \simeq V\oplus  \overline{V}$. Under the same 3d $\CN=2$ subalgebra as above \eqref{QHT-A}, the 3d $\CN=4$ vector multiplet decomposes into an $\CN=2$ vector multiplet and an adjoint chiral multiplet; and the hypermultiplet decomposes into $V$ and $\ol V$-valued chiral multiplets.

Continuing to work in the conventions described around \eqref{QHT-A}, we note that the 3d $\CN=4$ supercharges $Q_\alpha^{a\dot a}$ have charge $(a,\dot a)$ under a maximal torus $U(1)_H\times U(1)_C$ of R-symmetry group $SU(2)_H\times SU(2)_C$. The anti-diagonal $U(1)_\epsilon \subset U(1)_H\times U(1)_C$ with charge $\epsilon = H-C$ as in \eqref{eHC} act trivially on the $\CN=2$ subalgebra, and thus is a flavor symmetry from an $\CN=2$ perspective. In defining the holomorphic-topological twist, we must further choose a 3d $\CN=2$ R-symmetry, and we take it to be
\be U(1)_R  = U(1)_H \qquad \text{(\ie\  $R=H$)}\,, \ee
in order to be compatible with a further deformation to the A-twist.
(This may be contrasted with the choice $U(1)_R=U(1)_C$ above \eqref{HTB'}, compatible with a further deformation  to the B-twist.) This assures that the spins of various fields in the $\C_{z,\bar z}$ plane, defined in the HT twist as
\be J = \frac{R}{2}-J_0 \ee
(where $J_0$ is spin for the untwisted Lorentz group), will agree with spins in the A-twist.

To set up the BV formalism for the HT twist, we follow \cite{CDG}.
On a local spacetime $M=\C_{z,\bar z}\times \R_t$, we denote by $\bOmega^\bullet$ the Dolbeault complex of $\C_{z,\bar z}$ tensored with the de Rham complex of $\R_t$, so that
\be \bOmega^0 = C^\infty(M)\,,\qquad \bOmega^1 = C^\infty(M) \diff \bar z \oplus  C^\infty(M)\diff t\,, \qquad  \bOmega^2 = C^\infty(M)\diff \bar z \diff t\,,\ee
and denote by $\bOmega^{\bullet,(p)} := \bOmega^\bullet \diff z^p$ its twist by a $p$-th  power of the canonical bundle on $\C_{z,\bar z}$. In the HT twist, cohomological degree is identified with R-charge. The differentials $\diff \bar z,\diff t $ have R-charge 1; and further shifts in cohomological degree are denoted by ``$[r]$''.

The physical fields of a 3d $\CN=2$ theory are regrouped into twisted superfields --- various forms on $M$ --- according to their R-charge and spin. In particular, the fields of our 3d $\CN=4$ gauge theory regroup into the twisted superfields
\begin{itemize}
	\item $\bA \in \bOmega^{\bullet, (0)} \otimes \mathfrak{g}_\C[1]$,  whose leading component is a ghost $c$ for gauge transformations (such that $\pd_zc$ is cohomologous to a physical gaugino), and whose 1-form component contains the gauge connection $A_{\bar z}\diff \bar z+A_t \diff t$,  complexified in the $\diff t$ direction by the real vector multiplet scalar.
	\item $\bB \in \bOmega^{\bullet, (1)} \otimes \mathfrak{g}_\C^*$, whose lowest component $B$ contains the curvature $F_{zt}$ 
	\item $\bPhi \in \bOmega^{\bullet, (0)}\otimes \mathfrak g_\C$, containing the complex adjoint scalar $\phi$ from the 3d $\CN=2$ adjoint chiral multiplet
	\item $\bLambda \in \bOmega^{\bullet, (1)}\otimes\mathfrak{g}_\C^*[1]$, whose lowest component contains a gaugino $\lambda$ from the 3d $\CN=2$ adjoint chiral multiplet.
	\item $(\bX, \bY) \in \bOmega^{\bullet, (1/2)} \otimes (V\oplus \overline{V})[1]$, whose lowest components contain the complex hypermultiplet scalars $(X,Y)$
	\item $(\bPsi_{\bX}, \bPsi_{\bY}) \in \bOmega^{\bullet, (1/2)} \otimes (\overline{V}\oplus V)$, whose lowest  components contain fermions $\psi_X,\psi_Y$ from the hypermultiplet
\end{itemize}
For reference, the charges of the lowest components of various fields are
\be \label{YM-charges} \begin{array}{c|cccccccc}
& c&B&\phi&\lambda&X&Y&\psi_X&\psi_Y \\\hline
H & 1 & 0 & 0 & 1 & 1 & 1 & 0 & 0 \\
C & 1 & 0 & 2 & -1 & 0 & 0 &1 & 1 \\
\epsilon = H-C & 0 & 0 & -2 & 2 & 1 & 1 & -1 & -1 \\
J = \frac H2 - J_0 & 0&1&0&1&\frac12 & \frac12 & \frac12 & \frac12
\end{array} \ee

Using these superfields, one constructs a twisted action
\be S_{HT} = \int_M \bB F'(\bA) +\bLambda \diff'_{\bA}\bPhi + \bPsi_{\bX} \diff'_{\bA} \bX + \bPsi_{\bY} \diff'_{\bA} \bY + \bY\bPhi \bX\,, \label{YM-HT-action} \ee
where $\diff' = \pd_t \diff t + \pd_{\bar{z}} \diff \bar{z}$, $\pd = \pd_z \diff z$, $\diff'_{\bA} = \diff' + \bA$ (acting as a covariant derivative in appropriate representations), and $F'(\bA) = \diff' \bA + \bA^2$ is the corresponding curvature.
In general, the (classical) action of the BV-BRST differential $Q$ is directly tied to the action $S$ by the BV bracket $\{-,-\}_{BV}$ via $Q\CO = \{\CO, S\}_{BV}$. For the present situation, the BV brackets of the above twisted superfields are given by
\be \label{BVbracket-YM}
	\begin{array}{l}\{\bA(x), \bB(y)\}_{BV} =\{\bPhi(x), \bLambda_{\bX}(y)\}_{BV} \\[.1cm]
	 = \{\bX(x), \bPsi_{\bX}(y)\}_{BV} = \{\bY(x), \bPsi_{\bY}(y)\}_{BV} = \delta^{(3)}(x-y) \diff{\rm Vol}\,. \end{array}
\ee
Using these brackets and the relation $Q_{HT} = \{-,S_{HT}\}_{BV}$, the action of the HT supercharge 
on the fundamental fields is easily found to be
\be \label{QHT-YM}
\begin{aligned}
	Q_{HT}\, \bA & = F'(\bA) \qquad & Q_{HT}\, \bB & = \diff'_{\bA}\bB - \bLambda f \bPhi - \bPsi_{\bX}\tau \bX + \bY \tau \bPsi_{\bY}\\
	Q_{HT}\, \bPhi & = \diff'_{\bA} \bPhi & Q_{HT}\, \bLambda &=  \diff'_{\bA} \bLambda + \bY\tau\bX  \\
	Q_{HT}\, \bX & = \diff'_{\bA} \bX \qquad & Q_{HT}\, \bPsi_{\bX} & = \diff'_{\bA} \bPsi_{\bX} + \bY \bPhi\\
	Q_{HT}\, \bY & = \diff'_{\bA} \bY \qquad & Q_{HT}\, \bPsi_{\bY} & = \diff'_{\bA} \bPsi_{\bY} + \bPhi  \bX
\end{aligned}
\ee
Here `$\tau$' denotes the Lie-algebra generators in representation $V$; $Y\tau X=\mu_\C$ is the complex moment map for the $G$ action, and  $\bPsi_{\bX}\tau \bX -  \bY \tau \bPsi_{\bY}$ contains (in its 1-form component) a derivative of the real moment map. Similarly, `$f$' denotes the Lie-algebra generators in the adjoint representation, \emph{a.k.a.} the structure constants.

\emph{Perturbatively}, the Poisson vertex algebra $\CV$ of bulk local operators is strongly generated by the $z$-modes of the lowest components $B(z),\phi(z),\lambda(z), X(z), \psi_X(z), Y(z), \psi_Y(z)$ of the various twisted superfields, and the nonzero modes of $c(z)$,
with differential $Q_{HT}$ induced from \eqref{QHT-YM},
\be \label{QHT-YM-c}
\begin{aligned}
	Q_{HT}\, c & = \tfrac{1}{2} [c,c] \qquad & Q_{HT}\, B & = [c,B] + [\lambda,\phi] - \psi_X \tau X + Y \tau \psi_Y \\
	Q_{HT}\, \phi & = [c, \phi]  & Q_{HT}\, \lambda &=  [c,\lambda] + \mu_\C(X,Y)  \\
	Q_{HT}\, X & = c \cdot  X \qquad & Q_{HT}\, \psi_X & = c\cdot \psi_X + \phi \cdot Y\\
	Q_{HT}\, Y & = c \cdot Y \qquad & Q_{HT}\, \psi_Y & = c\cdot\psi_Y + \phi \cdot X\,,
\end{aligned}
\ee
a trivial OPE, and a descent bracket\footnote{As mentioned in \cite{CDG}, even though these components are \emph{not} $Q_{HT}$ closed these basic brackets can be used to determine brackets on honest cohomology classes.}
\be
	\{\!\{c,B\}\!\} = \{\!\{\phi,\lambda\}\!\} =  \{\!\{X,\psi_X\}\!\} = \{\!\{Y,\psi_Y\}\!\} = 1\,.
\ee
The secondary stress tensor $G(z)\in \CV$ is given by
\be  \label{G2-YM}
	G = -B \pd_z c + \lambda \pd_z\phi + \psi_X \pd_z X - Y \pd_z \psi_Y.
\ee
This description, however, is only perturbative. In the HT-twist (and the A-twist) of a 3d gauge theory, monopole operators will also be present.

We would now like to deform the HT-twist to the topological A-twist. The supercharge $Q_A' = Q_-^{-\dot +}$ that deforms the HT-twist to the A-twist in a supersymmetric setting is realized on the twisted theory above by BV bracket with
\be S_A':= \int_M \bB \bPhi  - \bPsi_{\bX} \bPsi_{\bY}\,,  \label{SA'-YM}\ee
in the sense that
\be Q_A'(\CO) = \{\CO, S_A'\}_{BV} \ee
for any local operator $\CO$. On lowest components, the nontrivial transformations are $Q_A'(c) =\phi$, $Q_A'(\lambda) = B$, $Q_A'(X)=-\psi_Y$, $Q_A'(Y)=\psi_X$. Note that $\{Q_{HT},Q_A'\} = Q_A'^2 = 0$, or equivalently $\{S_{HT},S_A'\}_{BV} = \{S_A',S_A'\}_{BV} = 0$.

To actually implement the A-twist in the BV formalism, we should add $S_A'$ to the action,
\be S_{A} = S_{HT}+S_A'= \int_M \bB F'(\bA) +\bLambda \diff'_{\bA}\bPhi + \bPsi_{\bX} \diff'_{\bA} \bX + \bPsi_{\bY} \diff'_{\bA} \bY + \bY\bPhi \bX+ \bB \bPhi  - \bPsi_{\bX} \bPsi_{\bY}\,, \label{YM-A-action} \ee
This may be interpreted as introducing an exotic superpotential.
The BV brackets \eqref{BVbracket-YM} are unchanged (since no derivatives appear in $S_A'$), but the BV differential is naturally modified to $Q_A =  \{-, S_A\}_{BV} = Q_{HT}+Q_A'$.

An important subtlety is that in the A-twist, cohomological degree is no longer identified with the original R-charge (\ie\  with $H$).
 It is easy to see that $H$ will not work, since $H(Q_{HT})=1$ (as appropriate for a differential) but $H(Q_A')=-1$. Related to this, the term $S'_A$ breaks both $U(1)_H$ and $U(1)_\epsilon$ symmetries, but preserves their anti-diagonal combination $U(1)_C$.
The correct cohomological degree in the A-twist is $U(1)_C$ charge
\be \text{cohomological degree in A twist:} \quad C = H-\epsilon\,,  \label{A-shift} \ee
for which we have
\be C(Q_{HT}) = C(Q_A') = C(Q_A) = 1\,, \ee
as desired. Note, however, that the spin of various fields is still given by $J$ in \eqref{YM-charges}.

The action of the $A$-twist supercharge $Q_A$ on twisted superfields becomes
\be
\label{eq:QA-YM}
\begin{aligned}
	Q_{A}\, \bA & = F'(\bA) + \bPhi\qquad & Q_{A}\, \bB & = \diff'_{\bA}\bB - \bLambda f \bPhi - \bPsi_{\bX}\tau \bX + \bY \tau \bPsi_{\bY}\\
	Q_{A}\, \bPhi & = \diff'_{\bA} \bPhi & Q_{A}\, \bLambda &=  \diff'_{\bA} \bLambda + \bY\tau\bX + \bB \\
	Q_{A}\, \bX & = \diff'_{\bA} \bX - \bPsi_{\bY} \qquad & Q_{A}\, \bPsi_{\bX} & = \diff'_{\bA} \bPsi_{\bX} + \bY \bPhi\\
	Q_{A}\, \bY & = \diff'_{\bA} \bY + \bPsi_{\bX} \qquad & Q_{A}\, \bPsi_{\bY} & = \diff'_{\bA} \bPsi_{\bY} + \bPhi  \bX
\end{aligned}
\ee
with corresponding transformations of the lowest components that generate the perturbative bulk operator algebra $\CV$. It is not hard to see that taking $Q_A$-cohomology kills all perturbative local operators, aside from gauge-invariant polynomials $p(\phi)$ formed from the zero-mode of $\phi$. Naively, the relation $Q_A'c = \phi$ would suggest that even such operators are exact; but as explained in \cite[Sec. 6.2]{CDG}, the zero-mode of $c$ should be removed from the operator algebra by hand.

Gauge-invariant polynomials $p(\phi)$ give the correct description of perturbative local operators in the topological A-twist. Their expectation values parameterize ``half'' of the Coulomb branch, the base of an integrable system \cite{CHZ, BDG, Nak, BFN}. Non-perturbative monopole operators should also be present in the A-twist, parameterizing the fibers of the Coulomb-branch integrable system.

Finally, we verify that the secondary stress tensor has become exact in the A-twist, at least classically. A simple calculation shows that 
\be
	G = Q_A(-\lambda\pd_z c + Y \pd_z X) \,.
\ee

\subsubsection{Chern-Simons gauging of hypermuliplets}
\label{sec:CShypers}

Next, let us describe the HT twist of 3d $\CN=4$ hypermultiplets in representation $T^*V$, gauged with a 3d $\CN=2$ vector multiplet at Chern-Simons level $k$, \ie\ with a bilinear form schematically written as $\frac{k}{4\pi}\text{Tr}(ab)$ in the Chern-Simons kinetic term.

In the BV formalism, the twisted superfields are
\be \begin{array}{l@{\qquad}l}
 \bA \in \bOmega^{\bullet, (0)} \otimes \mathfrak{g}_\C[1] & (\bX, \bY) \in \bOmega^{\bullet, (1/2)} \otimes (V\oplus \overline{V}) \\
 \bB \in \bOmega^{\bullet, (1)} \otimes \mathfrak{g}_\C^* & (\bPsi_{\bX}, \bPsi_{\bY}) \in \bOmega^{\bullet, (1/2)} \otimes (\overline{V}\oplus V)[1]\,, \end{array}
\ee
with exactly the same charges as in \eqref{YM-charges}. The only difference is that $\bPhi,\bLambda$, which came from the 3d $\CN=2$ adjoint multiplet, are missing. The HT-twisted action is now 
\be
\label{eq:HTtwistedactionCS}
	S_{HT} = \int \bB F'(\bA) + \bPsi_{\bX} \diff'_{\bA} \bX + \bPsi_{\bY} \diff'_{\bA} \bY + \tfrac{k}{4 \pi} \text{Tr}( \bA \pd \bA)\,,
\ee
with BV brackets
\be
	\{\bA(x), \bB(y)\}_{BV} = \{\bX(x), \bPsi_{\bX}(y)\}_{BV} = \{\bY(x), \bPsi_{\bY}(y)\}_{BV} = \delta^{(3)}(x-y) \diff{\rm Vol}\,,
\ee
inducing the action of $Q_{HT} = \{-, S_{HT}\}_{BV}$. Explicitly, we have
\be
\begin{aligned}
	Q_{HT}\, \bA & = F'(\bA) \qquad & Q_{HT}\, \bB & = \diff'_{\bA}\bB - \bPsi_{\bX}\tau \bX + \bY \tau \bPsi_{\bY} + \tfrac{k}{2\pi} \pd \bA\\
	Q_{HT}\, \bX & = \diff'_{\bA} \bX \qquad & Q_{HT}\, \bPsi_{\bX} & = \diff'_{\bA} \bPsi_{\bX}\\
	Q_{HT}\, \bY & = \diff'_{\bA} \bY \qquad & Q_{HT}\, \bPsi_{\bY} & = \diff'_{\bA} \bPsi_{\bY}\,,\\
\end{aligned}
\ee
corresponding to the transformations of lowest components
\be  \label{CS-HT-c}
\begin{aligned}
	Q_{HT}\, c & = \tfrac{1}{2}[c,c] \qquad & Q_{HT}\, B & = [c,B] - \psi_X\tau X + Y \tau \psi_Y + \tfrac{k}{2\pi}\pd_zc  \\
	Q_{HT}\, X & = c\cdot X \qquad & Q_{HT}\, \psi_X & = c\cdot \psi_X\\
	Q_{HT}\, Y & = c\cdot Y \qquad & Q_{HT}\, \psi_Y & = c\cdot\psi_Y\,.
\end{aligned}
\ee

Note that $U(1)_\epsilon$ remains a flavor symmetry, as the Chern-Simons term $\tfrac{k}{4 \pi} \text{Tr}( \bA \pd \bA)$ is invariant. More so, $U(1)_\epsilon$ acts trivially on the gauge multiplet $(\bA,\bB)$.

\subsubsection{$\CT_{n,k}$}
\label{sec:CSGLSM}

We finally return to the main theory of interest: $\CT_{n,k}^{UV}=T[SU(n)]^{UV}/SU(n)_k$. We construct its HT-twisted action by coupling an action for the 3d $\CN=4$ Yang-Mills gauge theory $T[SU(n)]$ (as in Section \ref{sec:YM-HT}) to a 3d $\CN=2$ Chern-Simons theory at level $k$ (as in Section \ref{sec:CShypers}).  We take the liberty of shifting the cohomological degree from $H$ to $C$ as in \eqref{A-shift}, from the beginning.

The gauge group is $SU(n)_k\times \prod_{a=1}^{n-1} U(a)$, giving rise to  superfields
\be \begin{array}{c}  (\bA_n,\bB_n)\in \bOmega^\bullet\otimes (\mathfrak{sl_n}[1]\oplus\mathfrak{sl}_n^*dz)  \\[.1cm]
 (\bA_a,\bB_a) \in \bOmega^\bullet(\mathfrak{gl}_a[1]\oplus\mathfrak{gl}_a^*dz)\,,\qquad  (\bPhi_a,\bLambda_a) \in \bOmega^\bullet \otimes(\mathfrak{gl}_a[2]\oplus\mathfrak{gl}_a^*[-1]dz)\,,\quad a=1,...,n-1\,.\end{array} \ee
The matter representation is $T^* \bigoplus_{a=1}^{n-1} V_a$ with $V_a:= \text{Hom}(\C^a,\C^{a+1})$ (\ie\ bifundamental hypermultiplets for each edge in the quiver \eqref{Tnk-quiver}), giving rise to
\be (\bX_a,\bY_a) \in \bOmega^\bullet dz^{\frac12}\otimes(V_a\oplus \overline V_a)\,,\qquad (\bPsi_{\bX,a},\bPsi_{\bY,a}) \in  \bOmega^\bullet dz^{\frac12}\otimes(\overline V_a\oplus  V_a)[1] \,,\quad a=1,...,n-1\,.\ee
The BV action is
\begin{align}
	S_{HT}[\CT_{n,k}^{UV}] & = \int_M \tfrac{k}{4\pi} \text{Tr}[\bA_n \pd \bA_n] + \sum_{a=1}^{n}\bB_a F'(\bA_a) + \sum_{a=1}^{n-1}\bigg( \bLambda_a \diff'_{\bA} \bPhi_a  + \bPsi_{\bX,a} \diff_{\bA} \bX_a + \bPsi_{\bY,a} \diff_{\bA} \bY_a  \notag \\
	& \hspace{1cm} + \Tr[\bX_a \bPhi_a \bY_a] - \Tr[\bY_{a} \bPhi_{a+1} \bX_{a}] \bigg)\,, 
	 \label{eq:HTtwistedactionFinal}
\end{align}
with the convention that $\bPhi_n= 0$. The BV brackets induce an action of $Q_{HT}=\{-, S_{HT}\}_{BV}$, whose explicit form is a straightforward generalization of previous formulas.

We now observe that 
\be S_A'[\CT_{n,k}^{UV}] := \int_M \sum_{a=1}^{n-1}\Big( {\bB}_a \bPhi_a - \bPsi_{\bX,a} \bPsi_{\bY,a} \Big)\,, \ee
which would have been the A-twist deformation for the Yang-Mills theory $T[SU(n)]^{UV}$ alone, still satisfies $\{S_A',S_A'\}_{BV} = \{S_A',S_{HT}\}_{BV} = 0$. It therefore defines the action of a second differential $Q_A'=\{-, S_A'\}_{BV}$ on the HT-twisted $\CT_{n,k}^{UV}$. We add it to the action, setting
\be S_A[\CT_{n,k}^{UV}] := S_{HT}[\CT_{n,k}^{UV}] + S_A'[\CT_{n,k}^{UV}]\,,  \label{eq:AtwistedactionFinal} \ee
in order to implement a twist by the combined differential $Q_A=Q_{HT}+Q_A'$. $U(1)_C$ (cohomological degree) and twisted spin $U(1)_J$ remain symmetries of this action.

As the notation indicates, we expect this to be the topological A-twist --- \emph{i.e.} that the resulting theory is equivalent to the topological A-twist of the infrared 3d $\CN=4$ theory $\CT_{n,k}$. 
We will check topological invariance of the A-twist \eqref{eq:AtwistedactionFinal} at the classical level, by showing that the secondary stress tensor is $Q_A$-exact.

The action of $Q_A=\{-, S_A\}_{BV}$ on superfields is given by 
\be
\begin{aligned}
	\label{eq:QACS}
	Q_A\, \bA_n & = F'(\bA_n)  \qquad & Q_A  \bB_n & = \diff'_{\bA_n} \bB_n - \mu_n +\tfrac{k}{2\pi} \pd \bA_n \\
	Q_A\, \bA_a & = F'(\bA_a) + \bPhi_a \qquad & Q_A  \bB_a & = \diff'_{\bA_a} \bB_a - \mu_a  \\
	Q_A\, \bPhi_a & = \diff'_{{\bA}_a} \bPhi_a \qquad & Q_A\, \bLambda_a & = \diff'_{{\bA}_a} \bLambda_a + \mu_{\C,a} + {\bB}_a\\
	Q_A\, \bX_{a} & = \diff'_{ \bA} \bX_{a} - \bPsi_{\bY,a}\qquad & Q_A\, \bPsi_{\bX,a} & = \diff'_{ \bA} \bPsi_{\bX,a} + \bPhi_{a} \bY_{a} - \bY_{a} \bPhi_{a+1}\\
	Q_A\, \bY_{a} & = \diff'_{ \bA} \bY_{a} +\bPsi_{\bX,a} \qquad & Q_A\, \bPsi_{\bY,a} & = \diff'_{ \bA} \bPsi_{\bY,a} + \bPhi_{a} \bX_{a} - \bPhi_{a+1} \bX_a
\end{aligned}
\ee
for $a = 1, ..., n-1$, where the moment maps $\mu_n, \mu_a, \mu_{\C,a}$ are given by
\be
\begin{aligned}
	\mu_n & = \bPsi_{\bX,n-1} \bX_{n-1}  - \bY_{n-1} \bPsi_{\bY,n-1} - \tfrac{1}{n} \id_{n \times n} \Tr(\bPsi_{\bX,n-1} \bX_{n-1}  - \bY_{n-1} \bPsi_{\bY,n-1}) \\
	\mu_a & = \bPsi_{\bY,a} \bY_{a} - \bX_{a}\bPsi_{\bX,a} + \bPsi_{\bX,a-1} \bX_{a-1}  - \bY_{a-1} \bPsi_{\bY,a-1} - [\bLambda_a, \bPhi_a]\\
	\mu_{\C,a} & = \bX_a \bY_a - \bY_{a-1} \bX_{a-1}\,,
\end{aligned}
\ee
with $\bX_0 = \bPhi_n = \bY_0 = 0$ and $\id_{n \times n}$ is the $n \times n$ identity matrix.

The bulk vertex algebra $\CV$ is generated perturbatively by modes of the lowest components of each superfield (removing the zero-mode of the ghosts $c_a$). The (classical) secondary stress tensor is a direct generalization of \eqref{G2-YM},
\be
	G = -\sum_{a=1}^n B_a \pd_z c_a + \sum_{a=1}^{n-1}\big[ \lambda_a \pd_z \phi_a + \psi_{X,a} \pd_z X_a - Y_a \pd_z \psi_{Y,a}\big]\,.
\ee
It is now $Q_A$-exact, with
\be G = Q_A\bigg[ -\frac\pi k \text{Tr}\big(B_n + \mu_{\C,n}\big)^2- \sum_{a=1}^{n-1} \big(\lambda_a\pd_z c_a+Y_a\pd_z X_a\big)\bigg]\,, \ee
where $\mu_{\C,n} = Y_{n-1}X_{n-1} - \tfrac{1}{n} \mathds{1}_{n \times n} \Tr(Y_{n-1} X_{n-1})$ is the complex moment map for the Chern-Simons gauge group.
In order to verify this identity, it is helpful to introduce $\widehat B = B_n + \mu_{\C,n}$, which satisfies $Q_A\,\widehat B = [c,\widehat B]+\frac{k}{2\pi}\pd_z c$. Then
\begin{align}
		 & Q_A\big[-\tfrac{\pi}{k} \text{Tr}(\widehat{B}^2 )- \lambda_a \pd_z c_a + Y_a \pd_z X_a\big] \notag \\
		 & \quad = - \widehat{B} \pd_z c_n - (f_a c_a \lambda_a + \mu_{\C,a} + B_a) \pd_z c_a + \lambda_a \pd_z(\tfrac{1}{2} f_a c_a^2 + \phi_a) \label{G-exact} \\ & \hspace{.5in}  + \psi_{X,a} \pd_z X_a - Y_a \pd_z \psi_{Y,a} + \mu_{\C,n} \pd_z c_n + \mu_{\C,a} \pd_z c_a \notag \\
		&\quad = - B_n \pd_z c_n -  B_a \pd_z c_a+ \lambda_a \pd_z {\phi}_a + \psi_{X,a} \pd_z X_a - Y_a \pd_z \psi_{Y,a} = G\,, \notag
\end{align}
with implied summations over $a=1,...,n-1$.

Rather surprisingly, the fundamental identity seems to be entirely bypassed in this derivation. This would have not been the case for the $B$-twist; the (classical) existence of both $A$ and $B$ twist deformations does require the fundamental identity \cite{topCSM}. On the other hand, \eqref{G-exact} only amounts to a classical verification that $G$ is $Q_A$-exact. It would be interesting (and important for further progress) to check whether quantum corrections modify \eqref{G-exact}, perhaps in a way that involves the fundamental identity. In an HT-twisted theory, there may be (only) one-loop perturbative corrections \cite{Williams-oneloop}, or non-perturbative corrections.

\subsubsection{Flavor symmetry and flat connections}
\label{sec:YMCSflav}
The theory $\CT_{n,k}$ has a $G^\vee=PSU(n)$ flavor symmetry, inherited from the ``topological'' $PSU(n)$ flavor symmetry of $T[SU(n)]$, that acts on its Coulomb branch \cite{GaiottoWitten-Sduality}. We correspondingly expect that the A-twisted theory $\CT_{n,k}^A$ can be deformed by complexified $PGL(n,\C)$ flat connections.

 In the UV gauge theory $\CT_{n,k}^{UV}$ with quiver \eqref{Tnk-quiver}, only a maximal torus $T\simeq U(1)^{n-1}\subset PSU(n)$ of this symmetry acts. Thus the Lagrangian A-twisted theory $\CT_{n,k}^{UV,A}$ should at least be deformable by complexified $T_\C=GL(1,\C)^{n-1}$ flat connections $\CA$. The real part of $\CA$ is a standard $U(1)^{n-1}$ connection, and the complexification comes from triples of 3d $\CN=4$ FI parameters that have been twisted into 1-forms. This description is 3d-mirror (and otherwise equivalent) to the complexification by mass parameters in \eqref{A-mass}.
 
We would like to explain how the deformation appears in the twisted BV action. Let $\CA^a$ $(a=1,...,n-1)$ denote complex abelian connections for each $GL(1,\C)$ factor in the maximal torus, and let us decompose
\be \CA^a = \CA^a_z \diff z + \CA^a{}'\,, \ee
with $\CA^a{}' = \CA^a_{\bar z}\diff \bar z+\CA^a_t \diff t$. Then we may further deform the action \eqref{eq:AtwistedactionFinal} by a term
\be S_{\rm flavor} =  \sum_{a=1}^{n-1} \int_M -\text{Tr}(\bPhi_a) \CA^a_z\,\diff z+ \text{Tr}(\bA_a)\pd \CA^a{}'\,. \label{S-flavor} \ee
It is clear that $\{S_{\rm flavor},S_{\rm flavor}\}_{BV}=0$; however
\begin{align} \{S_A,S_{\rm flavor}\}_{BV} &= \int_M -\diff'\text{Tr}(\bPhi_a)\CA^a_z\, \diff z + \big[\diff'\text{Tr}(\bA_a)+\text{Tr}(\bPhi_a)\big] \pd \CA^a{}'  \notag \\
& = \int_M \text{Tr}(\bPhi_a)\big[ \diff' \CA_z^a \diff z+\pd \CA^a{}'\big] - \pd \text{Tr}(\bA_a)\big[\diff' \CA^a{}'\big]
\end{align}
only vanishes once we impose the flatness condition $\CF_\CA = (\diff'\CA^a{}') + (\diff' \CA_z^a \diff z+\pd \CA^a{}')=0$. (The two parts of the curvature here have different form indices and must vanish separately.)

Thus, given a flat $GL(1,\C)^{n-1}$ connection, the total action $S_A[\CT_{n,k}^{UV}]+S_{\rm flavor}$ will induce a new nilpotent differential. The transformations in \eqref{eq:QACS} that are deformed are those involving the real and complex moment maps --- unsurprising, since $\CA$ involves FI parameters that should shift the moment maps. Namely, we now have
\be Q_A\, \bB_a = \diff'_{\bA_a} \bB_a - \mu_a + \id_{a\times a}\,\pd\CA^a{}'\,,\quad   Q_A\, \bLambda_a  = \diff'_{{\bA}_a} \bLambda_a + \mu_{\C,a} + {\bB}_a - \id_{a\times a}\, \CA_z^a\,\diff z\,, \label{QA-flavor} \ee
for $a=1,...,n-1$, where $\id_{a\times a}$ denotes the $a \times a$ identity matrix.

\subsection{A first look at line operators}
\label{sec:line1}

We will eventually construct the entire category of line operators in $\CT_{n,k}^A$ by using boundary and corner VOA's, and we will probe its Grothendieck ring using localization computations. Here we wish to describe where some of these line operators originate, given the 4d constructions of $\CT_{n,k}^A$ in Section \ref{sec:4d} and the conjectured Lagrangian description in Section \ref{sec:Tnk-BV}.

In the 4d setup on the left of \eqref{sandA}, involving the 4d A-twisted SYM theory with deformed Neumann ($B_{1,k}$) and $T[G]$ ($\wt{B}_{1,0}$) boundary conditions, we expect to find
\begin{itemize}
\item half-BPS 't Hooft lines in the bulk, which are the only bulk line operators preserved by the 4d A-twist \cite{KapustinWitten}
\item quarter-BPS Wilson lines trapped on the Neumann boundary condition, which have played a major role in 4d constructions of Chern-Simons theory and its analytic continuation \cite{Witten-anal, Witten-path, Witten-fivebranes}.
\item quarter-BPS vortex lines on the $T[G]$ boundary condition, which are preserved by the 3d A-twist \cite{AsselGomis, DGGH}.
\end{itemize}
The bulk 't Hooft lines  `act' on the categories of line operators on either boundary condition by collision.
They are only detectable in the final sandwiched theory $\CT_{G,k}^A$ via their images on the boundaries; thus it suffices to focus our attention on the boundary Wilson and vortex line operators.

From a purely 3d $\CN=4$ perspective, the presence of gauge Wilson lines may seem at odds with the A-twist, as one typically encounters vortex-line operators in A-twisted theories containing only $\CN=4$ vector multiplets. Nonetheless, due to the Chern-Simons gauge fields, $\CT_{n,k}$ admits the desired Wilson lines; see, \eg, \cite{CDTprofusion, DTThyperloops} or the review \cite{roadmap} and references therein for related examples. We may check using the Lagrangian construction of $\CT_{n,k}^A$ in Section \ref{sec:CSGLSM} that Wilson lines appear in the final A-twisted 3d field theory $\CT_{G,k}^A$. To this end, let us define a new twisted superfield
\be \widehat \bA_n := \bA_n +\tfrac{2\pi}{k}\big(\bB_n + \mu_{\C,n}\big)\,, \ee
where $\mu_{\C,n}=\bY_{n-1}\bX_{n-1} - \tfrac{1}{n} \id_{n \times n} \Tr(\bY_{n-1} \bX_{n-1})$ is the complex moment map for the $SU(n)$ Chern-Simons gauge group. The 1-form part of $\widehat \bA_n$ now has $\diff \bar z,\diff t$ \underline{and} $\diff z$ components; the $\diff z$ component is ``borrowed'' from $B$ and the moment map. Rather beautifully, this superfield satisfies $Q_A\, \widehat \bA_n = F(\widehat \bA_n)=\diff \widehat \bA_n + \widehat \bA_n^2$, containing all components of the curvature.

Given an arbitrary closed curve $\gamma\in M$ and a finite-dimensional representation $\rho$ of the Chern-Simons gauge group $SU(n)$, we may define a Wilson-loop operator in the twisted BV formalism as
\be W_\rho(\gamma) = \text{Tr}\Big[P\,\exp \oint_\gamma \rho(\widehat \bA_n)\Big]\,, \label{Wrho} \ee
using a chosen framing of $\gamma$ for a point-splitting regularization. Due to $Q_A\, \widehat \bA_n = F(\widehat \bA_n)$, this operator is not only $Q_A$-closed, but small variations of $\gamma$ will also be $Q_A$-exact. 
Wilson lines may be defined similarly for open curves $\gamma$, removing the trace, as long as gauge-invariant boundary conditions are imposed.

Wilson lines also survive deformation by background flat connections for the flavor symmetry of $\CT_{n,k}^A$. This follows (at least for abelian backgrounds) from the fact that the $Q_A$ transformations in \eqref{QA-flavor} do not affect the superfields $\bA_n,\bB_n,\bX_{n-1},\bY_{n-1}$ involved in $\widehat \bA_n$.

\subsubsection{Counting, one-form symmetry, and anomaly}
\label{sec:countW}

Due to the Chern-Simons kinetic term, we might expect to find only a finite set of Wilson lines in $\CT_{n,k}^A$.
Indeed, applying arguments of \cite{Witten-Jones,EMSS}, we would expect that large gauge transformations induce an equivalence of Wilson lines in irreducible representations labelled by dominant weights $\lambda,\lambda'$ of $SU(n)$ such that the Weyl orbits of $\lambda$ and $\lambda'$ differ by $k$ times a coroot.  (One uses the bilinear pairing from the Chern-Simons term to dualize coweights and coroots to weights.) In pure bosonic Chern-Simons theory this leads to equivalence classes of Wilson lines labelled by elements of the coset
\be \frac{P}{W\ltimes k Q^\vee} \ee
where $P$ is the weight lattice, $Q^\vee$ the coroot lattice, and $W$ the Weyl group.

In $\CT_{n,k}^A$, the presence of $T[SU(n)]$ ``matter'' modifies the equivalence relation among Wilson lines. The computation of the ring of Bethe roots (the Grothendieck ring of the category of line operators) in Sections \ref{sec:T2kGrothendieck} and \ref{sec:SUnBethe} suggests that equivalence classes of Wilson lines in $\CT_{n,k}^A$ are labelled by elements of %
\footnote{Note that each element of $P/kQ^\vee$ can be represented by a dominant weight, and thus associated with an irreducible representation of $SU(n)$ labelling a Wilson line.}
\be \frac{P}{kQ^\vee}\,. \label{count-PkQ} \ee
In other words, the coupling to $T[SU(n)]$ effectively undoes the Weyl-group quotient. 
This matches the counting of (say) simple modules in any stalk of $U_q(\sln)$-mod, as well as the counting of modules of the Feigin-Tipunin algebra $\FT_k(\sln)$ later in Section \ref{sec:monodromy-modules}.

Furthermore, the later Bethe root analysis suggests that the $\Z_n$ one-form symmetry is naturally realized as an action of the subgroup $kP/kQ^\vee\simeq \Z_n$ on $P/kQ^\vee$. (Though we emphasize that the one-form symmetry generators are \emph{not} themselves Wilson lines, as the latter would not be invertible.) We will find that the $\Z_n$ 1-form symmetry has an 't Hooft anomaly, determined by the property that its generator has self-braiding $e^{-2\pi i k/n}$ (closely related to examples in \cite[Sec. 5]{HsinLam-discrete}). This again matches the structure of braiding of invertible modules for $U_q(\sln)$ as well as invertible modules for the Feigin-Tipunin algebra $\FT_k(\sln)$ (see Sections \ref{sec:Z2-anomaly} and \ref{sec:SUnBethe}). The anomaly vanishes when $k\equiv 0$ (mod $n$).

\subsection{The boundary VOA $\CN_{n,k}$}
\label{sec:VOAnk}

We now turn to defining holomorphic boundary conditions $\CB^A_{n,k}$ for the 3d topological theories $\CT^A_{n,k}$, using the explicit Lagrangian formulation of $\CT^A_{n,k}$ from Section \ref{sec:CSGLSM}.

Our basic approach will be to begin with the boundary conditions introduced by  \cite{CostelloGaiotto} for the A-twist of $T[SU(n)]$, and then further modify them (following \cite{DGPdualbdys,CDG}) to accommodate an $SU(n)_k$ Chern-Simons gauging in the bulk. The result will be a family of Neumann-like boundary conditions --- in the sense that they preserve all the bulk gauge symmetry of (the Lagrangian formulation of) $\CT^A_{n,k}$, and impose Neumann b.c. on all bulk hypermultiplet scalars. We then compute the VOA supported on $\CB^A_{n,k}$ to be
\be \CN_{n,k} = \big[A(\sln) \otimes \FF(n(k-1))\big]^{SL(n,\C[\![z]\!])}\,, \label{Nnk-form1} \ee
where $A(\mathfrak{sl}_n)$ is the boundary VOA for $T[SU(n)]^A$,  $\FF(n(k-1))$ denotes $n(k-1)$ complex free fermions, and the superscript $SL(n,\C[\![z]\!])$ denotes taking derived $SL(n,\C[\![z]\!])$ invariants. See \cite{topCSM} for a discussion of more general theories. When \eqref{Nnk-form1} lies entirely in cohomological degree zero (which we expect, but do not prove), we show that the VOA can also be described as the coset of an affine VOA
\be \CN_{n,k}  \simeq \text{Com}\big(V^k(\mathfrak{sl}_n),\,A(\sln) \otimes \FF(n(k-1))\big)\,. \ee

We note that the corner configuration in  \eqref{brane-D} defines a \emph{second}, Dirichlet-like boundary condition for $\CT^A_{n,k}$. It supports a second family of VOA's $\CD_{n,k}$ that are introduced in Section~\ref{sec:D-FT}.

\subsubsection{Yang-Mills gauging of hypermultiplets}

We work again in steps.
We start by reviewing the construction from \cite{CostelloGaiotto} of holomorphic boundary conditions for A-twists of 3d $\CN=4$ gauge theories without Chern-Simons couplings.
We will recast the construction in the twisted BV formalism, following \cite{CDG}, thinking of the 3d A twist as a further deformation of the 3d HT twist.
This has two advantages. Practically, it makes it easy to generalize the construction to our Lagrangian definition of $\CT_{n,k}$ theories from Section \ref{sec:CSGLSM}. Moreover, subtle modifications employed by \cite{CostelloGaiotto} to render $\CN=(0,4)$ boundary conditions compatible with topological twists become exact in the HT twist, and no longer appear in the twisted Lagrangian.

Consider, then, the A-twist of 3d $\CN=4$ super Yang-Mills with gauge group $G$ and hypermultiplets transforming in  representation $T^*R$. The twisted superfields and A-twisted Lagrangian of this theory were written out in Section \ref{sec:YM-HT} (we follow the same notation). We impose the following boundary conditions on the bulk superfields:
\begin{itemize}
	\item Neumann boundary conditions for the (3d $\CN = 2$) vector multiplet: $\bB|_{\pd}=0$
	\item Dirichlet boundary conditions for the $\mathfrak{g}$-valued chiral multiplet: $\bPhi|_{\pd} = 0$ 
	\item Neumann boundary conditions for the $R$ and $R^*$ chiral multiplets: $\bPsi_{\bX}|_{\pd}=\bPsi_{\bY}|_{\pd} = 0$
\end{itemize}

Due to unbroken gauge symmetry at the boundary, we further need to couple to 2d degrees of freedom, in order to cancel a boundary gauge anomaly.  In the conventions of \cite{DGPdualbdys}, the bulk fields with the above choice of b.c. contribute $2h-T_R$ to the gauge anomaly, where $h$ is the dual Coxeter number of $G$ and $T_R$ is the quadratic index of the representation $R$, normalized so that $\Tr_R(F^2)= T_R \Tr(F^2)$, with `$\Tr$' the trace in the fundamental representation of $SU$ gauge groups. Boundary $\CN=(0,2)$ Fermi multiplets in a representation $V$ contribute $T_V$. Thus, as long as there is enough bulk matter, so that $T_R-2h\geq0$, we can cancel the boundary anomaly by adding Fermi multiplets in a representation $V$ such that $T_V=T_R-2h$.

In the twisted formalism, boundary Fermi multiplets are realized by a pair of superfields $\bGamma, \wt{\bGamma}$.
In their presence, the boundary condition for the bulk vector multiplet is modified to $\bB|_{\pd}=\wt{\bGamma} \sigma \bGamma$, where `$\sigma$' denotes the $\mathfrak{su}(n)$ generators in the representation $V$ and $\wt{\gamma} \sigma \gamma$ is the moment map for the $G$ action on $T^*V$. The remaining b.c. on bulk superfields are unchanged. We also note that since the superpotential in \eqref{YM-A-action} vanishes at the boundary, we do not need to include any $E$ or $J$ terms for the boundary Fermi multiplets. (The BV differential $Q_A=\{-,S_A\}_{BV}$ will automatically square to zero on a half-space.)

The analysis of \cite{CDG} (as well as \cite{CostelloGaiotto}) shows that the boundary VOA can now be computed perturbatively. (There are no monopole operators, and thus no nonperturbative corrections, on a Neumann boundary condition.)
It is generated by the leading components of all the superfields that are unconstrained at the boundary: the 2d fermions $\gamma(z),\wt\gamma(z)$, the bulk scalars $X(z),Y(z)$, and the bulk fermions $\lambda(z), c(z)$ ---
up to subtle feature, explained in \cite[Sec. 6.2.1]{CDG}, that the zero-mode of $c(z)$ does not enter the boundary VOA.%
\footnote{The field $c(z)$ is a ghost for holomorphic gauge transformations at the boundary, and one should not introduce a ghost for the gauge transformations that are constant along the boundary. Rather, one should restrict to invariants for constant $G$-valued gauge transformations by hand.} %
Scaling dimensions are given by twisted spin $J$: $\gamma,\wt\gamma$ and $X,Y$ have dimension $\frac12$, while $c$ and $\lambda$ have dimensions 0 and 1.

\begin{subequations}
	\label{eq:YMAbdyOPE}
	Among the generators $\gamma,\wt \gamma,X,Y, c,\lambda$, there are singular OPE's 
	\be
	\gamma^\alpha(z) \wt{\gamma}_\beta(w) \sim \frac{\delta^\alpha{}_\beta}{z-w}\,, \qquad (\alpha = 1, ..., \dim V)
	\ee
	(as usual for 2d fermions), as well as
	\be   c^a(z) \lambda_b(w) \sim \frac{\delta^a{}_b}{z-w} \qquad X^n(z) Y_m(w) \sim \frac{\delta^n{}_m}{z-w}\,, \ee
\end{subequations}
induced by second derivatives of the bulk superpotential $\bW = \bY \bPhi \bX + \bB \bPhi - \bPsi_{\bX} \bPsi_{\bY}$, evaluated at the boundary. In addition, there is a boundary differential coming from \eqref{eq:QA-YM}, restricted to the boundary:
\be
\label{eq:YMAQ}
\begin{aligned}
	Q_A c & = \tfrac{1}{2}[c,c] \qquad & Q_A \lambda & = \norm{[c, \lambda]} + \norm{Y\tau X} + \norm{\wt{\gamma} \sigma \gamma}\\
	Q_A X & = c \cdot X \qquad & Q_A Y & = c \cdot Y\\
	Q_A \gamma & = c \cdot \gamma \qquad & Q_A \wt{\gamma} & = c \cdot \wt{\gamma}\,.
\end{aligned}
\ee
The boundary vertex algebra is then constructed as
\be \label{VA-generic}
\CV^A[G,R;V] = H^\bullet \big(\langle \! \langle X, Y, \gamma, \wt{\gamma}, \lambda, c \,\big|\, \eqref{eq:YMAbdyOPE}\rangle \! \rangle^G, Q_A\big)\,,
\ee
taking the part of the algebra generated by $X, Y, \gamma, \wt{\gamma}, \lambda, c$ invariant under constant $G$ gauge transformations (then removing the zero-mode of $c$ by hand), and taking $Q_A$ cohomology. 

All in all, we may identify $\gamma,\wt\gamma$ as complex free fermions valued in $V$, $X,Y$ as symplectic bosons valued in $T^*R$, and $\lambda,c$ as a $\g$-valued $bc$ ghost system for an internal Kac-Moody symmetry.
Indeed, the differential $Q_A$ is just a standard BRST differential, whose action on any $G$-invariant operator $\CO(z)$ can be realized as
\be \label{Q-internal}
Q_A \CO(z) = \oint \frac{\diff w}{2 \pi i} \mathcal{Q}_A(w) \CO(z) \qquad \mathcal{Q}_A
= \tfrac{1}{2}\norm{\lambda c^2} + c\,(\norm{Y \tau X} + \norm{\wt{\gamma} \sigma \gamma})\,,
\ee
where $\norm{Y \tau X} + \norm{\wt{\gamma} \sigma \gamma}$ is a Kac-Moody current in the beta-gamma + free-fermion system. The cancellation of a boundary gauge anomaly ensures that the level of this Kac-Moody current is $-2h$, as required for BRST reduction. Altogether, one arrives at a simple description of the boundary VOA: it is a  BRST reduction of a $T^*R$-valued beta-gamma system, tensored with free fermions. This is of course precisely the description found in \cite{CostelloGaiotto} (where $\CV^A[G,R;V]$ was denoted $\mathfrak A_H[G,R;V]$).

\subsubsection{Boundary VOA for $T[SU(n)]^A$}

We may easily specialize \eqref{VA-generic} to the A-twist of $T[SU(n)]$, with its Lagrangian description. Following the notation of Section \ref{sec:CSGLSM}, the gauge group is $G=\prod_{a=1}^{n-1}U(a)$, with bifundamental matter $R=\oplus_{a=1}^{n-1}\text{Hom}(\C^a,\C^{a+1})$. 
The boundary 't Hooft anomalies for the non-abelian parts of each $U(a)$ exactly cancel, but the abelian anomalies are non-trivial and are given by the Cartan matrix of $SU(n)$. To cancel these anomalies, \cite{CostelloGaiotto} introduce $n$ boundary Fermi multiplets $\gamma_a, \wt{\gamma}_a$ of weights $(1,0,..., 0, 0), (-1,1,..., 0, 0), ..., (0,0,..., 0, -1)$ under these $U(1)$'s, \ie\, $\gamma_1$ transforms in the representation $\det_1$, $\gamma_a$ transforms in the representation $\det_a \times (\det_{a-1})^{-1}$ for $a = 2, ..., n-1$, and $\gamma_n$ transforms as $(\det_{n-1})^{-1}$.

In the notation of Section \ref{sec:CSGLSM}, the boundary VOA is built from the symplectic bosons $X_a, Y_a$, the boundary complex fermions $\gamma_a, \wt{\gamma}_a$, and the $bc$-ghosts $\pd c_a, \lambda_a$, subject to the differential
\be
\label{eq:TSUnAQ}
\begin{aligned}
	Q_A c_a & = \tfrac{1}{2}[c_a,c_a] \qquad & Q_A \lambda_a & = \norm{[c_a, \lambda_a]} + J_a\\
	Q_A X_a & = c_{a+1} X_a - X_a c_a \qquad & Q_A Y_a & = c_{a} Y_a - Y_a c_{a+1}\\
	Q_A \gamma_a & = (\Tr(c_a) - \Tr(c_{a-1})) \gamma_a \qquad & Q_A \wt{\gamma}_a & = (\Tr(c_{a-1}) - \Tr(c_{a})) \wt{\gamma}_a\\
\end{aligned}\,,
\ee
where $c_0 = c_n = 0$ and $J_a = \norm{X_a Y_a} - \norm{Y_{a-1} X_{a-1}} + \id_{a \times a} (\norm{\wt{\gamma}_{a} \gamma_{a}}-\norm{\wt{\gamma}_{a+1} \gamma_{a+1}})$ is the current generating the $U(a)$ gauge transformations on the symplectic bosons and boundary complex fermions. The boundary VOA is simply the $U(1) \times ... \times U(n-1)$-BRST reduction of this symplectic boson + free fermion system; the corresponding VOA is the ``Langlands duality kernel'' of \cite{CostelloGaiotto, CCG},
\be
	A(\sln) = \CV^A\bigg[\prod\limits_{a=1}^{n-1} U(a), T^*\bigoplus_{a=1}^{n-1} \Hom(\C^a, \C^{a+1}), V_n\bigg]\,,
\ee
where $V_n = \det_1 \oplus \det_2 \otimes (\det_1)^{-1} \oplus \cdots \oplus (\det_{n-1})^{-1}$. This VOA also arises as a large level limit of a corner VOA \cite{GaiottoRapcak} (after decoupling a large commutative subalgebra), as described later in Section \ref{sec:kernel}.  In the special case $n=2$, the VOA $A(\mathfrak{sl}_2)$ admits a concise reformulation as the affine superalgebra ${\rm psu}(2|2)_1$.

Of particular importance is the $U(1) \times ... \times U(n-1)$-invariant and $Q_A$-closed current
\be \label{J-TSUn}
	J_{\mathfrak{su}(n)} = \norm{Y_n X_n} - \tfrac{1}{n} \Tr(\norm{Y_n X_n})\,.
\ee
It generates an $\mathfrak{su}(n)$ current subalgebra of $A(\sln)$ at level $1-n$, realizing the $SU(n)$ Higgs-branch flavor symmetry of $T[SU(n)]$.

\subsubsection{Deformation by abelian flat connections}

Before moving to the more interesting case of $\CT^A_{n,k}$, we pause to describe how the above construction is deformed in the presence of a background flat connection for the topological flavor symmetry. (A similar analysis appears in \cite{Gaiotto-blocks}.) 

We saw how to deform the $A$-twist of an $\CN=4$ super Yang-Mills theory coupled to hypermultiplets by a background flat connection for (a maximal torus of) the topological flavor symmetry in Section \ref{sec:Tnk-BV}; if $\CA = \CA_z \diff z + \CA'$ is such an abelian flat connection, we can introduce the superpotential $-\Tr(\bPhi) \CA_z \diff z + \Tr(\bA) \pd \CA'$. We will work in a holomorphic gauge where the flat connection is simply $\CA = \CA_z(z) \diff z$.

The new superpotential $-\Tr(\bPhi) \CA_z \diff z$ does not introduce any additional OPE's among generators $X,Y,\gamma,\wt\gamma,\lambda,\pd c$ on top of those in \eqref{eq:YMAbdyOPE}. However, it modifies the action of 
$Q_A$ on~$\lambda$:
\be \label{QA-Az-def}
	Q_A \Tr(\lambda) = \Tr(J) \quad \rightsquigarrow \quad Q_A \Tr(\lambda) = \Tr(J) - \CA_z\,.
\ee
This will effectively deform the OPE's of $Q_A$-\emph{cohomology classes} in the final boundary VOA.

Several other descriptions of deformations by flat flavor connections appear throughout the paper. We give an explicit analysis of the deformation \eqref{QA-Az-def} for $A(\mathfrak{sl}_2)$ in Appendix \ref{app:TSU2}, and show that it is equivalent to coupling to a $PSU(2)$ Poisson vertex algebra as described later in Section \ref{sec:VOA}.

\subsubsection{Boundary VOA for $\CT^A_{n,k}$}
\label{sec:Tnk-VOA}

Now consider $\CT^A_{n,k}$, in its Lagrangian formulation from Section \ref{sec:CSGLSM}. We define a holomorphic boundary condition for it by choosing the same Neumann-like boundary conditions as above for the $T[SU(n)]$ subsector, and choosing Neumann boundary conditions for the new $SU(n)_k$ Chern-Simons fields. This will again require adding boundary degrees of freedom to cancel a gauge anomaly.

At this point, we must make a choice of (sign) convention regarding how bulk Chern-Simons levels contribute to a boundary anomaly. Adopting the conventions/formalism of \cite{DGPdualbdys}, we will assume that a bulk Chern-Simons level $k$ contributes $+k$ to the anomaly on a right boundary condition, and $-k$ to a left boundary condition. We further assume, as we do throughout the paper, that $k$ is positive. Then we place a Neumann-like boundary condition on the \emph{left}, so that the anomaly can be cancelled with the addition of boundary fermions.

In contrast, for positive $k$, we expect the Dirichlet-like boundary condition of Section \ref{sec:D-FT} to only make sense on the \emph{right}. The relative orientation of the current Neumann-like b.c. and the later Dirichlet-like b.c. is ultimately responsible for the respective categories of modules $\CN_{n,k}$-mod and $\CD_{n,k}$-mod having a braiding-\emph{reversed} equivalence.

Working with left boundary conditions, then, the bulk fields contribute a total of $-k + n - (n-1) = -(k-1)$ to the boundary anomaly for the $SU(n)$ symmetry.  Thus, we must introduce boundary degrees of freedom transforming in a representation $V$ with quadratic index $T_V = k - 1 > 0$. Since $k \geq n$, we introduce $k - 1$ boundary Fermi multiplets transforming in the fundamental representation of $SU(n)$, \ie\, $V = (\C^n)^{k-1}$.

Again following \cite[Sec 6.2]{CDG}, we expect that the corresponding boundary VOA can be obtained by taking derived $SL(n,\C[\![z]\!])$-invariants of the product of the $T[SU(n)]^A$ VOA (namely, $A(\sln)$) and the $k-1$ $\C^n$-valued complex fermions:
\be
	\CN_{n,k} = \big[A(\sln) \otimes \FF(n(k-1))\big]^{SL(n,\C[\![z]\!])}\,. \label{Cnk-derived}
\ee
Here ``derived $SL(n,\C[\![z]\!])$ invariants'' simply means taking $SU(n)$ invariants by hand, and introducing a $c$-ghost and corresponding differential to take invariants for non-constant boundary gauge transformations cohomologically. We can further incorporate deformations by (torus-valued) background flat connections by simply replacing $A(\sln)$ with a deformed version, as in \eqref{QA-Az-def}.

It is important to note that the derived invariants appearing in \eqref{Cnk-derived} are \emph{not} the same as a BRST reduction. Namely, there is no $\mathfrak{sl}_n$-valued gaugino `$\lambda_n$' to fill out the $bc$-ghost system and implement the vanishing of the $SU(n)$ moment map. The distinction can be thought of as a consequence of gauging $SU(n)_k$ with a 3d $\CN=2$ rather than 3d $\CN=4$ vectormultiplet. Nevertheless, in contrast to general 3d $\CN=2$ theories, there \emph{is} an internal Kac-Moody current associated to the $SU(n)$ action: in the product $A(\sln) \otimes \FF(n(k-1))$ it is given by 
\be
	J^{\rm tot}_{\mathfrak{su}(n)} = J_{\mathfrak{su}(n)} - \sum\limits_{i=1}^{k-1} \big(\norm{\rho_i \wt{\rho}_i} - \tfrac{1}{n}\Tr(\norm{\rho_i \wt{\rho}_i})\big)\,,
\ee
where $J_{\mathfrak{su}(n)}$ is the $T[SU(n)]$ current from \eqref{J-TSUn} and the $\rho_i, \wt{\rho_i}$ for $i = 1, ..., k-1$ are the leading components of the boundary Fermi multiplets. The current $J^{\rm tot}_{\mathfrak{su}(n)}$ generates an $\mathfrak{su}(n)_{k-n}$ Kac-Moody subalgebra.

The algebra \eqref{Cnk-derived} can now be described more explicitly as
\be \CN_{n,k} = H^\bullet\big( \big[A(\sln) \otimes \FF(n(k-1))\otimes \langle\!\langle \pd c\rangle\!\rangle\big]^{SU(n)}, Q_A' \big) \label{Cnk-current} \ee
where $[\cdots]^{SU(n)}$ denotes the $SU(n)$-invariant subalgebra, and the action of the differential $Q_A'$ (implementing derived invariants) is given on any $SU(n)$-invariant operator 
$\CO(z) \in A(\sln) \otimes \FF(n(k-1))$ by
\be
	Q_A' \CO(z) = \oint \frac{\diff w}{2 \pi i} c(w) \cdot J^{\rm tot}_{\mathfrak{su}(n)}(w) \CO(z)= \sum \limits_{\ell \geq 0}\frac{1}{\ell !} \pd^\ell c(z) \oint \frac{\diff w}{2 \pi i} (w-z)^\ell J^{\rm tot}_{\mathfrak{su}(n)}(w) \CO(z)\,.
\ee

One further useful reformulation is possible, under the nontrivial assumption that the cohomology \eqref{Cnk-current} is entirely supported in cohomological degree zero (\emph{i.e.} there is no higher cohomology). We will not prove this assumption here, though we do make some explicit verifications in Appendix \ref{sec:TSU2CS}. Since the differential is given by OPE's with the current $J^{\rm tot}_{\mathfrak{su}(n)}(w)$, the algebra $\CN_{n,k}$ in degree zero is just the coset (\emph{a.k.a.} commutant) of the internal $\mathfrak{su}(n)_{k-n}$ Kac-Moody symmetry. Assuming that there is no higher cohomology, we thus expect that
\be \label{Nnk-Com}
	\CN_{n,k} \overset{\rm conj.}{\cong} \frac{A(\sln) \otimes \FF(n(k-1))}{\mathfrak{su}(n)_{k-n}} = {\rm Com}\big(V^{k}\big(\sln),A(\sln) \otimes \FF(n(k-1))\big)\,.
\ee

\section{QFT computations and predictions}
\label{sec:Hilbert}

In this section, we apply various techniques to analyze quantitative features of the categories of line operators and state spaces in the topological QFT's $\CT_{G,k}^A$. We mainly focus on $G=SU(n)$, \ie\ the theories $\CT_{n,k}^A$.

We begin in Section \ref{sec:Bethe} by computing the ``Bethe roots'' of (the 3d $\CN=2$ precursor of) $\CT_{2,k}^A$ following \cite{NS-Bethe, NS-int}, and apply this in Section \ref{sec:T2kGrothendieck} to find the Grothendieck ring of the category of line operators, perfectly reproducing the quantum-group result from Section \ref{sec:Uqsl2}. We then employ methods of \cite{NS-index, BZ-Riemann, BZ-index, CK-comments} to compute the Euler characters (\emph{a.k.a.} indices) of state spaces of $\CT_{2,k}^A$ in all genera, again reproducing quantum-group results.

We extend the computation of Bethe roots and the character of the genus-one state space to $G=SO(3)$ in Section \ref{sec:SO3}, following \cite{EKSW, Willett-notes}. In the process, we will compute the 't Hooft anomaly in the one-form $\Z_2$ symmetry of $\CT_{2,k}^A$. (The way one obtains $\CT_{SO(3),k}^A$ is by gauging the one-form symmetry, which is only possible when the anomaly vanishes.)
We then extend to higher-rank $\CT_{n,k}^A$ theories in Section \ref{sec:SUnBethe}, again computing Bethe roots, the character of the genus-one state space, and the anomaly in the $\Z_n$ one-form symmetry. We briefly comment on the effect of gauging various subgroups of the $\Z_n$ symmetry, to obtain different global forms of the higher-rank theories.
 We hope that the results in these sections will provide guidance in comparing with other quantum-group and VOA categories in the future.

Finally, we include some general remarks on the expected algebraic structure of state spaces and the categories of line operators themselves (as opposed to indices and the Grothendieck ring) in Sections \ref{sec:T-Hilb} and \ref{sec:T-cat}. We explain the origin of the approximate factorization of state spaces \eqref{Hilb-fact} from the Introduction. We also speculate on a putative geometric model for the category of line operators in $\CT_{n,k}^A$, as  weakly equivariant D-modules on a loop space.

\subsection{Twisted superpotential and Bethe roots}
\label{sec:Bethe}

The first step toward analyzing expectation values of line operators and computing $\Sigma\times S^1$ partition functions is to determine the supersymmetric vacua of $\CT_{n,k}$ on a finite-size circle.  

We will work with the Lagrangian 3d $\CN=2$ theory $\CT^{UV}_{n,k} = T[SU(n)]^{UV}/SU(n)_k$ defined in Section \ref{sec:3d}, which flows in the infrared to $\CT_{n,k}$. Putting $\CT^{UV}_{n,k}$ on a finite-size circle defines an effective 2d $\CN=(2,2)$ theory $\CT^{UV}_{n,k}[S^1]$. Furthermore, upon introducing generic real masses and background connections (around $S^1$) for the flavor symmetries of $\CT^{UV}_{n,k}$, the 2d theory $\CT^{UV}_{n,k}[S^1]$ will become fully massive. Its supersymmetric vacua are given by critical points of an effective twisted superpotential $\CW$ on its Coulomb branch. The effective twisted superpotential of $\CT^{UV}_{n,k}[S^1]$ is easily computed using methods developed by \cite{NS-Bethe, NS-int} in the context of the Bethe/Gauge correspondence.

We specialize to $n=2$ for now, and use the following conventions/notation.
 As an $\CN=2$ theory, $\CT^{UV}_{2,k}$ has gauge group $SU(2)\times U(1)$ and five chiral multiplets, with charges:
\be \label{T2-charges} \begin{array}{c|c|c|ccccc}
& \text{charge} & \text{fugacity} & (X_1)^1 & (X_1)^2 & (Y_1)_1 & (Y_1)_2 & \phi_1 \\\hline
U(1) && x & -1 & -1 & 1 & 1 & 0 \\
U(1)\subset SU(2) && z& 1 & -1 & -1 & 1 & 0\\  \hline
U(1)_R=U(1)_H & R & (-1) & 1 & 1 & 1 & 1 & 0 \\
U(1)_\epsilon & \epsilon & t & 1 & 1 & 1 & 1 & -2 \\
U(1)_{\rm top} \subset G^\vee & f & y & 0&0&0&0&0
 \end{array} \ee
Here the top two rows are the gauge charges. The third row contains the $\CN=2$ $R$-charge, which we have chosen to coincide with the maximal torus of the 3d $\CN=4$ R-symmetry $SU(2)_H$ that emerges in the IR. The fourth row is the $\CN=2$ flavor symmetry that corresponds to a difference $\epsilon=H-C$ of 3d $\CN=4$ R-charges. The final row is the maximal torus of the topological flavor symmetry $G^\vee=PSU(2)$ that emerges in the IR. The latter is the symmetry that allows coupling to background flat $PGL(2,\C)$ connections.

The ``fugacity'' column here lists the variables that we will use to denote respective fugacities in supersymmetric indices. Alternatively, these variables represent holonomies of connections for the various symmetries around $S^1$, complexified by appropriate real masses. Thus $y = \exp\big[\oint_{S^1} (iA_{\rm top}+t_\R)\big]$ where $t_\R$ is the  real FI parameter; $x = \exp\big[\oint_{S^1}(iA_{U(1)}+\sigma)\big]$ where $\sigma$ is the $U(1)$ vectormultiplet scalar; etc.

With this notation, we can write the effective twisted superpotential \cite{NS-Bethe, NS-int} as
\be \CW = k(\log z)^2 + \log y\log x + L_2(txz)+L_2(tx/z)+L_2(tz/x)+L_2(t/(xz)) + L_2(-t^{-2})\,,  \label{WSU2} \ee
where $L_2(u):= \text{Li}_2(u)+\frac14(\log u)^2$ is a function that satisfies
\be \exp \big( u\pd_u L_2(u)\big) = \frac{u^{1/2}}{1-u}  =  \frac{1}{u^{-1/2}-u^{1/2}}\,. \ee
The first two terms in $\CW$ come from classical contributions from the Chern-Simons terms for $SU(2)$ and for the mixed $U(1)$-$U(1)_{\rm top}$ Chern-Simons term, respectively, while the last 5 terms come from 1-loop corrections due to the chiral multiplets.

The supersymmetric vacua of the 3d theory on $S^1$ are the solutions of
\be P_x(x,z) := \exp( x\pd_x \CW) = 1\,,\qquad P_z(x,z) := \exp( z\pd_z \CW) = 1\,, \label{BetheP} \ee
modulo the action of the Weyl group. Explicitly,
\be P_x = y \frac{xz-t}{1-t xz}\frac{x/z-t}{1-t x/z} \,,\qquad P_z = z^{2k} \frac{xz-t}{1-txz}\frac{1-t x/z}{x/z-t}\,.  \label{Pxz} \ee
Note that, while $\CW$ suffers from several multi-log branch-cut ambiguities, $P_x$ and $P_z$ do not. The Weyl group $S_2=\Z_2$ of the $SU(2)$ gauge symmetry acts on the set of solutions by sending $z\mapsto z^{-1}$. 
By a tried and tested prescription%
\footnote{This prescription could still benefit from a complete physical derivation.}, %
one should not associate fixed points of the Weyl group (solutions with $z=\pm 1$, often called degenerate vacua) with true supersymmetric vacua. The remaining solutions fall into orbits of size two; we denote the set of orbits
\be \CB := \{(x,z)\,|\, P_x=P_z=1\,,\; z\neq \pm 1\}/S_2\,, \label{Bethe} \ee
and, following \cite{NS-Bethe, NS-int}, call the elements of $\CB$ the ``Bethe vacua."

Various computations of expectation values and indices/partition functions based on these Bethe vacua can alternatively be thought of as computations in the holomorphic-topological (HT) twist of a 3d $\CN=2$ theory. 
The choice of 3d $\CN=2$ R-symmetry in \eqref{T2-charges} is the one compatible with a further deformation of the HT twist to a 3d topological A-twist, as outlined abstractly in Section \ref{sec:twistedindex}, and explicitly for $\CT_{n,k}$ in Section \ref{sec:Tnk-BV}. We will perform various calculations at generic fugacity $t$ for the anti-diagonal subgroup $U(1)_\epsilon \subset U(1)_H \times U(1)_C$, and then take the limit
\be \text{HT $\leadsto$ A-twist}:\quad t \to -1\,. \ee
to implement the HT $\leadsto$ A deformation, and ensure that we use the correct cohomological grading for the 3d topological A-twist.

\subsection{Grothendieck ring}
\label{sec:T2kGrothendieck}

Let $\CC^{(n,k)}$ denote the dg category of line operators in $\CT_{n,k}^A$. We expect its Grothendieck ring $K_0(\CC^{(n,k)})$ coincides with the algebraic ring in which the Bethe vacua of $\CT_{n,k}^{UV}[S^1]$ are defined  --- essentially the Jacobian ring of $\CW$ --- in the limit $t\to -1$. This is the twisted chiral ring of $\CT_{n,k}^{UV}[S^1]$. 

One (standard) way to understand this statement is the following. The $\C\times S^1$ geometry used in compactifying $\CT_{n,k}$ has asymptotic boundary $\pd(\C\times S^1) \simeq T^2$.
The Bethe vacua can be identified with the supersymmetric ground states of $\CT_{n,k}$ on this asymptotic 2-torus, in the presence of flat background connections as specified by flavor fugacities. For example, for $n=2$, $U(1)_{\rm top}$ fugacity $y$ corresponds to a $PGL(2,\C)$ connection with holonomy $\text{diag}(y,1)$ along the non-contractible $S^1$. As $t\to -1$, we expect these to become the states of the topologically twisted theory $\CT_{n,k}^A$ on $T^2$, in the presence of a flat connection with generic holonomy along one of the cycles.

In principle, additional differentials could arise when implementing the topological A-twist at $t=-1$. However, the assumption that $\CT_{n,k}[S^1]$ can be deformed to be fully massive (which we verify by explicit computations of $\CW$) implies with this amount of supersymmetry that all its supersymmetric vacua will lie in cohomological degree zero, and precludes the existence of additional differentials. A corollary is that whenever $\CT_{n,k}$ can be deformed to be fully massive, the $T^2$ state space of $\CT_{n,k}^A$ with generic flat connection is guaranteed to lie in cohomological degree zero, with
\be \text{dim} \CH(T^2,\CA_{\rm generic}) = \#\,\text{Bethe vacua}\,. \label{H-Bethe} \ee

Now, any line operator preserving the HT twist supercharge can be inserted at  $\{0\}\times S^1\subset \C\times S^1$ to define a ground state on the asymptotic $T^2$, and every asymptotic ground state should arise from such a line-operator insertion. (Analogous statements about chiral rings in 2d $\CN=(2,2)$ theories go back to \cite{CV-tt*,CV-class}.) This should define an isomorphism between the Grothendieck group of the category of line operators in the HT twist and the space of Bethe vacua. As $t\to -1$, we expect this to become an isomorphism between the Grothendieck group of $\CC^{(n,k)}$ and the Bethe vacua. Moreover, just as with chiral rings in 2d $\CN=(2,2)$ theories, the product on $K_0(\CC^{(n,k)})$ (induced by non-singular OPE of line operators preserving the A-twist) should coincide with the ring structure of Bethe vacua.

Let's apply this to $n=2$. The algebraic ring of functions on the Bethe vacua takes the form
\be \label{ring-2} \CR_{2,k}:= \C[\CB]= \Big(\C(y,t)[x^{\pm 1},z^{\pm 1},\alpha]/\big(\wt P_x,\wt P_z,\alpha(z^2-1)-1\big)\Big)^{S_2}\,. \ee
Here we work over Laurent polynomials in the gauge fugacities, but arbitrary rational functions in the flavor fugacities, corresponding to the fact that we assume the flavor fugacities to be generic (giving us a massive effective 2d theory). We define
\be \begin{array}{l} \wt P_x := y(xz-t)(x/z-t)-(1-txz)(1-tx/z)\,,\\ \wt P_z := z^{2k}(xz-t)(1-tx/z)-(1-txz)(x/z-t) \end{array}\ee
as the denominator-cleared forms of $P_x-1,P_z-1$; add an extra variable $\alpha$ to implement the condition $z\neq \pm 1$; and take $S_2$ Weyl invariants at the end.

As a ring over $\C(y,t)$, $\CR_{2,k}$ has dimension $2k$. It can be further simplified in a physically meaningful way by introducing
\be \nu := z+z^{-1}\,, \ee
which represents the fundamental Wilson line for the $SU(2)$ gauge symmetry of $\CT_{2,k}$. Then we find (by computing Gr\"obner bases) that 
\be \CR_{2,k} \simeq \C(y,t)[\nu] / F_k(\nu,y^\pm,t^\pm)\,, \ee
where $F_k$ is a polynomial of degree $2k$ in $\nu$ that remains finite as $t\to -1$. For example:
\be \begin{array}{l}
	F_1 = \nu^2-t^2(y+2+y^{-1}) \\
	F_2 = (\nu^2-t^2-1)^2-t^2(y+2+y^{-1})  \\
	F_3 = \nu^2(\nu^2-t^2-2)^2-t^2(y+2+y^{-1})  \\
	F_4 = (\nu^4-(t^2+3)\nu^2+t^2+1)^2-t^2(y+2+y^{-1})  \\ \ldots
\end{array} \ee
Sending $t\to -1$ (which algebraically requires some care) we obtain 
\be  \CR_{2,k} \big|_{t\to-1} \simeq \C(y)[\nu] / f_k(\nu,y^\pm)\,,\ee
with
\be f_k(\nu,y) = (-1)^k L_{2k}(i\nu)-(y+y^{-1})\,,\ee
where $L_{2k}(x)$ is the $2k$-th Lucas polynomial, defined by $L_{2k}(i(z+z^{-1})) = (-1)^k(z^{2k} + z^{-2k})$. For example, 
\be \begin{array}{l}
	f_2 = (\nu^2-2)^2 - (y^{-1}+2+y)  \\
	f_3 = \nu^2(\nu^2-3)^2-(y^{-1}+2+y)  \\
	f_4 = (\nu^4-4\nu^2+2)^2 - (y^{-1}+2+y)  \\
	\ldots \end{array}
\ee

Further setting $y\to 1$ to ignore $PGL(2)$ equivariance, we may use the fact that $L_{2k}(i(z+z^{-1})) - 2 = (z^k + z^{-k})^2$ has zeroes at the $2k$-th roots of $-1$ to factor $f_k(\nu,1)$ as
\be
	f_k(\nu,1) = \prod_{\ell = 1}^{2k}\big(\nu-2\cos\tfrac{\pi \ell}{k}\big)^2\,.
\ee
This perfectly reproduces the Grothendieck ring for the small quantum group $u_q(\mathfrak{sl}_2)$ at even root of unity $q=e^{i\pi/k}$, as discussed in Section \ref{sec:V-1} and in particular \eqref{U-ring}.

The $y$ dependence can also be easily interpreted if we take seriously the relation between line operators and Bethe vacua above, and the general construction of the torus state space from line operators in a twisted QFT reviewed in Section \ref{sec:torus-lines}. We are producing states in a $T^2$ state space by inserting line operators along the core of a solid torus in the presence of a background $PGL(2,\C)$ connection with holonomy $g=\text{diag}(y,1)$ along the non-contractible cycle. Then we would actually expect the ring of Bethe vacua to correspond to twisted Hochschild homology
\be  \CR_{2,k} \big|_{t\to-1} \simeq 
 HH_\bullet^g(\CC^{(2,k)})\,. \label{R-HH} \ee
This depends on the parameter $y$, and for generic $y$ should be supported entirely in degree zero, with the property that $\lim_{y\to 1} HH_\bullet^g(\CC^{(2,k)}) \simeq  K_0(\CC^{(2,k)})$. It would be interesting to verify \eqref{R-HH} with a quantum-group computation.

\subsection{Characters of state spaces on $\Sigma_g$}
\label{sec:T-char-g}

Next, we determine the Euler characters of state spaces on smooth genus-$g$ surfaces $\Sigma_g$. We continue to specialize to the case $n=2$. We follow the approach outlined in Section~\ref{sec:twistedindex}: treating $\CT_{2,k}$ as a 3d $\CN=2$ theory, we compute its genus-$g$ twisted index --- \emph{a.k.a.} partition function on $\Sigma_g\times S^1$ in the HT twist --- and then specialize $t\to -1$ to obtain the genus-$g$ index in the 3d topological A-twist.

To compute the twisted index, we apply the analysis of \cite{NS-index, BZ-Riemann, BZ-index, CK-comments} to $\CT_{2,k}^{UV}$. The twisted index takes the form:
\be \CI_{\CT_{2,k}}(\Sigma_g) = \text{Tr}_{\CH(\Sigma_g)} (-1)^R t^\epsilon y^f 
 = \sum_{(x,z)\in \CB} H(x,z,y,t)^{g-1}\,. \ee
Here $H(x,z,y,t)$ is the ``handle gluing operator,'' given by the Hessian of the twisted superpotential $\CW$ 
times the exponential of the effective dilaton $\Omega$, which controls the coupling to the curvature of $\Sigma_g$ \cite{NS-index}, 
\be H(x,z,y,t) = e^{\Omega} \det \begin{pmatrix} (x\pd_x)^2 \CW & x\pd_x\, z\pd_z \CW \\ z\pd_z\,x\pd_x \CW & (z\pd_z)^2 \CW \end{pmatrix}\,. \ee
For $\CT^{UV}_{2,k}$, the effective dilaton $\Omega$ is given by%
\be \Omega = \log t + \log(1-t^{-2}) - \log(1-z^{2}) - \log(1-z^{-2})\,.\ee
The first term comes from mixed $U(1)_\varepsilon$-$U(1)_H$ Chern-Simons terms, the second from 1-loop corrections due to the chiral multiplet with the scalar $\phi_1$, and the last two terms come from 1-loop corrections due to $W$-bosons. Just as with $P_x, P_z$, the handle operator $H(x,z,y,t)$ does not suffer from branch-cut ambiguities. 

After evaluating $\CI_{\CT_{2,k}}(\Sigma_g)$ for generic $t$, we should send $t\to -1$ to obtain the answer relevant for the topological A-twist. At $g=0$, \emph{i.e.} for $\Sigma_0 = S^2$, we find
\be\begin{aligned} k=1:& \quad \chi\big[\CH(S^2)\big] = -t\big|_{t=-1}= 1\,,\\
	 k \geq 2:& \quad \chi\big[\CH(S^2)\big] = \frac{-yt(1+t^2)}{(y-t^2)(1-yt^2)}\Big|_{t=-1} = \frac{2}{(1-y)(1-y^{-1})}\,.
\end{aligned} \label{char-g0} \ee
Importantly, for $k \geq 2$ we find that the $t \to -1$ limit of the twisted index on $S^2$ agrees with the Euler character of the $S^2$ state space found in Section \ref{sec:U-Hilbert}, \cf\, \eqref{P-gen0} at $t=-1$.

The Euler character of the state space $\CH(T^2)$ in genus 1 is just the number of Bethe vacua. There are exactly $2k$ Bethe vacua, and so we find
\be   \chi\big[\CH(T^2)\big] = \text{Tr}_{\CH(T^2)} (-1)^R t^\epsilon y^f =  |\CB| = 2k\,. \ee
For $g=2$, we find
\be \chi\big[\CH(\Sigma_2)\big] = 4k^3\,, \ee
independent of $y$.
For $g\geq 3$, we encounter increasingly complicated dependence on $y$. Nonetheless, by analyzing the handle-gluing operator, we find an all-genus formula for the equivariant character given by%
\be \label{eqchar}
	\chi\big[\CH(\Sigma_g)\big] = \sum\limits_{\ell = 0}^{2k-1}\Bigg[\frac{1}{2k}\bigg(\frac{\zeta^\ell y^{\tfrac{1}{2k}} - \zeta^{-\ell} y^{-\tfrac{1}{2k}}}{y^{\tfrac{1}{2}}-y^{-\tfrac{1}{2}}}\bigg)^2\Bigg]^{1-g}\,,
\ee
where $\zeta$ is any primitive $2k$-th root of unity.%
\footnote{This expression does not depend on a choice of $2k$-th root of $y$, which follows from the fact that the sum is invariant under $y^{\tfrac{1}{2k}} \to \zeta y^{\tfrac{1}{2k}}$.}
We will reproduce this answer from a VOA perspective in Section \ref{sec:VOA} (albeit also somewhat experimentally). After setting $y=1$ (ignoring $PGL(2,\C)$ equivariance), we arrive at the simple formula for higher-genus Euler characters:
\be \chi\big[\CH(\Sigma_g)\big]\big|_{y=1} = 2^g k^{3g-3}\,,\qquad (g\geq 2)\,. \ee

Summarizing, for any $k\geq 2$:
\begin{itemize}
\item The character of the genus $g=0$ state space matches the derived computation in \eqref{P-gen0}. The CGP TQFT just assigns a 1-dimensional state space to $S^2$, which is the subspace of $\CH(S^2)$ in cohomological degree zero. It is not possible to see the degree-zero subspace in the character (though it is clear from \eqref{P-gen0}).
\item In genus $g=1$, the character $\chi[\CH(T^2)]$ matches the derived computation of Section \ref{sec:U-1}, as well as the dimension of the CGP state space after deforming by a generic flat connection (Sections \ref{sec:U-g} and \ref{sec:CGP-torus}).
\item In genus $g\geq 2$, the character $\chi[\CH(\Sigma_g)]$ matches the dimensions of quantum-group spaces after deformation by a generic flat connection, either computed naively as in Section \ref{sec:U-g} or in the CGP TQFT as in Section \ref{sec:CGP-0}.
\end{itemize}

\subsection{$\Z_2$ anomaly and SO(3) theory}
\label{sec:SO3}

Although in this paper we mainly focus on theories $\CT_{G,k}^A$ for simply-connected groups $G=SU(n)$, the computations of indices and the Grothendieck ring described above are fairly accessible in other types as well. We illustrate this briefly here for $G=SO(3)$, \emph{i.e.} for the theory $\CT_{SO(3),k} = T[SU(2)]/SO(3)_k$, using methods developed in \cite{EKSW} and \cite{Willett-notes}.%
\footnote{We thank B. Willett, H.Y. Kim, and S. Sch\"afer-Nameki for enlightening discussions on global forms of gauge groups and related anomalies, and especially thank B. Willett for sharing his unpublished notes \cite{Willett-notes}.} %
We hope that these computations can be suitably matched with quantum group and VOA perspectives in the future.

\subsubsection{One-form anomaly}
\label{sec:Z2-anomaly}

A useful way to construct $\CT_{SO(3),k}$ is by gauging the one-form symmetry of $\CT_{SU(2),k}$. We recall from Section \ref{sec:global} that $\CT_{SU(2),k} \simeq T[SU(2)]/SU(2)_k$ has a $\Z_2$ one-form symmetry that arises when the center of $SU(2)$ is gauged. The generator of the $\Z_2$ symmetry is a topological line operator $\omega$ (topological even in the full physical theory) that satisfies $\omega\otimes \omega=\id$. The $\Z_2$ action on the category of line operators comes from collision with $\omega$,
\be L \mapsto \omega\otimes L \qquad \text{(any line operator $L$)}\,. \ee

Gauging the $\Z_2$ symmetry amounts to inserting the projection line operator $\id \oplus \omega$ in all possible configurations in any correlation function \cite{GKSW}. This is a well defined operation if and only if $\omega$ double-braids trivially with itself, so that configurations
\be \includegraphics[width=1.7in]{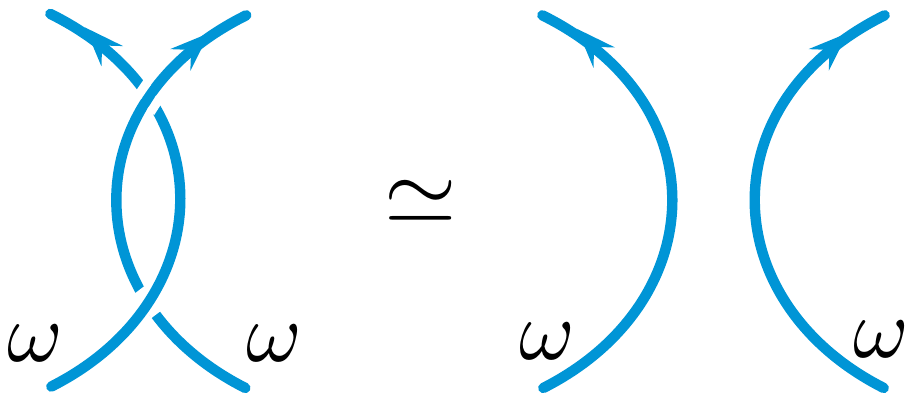} \ee
are indistinguishable.
 If $\omega$ does not double-braid trivially with itself, one says that the $\Z_2$ symmetry has an anomaly, and cannot be gauged. In the case of $\CT_{SU(2),k}$, it turns out (and we shall check momentarily) that $\omega$ braids trivially with itself if and only if the level $k$ is even. Thus we only expect to be able to define $\CT_{SO(3),k}\simeq\CT_{SU(2),k}/\Z_2 $ when $k$ is even.

Let us review how to detect a potential anomaly. The self-double-braiding of $\omega$ (as for any invertible line operator) is just given by a constant. In other words, the local operator $\mu_\omega \in \text{End}(\omega\otimes \omega)$ represented by
\be \includegraphics[width=2.2in]{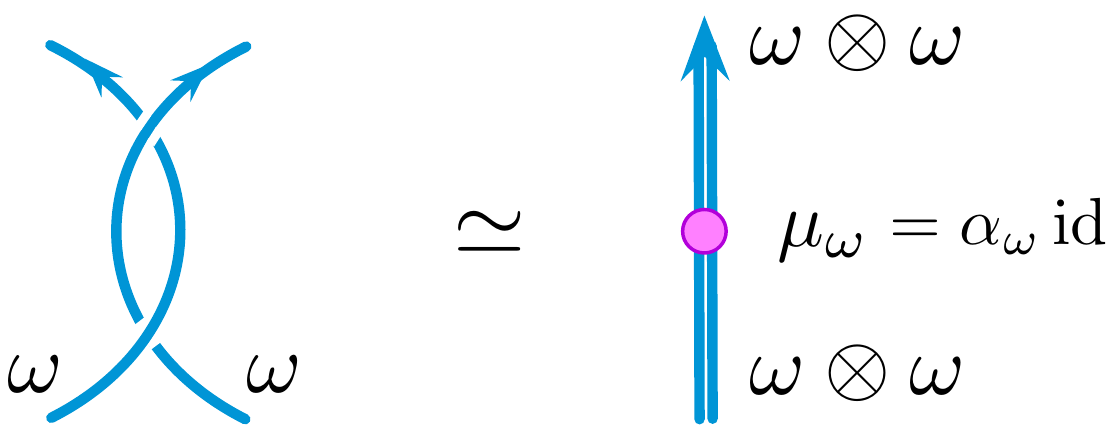}  \ee
can at most be a constant $\alpha_\omega$ times the identity. Moreover, the double-braiding furnishes a representation of the $\Z_2$ symmetry, so $\alpha_\omega^2=1$.

To determine $\alpha_\omega$, one may consider the action of $\omega$ on the torus state space $\CH(T^2)$, either in the full physical theory or in any twist. Let $\omega_A$ denote the line operator $\omega$ wrapping an ``A'' cycle of $T^2$ and let $\omega_B$ denote $\omega$ wrapping a dual ``B'' cycle. Then the operators $\omega_A:\CH(T^2)\to \CH(T^2)$ and $\omega_B:\CH(T^2)\to \CH(T^2)$ generate \emph{two} $\Z_2$ actions on $\CH(T^2)$, and must satisfy
\be \omega_A\omega_B=\alpha_\omega\, \omega_B\omega_A\,. \ee
In other words, a potential projective factor in the full $\Z_2\times \Z_2$ action on $\CH(T^2)$ is controlled by the double-braiding, \emph{a.k.a.} the anomaly.

In the topological A-twist, with generic abelian background connection, the torus state space $\CH(T^2)$ has a basis indexed by Bethe roots, as discussed below \eqref{H-Bethe}. It was explained in \cite{EKSW,Willett-notes} how $\omega_A$ and $\omega_B$ act in this basis. In the theory $\CT_{SU(2),k}^A$, $\omega_A$ acts on a state $|x,z\rangle$ labelled by a solution to \eqref{BetheP} as
\be \omega_A|x,z\rangle = |-x,-z\rangle\,.\ee
This is just the action of the $\Z_2$ center of $SU(2)$ on its $z$ fugacity, combined with a compensating transformation of the $U(1)$ fugacity $x$. The compensating transformation is needed to keep $P_x, P_z$ invariant; it ultimately arises because the hypermultiplets are in fundamental representations of $SU(2)$, on which the center of $SU(2)$ acts trivially only up to an accompanying $U(1)$ gauge transformation. Put differently, only a diagonal torus $U(1)_Z\subset SU(2)\times U(1)$ contains a $\Z_2$ that acts trivially on all the matter fields.
The other operator $\omega_B$ acts diagonally
\be \omega_B|x,z\rangle = P_Z^{1/2}(x,z)|x,z\rangle\,, \ee
with eigenvalues given by any fixed square root of $P_Z=\exp(Z\pd_Z\CW)$, where $Z$ is a fugacity for the diagonal $U(1)_Z$.

To make this more explicit, let us change fugacity variables $X = x/z, Z = z$ in the twisted superpotential \eqref{WSU2}. This obfuscates the Weyl symmetry $(X,Z) \mapsto (XZ^2,Z^{-1})$,
but simplifies the action of the $\Z_2$ center, $\omega_A:(X,Z) \mapsto (X,-Z)$.
We may define $P_X = \exp(X\pd_X \CW)$ and $P_Z =\exp(Z\pd_Z \CW)$, finding
\be P_X = y \frac{XZ^2-t}{1-t XZ^2}\frac{X-t}{1-tX}\,,\qquad P_Z = Z^{2k}y\left(\frac{XZ^2-t}{1-tXZ^2}\right)^2\,,\ee
and we choose the root
\be P_Z^{1/2} := Z^k y^{1/2} \frac{XZ^2-t}{1-t XZ^2} \ee
(which is necessarily algebraic in all variables except potentially $y$). Due to the $Z^k$ factor in $P_Z^{1/2}$, it is easy to see that
\be \omega_A\omega_B = (-1)^k\omega_B\omega_A\,, \ee
whence the anomaly in the one-form symmetry is $\alpha_\omega=(-1)^k$. It vanishes when $k$ is even.

This one-form anomaly beautifully matches the double-braiding of $\Z_2$ generators from quantum-group and VOA perspectives. We saw in Section \ref{sec:V-1} that the $\Z_2$ symmetry generator $\omega$ is identified with the one-dimensional module $S_1^-$ in $u_q(\mathfrak{sl}_2)$-mod, at $q=e^{i\pi/k}$. In the CGP TQFT of Section \ref{sec:defCH}, which ``passes through'' representations of the unrolled quantum group in order to define braiding, the $\Z_2$ generator is identified as $\omega=\C_{mk}^H$ for any odd $m$, \emph{i.e.} the one-dimensional module on which $H$ act by $mk$. Braiding of one-dimensional modules is extremely simple, as it is given purely by the Cartan part of the R-matrix $q^{\frac12H\otimes H}$. Thus, the single braiding of $\omega$ is
\be q^{\frac12H\otimes H} \omega\otimes \omega = q^{\frac12 mk\cdot m'k}\,\text{id} = e^{\frac{i\pi k}{2} m m'}\,\text{id} \qquad \text{($m,m'$ odd)} \ee
and the double-braiding is
\be \alpha_\omega = \big(e^{\frac{i\pi k}{2} m m'}\big)^2 = (-1)^k\,, \ee
exactly matching the quantum group result, \cf\, Eq. \eqref{E:Braiding-sigma}.

The same double-braiding also arises from a VOA perspective. Let $P,Q$ denote the weight and root lattices of $\sln$. For a Feigin-Tipunin algebra $\FT_k(\sln)$, with an associated lattice VOA $V_{\sqrt k Q}$ (see Remark \ref{rmk:FT} and also Section \ref{sec:FT}), the invertible modules generating a $\Z_n$ symmetry are labelled by elements of $\sqrt k P / \sqrt k Q \simeq \Z_n$. Their braiding coincides with the braiding of corresponding $V_{\sqrt k Q}$ modules; in particular for weights $\lambda,\lambda'\in P$, the double braiding is given by a constant $\alpha_{\lambda,\lambda'}=e^{2\pi i k (\lambda,\lambda')}$, with $(\;,\;)$ the Cartan pairing on $P$. The $\Z_n$ symmetry is generated by a module $\omega$ labelled by the highest weight of the fundamental representation $\lambda_1$, which has $(\lambda_1,\lambda_1)=\frac{n-1}{n}$, and so
\be \alpha_{\lambda_1,\lambda_1} = e^{2\pi i k \frac{n-1}{n}} = e^{-\frac{2\pi  i k}{n}}\,. \label{braid-n} \ee
For $n=2$, this becomes $(-1)^k$ as above.

\subsubsection{Gauging the one-form symmetry}
\label{sec:Z2-gauge}

Now, let us assume that $k=2p$ is  indeed even, and gauge the $\Z_2$ symmetry to obtain $\CT_{SO(3),k}$. The Bethe roots relevant for $SO(3)$ are the solutions to $P_X=1$ and $P_Z^{1/2}=1$ that are not fixed points of the Weyl symmetry:
\be \CB_{SO(3)} = \{(X,Z)\,|\, P_X=P_Z^{1/2}=1\,,\; (X,Z)\neq (XZ^2,Z^{-1})\} / S_2\,.\ee
This is a subset of the $SU(2)$ Bethe roots --- it's the $SU(2)$ roots that satisfy $P_Z^{1/2}=+1$.
Note that $\Z_2$ acts on $\CB_{SO(3)}$.  Potential twisted sectors that complicate the analysis of state spaces come from $\Z_2$-fixed points of $\CB_{SO(3)}$, \emph{i.e.} solutions with $Z=\pm i$. It is straightforward to see that there are no such solutions, and thankfully no twisted sectors to worry about.

The index of the state space on $\Sigma_g$ is simply obtained as
\be \chi\big(\CH(\Sigma_g)\big) = \frac{1}{|\Z_2|^{2g}} \sum_{(X,Z)\in \CB_{SO(3)}}H(X,Z,y,t)^{g-1}\Big|_{t=-1}\,, \ee
with the $SU(2)$ handle-gluing operator $H$ appearing, and a prefactor $\frac{1}{|\Z_2|^{2g}}=\frac{1}{2^{2g}}$ coming from projections to invariants of the $\Z_2$ 1-form symmetry on each cycle of $\Sigma_g$. We simply find $2^{-2g-1}$ times the $SU(2)$ result; in particular, setting $y=1$, we get
\be \chi\big(\CH(\Sigma_g)\big)\big|_{y=1} =  \frac{1}{2^{2g+1}} \begin{cases} 2k & g=1 \\ 2^g k^{3g-3} & g \geq 2 \end{cases}
 = \begin{cases} k/2 & g=1 \\ k^{3g-3}/2^{g+1} & g\geq 2\end{cases} = \begin{cases} p & g=1 \\ 2^{2g-4}p^{3g-3} & g\geq 2 \end{cases} \ee
 
We compute the Grothendieck ring of the category of Wilson lines by looking for the minimal polynomial satisfied by $\xi = Z^2+1+Z^{-2} =\chi_2(Z) = \chi(Z)^2-1$ (the character of the adjoint rep) when evaluated at the Bethe roots. We find the ring to be $\C(y)[\xi]/h_p(\xi,y,-1)$ with $h_p(\xi,y,-1) = L_{4p}(i \sqrt{\xi+1}) - (y^{1/2} + y^{-1/2}).$ For example, 
\be \begin{array}{l}
h_1 = (\xi+1-(y^{1/2}+2+y^{-1/2})) \\
h_2 = ((\xi-1)^2-(y^{1/2}+2+y^{-1/2})) \\
h_3 = ((\xi-2)^2(\xi+1)-(y^{1/2}+2+y^{-1/2})) \\
h_4 = ((\xi^2-2\xi-1)^2-(y^{1/2}+2+y^{-1/2})) \\
h_5 = ((\xi^3-3\xi+1)^2(\xi+1)-(y^{1/2}+2+y^{-1/2}))
\end{array}
\ee
Using $\xi = \chi^2-1$, this is very close to the $SU(2)$ Grothendieck ring:
\be  h_p(\chi^2-1,y,-1) = f_p(\chi, y^{1/2})\,. \ee

\subsubsection{$SO(3)$ at $k=2$ and a B-twisted hypermultiplet}
\label{sec:SO3-ferm}

We strongly suspect that the theories $\CT_{SO(3),2}^A$ and $\CT_{\rm hyper}^B$ are equivalent, as anticipated back in \eqref{hyper-SO(3)}. First, we can check their 0-form and 1-form global symmetries. Gauging of the (electric) $\Z_2$ 1-form symmetry of $\CT_{SU(2),2}^A$ to go to $\CT_{SO(3),2}^A$ trades the $\Z_2$ 1-form symmetry for a $\Z_2$ 0-form (magnetic) symmetry extending the topological flavor symmetry $G^\vee = PSU(2)$ to $SU(2)$ \cite{GKSW}. Thus, $\CT_{SO(3),2}^A$ has a $SU(2)$ 0-form symmetry and trivial 1-form symmetry. Moreover, the $SU(2)$ topological flavor symmetry can be used to deform $\CT_{SO(3),2}^A$ by flat background $SU(2)$ connections, exactly matching $\CT_{\rm hyper}^B$.

The Bethe root analysis above provides a more robust check of our proposed equivalence. There is a single Bethe root $\CB_{SO(3)}|_{k=2} = 1$, compatible with the observation that a $B$-twisted hypermultiplet has a single line operator (the trivial line) compatible with a generic, flat background $SU(2)$ connection. More generally, we find an exact match of characters: for $k = 2$, and arbitrary $y$ we have
\be
	\chi_{\CT_{SO(3),2}^A}\big(\CH(\Sigma_g)\big) = \big((1+\wt{y})(1+\wt{y}^{-1})\big)^{g-1}\,,
\ee
where $\wt{y} = \sqrt{y}$ is the fugacity for a maximal torus of the topological $SU(2)$ flavor group, \cf\, Eq. \eqref{P-hyper}.

The proposal is also supported from the perspective of boundary VOA's. We will argue in Section \ref{sec:VOA} that $\CT_{SU(2),2}^A$ supports the triplet VOA $\FT_{2}(\mathfrak{sl}_2)$ on its boundary, with its category of line operators equivalent to the derived category of $\FT_{2}(\mathfrak{sl}_2)$-mod. In turn, gauging the bulk $\Z_2$ one-form symmetry should induce a simple current extension of $\FT_{2}(\mathfrak{sl}_2)$, which is well known \cite{Kausch:2000fu} to be isomorphic to symplectic fermions, \emph{a.k.a.} the boundary VOA of $\CT_{\rm hyper}^B$.

Finally, the effect of gauging the $\Z_2$ one-form symmetry on the category $u_q(\mathfrak{sl}_2)$-mod is discussed in Appendix \ref{app:U}. (This is not really independent of the preceding comments about VOA's, given the equivalance $\FT_{2}(\mathfrak{sl}_2)\text{-mod}\simeq u_q(\mathfrak{sl}_2)$-mod.) Gauging the $\Z_2$ symmetry has the effect of quotienting derived endomorphism algebras of simple (respectively, projective) modules of $u_i(\mathfrak{sl}_2)$, in such a way that they reduce to the endomorphism algebras $\C[X,Y]$ (respectively, $\C[\xi,\psi]$) found on $\CT_{\rm hyper}^B$ line operators in Section \ref{sec:hyper-lines}.

\subsection{Bethe vacua for higher rank and the $\Z_n$ anomaly}
\label{sec:SUnBethe}

We now generalize to arbitrary rank (any $n$). We will just describe a few properties of Bethe vacua, first for theories $\CT_{n,k} = T[SU(n)]/SU(n)_k$, and then for other global forms $T[SU(n)]/G_k$. This is sufficient for producing the characters of $T^2$ state spaces. We hope to discuss characters in higher genus and Grothendieck rings in future work.

The Bethe vacua become increasingly complicated for $n > 2$. Nonetheless, since we are only interested in the limit $t \to -1$, we can work in a formal neighborhood of $t = -1$, \ie\, with formal series in $t+1$, to significantly simplify the analysis. We will also split the computation into two steps, first considering the theory $T[SU(n)]$, and then gauging its $SU(n)_k$ (or $G_k$) flavor symmetry.

For $T[SU(n)]^{UV}$, defined by the quiver \eqref{TSUn-UV}, we use fugacities  $x_{a,j}$ ($j=1,...,a$) for the maximal torus of each gauge group $U(a)$ ($a=1,...,n-1$); fugacities $z_j$ for the ordinary $SU(n)$ flavor symmetry; and fugacities $y_a$ for the maximal torus of the topological flavor symmetry.
The Bethe equations for $T[SU(n)]^{UV}$ are given by
\be
	P_{a,j}(x, z) = 1 \qquad a = 1, ..., n-1\,,\qquad  j = 1, ..., a\,,
\ee
with rational functions
\be
\label{eq:TSUnBethe}
\begin{aligned}
	P_{a,j}(x, z) & := y_a \Bigg(\prod_{j_-=1}^{n-1}\frac{x_{a,j} - t x_{a-1,j_-}}{t x_{a,j} - x_{a-1,j_-}}\Bigg)\Bigg(\prod_{j' \neq j}\frac{x_{a,j} - t^2 x_{a, j'}}{t^2 x_{a,j} - x_{a, j'}}\Bigg)\Bigg(\prod_{j_+=1}^{n+1}\frac{x_{a,j} - t x_{a+1, j_+}}{t x_{a,j} - x_{a+1, j_+}}\Bigg)
\end{aligned}\,
\ee
where $x_{n, j} := z_{j} z_{j-1}^{-1}$ for $j = 1, ..., n$, with the convention that $z_0 = z_n = 1$.

If we work in a formal neighborhood of $t = -1$, the Bethe vacua correspond to choices $x_{a,j} = - x_{a+1, \iota_a(j)} + O(t+1)$ such that $x_{a,j} \neq x_{a,j'}$ unless $j = j'$, \ie\, we require $\iota_a: \{1,..., a\} \to \{1, ..., a+1\}$ is an injection. We can fix the action of the full Weyl group by choosing $\iota_a(j) = j$, and thereby identify the $n!$ non-degenerate Bethe vacua as the $S_n$ orbit of the solution
\be
	x_{a,j} = - x_{a+1,j} + O(t+1) \quad \Rightarrow \quad x_{a,j} = (-1)^{n-a} z_j z_{j-1}^{-1} + O(t+1)\,,
\ee
where $S_n$ acts on the $z_{j}$ or, equivalently, the $x_{n,j}$. The equations for the higher order terms in the $x_{a,j}$ are linear and admit a unique solution.

The additional Bethe equations introduced by gauging with $SU(n)_k$ are
\be
\label{eq:SUNkBethe}
	P_j(x,z) := \frac{z_j^{2k}}{z_{j - 1}^k z_{j + 1}^k} \prod_{j_-=1}^{n-1}\frac{(z_j z_{j-1}^{-1} - t x_{n-1, j_-})(t z_{j+1} z_{j}^{-1} - x_{n-1, j_-})}{(t z_j z_{j-1}^{-1} - x_{n-1, j_-})(z_{j+1} z_{j}^{-1} - t x_{n-1, j_-})} = 1\,,
\ee
which can again be analyzed in a formal neighborhood of $t = -1$. The Weyl group of $SU(n)$ acts by permuting the $z_j$, thus it suffices to consider the non-degenerate roots to these equations such that $x$ is specialized to a single $T[SU(n)]$ vacuum, \eg\,, the $T[SU(n)]$ vacuum corresponding the fundamental solution discussed above. To lowest order in $t+1$ (which requires knowledge of first correction in $t+1$ to the vacua for $T[SU(n)]$), we find that the Bethe vacua correspond to
\be
	z_j\big[\vec{\lambda}\big] = e^{\frac{2 \pi i\lambda_j}{nk}}\prod_{a = 1}^{j} y_a^{\frac{a(j-n)}{nk}} \prod_{b = j+1}^{n-1} y_b^{\frac{j(b-n)}{nk}} + O(t+1)\,,
\ee
where $\lambda_j \in \Z_{nk}$ satisfy the equations
\be
	2 \lambda_j - \lambda_{j-1} - \lambda_{j+1} \equiv 0 \mod n
\ee
with $\lambda_0 \equiv \lambda_n \equiv 0 \mod nk$. For fixed $\lambda_{n-1}$, the remaining $\lambda_j$ are determined $\! \! \mod n$ by these equations and thus there are $(nk) k^{n-2} = n k^{n-1}$ Bethe vacua. Thus,
\be \label{eq:genericindextypeA}
	\chi\big[\CH(T^2)\big] = n k^{n-1}\,.
\ee

We expect the number $nk^{n-1}$ to agree with the number of distinct simple objects in a generic stalk of the category $U_q(\sln)$-mod at $q = e^{\frac{i \pi}{k}}$. We will also find in Section \ref{sec:FTtwisted} that it coincides with a conjectural computation of the number of twisted modules for the Feigin-Tipunin algebra $\FT_k(\sln)$, at generic twist.

We additionally expect $\CT_{n,k}$ to have a $\Z_n$ one-form symmetry. Its potential 't Hooft anomaly may be computed by following the same procedure described in Section \ref{sec:Z2-anomaly}. Let $\omega$ be a generator of the 1-form symmetry with the property that the corresponding A-cycle operator $\omega_A$ acts on $SU(n)$ fugacities as $x_{n,j}\mapsto e^{\frac{2\pi i}{n}} x_{n,j}$ (and thus $z_j \to e^{\frac{2 \pi i j}{n}} z_j$), as appropriate for a central $SU(n)$ element with diagonal entries $e^{\frac{2\pi i}{n}}$. This must be accompanied by a compensating transformation $x_{a,j}\mapsto e^{\frac{2\pi i}{n}}x_{a,j}$ of each $U(a)$ fugacity. Therefore, $\omega_A$ acts on Bethe vacua labelled by $\vec\lambda$ as
\be \omega_A|\vec\lambda\rangle = |\vec{\lambda} + k(1,..., n-1)\rangle\,. \ee
The ``conjugate momentum'' with respect to this gauge transformation, 
 \ie\, the generalization of $P^{1/2}_Z$, is given by
\be
	P_{\rm center}^{1/n} = z_{n-1}^{k} \bigg(\prod_{a=1}^{n-1} y_a^{\frac{a}{N}}\bigg) \bigg(\prod \limits_{j_-=1}^{n-1} \frac{t z_{n-1}^{-1}-x_{n-1,j_-}}{z_{n-1}^{-1}-t x_{n-1,j_-}}\bigg)\,,
\ee
and evaluates to $P_{\rm center}^{1/n}[\vec{\lambda}] = e^{\frac{2 \pi i\lambda_{n-1}}{n}}$
on the Bethe vacuum labeled by $\vec{\lambda}$. Therefore, the dual B-cycle operator acts as
\be \omega_B |\vec\lambda\rangle = e^{\frac{2 \pi i \lambda_{n-1}}{n}} \ket{\vec{\lambda}} \ee
The anomaly is captured by the commutation relation of these generators,
\be \omega_A\omega_B=e^{\frac{2\pi i k}{n}} \omega_B\omega_A\qquad\Rightarrow\qquad \alpha_{\omega} = e^{\frac{2\pi i k}{n}}\,. \label{anom-n} \ee

Note that this beautifully matches the VOA result \eqref{braid-n} (up to a minus sign that may be absorbed in the orientation of lines on $T^2$). It also matches the generalization of the Cartan part of the R-matrix $q^{ (H\otimes H)}$ (involving the Cartan pairing) to unrolled $U_q^H(\sln)$.

The upshot of \eqref{anom-n} is that,  in general, only  a $\Z_{\gcd(k,n)}$ subgroup of the $\Z_n$ 1-form symmetry is non-anomalous.  This should allow us to define the theory associated to gauging different global forms $SU(n)/\Z_m$, where $m$ divides $\gcd(k,n)$. The Bethe vacua that survive gauging the $\Z_m$ 1-form symmetry are those with eigenvalue $1$ under both $(\omega_A)^{\frac{n}{m}}$ and $(\omega_B)^{\frac{n}{m}}$. In particular, they are labeled by $\vec{\lambda}$ with $\lambda_{n-1} \equiv 0 \mod m$ modulo shifts by $\tfrac{k n}{m}(1,..., n-1)$, resulting in a total of $\frac{n k^{n-1}}{m^2}$ vacua.

\subsection{Full state spaces}
\label{sec:T-Hilb}

Finally, we would like to describe some of the structure of the actual state spaces (as opposed to their characters) and the full dg category of line operators (as opposed to its Grothendieck ring) in theories $\CT_{G,k}^A$. We will only begin the analysis here; this section and the next are somewhat speculative, and we hope they will lead to interesting future work. We can work with general group $G$ for the moment.

We expect the derived state spaces of $\CT_{G,k}^A$ to have a geometric description that generalizes classic work of \cite{Witten-GQ} on geometric quantization in Chern-Simons theory. Recall that in ordinary Chern-Simons theory with group $G$ at level $k-h^\vee$, the state space on a Riemann surface $\Sigma$ (a smooth surface endowed with an algebraic structure) may be described as sheaf cohomology
\be \CH^{CS}_{G,k}(\Sigma) = H^\bullet(\text{Bun}_{G_\C}(\Sigma), \CL^{\otimes k} )\,, \label{CS-Hilb} \ee
where $\text{Bun}_{G_\C}(\Sigma_g)$ is the moduli space of algebraic (\emph{a.k.a.} holomorphic) $G_\C$-bundles on $\Sigma$ and $\CL$ is a line bundle whose first Chern class generates $H^2(\text{Bun}_{G_\C},\Z)$.

A geometric-quantization-like approach for constructing state spaces of general 3d $\CN=2$ theories in the HT twist was developed in \cite{BF-Hilb}, and extended to twists of 3d $\CN=4$ theories in \cite{BFK-Hilb}. Some abstract properties of state spaces in the topological twists of 3d $\CN=4$ theories were derived in \cite{Gaiotto-blocks}. We can use these results to give several --- still somewhat abstract --- descriptions of the state spaces of $\CT_{G,k}^A$.

Let us use the definition $\CT_{G,k} = T[G]/G_k$ and work in steps, beginning with the state space of $T[G]$. The theory $T[G]$ has $G\times G^\vee$ symmetry. In the 3d topological A-twist, its state space on a Riemann surface $\Sigma$ depends on the choice of 1) an algebraic $G_\C$ bundle on $\Sigma$, which may be encoded in the choice of an algebraic connection $\CA'$; and 2) a flat $G_\C^\vee$ bundle (\emph{a.k.a.} local system) on $\Sigma$, encoded in the choice of a flat connection $\CA$. Globally, these state spaces $\CH_{T[G]^A}(\Sigma;\CA',\CA)$ become the stalks of a sheaf 
\be \CE_{T[G]^A}(\Sigma) \to \text{Bun}_{G_\C}(\Sigma)\times \text{Loc}_{G_\C^\vee}(\Sigma) \label{TG-sheaf} \ee
argued in \cite{Gaiotto-blocks} to be a flat sheaf (local system) on the first factor and a coherent sheaf on the second. Thus, $\CH_{T[G]^A}(\Sigma;\CA',\CA)$ is the stalk $\CE_{T[G]^A}(\Sigma)\big|_{\CA',\CA}$ at $\CA'\in \text{Bun}_{G_\C}(\Sigma)$ and $\CA\in \text{Loc}_{G_\C^\vee}(\Sigma)$.

For $G=SU(n)$ and abelian $\CA$ the stalks of the sheaf \eqref{TG-sheaf} can in principle be constructed by applying the methods of \cite{BF-Hilb,BFK-Hilb} to the UV Lagrangian gauge theory $\CT[SU(n)]^{UV}$. The construction is explicit but not yet practical for computations.

Further gauging the $G$ symmetry of $T[G]$ with an $\CN=2$ vectormultiplet at Chern-Simons level $k$ translates to taking derived global sections of the sheaf \eqref{TG-sheaf}, tensored with $\CL^k$, over $\text{Bun}_{G_\C}(\Sigma)$ \cite{BF-Hilb}. For each choice of flat $G_\C^\vee$ connection $\CA$, the state space of $\CT_{G,k}^A$ then takes the form
\be \CH(\Sigma;\CA)  = H^\bullet(\text{Bun}_{G_\C}(\Sigma), \CE_{T[G]^A}(\Sigma)\big|_\CA \otimes \CL^k)\,.  \label{Hilb-TGk} \ee
Note that once we fix $\CA$, we may interpret $\CE_{T[G]^A}(\Sigma)\big|_\CA$ as a sheaf over $\text{Bun}_{G_\C}(\Sigma)$ alone.

We expect that \eqref{Hilb-TGk} may be computed by a spectral sequence, whose first page is $H^\bullet(\text{Bun}_{G_\C}(\Sigma), \CL^k)\otimes \CE_{T[G]^A}(\Sigma)\big|_{\CA'=*,\CA}$. Here $\CE_{T[G]^A}(\Sigma)\big|_{\CA'=*,\CA}$ denotes the stalk of the local system $\CE_{T[G]^A}(\Sigma)\big|_{\CA}$ over any point $\CA'\in \text{Bun}_{G_\C}(\Sigma)$. This leads to an approximation
\be \CH(\Sigma;\CA) \approx \CH^{\rm CS}_{G,k}(\Sigma) \otimes \CH_{T[G]^A}(\Sigma;\CA'=*,\CA)\,, \label{Hilb-approx} \ee
relating state spaces of $\CT_{G,k}^A$ to state spaces in Chern-Simons theory and in A-twisted $T[G]$. Of course, there may be further differentials in the spectral sequence, correcting \eqref{Hilb-approx}. Nevertheless, the approximation turns out to be remarkably useful and accurate, as we now explore.

\subsubsection{Approximate state spaces at $\CA=0$}

The stalks of $\CE_{T[G]^A}(\Sigma)$ at trivial $\CA=0$ (and any $\CA'$) will be infinite-dimensional. We expect them to be dg vector spaces with non-negative, unbounded cohomological degrees and finite cohomology in each degree --- just like the state spaces of $\CT_{G,k}^A$. More so, we can introduce a real mass deformation to resolve the Coulomb branch $\CM_{\rm Coul}$ of $T[G]$, which should not affect its A-twisted state spaces. Then 3d mirror symmetry predicts that the state spaces of $T[G]^A$ will be equivalent to state spaces of a B-twisted sigma model (\emph{a.k.a.} Rozansky-Witten theory) whose target is $\CM_{\rm Coul}$. The resolved Coulomb branch is a cotangent bundle
\be \CM_{\rm Coul} = T^*[2]\CF^\vee\,,\qquad \CF^\vee = G_\C^\vee/B^\vee\,,\ee
where $\CF^\vee$ is the Langlands-dual flag manifold, and we have introduced a degree shift to correctly account for cohomological degree ( = $U(1)_C$ charge). Then \cite{RW, KRS} imply that in genus $g$,
\be \CE_{T[G]^A}(\Sigma_g)\big|_{\CA'=*,\CA=0} \simeq H^\bullet(\CM_{\rm Coul},\Lambda^\bullet (T^*)^{\oplus g})\,, \ee
where $H^\bullet$ denotes sheaf cohomology and $T^*$ is the holomorphic cotangent bundle. (All degrees add to give the total cohomological degree on the LHS.) Special cases include
\be \begin{array}{l} \CE_{T[G]^A}(S^2) \simeq \C[T^*[2]\CF^\vee] \qquad \text{(ring of algebraic functions)} \\[.2cm]
 \CE_{T[G]^A}(T^2) \simeq H^{\bullet,\bullet}_{\bar\pd}(T^*[2]\CF^\vee)\qquad \text{(total algebraic Dolbeault cohomology)} \end{array} \ee
These are both infinite-dimensional. (They were explicitly described in Sections \ref{sec:U-0}--\ref{sec:U-1} for $G=SU(2)$, where $\CF^\vee=\mathbb P^1$.)

Now consider the approximation \eqref{Hilb-approx} for state spaces of $\CT_{G,k}^A$. The Chern-Simons state space in genus zero is always one-dimensional, so the approximation simply takes the form
\be \CH(S^2) \approx  \C[T^*[2]\CF^\vee]\,. \label{approx-g0} \ee
For $G=SU(2)$, the character computed in \eqref{char-g0} agrees perfectly with the character of the ring of functions on $T^*[2]\mathbb P^1$. This suggests that the approximation \eqref{approx-g0} is actually \emph{exact}.

Note that having an exact equality in \eqref{approx-g0} is also consistent with our conjectured equivalence of line operators in $\CT_{G,k}^A$ with the category of modules for the reduced quantum group $u_q(\g)$ at $q=e^{i\pi/k}$, for any $G=SU(n)$. The results of \cite{ABG,BL} identify the principal block of $D^b\,u_q(\g)$-mod with coherent sheaves on the flag variety $T^*[2]\CF$, which is isomorphic to  $T^*[2]\CF^\vee$ when $G=SU(n)$. Then a generalization of the computation of endomorphisms of the trivial line from Section \ref{sec:U-0} leads directly to the RHS of \eqref{approx-g0}.

In genus one, the Chern-Simons state space has a basis corresponding to elements of the quotient lattice
\be \frac{P}{W\ltimes (k-h^\vee) Q^\vee}\,, \label{CS-lattice} \ee
where $P$ is the weight lattice of $G$, $Q^\vee$ is the coroot lattice, and $W$ is the Weyl group. Let $d_{G,k}=\Big| \frac{P}{W\ltimes (k-h^\vee) Q^\vee}\Big|$ be its dimension. Then the approximation \eqref{Hilb-approx} predicts
\be \CH(T^2)\approx \C^{d_{G,k}} \otimes H^{\bullet,\bullet}_{\bar\pd}(T^*[2]\CF^\vee)\,. \label{approx-g1} \ee
For $G=SU(2)$, we have $d_{G,k} = k-1$ and $H^{\bullet,\bullet}_{\bar\pd}(T^*[2]\CF^\vee)$ given by \eqref{dol-P1}. The prediction \eqref{approx-g1} differs from the exact quantum-group calculation \eqref{Hilb-T2} by a single $\C^2$ summand.

\subsubsection{Approximate state spaces at generic abelian $\CA$}

Having generic abelian (diagonal) $\CA$ corresponds to introducing generic complex FI parameters that should localize the theory to fixed points of the torus $T^\vee\subset G^\vee$ acting on the Coulomb branch. Some aspects of this localization were discussed in \cite{RW-Coulomb}. Since the number of fixed points is the order of the Weyl group $|W|$, we would expect the stalks for generic $\CA$ to be $|W|$-dimensional, supported in cohomological degree zero,
\be \CE_{T[G]^A}(\Sigma_g)\big|_{\CA'=*,\CA\,\text{generic}} \simeq \C^{|W|}\,. \ee

Now the approximation \eqref{Hilb-approx} implies
\be \CH(\Sigma;\CA_{\rm generic}) \approx \CH_{G,k}^{\rm CS}(\Sigma) \otimes  \C^{|W|}\,.\ee
In genus one, one might expect that tensoring with $\C^{|W|}$ would undo the Weyl-group quotient in \eqref{CS-lattice}, leaving behind a space of dimension
\be \text{dim}\, \CH(T^2;\CA_{\rm generic}) \approx \big| P/(k-h^\vee)Q^\vee\big|\,.\ee
For $G=SU(n)$, the RHS is $n(k-n)^{n-1}$, in agreement with \eqref{eq:genericindextypeA} at large $k$.

\subsection{Category of line operators}
\label{sec:T-cat}

In Section \ref{sec:VOA}, we will access the category of line operators in $\CT_{n,k}^A$ by relating it to module categories for boundary VOA's. However, the category should also have an intrinsic description that depends only on the bulk field content of $\CT_{n,k}^A$ --- analogous to the analysis of Section \ref{sec:hyper-lines} for the B-twist of a free hypermultiplet.
Recent developments in the structure of twists of 3d $\CN=2$ and $\CN=4$ theories lead to a prediction that we outline here, mainly for theoretical interest, and as a starting point for further investigations.
Further work is needed to make the prediction mathematically and physically precise, and to do meaningful computations.

\subsubsection{Line operators in A-twisted gauge theory}
\label{sec:support}

We first recall that in the A-twist of a standard 3d $\CN=4$ gauge theory with matter $T^*V$ and gauge group $G$, the category of line operators is a version of
\be \CC_{G,V}^A = \text{D-mod}\big(V(\!(z)\!)/G(\!(z)\!)\big)\,, \label{D-modVG} \ee
the dg category of D-modules on the dg ind-scheme $V(\!(z)\!)/G(\!(z)\!)$, where $V(\!(z))\!)$ denotes the algebraic loop space of $V$  (\ie\ the space of $V$-valued Laurent series) and $G(\!(z)\!)$ denotes the algebraic loop group (\ie\ the complexified group $G_\C$ defined over Laurent series). This category was first proposed in unpublished work of J. Hilburn and P. Yoo, and has been discussed and explored from various perspectives, including \cite{BF-lines,CostelloGaiotto,Webster-tilting,DGGH,HilburnRaskin}.

The category \eqref{D-modVG} can be described a bit more explicitly. We'll approach it in several steps. D-modules on a vector space~$V$, denoted $\text{D-mod}(V)$, are (by definition) modules for the algebra of differential operators on $V$, \emph{a.k.a.} modules for a deformation quantization of functions on $T^*V$. Let $x^i$ be coordinates on $V$ and $y_i$ dual coordinates on the cotangent fiber. Their quantization leads to an algebra $\C[x,y]$ with $[x^i,y_j]=\delta^i{}_j$, whence
\be  \text{D-mod}(V) := \C[x,y]\text{-mod}\,. \ee

Next, suppose that $V$ has an algebraic action of a complex Lie algebra $\g$. Then D-modules on the corresponding stack $\text{D-mod}(V/\g)$, also  known as strongly $\g$-equivariant D-modules on $V$,  are modules for the dg algebra generated by $x,y$ along with fermionic generators $c\in \mathfrak g$, $b\in \mathfrak g^*$, with
\be \label{V/g} \begin{array}{r@{\qquad}l}
\text{degrees:} & |x|=0\,,\quad |y|=2\,,\qquad |c|=|b|=1\,, \\
\text{commutators:} &  [x^i,y_j]=\delta^i{}_j\,,\qquad [b_a,c^{a'}] = \delta_a{}^{a'}\quad \text{(rest trivial)} \\
\text{differential:} & Q\,x = c\cdot x\,,\quad Q\,y = c\cdot y\,,\quad Q\,c= \tfrac12[c,c]\,,\quad  Q\,b = \mu(x,y)+c\cdot b\,,
\end{array}
\ee 
where in the last line `$\cdot$' denotes the action of $\g$ in the appropriate representation, and $\mu(x,y)$ is the normal-ordered moment-map operator. (Note that the $xy$ commutator is an ordinary commutator, while the $bc$ commutator is an anti-commutator, as these generators are fermionic.) Let's simply denote the algebra \eqref{V/g}  as $\C[x,y,b,c]$. Note that this is the algebra that computes BRST cohomology of $\C[x,y]$, \emph{a.k.a.} functions on the derived quantum symplectic quotient of $T^*V$.
 Then
\be \text{D-mod}(V/\g) \simeq \C[x,y,b,c]\text{-mod}\,. \ee

Now consider the loop space $V(\!(z)\!)$. Its coordinates are the modes $x^i_n$ of Laurent series $x(z) = \sum_{n\in \Z} x^i_n  z^{-n}$, while coordinates on the cotangent fibers are the modes $y_{i,n}$ of Laurent series $y(z) = \sum_{n\in \Z} y_{i,n} z^{-n-1}$. They generate an algebra $\C[x(z),y(z)]$ with commutation relations $[x^i_n,y_{j,m}] = \delta^i{}_j\delta_{n+m,0}$. Similarly, for the loop algebra we introduce $c(z) = \sum_{n\in \Z} c_n z^{-n-1} \in \g(\!(z)\!)$, $b(z) =  \sum_{n\in \Z} b_n z^{-n}\in \g^*(\!(z)\!)$, with anti-commutators $[b_{a,n},c^{a',m}] = \delta_a{}^{a'}\delta_{m+n,0}$. Altogether, the modes generate the BRST algebra $\C[x(z),y(z),c(z),b(z)]$ with
\be \label{Vz/g} \begin{array}{r@{\quad}l}
\text{degrees:} & |x(z)|=0\,,\quad |y(z)|=2\,,\qquad |c(z)|=|b(z)|=1\,, \\
\text{commutators:} &  [x^i(z),y_j(w)]=\delta(z-w)\,,\qquad [b_a(z),c^{a'}(w)] = \delta_a{}^{a'}\delta(z-w)\\
\text{differential:} & Q\,x(z) = c(z)\cdot x(z)\,,\quad Q\,y(z) = c(z)\cdot y(z)\,,\\ &\hspace{.7in} Q\,c(z)= \tfrac12[c(z),c(z)]\,,\quad  Q\,b(z) = \mu(x,y)(z)+c(z)\cdot b(z)\,.
\end{array}
\ee 
Then
\be \text{D-mod}\big(V(\!(z)\!)/\g(\!(z)\!)\big) \simeq \C[x(z),y(z),b(z),c(z)]\text{-mod}\,. \label{D-modVg} \ee

Finally, there is the category $\text{D-mod}\big(V(\!(z)\!)/G_\C(\!(z)\!)\big)$ that actually appears in \eqref{D-modVG}. It differs from \eqref{D-modVg} in subtle ways, related to the fact that, when $G_\C$ is reductive, derived $G_\C(\!(z)\!)$ invariants and derived $\mathfrak g(\!(z)\!)$ invariants are not quite the same. This difference was discussed in \cite{CostelloGaiotto, CDG}, but is beyond our level of sophistication in the current paper. 

Both categories \eqref{D-modVg}, \eqref{D-modVG} also still require further restrictions on their objects in order for morphisms and their compositions, and (optimistically) tensor products and braiding, to be well defined. This is directly due to the infinite-dimensionality of loop spaces and loop groups. In particular, subtle choices must be made for the allowed supports of objects, and these choices are physically meaningful. Some options are carefully considered in \cite{HilburnRaskin}, and further generalized in \cite{BN-betagamma}. 

For example, the trivial/identity line operator $\id$ corresponds to the module generated by a vector on which all negative modes $\{x_{n+1},y_n,c_n,b_{n+1}\}_{n\geq 0}$ act as zero. One may consider a category containing all modules whose support is finitely far away from $\id$, in the sense that all  $\{x_n,y_n,c_n,b_n\}_{n\geq N}$ act as zero for sufficiently large $N$. This seems to contain line operators in the topological QFT that descend from physical line operators in the untwisted QFT.  These are generally vortex lines: the Laurent series $x(z),y(z)$ represent meromorphic profiles of hypermultiplet matter fields in the neighborhood of a line operator, and setting various modes to be zero described the profile of a vortex. (See \cite{DGGH} for extended discussion.)

Alternatively, one may also extend to a larger category such that (say) sufficiently large $\{x_n,c_n,b_n\}_{n\geq N}$ act as zero, or $\{y_n,c_n,b_n\}_{n\geq N}$ act as zero. These categories contain some line operators in the topological QFT that descend from boundary conditions wrapped on a circle in the untwisted QFT; however, they do not appear to be braided. To the best of our knowledge, neither the full set of consistent choices nor their physical interpretations have yet been carefully studied.

These choices/subtleties are the reason we said that above the category of line operators is a \emph{version} of \eqref{D-modVG}. We will not elaborate further here.

\subsubsection{Chern-Simons gauging}

We now return to the theory $\CT_{n,k}^A$, obtained by A-twisting $T[SU(n)]^{UV}/SU(n)_k$ (following an HT twist, as described in Section \ref{sec:Tnk-BV}). The quiver gauge theory $T[SU(n)]^{UV}$ is an ordinary 3d $\CN=4$ gauge theory, with group $\Gamma = \prod_{a=1}^{n-1} U(a)$ and representation $V=\prod_{a=1}^{n-1} \text{Hom}(\C^a,\C^{a+1})$. Thus, in the A twist, its category of line operators is a version of
\be \CC_{\Gamma,V}^A = \text{D-mod}\big( V(\!(z)\!)/\Gamma(\!(z)\!)\big)\,, \ee
\emph{a.k.a.} strongly $\Gamma(\!(z)\!)$-equivariant D-modules on $V(\!(z)\!)$. We expect that gauging the additional $SU(n)$ symmetry at Chern-Simons level $k$ will modify this to what are known as \emph{weakly} equivariant D-modules for the centrally extended algebraic loop group $SL_n(\!(z)\!)_{k-n}$, denoted
\be \CC^{(n,k)}  \overset{?}= \text{D-mod}_{SL_n(\!(z)\!)_{k-n}}\big( V(\!(z)\!)/\Gamma(\!(z)\!)\big)\,. \label{weakD} \ee

Let us explain this briefly. Neglecting the difference between Lie groups and Lie algebras, we saw above that  $\text{D-mod}\big( V(\!(z)\!)/\Gamma(\!(z)\!)\big)$ is equivalent to modules for a dg algebra $\C[x(z),y(z),c(z),b(z)]$ as in \eqref{Vz/g} (for the appropriate group $\Gamma$ and representation $V$). This dg algebra has an action of the affine algebra $\mathfrak{sl}_n(\!(z)\!)_\kappa$ 
at any level $\kappa$.%
\footnote{However, it is not an internal action: it is not generated by a current formed from $x,y,c,b$ themselves.} %
If $\kappa$ is an integer, the action integrates to one of the loop group $SL_n(\!(z)\!)_\kappa$.
Then \eqref{weakD} is the category of $SL_n(\!(z)\!)_\kappa$-equivariant modules for the algebra $\C[x(z),y(z),c(z),b(z)]$.

Note that the difference between an $\CN=4$ and $\CN=2$ gauging amounts to strong \emph{vs.} weak equivariance for D-modules. Alternatively, we have a symplectic (BRST) quotient vs. an ordinary quotient at the level of stacks.

We remark that if the stack $V(\!(z)\!)/\Gamma(\!(z)\!)$ in \eqref{weakD} is replaced by a point, we recover the standard category of line operators in Chern-Simons theory, namely,
\begin{align} \text{D-mod}_{SL_n(\!(z)\!)_{k-n}} (\text{pt})  &=  SL_n(\!(z)\!)_{k-n}\text{-mod}  \\
 &= \text{line ops in $SU(n)$ CS at level $k-n$}\,. \notag \end{align}
Including the stack $V(\!(z)\!)/\Gamma(\!(z)\!)$ accounts for coupling Chern-Simons theory to $T[SU(n)]$. The approximate factorization \eqref{Hilb-approx} of state spaces is reflected at the level of categories in an approximate factorization
\be \text{D-mod}_{SL_n(\!(z)\!)_{k-n}}\big( V(\!(z)\!)/\Gamma(\!(z)\!)\big) \overset{?}{\approx}  \big[\text{D-mod}_{SL_n(\!(z)\!)_{k-n}} (\text{pt})\big] \boxtimes \big[\text{D-mod}\big( V(\!(z)\!)/\Gamma(\!(z)\!)\big)\big]\,. \ee

It should be extremely interesting to explore the category \eqref{weakD} and computations within it further. The subtle choices of support discussed in Section \ref{sec:support} must be carefully specified for \eqref{weakD} as well.

\subsubsection{Comparison to boundary VOA}

The algebras appearing above should be highly reminiscent of mode algebras of VOA's.

For example, the algebra $\C[x(z),y(z)]$ whose modules are D-modules on $V(\!(z)\!)$ is just the mode algebra of a beta-gamma system valued in $V$, in obvious way: $x,y$ are simply identified with the usual $\beta,\gamma$ fields. This is no coincidence. The beta-gamma VOA is the boundary VOA of a boundary condition for free 3d $\CN=4$ hypermultiplets that is compatible with the A-twist \cite{CostelloGaiotto}. In this case, the category $\text{D-mod}\big(V(\!(z)\!)\big)$ and the category of boundary-VOA modules are identical, up to subtle issues of support.

Going further, the algebra $\C[x(z),y(z),c(z),b(z)]$ is reminiscent of a BRST reduction of a beta-gamma VOA valued in $V$. Here, however, there is an important difference: the differential on $\C[x(z),y(z),c(z),b(z)]$ does \emph{not} come from commutation with an operator $\oint b(z)J(z)$ built from an internal current $J(z)$, as would be the case in BRST reduction of VOA's, \emph{cf.} \cite[Sec. 4]{Arakawa-W}. The problem is that the putative current does not have the correct level to make a VOA-style reduction possible.

Similarly, it seems that the category \eqref{weakD} might be realized as modules for an $\mathfrak{sl}_n(\!(z)\!)_{k-n}$ coset of the algebra $\C[x(z),y(z),c(z),b(z)]$. Here again the problem is that the $\mathfrak{sl}_n(\!(z)\!)_{k-n}$ action on $\C[x(z),y(z),c(z),b(z)]$ cannot be generated by an internal current at level $k-n$, for arbitrary $k$.

These differences/discrepancies can all be fixed by tensoring the algebra $\C[x(z),y(z),c(z),b(z)]$ to free-fermion VOA's, in the correct representations to allow the existence of (1) an internal $U(1)\times\cdots U(n-1)$ current at the right level to define a nilpotent BRST operator $\oint b(z)J(z)$; and (2) an internal $\mathfrak{sl}_n(\!(z)\!)_{k-n}$ Kac-Moody symmetry, whose coset can be taken. This is precisely what happens in the correct construction of the anomaly-free boundary condition of Section \eqref{sec:VOAnk} that leads to the boundary VOA $\CN_{n,k}$. Since free fermions have a trivial representation category, we would expect the category of $\CN_{n,k}$ modules to be equivalent to \eqref{weakD} --- modulo a proper mathematical definition of \eqref{weakD} and the resolution of the usual issues of support.

\section{Vertex operator algebras}
\label{sec:VOA}

In this section, we specialize to $G=SU(n)$, and study two VOA's supported on boundary conditions for the 3d theory $\CT_{n,k}^A$. One of these VOA's was constructed explicitly in Section~\ref{sec:VOAnk}, using the Lagrangian definition of $\CT_{n,k}^A$, in the twisted BV formalism. It was denoted $\CN_{n,k}$, corresponding to the fact that it is supported on a \underline{N}eumann-like boundary condition. The other VOA, which we denote $\CD_{n,k}$, is supported on a \underline{D}irichlet-like boundary condition. This second condition is implicitly defined by the brane configuration in \eqref{brane-D}. We won't construct it explicitly in field theory in this paper.

We begin by arguing in Section \ref{sec:D-FT} that $\CD_{n,k}$ is equivalent to the Feigin-Tipunin algebra $\FT_k(\mathfrak{sl}_n)$. We do this in part by using the corner-VOA manipulations/identities of \cite{GaiottoRapcak,CreutzigGaiotto-S}. These corner-VOA methods are best understood when 4d Langlands-twist parameters are generic, leading not exactly to the Feigin-Tipunin algebra, but to a deformation thereof: a deformable family of VOA's defined over the field $\C(\psi)$. We explain how the large-$\psi$ limit is expected to reproduce the Feigin-Tipunin VOA, using a decomposition theorem recently proven by Sugimoto \cite{sugimoto2021feigintipunin, sugimoto2021simplicities}.

The results of Section \ref{sec:D-FT} complete a physics proof of Theorem \ref{th:1} from the introduction, by combining \vspace{-.2cm}
\begin{enumerate}[leftmargin=.5cm]
\setlength{\itemsep}{-.1cm}
\item The bulk-boundary functor $\CC_{g=1}^{n,k}\to  D^b\big(\CD_{n,k}\text{-mod}\big)$, assumed to be an equivalence relating the category of line operators in $\CT_{n,k}^A$ and modules for the boundary VOA $\CD_{n,k}$, as discussed in Section \ref{sec:bdy-VOA}.
\item The identification $\CD_{n,k}\simeq \FT_k(\mathfrak{sl}_n)$ of Section \ref{sec:D-FT}.
\item The logarithmic Kazhdan-Lusztig correspondence of \cite{Creutzig:2021cpu,GannonNegron} (establishing an equivalence of abelian braided tensor categories 
 $\FT_k(\mathfrak{sl}_2)\text{-mod}\simeq u_q(\mathfrak{sl}_2)\text{-mod}$, with monoidal structure on the quantum-group side given by \cite{Gainutdinov:2015lja,Creutzig:2017khq, Creutzig:2020jxj, Gainutdinov:2018pni}).
\end{enumerate}

In Section \ref{sec:FT}, we then discuss some representation theory of Feigin-Tipunin algebras, and in particular the triplet algebras $\FT_k(\mathfrak{sl}_2)$. We review the well-known correspondence between triplet modules and $u_q(\mathfrak{sl}_2)$ modules, and the construction of (underived) state spaces/conformal blocks. We also propose a somewhat experimental procedure for computing characters of spaces of derived conformal blocks. We then consider deformations of $\FT_k(\mathfrak{sl}_n)$ by generic, diagonal flat connections, and the effect this has on module categories.
 
In Section \ref{sec:A-N} we revisit the Neumann-like VOA's $\CN_{n,k}$, promoting them to a deformable family $\CN_{n,k}^\psi$ as well. We derive a formula for $\CN_{n,k}^\psi$  as an iterated extension of elementary corner VOA's (W algebras and affine algebras). This sets us up in Section \ref{sec:levelrank} to formulate Conjecture~\ref{conj:2} from the introduction, on a new logarithmic level-rank duality relating $\CD_{n,k}\simeq\FT_k(\sln)$ and a slight modification $\wt \CN_{n,k}$ of $\CN_{n,k}$. In particular, we conjecture a braid-reversed equivalence of (abelian) braided tensor categories $\CD_{n,k}\text{-mod}\simeq \wt\CN_{n,k}\text{-mod}$, induced by an embedding
\be \CD_{n,k} \hookrightarrow \FF(nk) \hookleftarrow \wt\CN_{n.k} \ee
as mutual commutants inside $nk$ complex free fermions. We establish in Section \ref{sec:dual-def}  the corresponding duality of deformable families. 
The remainder of the Section \ref{sec:levelrank} explains categorical background on which the main conjecture relies, and presents explicit computations that support the conjecture in the case $n=2$.
We finish in Section \ref{sec:rect} with some brief comments about a relation between $\CN_{n,k}$ and rectangular W-algebras, which we hope to explore further in future work.

\subsection{Notation}

Throughout this section, we work with the complex Lie algebra $\mathfrak g = \mathfrak{sl}_n$, and denote its root lattice by $Q$, its weight lattice by $P$, and dominant weights and roots by $P^+$, $Q^+=P^+\cap Q$. We denote finite-dimensional irreducible $\g$-modules with highest weight $\lambda$ as $R_\lambda$.  We also use the following notation for some standard VOA's and their modules:

\begin{itemize}[leftmargin=1.5cm]
\item[$V^\psi(\mathfrak g)$:] The universal affine vertex algebra of $\g$ at critically-shifted level $\psi$ (\ie\ at level $\psi-h^\vee$, where $h^\vee=n$ is the dual Coxeter number of $\sln$). $V^\psi(\mathfrak g)$ is simple for generic $\psi$.

\item[$V^\psi_\lambda$:] the Weyl module of $V^\psi(\mathfrak g)$ of highest-weight $\lambda$. The top level of $V^\psi_\lambda$ is the irreducible highest-weight representation of $\g$ of highest-weight $\lambda$. $V^\psi_\lambda$ is simple for generic $\psi$.

\item[$Z(\g)$:] The vertex Poisson algebra arising as a (particular) $\psi\to\infty$ limit of $V^\psi(\g)$.

\item[$W^\psi(\mathfrak g)$:] The universal principal W-algebra of $\g$ at critically-shifted level $\psi$, \ie\ the quantum Hamiltonian reduction of $V^\psi(\mathfrak g)$ with principal nilpotent element $f$ \cite{Arakawa-W}. $W^\psi(\mathfrak g)$ is simple for generic $\psi$.

\item[$\M_{\lambda,\mu}^\psi$:] The simple quotient of the  Verma module for $W^\psi(\sln)$ associated to the weight $\lambda - \psi \mu$. 

(Note that a Verma module is characterized by the action of Zhu's algebra on the top level and modules of Zhu's algebra can be labelled by elements in the dual of the Cartan subalgebra.)

\item[$\pi^\psi$:] The Heisenberg VOA of level $\psi$. It is generated by a field $X(z)$ with OPE $X(z)X(w) = \psi/ (z-w)^{2}$.
\item[$\pi^\psi_\lambda$:] The Fock module of $\pi^\psi$ of highest-weight $\lambda \in \mathbb C$.

\item[$V_\Lambda$:] The lattice VOA associated to the lattice $\Lambda$.  We will often consider the case $\Lambda = Q$ the root lattice of $\g$. 
\item[$V_{\Lambda + \mu}$:] The lattice VOA module corresponding to the coset $\mu + \Lambda$. This is a local module if and only if $\mu$ is in the dual lattice: $\mu \in \Lambda'$.

\item[$L^{(n)}$ and $L^{(n)}_s$:] Shorthand notation for the following lattice VOA's and their modules, which appear throughout: $L^{(n)}$ denotes the lattice VOA of the lattice $\sqrt{n} \mathbb Z$, and $L^{(n)}_s$ denotes the module corresponding to the coset $\frac{s}{\sqrt{n}}+ \sqrt{n}\mathbb Z$.
 
\item[$\FF(m)$:] The  VOA of $m$ complex free fermions. By bozonization, $\FF(m)\simeq V_{\Z^m}$.
\end{itemize}

In the case of affine and W algebras (and their modules), the above notation may refer to \emph{either} a deformable family of VOA's, defined over the field $\C(\psi)$; or a VOA at fixed parameter $\psi$, defined over $\C$. This will be clear depending on context.  

The various Verma modules above are automatically simple over deformable families of VOA's, and will stay simple if the levels are specialized to generic complex numbers. However, the Verma modules will not necessarily stay simple at fixed rational $\psi$, whereupon the above notation denotes their simple quotients.

\subsubsection{Deformable families}
\label{sec:def-fam}

Usually one considers VOA's over the complex numbers. However, in some instances, it is instructive to enlarge the underlying field to a larger field or even just a commutative ring. One such enlargement is a deformable family of VOA's.
The idea of deformable families was introduced in order to understand cosets of VOA's by affine subalgebras \cite{Creutzig:2012sf,Creutzig:2014lsa}. It is defined as follows. Let $K$ be an at most countable subset of the complex numbers and let $F_K$ be the $\mathbb C$-algebra of rational functions in a formal variable $\kappa$ of the form $\frac{p(\kappa)}{q(\kappa)}$ with the roots of $q(\kappa)$ lying in $K$ and the degree of $p(\kappa)$ being at most the degree of $q(\kappa)$. Then a deformable family is a vertex algebra over $F_K$; see Section 3 of \cite{Creutzig:2014lsa} for complete details. 
One can then specialize $\kappa$ to any number in $\mathbb C \setminus K$; in particular, since the degree of elements in $F_K$ is at most zero 
one can take the limit $\kappa \rightarrow \infty$.  (There \emph{may} be multiple ways to take limits/specializations in $\kappa$, requiring additional choices to be made. We will see this below.)

As a simple example, the universal affine vertex operator algebra of a simple Lie algebra $\mathfrak g$ at level $k$ has generators $J^x$ for $x \in \mathfrak g$ and OPE 
\be
J^x(z) J^y(w) = \frac{k \kappa(x, y)}{(z-w)^2} + \frac{J^{[x, y]}(w)}{(z-w)}
\ee
with $\kappa$ the Killing form on $\mathfrak g$. We see that coefficients appearing in the OPE are polynomials in $k$, and so it also makes sense to consider the affine vertex  algebra of $\mathfrak g$ over $\mathbb C[k]$ where $k$ is now an indeterminate. One can then enlarge further to the field of rational functions $\mathbb C(k)$ in order to accommodate the Sugawara stress tensor (which involves a factor of $1/(k+h^\vee)$).
Moreover, it turns out to be convenient to replace $k$ by $\psi = k+ h^\vee$ with $h^\vee$ the dual Coxeter number of $\mathfrak g$.
Rescale the generators of the affine vertex algebra by $\frac{1}{k}$, that is set $I^x := \frac{J^x}{k}$. The OPE is then
\be\label{eq:scalingOPE} 
\begin{split}
I^x(z) I^y(w) &=  \frac{1}{k} \frac{ \kappa(x, y)}{(z-w)^2} +  \frac{1}{k}  \frac{ I^{[x, y]}(w)}{(z-w)} \\
&= \frac{1}{\psi - h^\vee} \frac{ \kappa(x, y)}{(z-w)^2} + \frac{1}{\psi - h^\vee}\frac{( I^{[x, y]}(w)}{(z-w)}
\end{split}
\ee
in particular the OPE coefficients are rational functions in both $k$ and $\psi$ of degree minus one. Thus the rescaled generators form a deformable family in the formal variable $\psi$ with $K = \{ 0, h^\vee\}$. Here we also exclude $0$ in order to have a Virasoro field given by the Sugawara stress tensor.

\subsection{$\CD_{n,k}$ and Feigin-Tipunin algebras}
\label{sec:D-FT}

\subsubsection{Corner VOA's and decompositions}
\label{sec:corners}

We recall a bit of background on VOA's supported at corners of topologically twisted 4d $\CN=4$ Yang-Mills theory, following \cite{GaiottoRapcak,CreutzigGaiotto-S}. (See also \cite{FrenkelGaiotto} for further developments.)

As already discussed in Section \ref{sec:4d}, 4d $\CN=4$ Yang-Mills with gauge group $G=SU(n)$ or $PSU(n)$ admits families of half-BPS boundary conditions $B_{p,q}$, $\wt B_{p,q}$, labelled by coprime integers $(p,q)$. In a IIB brane construction, $B_{p,q}$  (resp. $\wt B_{p,q}$) are engineered by $n$ D3 branes ending on a single $(p,q)$ 5-brane (resp. ending one-by-one on a stack of  $n$ $(p,q)$ 5-branes). These boundary conditions go back to the series of papers \cite{GaiottoWitten-Janus,GaiottoWitten-boundary,GaiottoWitten-Sduality}. Special cases include Neumann $B_{1,0}$; principal Nahm pole $B_{0,1}$; Dirichlet $\wt B_{0,1}$; and Neumann coupled to $T[G]$ $\wt B_{1,0}$.

Somewhat nontrivially, the various $B_{p,q},\wt B_{p,q}$ boundary conditions can be deformed to be compatible with generic bulk geometric-Langlands twist parameter $\psi$ --- the ``canonical parameter'' of \cite{KapustinWitten}. In a topologically twisted theory, the duality group $SL(2,\Z)$ then acts simultaneously on $\psi$; the global form of the group $G$ (with $S$ sending $G\to G^\vee$, etc.); and the labels $(p,q)$. Explicitly, the action is such that
\be g = \bp a&b\\c&d \ep\in SL(2,\Z)\,:\quad \psi \mapsto \frac{a\psi+b}{c\psi+d}\,,\qquad \bp q \\ p\ep \mapsto g \bp q \\ p\ep = \bp aq+bp\\ cq+dp \ep\,. \ee

Different half-BPS boundary conditions can intersect to form quarter-BPS corners. Suitable deformations of these corner configurations also preserve twists with generic parameters $\psi$. However, the 4d bulk topological twist induces a \emph{holomorphic}, non-topological twist at the 2d corners. The corners thus support vertex algebras. Since both the 4d bulk and 3d boundaries are fully topological (only the corners are holomorphic), the vertex algebras are expected to have a conserved holomorphic stress tensor, \emph{i.e.} to be VOA's.

All corner VOA's can ultimately be constructed from the three fundamental corners
\be \label{corner-fund} \raisebox{-.5in}{\includegraphics[width=5in]{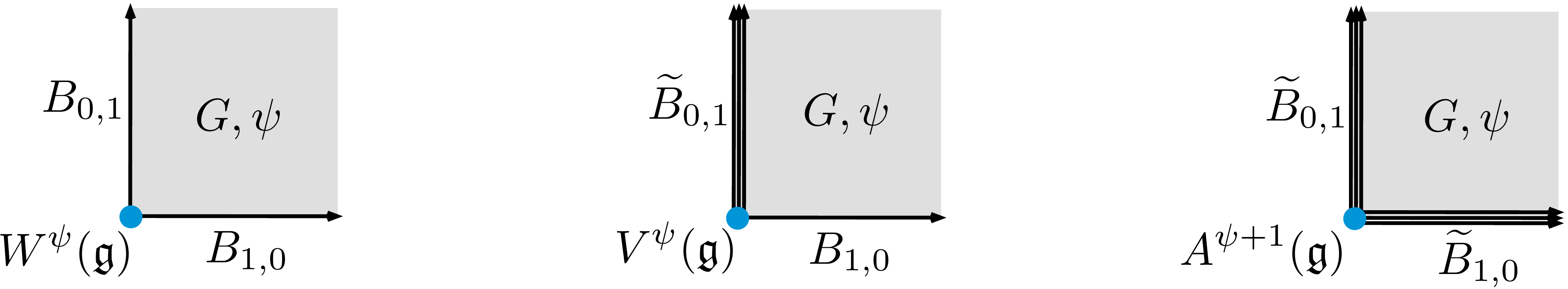}} \ee
supporting a principal W-algebra, an affine algebra, and the ``quantum geometric Langlands duality kernel'' VOA, respectively. The critically-shifted levels of the W and affine algebras match the bulk parameters $\psi$; while the level-like parameter of the Langlands kernel $A^{\psi+1}(\mathfrak g)$ is (by convention) shifted by 1. We will revisit the Langlands kernel in Section \ref{sec:A-N}. The first and third configurations are essentially invariant under S-duality (up to replacing $G\leftrightarrow G^\vee$ and flipping some orientations), which is reflected in Feigen-Frenkel duality of the principle W-algebra, and self-duality of the Langlands kernel:
\be W^{\psi}(\g) \simeq W^{\frac1\psi}(\g)\,,\qquad A^{\psi+1}(\g) \simeq A^{\frac1\psi+1}(\g)\,. \label{FeiginFrenkel} \ee

\subsubsection{The $\CD_{n,k}$ corner}
\label{sec:Dnk}

Now consider the sandwich of boundary conditions $B_{1,k},\wt B_{1,0}$ for 4d $G=SU(n)$ super-Yang-Mills that engineered our 3d $\CN=4$ theory $\CT_{n,k}^A$, as on the LHS of  \eqref{sandA}. Its 3d A-twist is induced by a 4d twist with parameter $\psi=0$. In a quarter-BPS brane construction, the branes representing the boundary conditions $B_{1,k},\wt B_{1,0}$ must intersect, and their intersection implicitly defines a boundary condition for $\CT_{n,k}$. In the $\psi=0$ twist, it supports a VOA that we denote $\CD_{n,k}$.

\begin{figure}[htb]
\centering
\includegraphics[width=5.5in]{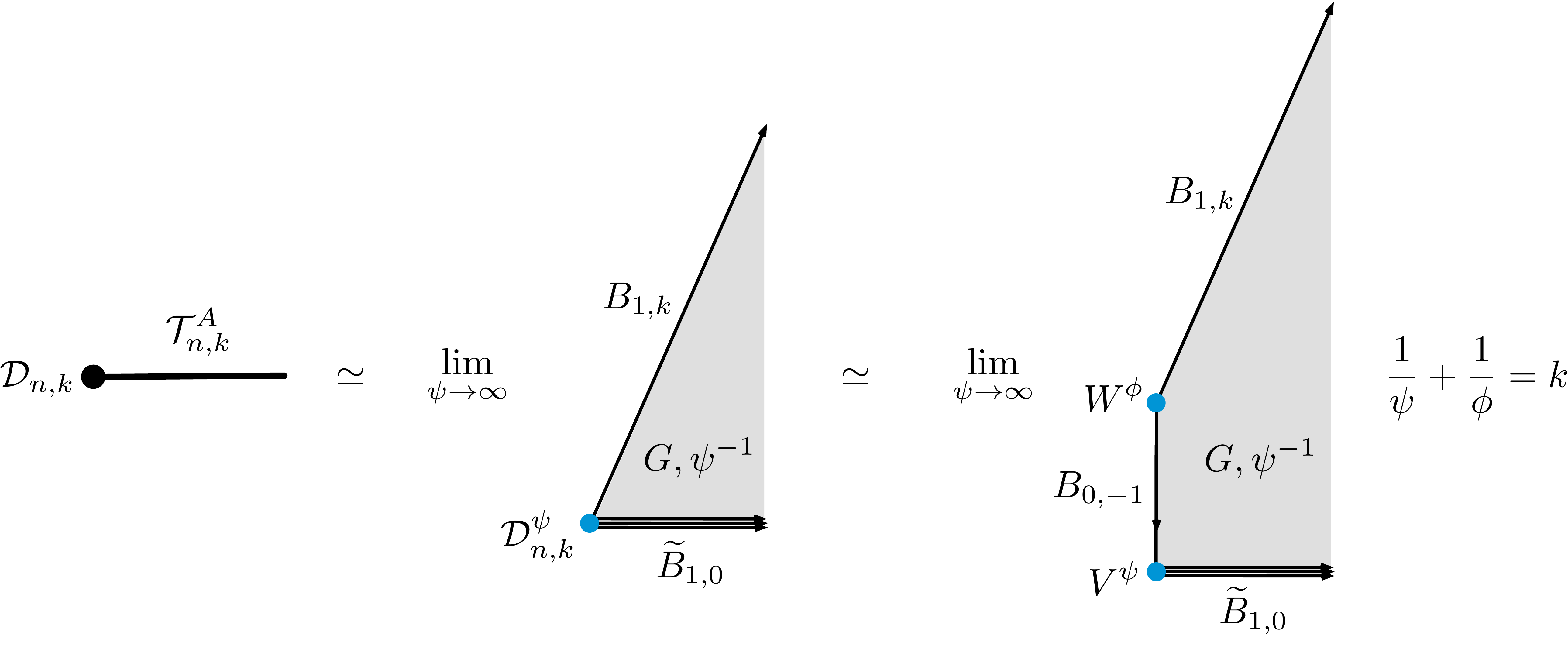}
\caption{The boundary VOA $\CD_{n,k}$ as a limit of a deformable family $\CD^\psi_{n,k}$.}
\label{fig:corner-FTres}
\end{figure}

We may enhance $\CD_{n,k}$ to a deformable family of VOA's $\CD^\psi_{n,k}$, over $\C(\psi)$, implicitly defined by the same corner configuration, but with bulk twist parameter $\psi^{-1}$. This depicted in the middle of Figure \ref{fig:corner-FTres}. We expect that $\CD_{n,k}$ will be recovered as an appropriate large-level limit: $\CD_{n,k} \overset{?}= \lim_{\psi\to\infty} \CD^\psi_{n,k}$.

To understand this better, we ``resolve'' the corner in the middle of Figure \ref{fig:corner-FTres} by intersecting with a third boundary condition $B_{0,-1}$. The boundary $B_{0,-1}$ is an (anti-)D5 brane, implementing a principal-Nahm-pole boundary condition on the bulk 4d theory. We expect that it will induce a Dirichlet-like boundary condition for the 3d theory $\CT_{n,k}$, in which some fields are given singular profiles. This is why we have labelled the associated boundary VOA's as $\CD_{n,k}$.

The virtue of introducing the extra $B_{0,-1}$ is that each pair of integers $(p,q)$, $(p',q')$ labelling consecutive boundary conditions now satisfy $\Big|\det \bsp p&q\\p'&q'\esp\Big| =1$. This means that there is an $SL(2,\Z)$ duality transformation relating each of the two corners on the RHS of Figure \ref{fig:corner-FTres} to one of the fundamental corners of \eqref{corner-fund}, up to orientation. Explicitly this is accomplished by the two transformations:
\be \raisebox{-.5in}{\includegraphics[width=5.5in]{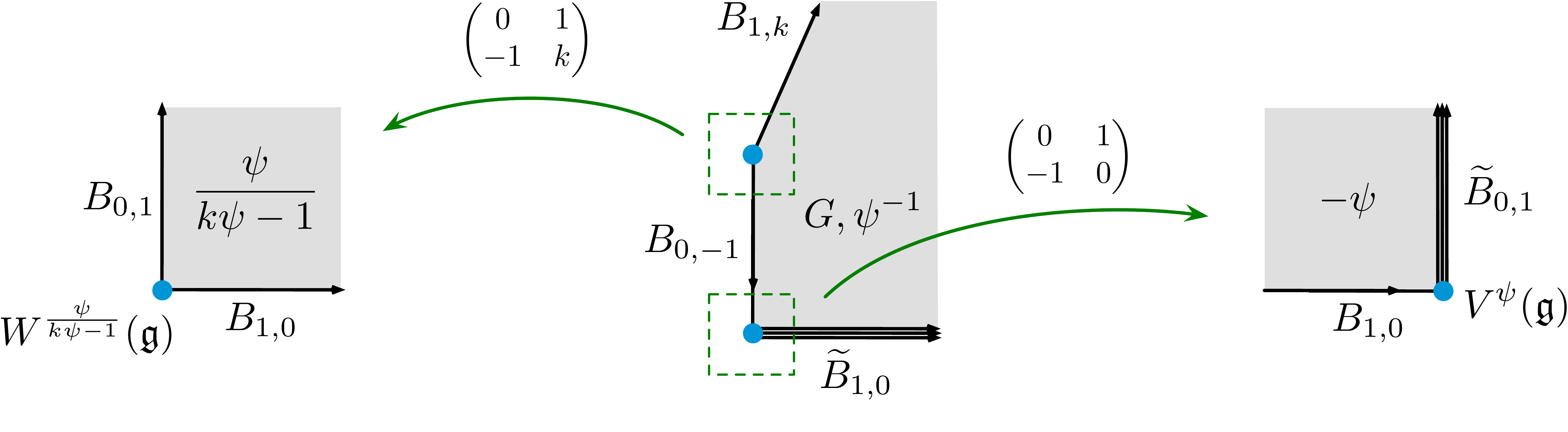}} \label{corner-FT2} \ee
We thus find that the lower corner supports an affine algebra of $\g=\sln$ at (critically shifted) level $\psi$ (negated relative to the bulk parameter $-\psi$ due to orientation) and that the upper corner supports a W-algebra at level $\phi=1/(-\psi^{-1}+k)=\frac{\psi}{k\psi-1}$. More succinctly, these two levels are related by $\psi^{-1}+\phi^{-1}=k$.

Our desired family $\CD_{n,k}^\psi$ should now arise as the tensor product of VOA's $V^\psi(\sln)\otimes W^\phi(\sln)$, \emph{extended} by a certain set of common modules. Physically, these modules are line operators on the 3d boundary $B_{0,-1}$, stretching between the $V^\psi$ and $W^\phi$ corners. The required extension was described in \cite{CreutzigGaiotto-S}. One finds
\be \CD^\psi_{n,k} = \bigoplus_{\lambda\in Q^+} V^\psi_\lambda \otimes \M^\phi_{\lambda,0}\,. \ee

The large-$\psi$ limit can now be taken in such a way that $V^\psi(\mathfrak{sl}_n)$ becomes a commutative vertex algebra $Z(\sln)$. 
This means we scale the fields of $V^\psi(\sln)$ as described in \eqref{eq:scalingOPE}. Other fields should however remain non-commutative in the limit. We illustrate several possible scalings and thus limits in instructive examples in sections \ref{sec:abelian} and \ref{sec:osp}.
$Z(\mathfrak{sl}_n)$ is a commutative vertex algebra that inherits a Poisson vertex algebra structure from  $V^\psi(\mathfrak{sl}_n)$. In this particular limit, we now find
\be \label{lim-FT} \lim_{\psi\to\infty} \CD^\psi_{n,k} = Z(\mathfrak{sl}_n)\otimes \bigoplus_{\lambda\in Q^+}  R_\lambda \otimes \M^{1/k}_{\lambda,0}\,. \ee
$ \CD^\psi_{n,1}$ is nothing by $V^{\psi-1}(\sln) \otimes V_Q$  by \cite{Arakawa:2018iyk}. We discuss $ \CD^\psi_{2,2}$ in Section \ref{sec:osp}. The existence of the $\CD_{n,k}^\psi$ for general $n, k$ will be proven in \cite{CN}.
The limit of $ \CD^\psi_{n,1}$ is nothing but the lattice VOA $V_Q$ and the limit of $ \CD^\psi_{2,2}$ is the even subalgebra of a pair of symplectic fermions, see also Section \ref{sec:osp}. The case of general $n, k$ remains conjectural.
We expect that the corner VOA $\CD_{n,k}$ defined by the physical setup at bulk twist $\psi=0$ corresponds to \eqref{lim-FT} with the large center $Z(\sln)$ factored out.%
\footnote{It is a general expectation that all vertex algebras sitting at a holomorphic corner of a purely topological 3d/4d configuration will have holomorphic stress tensors. There is not yet a proof --- some relevant discussion appears in \cite{CDG}. Assuming that $\CD_{n,k}$ has a stress tensor and is indeed a VOA, no commutative vertex algebra such as $Z(\sln)$ can appear in $\CD_{n,k}$: the holomorphic $\pd_z$ derivatives of every field in $\CD_{n,k}$ must be generated by OPE with the stress tensor, but the stress tensor would have trivial OPE's with $Z(\sln)$.} %
In other words, $\CD_{n,k}$ is the extension of the W-algebra $W^{1/k}(\sln)$ given by
\be \CD_{n,k} = \bigoplus_{\lambda\in Q^+}  R_\lambda \otimes \M^{1/k}_{\lambda,0}\,. \label{D-decomp} \ee

The RHS of \eqref{D-decomp} is precisely the decomposition of the Feigin-Tipunin algebra $\FT_k(\sln)$ conjectured by Feigin and Tipunin \cite{Feigin:2010xv} and recently proven by S. Sugimoto \cite{sugimoto2021feigintipunin, sugimoto2021simplicities}. Thus we find, subject to several assumptions/expectations noted above, that $\CD_{n,k}\simeq \FT_k(\sln)$.
In fact, we assume more. We assume that there exists several possibilities of scaling of the fields of $\CD^\psi_{n,k}$ such that the limit exists and the structure of the limiting algebra depends on the chosen scaling. In particular we expect that there exist limits where the center appears in the OPE algebra. This means one gets a variant of the Feigin-Tipunin algebra coupled to the large center, which we interpret as $\FT_k(\sln)$ deformed by a flat connection. (We discuss the interpretation in terms of flat connections further in Section \ref{sec:FTtwisted}.)

\subsubsection{Abelian example}
\label{sec:abelian}

The simplest example is the abelian case $\mathfrak{g} = \mathfrak{gl}_1$. The affine vertex algebra of $\mathfrak{gl}_1$ at level $\psi$ is nothing but the free boson or Heisenberg VOA $\pi^\psi$.
The dual Coxeter number is zero and so $\psi$ is already the shifted level. 
 The free boson is generated by a single field $X$ with OPE
\be 
X(z) X(w) = \frac{\psi}{(z-w)^2}\,.
\ee
For a complex number $\lambda$ we denote by $\pi_\lambda^\psi$ the Fock module of $\pi^\psi$ of highest weight $\lambda$. In particular there is a field $\phi^\psi_\lambda$ associated to the highest-weight vector. The OPE with the free boson is
\be
X(z) \phi^\psi_\lambda(w) = \frac{\lambda \phi^\psi_\lambda(w)}{(z-w)}\,.
\ee
The  fusion rule $\pi^\psi_\nu \boxtimes \pi^\psi_\mu = \pi^\psi_{\nu+\mu}$ corresponds to the OPE \cite{FBZ}
\begin{equation}
\begin{split}
\phi^\psi_\nu(z) \phi^\psi_\mu(w) &= (z-w)^{\frac{\nu \mu}{\psi}} :\!\phi^\psi_\lambda(z) \phi^\psi_\mu(w)\!:  \\ &= (z-w)^{\frac{\lambda \mu}{\psi}} \phi^\psi_{\lambda +\mu}\left(1 + (z-w) \frac{\lambda}{\psi} X(w) + \dots  \right)\,.
\end{split}
\end{equation}

Let $k$ be a positive integer and consider $\pi^\psi \otimes \pi^\kappa$ and denote the generators $X$ and  $Y$.  
Require that $\psi$ and $\kappa$ are related via
\be
\frac{1}{\psi} + \frac{1}{\kappa} = k\,.
\ee
The object 
\be
A^k := \bigoplus_{\lambda \in \mathbb Z} \pi^\psi_\lambda \otimes \pi_\lambda^\kappa
\ee
is easily seen to carry the structure of a simple vertex operator superalgebra isomorphic to $V_{\sqrt{k}\mathbb Z} \otimes \pi^{\psi + \kappa}$ and $ \pi^{\psi + \kappa}$ is generated by the field $X-Y$. Let $W_\lambda := \phi^\psi_\lambda \otimes \phi^\kappa_\lambda$ so that
\be
W_\lambda(z)W_\mu(w) = (z-w)^{k\lambda\mu} W_{\lambda + \mu}(w) \left[ 1 + (z-w) \left(\frac{\lambda}{\psi}X(w) + \frac{\lambda}{\kappa} Y(w) \right) + \dots \right]\,.
\ee
Setting $Z = \frac{X}{\psi} + \frac{Y}{\kappa}$ this OPE becomes 
\be
W_\lambda(z)W_\mu(w) = (z-w)^{k\lambda\mu} W_{\lambda + \mu}(w) \big[ 1 + (z-w) \lambda Z + \dots \big]\,.
\ee
and in particular we can  consider $A^k$ as a deformable family in the formal variable $\kappa$ and generated by $Z$, the $W_\lambda$ and $Z' = \frac{X-Y}{\kappa}$ then 
\be
\lim_{\kappa \rightarrow \infty}^{\qquad (1)} A^k \cong V_{\sqrt{k}\mathbb Z} \otimes \pi^0
\ee
is nothing but the lattice VOA $V_{\sqrt{k}\mathbb Z}$ times a commutative (that  is level zero) free boson $\pi^0$.  $\pi^0$ is generated by $Z'$. 

We now want to couple the lattice VOA to the abelian algebra. For this we consider $A^k$ as a deformable family generated by $X$, the $W_\lambda$ and $Z = \frac{Y}{\kappa}$. Note that the OPE
\be
Z(z) Y(w) = \frac{1}{(z-w)^2}
\ee
is then finite and independent of $\kappa$.
The OPE of the $W_\lambda$ becomes
\be
\begin{split}
W_\lambda(z)W_\mu(w) = (z-w)^{k\lambda\mu} W_{\lambda + \mu}(w) \left[ 1 + (z-w) \left(\frac{\lambda}{\psi}X(w) + \lambda Z(w) \right) + \dots \right]\,. \\
\lim_{\kappa \rightarrow \infty}^{\qquad (2)} W_\lambda(z)W_\mu(w) = (z-w)^{k\lambda\mu} W_{\lambda + \mu}(w) \big[ 1 + (z-w) \left(k \lambda X(w) + \lambda Z(w) \right) + \dots \big]\,.
\end{split} 
\ee
and so we see that the abelian free boson $Z$ appears in the OPE. 

Consider now the $A^k$-module 
\[
M^k_\mu :=\bigoplus_{\lambda \in \mathbb Z} \pi^\psi_{\lambda + \psi \mu} \otimes \pi^\kappa_{\lambda- \kappa \mu}.
\]
This is a local module, i.e. a VOA module. 
It satisfues 
\be
M^k_\mu = V_{\sqrt{k}\mathbb Z} \otimes \pi^{\frac{\psi k}{\kappa}}_\mu
\ee
as a $V_{\sqrt{k}\mathbb Z} \otimes \pi^{\frac{\psi k}{\kappa}}$-module, where the algebra $\pi^{\frac{\psi k}{\kappa}}$ is generated by $Z' = \frac{X-Y}{\kappa}$. 
In particular in the large $\kappa$ limit corresponding to our first choice of scaling 
\be
\lim_{\kappa \rightarrow \infty }^{\qquad (1)} M^k_\mu = V_{\sqrt{k}\mathbb Z} \otimes \pi^{0}_\mu.
\ee
The second choice of scaling is on the other hand 
\be
\lim_{\kappa \rightarrow \infty }^{\qquad (2)} M^k_\mu = \bigoplus_{\lambda \in \mathbb Z} \pi^\psi_{\lambda + \frac{ \mu}{k}} \otimes \pi^0_{- \mu}\ee
as a module for the vertex algebra generated by $X$ and $Z = \frac{Y}{\kappa}$. Recalling that $Z$ is abelian we can quotient by the action of $Z$ to get
\be 
\bigoplus_{\lambda \in \mathbb Z} \pi^\psi_{\lambda + \frac{ \mu}{k}}  = V_{\sqrt{k}\mathbb Z + \mu}
\ee
which is a twisted module for the lattice VOA  $V_{\sqrt{k}\mathbb Z}$ if $\mu \notin \frac{1}{\sqrt{k}}\mathbb Z$. 
Another way to phrase this is as follows. Fock modules of the abelian free bosons can be naturally identified with modules for the commutative ring $S := \mathbb C[Z_0, Z_{-1}, Z_{-2}, \dots ]$ which is a subring of $R := \mathbb C[[Z_n | n \in \mathbb Z]]$ and OPE coefficients are in $R$. Hence  
 the second limit is a twisted module for $V_{\sqrt{k}\mathbb Z}\otimes_{\mathbb C}R$. 
This is a key observation that we summarize:
\begin{enumerate}
\item The first limit is a lattice VOA times a commutative free boson
\item The second limit couples the lattice VOA to the commutative free boson 
\item The second limit of a generic module is a twisted module for the lattice VOA
\end{enumerate}

\subsubsection{The example of $\CD^\psi_{2,2}$ via $\mathfrak{osp}(1|2)$}
\label{sec:osp}

Another simple example of the large-level limits that we use is the case $k=2$ for  $\mathfrak{sl}_2$. We now present this example in detail, in particular we illustrate the subtlety of different large-level limits depending on a scaling. 
(Another interesting and more involved example of large-level behavior is the large $N=4$ superconformal algebra at central charge $-6$, developed in \cite{Creutzig:2018ltv}.)

Let $\tilde\psi$, $\psi$
be related via 
\be
\frac{1}{\tilde\psi} + \frac{1}{\psi}  = 2.
\ee
The basic example is $V^{\tilde\psi -\frac{1}{2}}(\mathfrak{osp}(1|2))$,  with the very nice decomposition \cite{CreutzigGaiotto-S}
\be
V^{\tilde\psi -\frac{1}{2}}(\mathfrak{osp}(1|2)) \cong  \bigoplus_{\lambda \in P^+} V^{\tilde\psi}_\lambda  \otimes W^{\psi}_{\lambda, 0}
\ee
as $V^{\tilde \psi}(\mathfrak{sl}_2) \otimes W^{\psi}(\mathfrak{sl}_2)$-modules, where of course $W^{\psi}(\mathfrak{sl}_2)$ is just the Virasoro algebra of central charge $13- \left( \psi+ \frac{1}{\psi}\right)$. 
The even subalgebra is
\be
V^{\tilde\psi -\frac{1}{2}}(\mathfrak{osp}(1|2))_{\text{even}}  \cong  \bigoplus_{\lambda \in Q^+} V^{\tilde\psi}_\lambda  \otimes W^{\psi}_{\lambda, 0} =\CD^\psi_{2, 2}.
\ee
We will now explain four choices of large $\tilde \psi$-limits. One of them will be symplectic fermions times a commutative vertex algebra and a second one will be symplectic fermions coupled to the commutative vertex algebra. We note that the even subalgebra of symplectic fermions times (resp. coupled to) the commutative vertex algebra appears then as the limit of the even subalgebra of $V^{\tilde\psi -\frac{1}{2}}(\mathfrak{osp}(1|2))$. This is good news as it is well-known that  the Feigin-Tipunin algebra $\FT_2(\mathfrak{sl}_2)$ is nothing but the even subalgebra of symplectic fermions. In particular we have two important limits of 
$\CD^\psi_{2, 2}$; one limit is $\FT_2(\mathfrak{sl}_2)$ times a commutative vertex algebra and another one is $\FT_2(\mathfrak{sl}_2)$ coupled to a commutative vertex algebra.

The algebra $V^{\tilde \psi - \frac{1}{2}}(\mathfrak{osp}(1|2))$ is strongly generated by even fields $e, h, f$ and odd fields $x, y$ with OPE's
\begin{equation}
	\begin{split}
		e(z)f(w) &\sim (\tilde \psi -2)(z-w)^{-2} + h(w)(z-w)^{-1}, \qquad  \\ h(z)h(w) &\sim 2 (\tilde \psi -2)(z-w)^{-2}, \\
		h(z)e(w) &\sim  2e(w)(z-w)^{-1}, \qquad  \\ h(z)f(w) &\sim  -2f(w)(z-w)^{-1},\\
		h(z)x(w) &\sim  x(w)(z-w)^{-1}, \qquad \\ h(z)y(w) &\sim  -y(w)(z-w)^{-1},\\
		e(z) y(w) &\sim x(w)(z-w)^{-1},\qquad \\ f(z) x(w) &\sim y(w)(z-w)^{-1},\\
		x(z)y(w) &\sim  (\tilde \psi -2)(z-w)^{-2} + \frac{h(w)}{2}(z-w)^{-1}, \qquad \\ x(z)x(w) &\sim -e(w)(z-w), \\  y(z)y(w) &\sim f(w)(z-w).\\
	\end{split}
\end{equation}
We want to take the limit $\tilde \psi \rightarrow \infty$. For this we have to first rescale fields and the limit depends on our rescaling.
Let us consider two rational functions $f, g$ and set
\begin{equation}
	\begin{split}
		h(\tilde\psi) :&= \frac{g(\tilde\psi)^2}{f(\tilde\psi)}\\
		e_f(z) :&= f(\tilde\psi) e(z), \qquad h_f(z):= f(\tilde\psi) h(z), \qquad f_f(z):= f(\tilde\psi) f(z),\\
		x_g(z) :&= g(\tilde\psi) x(z), \qquad y_g(z):= g(\tilde\psi) y(z).
	\end{split}
\end{equation}
Of course we can scale each field differently but for our purposes the above set-up is sufficient. OPE's are then
\begin{equation}
	\begin{split}
		e_f(z)f_f(w) &\sim f(\tilde\psi)^2(\tilde \psi -2)(z-w)^{-2} + f(\tilde\psi) h_f(w)(z-w)^{-1}, \qquad  \\ 
		h_f(z)h_f(w) &\sim 2 f(\tilde\psi)^2(\tilde \psi -2)(z-w)^{-2}, \\
		h_f(z)e_f(w) &\sim  2f(\tilde\psi)e_f(w)(z-w)^{-1}, \qquad  \\ h_f(z)f_f(w) &\sim  -2f(\tilde\psi)f_f(w)(z-w)^{-1},\\
		h_f(z)x_g(w) &\sim  f(\tilde\psi) x_g(w)(z-w)^{-1}, \qquad \\ h_f(z)y_g(w) &\sim  -f(\tilde\psi)y_g(w)(z-w)^{-1},\\
		e_f(z) y_g(w) &\sim f(\tilde\psi)x_g(w)(z-w)^{-1},\qquad \\ f_f(z) x_g(w) &\sim f(\tilde\psi)y_g(w)(z-w)^{-1},\\
		x_g(z)y_g(w) &\sim  g(\tilde\psi)^2(\tilde \psi -2)(z-w)^{-2} + h(\tilde\psi)\frac{h_f(w)}{2}(z-w)^{-1}, \qquad \\ x_g(z)x_g(w) &\sim -h(\tilde\psi)e_f(w)(z-w), \\  y_g(z)y_g(w) &\sim h(\tilde\psi)f_f(w)(z-w).\\
	\end{split}
\end{equation}

Here are a few cases:
\begin{enumerate}
	\item {\bf Commutative Limit}: Set $f(\tilde\psi) = g(\tilde\psi) = \frac{1}{\tilde\psi}$. Then in the large-$\tilde\psi$ limit all OPE's are regular, \emph{i.e.} we get a large commutative superVOA. This is too much as we want to keep a non-commutative part.
	\item  {\bf Free Field Limit}: Set $f(\tilde\psi) = g(\tilde\psi) = \frac{1}{\sqrt{\tilde\psi}}$. Then in the large-$\tilde\psi$ limit the non-regular OPE's are
	\begin{equation}
		\begin{split}
			e_f(z)f_f(w) &\sim (z-w)^{-2}, \qquad  \\ 
			h_f(z)h_f(w) &\sim 2 (z-w)^{-2}, \\
			x_g(z)y_g(w) &\sim  (z-w)^{-2} .\\
		\end{split}
	\end{equation}
	This is just the free field algebra of a rank three Heisenberg VOA times a pair of symplectic fermions. This is not enough as we also want to get a commutative part. 
	\item {\bf Mixed limit} Set $f(\tilde\psi) =  \frac{1}{\tilde\psi^{3/4}}$ and $g(\tilde\psi) = \frac{1}{\sqrt{\tilde\psi}}$ so that $h(\tilde\psi) =\frac{1}{\tilde \psi^{1/4}}$, then there is a single non-regular OPE
	\begin{equation}
	\begin{split}
		x_g(z)y_g(w) &\sim  (z-w)^{-2} 
			\end{split}
\end{equation}
The large $\tilde\psi$-limit is a pair of symplectic fermions (generated by $x_g, y_g$) times a large commutative vertex algebra (generated by $e_f, h_f, f_f$).
Note that the limit of the even subalgebra is then the even subalgebra of symplectic fermions times the commutative vertex algebra. This is exactly as desired, since the even subalgebra of symplectic fermions is the triplet vertex algebra at $k=2$. The triplet vertex algebra at $k=2$ is the Feigin-Tipunin algebra $\FT_2(\mathfrak{sl}_2)$.

	\item {\bf Coupling to SU(2) Poisson VA}: Set $f(\tilde\psi) =  \frac{1}{\tilde\psi}$ and $g(\tilde\psi) = \frac{1}{\sqrt{\tilde\psi}}$ so that $h(\tilde\psi) =1$, and in the large-$\tilde\psi$ limit the non-regular OPE's are
	\begin{equation}
		\begin{split}
			x_g(z)y_g(w) &\sim  (z-w)^{-2} + \frac{h_f(w)}{2}(z-w)^{-1}, \qquad \\ x_g(z)x_g(w) &\sim -e_f(w)(z-w), \\  y_g(z)y_g(w) &\sim f_f(w)(z-w).\\
		\end{split}
	\end{equation}
	This is the interesting limit. The $e_f, h_f, f_f$ form a commutative VOA that can be given the structure of a Poisson vertex algebra, while the $x_g, y_g$ obey the OPE relations of symplectic fermions coupled to this commutative VOA.
\end{enumerate}
The Poisson VA structure appears as follows. We stay with the last case and only consider the subalgebra generated by $e_f, h_f, f_f$. Set $\epsilon=\frac{1}{\tilde\psi}$. We consider large $\tilde\psi$, \ie\ small $\epsilon$. Then to leading order in $\epsilon$ the OPE's are
\begin{equation}
	\begin{split}
		e_f(z)f_f(w) &\sim \epsilon(z-w)^{-2} +\epsilon h_f(w)(z-w)^{-1}, \qquad  \\ 
		h_f(z)h_f(w) &\sim 2 \epsilon(z-w)^{-2}, \\
		h_f(z)e_f(w) &\sim  2\epsilon e_f(w)(z-w)^{-1}, \qquad  \\ h_f(z)f_f(w) &\sim  -2\epsilon f_f(w)(z-w)^{-1}.\\
	\end{split}
\end{equation}
Let
\be
e_f(z) = \sum_{n\in \mathbb Z} e_n^\epsilon z^{-n-1}, \qquad h_f(z) = \sum_{n\in \mathbb Z} h_n^\epsilon z^{-n-1}, \qquad f_f(z) = \sum_{n\in \mathbb Z} f_n^\epsilon z^{-n-1} 
\ee
be the mode expansion then the OPE's translate into the commutation relations
\begin{equation}
	\begin{split}
		[e^\epsilon_n, f^\epsilon_m] &= \epsilon \left(h^\epsilon_{n+m}+ n \delta_{n+m, 0}\right) \\
		[h^\epsilon_n, h^\epsilon_m] &= \epsilon  n \delta_{n+m, 0} \\
		[h^\epsilon_n, e^\epsilon_m] &= \epsilon 2e^\epsilon_{n+m} \\
		[h^\epsilon_n, f^\epsilon_m] &= \epsilon \left(-2 f^\epsilon_{n+m}\right) \\
	\end{split}
\end{equation}
the Poisson vertex algebra structure inherited in the $\epsilon$ to zero limit is then given by 
\begin{equation}
	\begin{split}
		\{e_n, f_m \}&=  h_{n+m}+ n \delta_{n+m, 0} \\
		\{h_n, h_m\} &=   n \delta_{n+m, 0} \\
		\{h_n, e_m\} &=  2e_{n+m} \\
		\{ h_n, f_m\} &= -2 f_{n+m} \\
	\end{split}
\end{equation}

\subsection{Representation theory of Feigin-Tipunin algebras and the triplet}
\label{sec:FT}

By a logarithmic vertex algebra one means a vertex algebra that has correlation functions with logarithmic singularities. This can only happen if the zero-mode of the Virasoro algebra acts non-semisimply (non-diagonalizably), and it leads to non-semisimple abelian categories of VOA modules. See \cite{Creutzig:2013hma} for an introduction.
 
Few examples are known and the best understood series are the triplet algebras that we discuss in a moment. These triplet algebras fall into a larger class of algebras, the Feigin-Tipunin algebras \cite{Feigin:2010xv}. The Feigin-Tipunin algebra $\FT_k(\g)$ is realized as a subalgebra of $V_{\sqrt{k}Q}$, the lattice VOA of the root lattice $Q$ of a simply-laced simple Lie algebra $\mathfrak{g}$, with $k \in \mathbb Z_{\geq 2}$. It is constructed as global sections of a certain bundle with fibers $V_{\sqrt{k}Q}$ over the flag variety $G_\C/B$.

Feigin and Tipunin  conjectured various nice properties of their algebras that by now have partially been proven by Shoma Sugimoto \cite{sugimoto2021feigintipunin, sugimoto2021simplicities}.
The most important for us is the decomposition (proven for $k \geq h^\vee-1$) of $\FT_k(\g)$ as a $G_\C \otimes W^{1/k}(\g)$ module,
\be
\FT_k(\mathfrak{g}) = \bigoplus_{\lambda \in Q^+} R_\lambda \otimes \M^{1/k}_{\lambda, 0}\,.
\ee
(Since only representations with weights in the root lattice are involved, $G_\C$ may be taken as the adjoint form of the complex reductive group with Lie algebra $\g$. The case relevant for the rest of this section is $G_\C = PGL(n,\C)$.)

\begin{rmk}\label{rmk:FT}
Let $\g$ be a simply-laced simple Lie algebra of rank $n$. Let $Q$ be its root lattice, $P$ be its weight lattice, and $\kappa$ the Killing form. Denote by $\alpha_1, \dots, \alpha_n$ the positive simple roots of $\g$. One then associates the rank $n$ Heisenberg vertex algebra to the Cartan subalgebra of $\g$. It is generated by fields $\alpha_1(z), \dots, \alpha_n(z)$ with OPE's
\begin{equation}
\alpha_i(z) \alpha_j(w) = \frac{A_{i, j}}{(z-w)^2}
\end{equation}
and $(A_{i, j})$ the Cartan matrix of $\g$. 
Denote by $\pi_\lambda$ the Fock module of highest-weight $\lambda$ and by $\phi_\lambda$ the field associated to the highest-weight vector. It is an intertwiner from $\pi_\mu$ to $\pi_{\mu+\lambda}$ for any weight $\mu$.
Denote the zero-mode of $\phi_\lambda$ by $\oint \phi_\lambda$, then 
\be
\FT_k(\mathfrak{g}) =  \bigcap_{i=1}^n \text{ker}\left( \oint \phi_{-\alpha_i/\sqrt{k}}: V_{\sqrt{k}Q} \rightarrow V_{\sqrt{k}Q -\alpha_i/\sqrt{k}} \right)
\ee
Consider $\sqrt{k}P$, which is a not necessarily integral lattice and so the lattice algebra $V_{\sqrt{k}P}$ is not quite a vertex algebra, but an abelian intertwining algebra. It follows that 
\be
A(\FT_k(\mathfrak{g}) ) :=  \bigcap_{i=1}^n \text{ker}\left( \oint \phi_{-\alpha_i/\sqrt{k}}: V_{\sqrt{k}P} \rightarrow V_{\sqrt{k}P -\alpha_i/\sqrt{k}} \right)
\ee
is an extension of $\FT_k(\mathfrak{g}) $ to some abelian intertwining algebra. In the case of $\g=\mathfrak{sl}_2$ this one is called the doublet algebra \cite{Ada-doublet, Adamovic:2020lvj}, e.g. in the special case $k=2$ this is nothing but the symplectic fermion vertex superalgebra. 

The algebra $V_{\sqrt{k}P}$ carries an action of $\sqrt{k}P/\sqrt{k}Q \cong P/Q$ via automorphism and the subalgebra  $A(\FT_k(\mathfrak{g}) )$ inherits this action. In particular one has the decomposition as $V_{\sqrt{k}Q}$ and $\FT_k(\mathfrak{g})$-modules
\begin{equation}
\begin{split}
V_{\sqrt{k}P}  &= \bigoplus_{\lambda \in \sqrt{k}P/\sqrt{k}Q}  \ V_{\lambda + \sqrt{k}Q} \\
A(\FT_k(\mathfrak{g}) ) &= \bigoplus_{\lambda \in \sqrt{k}P/\sqrt{k}Q}  \  \bigcap_{i=1}^n \text{ker}\left( \oint \phi_{-\alpha_i/\sqrt{k}}:  V_{\lambda + \sqrt{k}Q} \rightarrow  V_{\lambda -\alpha_i/\sqrt{k} + \sqrt{k}Q}\right)
\end{split}
\end{equation}
This is a decomposition of $A(\FT_k(\mathfrak{g}) )$ into simple currents by \cite{ Creutzig:2016ehb, McRae:2018wpu}, provided $\FT_k(\mathfrak{g}) $ is $C_2$-cofinite\footnote{The $C_2$-cofiniteness of $\FT_k(\mathfrak{g})$ is known to be true for $\g = \mathfrak{sl}_2$ \cite{Adamovic:2007er, Carqueville:2005nu} and our understanding is that Shoma Sugimoto is making progress towards the proof for some higher rank cases. }. 
Since the intertwining operators of $V_{\sqrt{k}P}$ restrict to intertwining operators of 
$A(\FT_k(\mathfrak{g}) )$ these simple currents must braid according to the quadratic form of the lattice $\sqrt{k}P$. 
\end{rmk}

The connection between Feigin-Tipunin algebras and quantum groups has been studied in \cite{Creutzig:2020jxj}; in particular quasi-Hopf modifications of quantum groups were constructed that have conjecturally equivalent representation categories as the Feigin-Tipunin algebras. Unrolled versions of these quantum groups \cite{rupert2020categories} should correspondingly be related to the orbifolds of the Feigin-Tipunin algebras of \cite{Creutzig:2016uqu}.
We also mention that these algebras are closely related to chiral algebras of Argyres-Douglas theories associated to $Q$
\cite{Creutzig:2018lbc}.  We turn now to the triplet algebra, \ie\ the case $n=2$, or $Q=A_1$.

\subsubsection{Triplet algebras $\FT_k(\mathfrak{sl}_2)$ and modules}
\label{sec:triplet-mod}

The triplet algebras  $\FT_k(\mathfrak{sl}_2)$ for $k \in \mathbb Z_{\geq 2}$ are the best studied family of logarithmic VOA's \cite{Tsuchiya:2012ru, Adamovic:2007er, Adamovic:2009xn,mcrae2021structure}. 
We review some of their representation theory, largely following work of Gannon and the first author \cite{Creutzig:2016fms}, which developed Verlinde's formula for the triplet. We would like to explain how some structure of the category of line operators and state spaces found in Sections \ref{sec:Uqsl2}, \ref{sec:Hilbert} appears from the perspective of the triplet.

We fix the integer $k\geq 2$. Let $\hat\CC$ denote the category of local $\FT_k(\mathfrak{sl}_2)$-modules.
We recall that according to \cite{GannonNegron, Creutzig:2021cpu} this category is equivalent to $u_q(\mathfrak{sl}_2)$-mod as a braided tensor category.
It has simple objects $S^\pm_s$ for $s =1, \dots, k$. The modules $S^\pm_k$ are projective, while the other ones have projective covers $P^\pm_s$ described by the following Loewy diagram:
\be \raisebox{-.6in}{
	\begin{tikzpicture}[thick,>=latex,nom/.style={circle,draw=black!20,fill=black!20,inner sep=1pt}]
		\node (top1) at (5,1.5) [] {$S_s^\pm$};
		\node (left1) at (3.5,0) [] {$S_{k-s}^\mp$};
		\node (right1) at (6.5,0) [] {$S_{k-s}^\mp$};
		\node (bot1) at (5,-1.5) [] {$S_s^\pm$};
		\node at (5,0) [nom] {$P^\pm_s$};
		\draw [->] (top1) -- (left1);
		\draw [->] (top1) -- (right1);
		\draw [->] (left1) -- (bot1);
		\draw [->] (right1) -- (bot1);		
	\end{tikzpicture}}
\ee
This means that $P^\pm_s$ has a socle series (\emph{a.k.a.} composition series)
\begin{equation}
 0 \leq S^\pm_s \leq S^\mp_{k-s} \oplus S^\mp_{k-s} \leq S^\pm_s.
\end{equation}
The endomorphisms of $P^\pm_s$ are spanned by the identity and the projection of the top component onto the socle (which squares to zero). This clearly matches the structure of the abelian category $u_q(\mathfrak{sl}_2)$-mod from Section \ref{sec:V-1}. (The projection endomorphism is denoted $\tau$ in Section \ref{sec:CGP-0} and in Appendix \ref{app:U}.)

The fusion rules for triplet modules may be determined by using the fact that the triplet algebra embeds in the lattice VOA $V_{\sqrt{2k}\mathbb Z}$. We introduce notation commonly used in the literature. Set
\begin{equation} \label{tensorproducts}
\alpha_{r, s} = \frac{1-r}{2}\alpha_+ + \frac{1-s}{2}\alpha_-\,, \qquad \alpha_+= \sqrt{2k}\,, \qquad \alpha_- = - \sqrt{\frac{2}{k}}\,.
\end{equation}
Then we write $V^+_{s}$ for the module $V_{\alpha_{1, s} + \alpha_+\mathbb Z}$ and $V^-_{s}$ for the module $V_{\alpha_{2, s} + \alpha_+\mathbb Z}$. Also introduce the notation $V_s$ for $V_{\frac{s}{\alpha_+} + \alpha_+ \mathbb Z}$.
They are indecomposable modules for the triplet algebra and satisfy
\begin{equation}
 0 \rightarrow  S^\pm_s \rightarrow V^\pm_s \rightarrow S^\mp_{k-s} \rightarrow 0\,. 
\end{equation}
(They match the Verma modules of Section \ref{sec:U-modules}.)
The fusion rules are completely determined by associativity, commutativity, rigidity and
\begin{equation}
	\begin{split}
	S^+_2 \boxtimes S^+_{s} &= \begin{cases}  S^+_{2} & \quad s=1 \\ S^+_{s-1} \oplus S^+_{s+1} &\quad 1<s<k \\ P^+_{k-1} &\quad s=k \end{cases} \\
	S^+_2 \boxtimes P^+_{s} &= \begin{cases}  P^+_{2} \oplus 2P^-_{k} & \quad s=1 \\ P^+_{s-1} \oplus P^+_{s+1} &\quad 1<s<k-1 \\ P^+_{k-2} \oplus 2P^+_{k} &\quad s=k-1 \end{cases} \\
	S^-_1 \boxtimes S^+_s &= S^-_s \\ 
	S^-_1 \boxtimes P^+_s &= P^-_s.
	\end{split}
	\end{equation}

\subsubsection{Characters of derived state spaces}
\label{sec:triplet-Hilb}

The vertex tensor category of the triplet algebra is a finite tensor category, so one has a modular functor that in particular assigns to marked Riemann surfaces finite dimensional vector spaces, \cite{Turaev:1994xb, Lyubashenko:1994tm}. These should of course be the spaces of conformal blocks.
They are expected to be the same as the spaces obtained by the CGP construction for $u_q(\mathfrak{sl}_2)$, outlined in Section \ref{sec:decTQFT}.

We further expect the full state space of the topological QFT $\CT_{n,k}^A$ on a Riemann surface $\Sigma_g$, with trivial flat connection, to correspond to the infinite-dimensional space of \emph{derived} conformal blocks $\Psi^{\rm der}_{n,k}(\Sigma_g)$ of the Feigin-Tipunin algebra $\FT_k(\sln)$. This is significantly harder to access at the moment. In the remainder of this section, we explore a somewhat experimental method for computing its graded Euler character
\be \chi\big[\Psi^{\rm der}_{n,k}(\Sigma)\big](y)=\text{Tr}_{\Psi^{\rm der}_{n,k}(\Sigma)} (-1)^H y^e\,, \ee
where (as usual) $H$ is homological degree, $e$ is weight for the $PGL(n,\C)$ automorphism of $\FT_k(\sln)$, and the fugacity $y$ is a cocharacter of $PGL(n,\C)$. We focus on the triplet, and aim to match the QFT computation \eqref{eqchar}.

Let us begin abstractly with the data $\mathbf t=\{g; (W_1, \nu_1), \dots, (W_m, \nu_m)\}$ of a genus-$g$ surface marked at $m$ points by  the modules $W_1, \dots, W_m$ with orientation $\nu_1, \dots, \nu_m\in\{\pm\}$. We set $W^+ =W$ and $W^- = W^*$, the dual module of $W$. 
For a semisimple modular tensor category $\CC$, one defines 
\begin{equation}
\Phi(\mathbf t) := W_1^{\nu_1} \otimes\dots \otimes W_m^{\nu_m} \otimes R^g, \qquad R = \bigoplus_{i \in I} S_{i} \otimes S_{i}^*\,,  
\end{equation}	
where  $I$ is the set of inequivalent simple modules. The vector space associated to the data $\mathbf t$ is then 
\begin{equation}
\Psi(\mathbf t) =  \text{Hom}_\CC( \id, \Phi(\mathbf t)), \label{HomPhi}
\end{equation}
where $\id$ is the tensor identity.  If $\CC$ is finite but not necessarily semisimple then one replaces $R$ by a coend. This coend satisfies 
\begin{equation}\label{eq:coend}
[R] = \sum_{i\in I} [P_i^* \otimes S_i]
\end{equation}
in the Grothendieck ring of $\CC$ by Theorem 4.11 of \cite{Shimizu17} (where $S_i$ are the simples and  $P_i$ are their projective covers). Moreover, to obtain a derived space of states, one would expect to replace \eqref{HomPhi} with a derived Hom, \emph{a.k.a.} $\text{Ext}^\bullet$.

\begin{figure}[htb]
\centering
\includegraphics[width=4.5in]{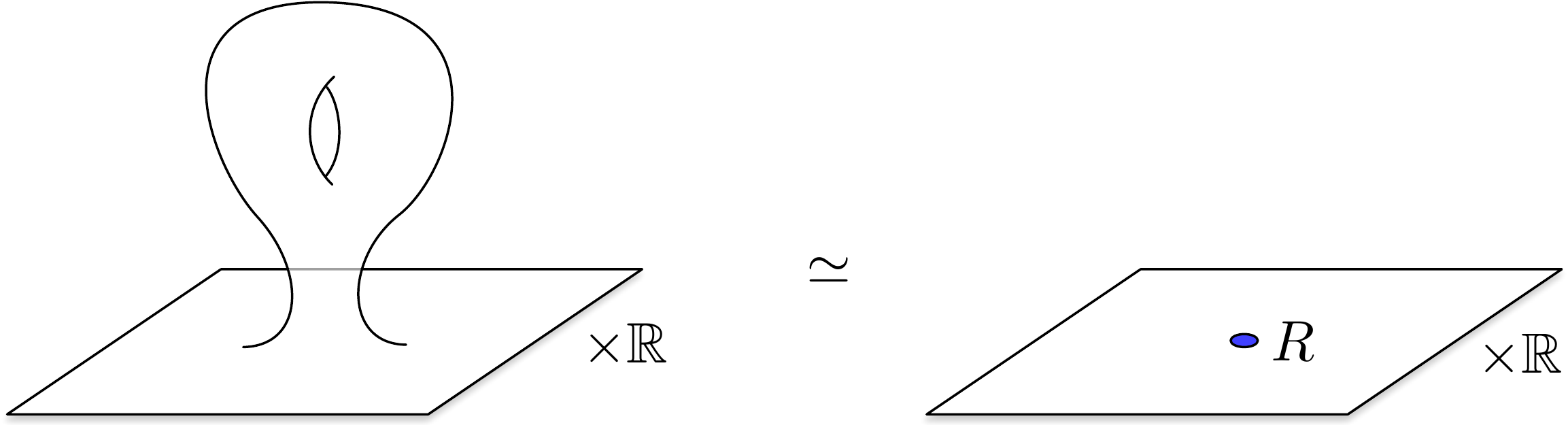}
\caption{The 3d geometry (one holed-torus)$\times\R$ defines a line operator $R$, the coend.}
\label{fig:coend}
\end{figure}

Note that the coend $R$ above may be interpreted as a ``handle-gluing'' object. From the perspective of 3d topological QFT, it is the line operator implicitly defined by placing the theory on a one-holded torus and shrinking the torus to infinitesimal size, as in Figure \ref{fig:coend}.

If $\CC$ is a semisimple modular tensor category, then the dimension of the state space is simply given by
\begin{equation}
\text{dim}\, \Psi(\mathbf t) = \text{dim}(R)^{g-1} \sum_{j \in I} \text{dim}(W_j)^{2(1-g)-m} \prod_{n=1}^m S_{W_n, W_j}\,, \label{Verlinde}
\end{equation}	
where $S_{\bullet, \bullet}$ is the Hopf-link invariant.  (This is often referred to as the Verlinde formula. It is a direct consequence of Verlinde's original formula for fusion rules \cite{Verlinde}.)
More generally, if $\CC$ is finite and allows for a sensible notion of dimension --- in particular, such that the coend $R$ has non-zero dimension --- then one might conjecture that a similar dimension formula still holds for $\Psi(\mathbf t)$. We will now explore this idea a little bit for the triplet.

In general, by ``dimension'' in a vertex tensor category $\CC$, we mean any ring homomorphism from the Grothendieck ring (\emph{a.k.a.} fusion ring) $K_0(\CC)$ to $\C$, or some other algebraic ring.
We have one way to assign abstract dimensions to modules of the triplet algebra, by using its relation to the lattice VOA $V_{\alpha_+\mathbb Z}$. 
Since $V_{\alpha_+\mathbb Z}$ is a vertex-algebra extension of the triplet algebra, there exists a functor $\CF$ from the category of triplet-algebra modules $\hat\CC$ to a category of not-necessarily-local lattice VOA modules.%
\footnote{See \cite{Creutzig:2021cpu} for the use of this functor in order to relate the triplet to a quasi-Hopf modification of the quantum group.}

Let $\CC_V$ be the category of not-necessary-local $V_{\alpha_+\mathbb Z}$-modules. The simple objects are the $V_s$ for integer $s$ with $V_s \cong V_{s+2k}$.  The triplet fusion rule of $V_{\alpha_+\mathbb Z}$ with $S^\pm_s$ descends in the Grothendieck ring $K_0(\CC_V)$ to
\begin{equation}
\begin{split}
[V_{\alpha_+\mathbb Z} \otimes S^+_s] &= [V_{1-s}] \oplus [V_{3-s}] \oplus \dots \oplus [V_{s-1}]\,, \\
[V_{\alpha_+\mathbb Z} \otimes S^-_s] &= [V_{k+1-s}] \oplus [V_{k+3-s}] \oplus \dots \oplus [V_{k+s-1}]\,. 
\end{split}
\end{equation}
This means that the composition factors of the induction of $S^+_s$ are  $V_{1-s}, V_{3-s}, \dots, V_{s-1}$ and similarly for $S^-_s$. The functor $\CF$ is monoidal \cite{Creutzig:2017anl}, so the definition $\text{dim}\,S^\pm_s := \text{dim} [\CF(S^\pm_s)]=s$ is a suitable (but not very sophisticated) candidate for a dimension.

To improve the situation, we deform the dimension.
 Let $x$ be a formal variable, and define a function on the simple and projective objects, valued in $\C[x]$, by
\begin{equation} \label{dimx}
\text{dim}_x(S^\pm_s) := \frac{x^s-x^{-s}}{x-x^{-1}}, \qquad \text{dim}_x(P^\pm_s) := (x^{k-s} + x^{-(k-s)}) \text{dim}_x(S^\pm_k)\,. 
\end{equation}
This does \emph{not} define a ring homomorphism $K_0(\hat\CC)\to \C[x,x^{-1}]$, unless extra relations are imposed. There are several options:
\begin{itemize}
\item Setting $x=1$ recovers the original dimension. It coincides with ordinary dimensions of $u_q(\mathfrak{sl}_2)$ modules.
\item Setting $x$ to be a primitive $2k$-th root of unity recovers the modified quantum dimensions of $u_q(\mathfrak{sl}_2)$ modules introduced by \cite{CGP2}, related to open Hopf-link invariants. This specialization was also used in \cite{Creutzig:2016fms} to derive a logarithmic analogue of Verlinde's formula for fusion rules, and modular S and T transformations of genus-1 conformal blocks. Unfortunately, this gives $\dim_x R=0$.
\item More generally, we may impose the weaker condition $(x^k-x^{-k})^2 \text{dim}_x(S^\pm_k) =0$, \emph{i.e.} $(x^k-x^{-k})^3=0$, to obtain a ring homomorphism
\be \text{dim}_x: \, K_0(\hat \CC) \to \C[x,x^{-1}]/((x^k-x^{-k})^3) \ee
Now we compute $\text{dim}_x(S^\pm_s)\text{dim}_x(P^\pm_s)+\text{dim}_x(S^\pm_{k-s})\text{dim}_x(P^\pm_{k-s})= 2 \text{dim}_x(S^\pm_k)$, and happily find that
\begin{equation}
\begin{split}
\text{dim}_x(R) &= \sum_{s=1}^k \text{dim}_x(S^+_s)\text{dim}_x(P^+_s) + \text{dim}_x(S^-_s)\text{dim}_x(P^-_s) \\ &= 2k (\text{dim}_x(S^\pm_k))^2=2k \left( \frac{x^k-x^{-k}}{x-x^{-1}}\right)^2 \neq 0\,.
\end{split}
\end{equation}
\end{itemize}
With this final, weakest specialization, we find that the ``dimension'' of the space associated to a smooth (unmarked) genus-$g$ surface is
\begin{equation}
\begin{split}
\dim_x\, \Psi(\Sigma_g) &= \text{dim}_x(R)^{g-1} \sum_{\substack{ s =1 \\ \epsilon \, \in \, \{\pm\}}}^{k} \text{dim}(S_s^\epsilon)^{2(1-g)} 
= 2\sum_{s =1}^{k}  \left[ \frac{1}{2k} \left( \frac{x^s-x^{-s}}{x^k-x^{-k}}\right)^2\right]^{(1-g)}\,.
\end{split}
\end{equation}

This final formula resembles \eqref{eqchar} for the Euler character of the QFT state space, but it does not coincide with it. We can obtain a perfect match by implementing one final modification: we replace the sum over dimensions of simple modules by a sum over pivotal structures, \ie\ a sum over all possible choices of traces and thus of dimensions. 

Any pivotal structure in a tensor category gives rise to a trace and hence a notion of a dimension. Any $2k$-th root of unity $\zeta$ defines a pivotal structure $a_\zeta$ that acts as multiplication by $\zeta^s$ on $V_s$ and hence changes the dimension of $V_s$ by the factor $\zeta^s$.  Let us denote the dimension associated to $a_\zeta$ by $\text{dim}_{x, \zeta}$. Then 
\begin{equation}
\begin{split}
\text{dim}_{x, \zeta}(S^+_s)&= \frac{x^s\zeta^s-x^{-s}\zeta^{-s}}{x\zeta-x^{-1}\zeta^{-1}}, \\ \text{dim}_{x, \zeta}(P^+_s) &= (x^{k-s}\zeta^{k-s} + x^{-(k-s)}\zeta^{-(k-s)} ) \text{dim}_{x, \zeta}(S^+_k), \qquad  \\
\text{dim}_{x, \zeta}(S^-_s) &= \zeta^k \text{dim}_{x, \zeta}(S^+_s), \qquad \text{dim}_{x, \zeta}(P^-_s) = \zeta^k\text{dim}_{x, \zeta}(P^+_s).
\end{split}
\end{equation}
We then have
\begin{equation}
\begin{split}
\text{dim}_{x, \zeta}(R) &= \sum_{s=1}^p \text{dim}_{x, \zeta}(S^+_s)\text{dim}_{x, \zeta}(P^+_s) + \text{dim}_{x, \zeta}(S^-_s)\text{dim}_{x, \zeta}(P^-_s) \\ &= 2k (\text{dim}_{x, \zeta}(S^\pm_k))^2=2k (S^x_{S^\pm_k, V_r})^2 = 2k \left( \frac{x^k-x^{-k}}{x\zeta-x^{-1}\zeta^{-1}}\right)^2.
\end{split}
\end{equation}
and hence the sum over $(\text{dim}_{x, \zeta}(R))^{g-1}$ for all possible $\zeta$ gives
\begin{equation}
\label{eqchar2}
\sum_{\{ \zeta\,|\, \zeta^{2k} = 1\}}(\text{dim}_{x, \zeta}(R))^{g-1} = \sum_{\{ \zeta\,|\, \zeta^{2k} = 1\}} \left[\frac{1}{2k}  \left(\frac{\zeta x - x^{-1} \zeta^{-1}   }{x^k-x^{-k}}\right)^2 \right]^{1-g}.
\end{equation}
Now if we identify $x^k = y^{\frac{1}{2}}$, we recover \eqref{eqchar}.

In summary we have found a quantity that reproduces the  Euler character of the fully derived space of states on $\Sigma_g$. At the moment we view this as an experimental observation that deserves further investigations. We note that $\text{dim}_x$ can be motivated using resolutions and so the defromation by $x$ somehow accesses derived structure.

\subsubsection{Flat connections and twisted module categories}
\label{sec:FTtwisted}

We now return to the general case $\g=\sln$. 
We saw that in the field theory $\CT_{n,k}^A$, the category of line operators could be deformed by a $PGL(n,\C)$ flat connection on an infinitesimal punctured disc. The category itself (ignoring braided tensor structure) only depends on the conjugacy class of the holonomy, due to \eqref{C-iso} in Section \ref{sec:flavor-line}. Moreover, for generic holonomy --- diagonalizable with distinct eigenvalues --- we expect the category to become semisimple.

We'd like to match this structure from a VOA perspective. We explained in Section \ref{sec:VOAnk} how a deformation by a diagonal connection manifests in the boundary VOA $\CN_{n,k}$, deforming its OPE's. Here we'll consider the boundary VOA $\CD_{n,k} \simeq \FT_k(\sln)$ instead.

It is known (see \emph{e.g.} Section 16 of \cite{FBZ}), that the dual of an affine Lie algebra is isomorphic to the space of connections on the trivial $G$-bundle on the punctured disc $\text{Spec}\,\C(\!(z)\!)$ \cite[Lemma 16.4.3]{FBZ}. Moreover, the large level limit of an affine vertex algebra can be taken in such a way that the limit is a commutative vertex algebra, which inherits a vertex Poisson algebra structure from the affine VOA. There is a notion of local Lie algebra attached to this Poisson vertex algebra and this local Lie algebra is isomorphic to local functionals on connections on $\text{Spec}\,\C(\!(z)\!)$ \cite[Lemma 16.4.5]{FBZ}. 

As explained in the examples of $\g = \mathfrak{gl}_1$ and $\g = \mathfrak{sl}_2$ (with $k=2$) in Sections \ref{sec:abelian} and \ref{sec:osp} the large level limit of the VOA's  $\CD^\psi_{n,k}$ can be taken in a way such that the commutative vertex algebra appears in the operator products of the limit VOA. Quotienting by the large center (the commutative vertex algebra) yields conjecturally the Feigin-Tipunin algebras $ \FT_k(\sln)$. 
Given that the Poisson vertex algebra structure on the commutative vertex algebra is identified with functionals on $G$-connections on the punctured disc 
 we view these limits as Feigin-Tipunin algebras $\FT_k(\sln)$ coupled to flat connections.
Modules of $\CD^\psi_{n,k}$ become modules of the Feigin-Tipunin algebras $\FT_k(\sln)$ coupled to flat connections in the large level limit. 

The holonomy $g$ of a flat connection should be identified with an inner automorphism of the center and thus should give rise to modules of the center that have  monodromy $g$. These modules should be coupled to $g$-twisted modules of the Feigin-Tipunin algebras $\FT_k(\sln)$ in such a way that monodromies cancel, i.e. the resulting module should be a local module for  the Feigin-Tipunin algebras $\FT_k(\sln)$ coupled to flat connections. These are expectations that we don't understand in detail yet, but they are motivated from our abelian example of Section \ref{sec:abelian}.

\subsubsection{Modules in the presence of abelian flat connections}
\label{sec:monodromy-modules}

Modules of VOA's are local, in the sense that the OPE of a field $V(z)$ of the VOA $V$ with a field $M(w)$ of the module $M$ is a Laurent series in $(z-w)$, of the form
\begin{equation}
V(z) M(w) = \sum_{n \geq -N} M_n(w) (z-w)^n
\end{equation}
for certain fields $M_n(w)$ of the module $M$. In particular for $\gamma_z : [0, 1] \rightarrow S^1_z$ a  loop around $z$ with $\gamma_z(0) = \gamma_z(1) = w$ the OPE doesn't have any monodromy, that is 
\begin{equation}
\lim_{x \rightarrow 0} V(z) M(\gamma_z(x)) = \lim_{x \rightarrow 1} V(z) M(\gamma_z(x))\,.
\end{equation} 

Let $\mu \in \mathbb C/\mathbb Z$ and $\lambda = e^{2\pi i \mu}$. Then a $\lambda$-twisted module $M$ has fields $M(w)$
with OPE's of the form 
\begin{equation}
V(z) M(w) = \sum_{n \geq -N} M_n(w) (z-w)^{n+\mu}\,,
\end{equation}
so that in this case there is a monodromy of $\lambda^\pm$, depending on the orientation of the loop:
\begin{equation}
\lambda^{\pm 1}\lim_{x \rightarrow 0} V(z) M(\gamma_z(x)) = \lim_{x \rightarrow 1} V(z) M(\gamma_z(x))\,.
\end{equation} 
We fix the orientation so that the monodromy is $\lambda$. This is an example of a $U(1)$-monodromy and as we will now see in the example of twisted modules of lattice VOA's, the monodromy can also depend on the chosen field $M(w)$ in $M$.

The simplest example is a lattice VOA $V=V_\Lambda$ for some even lattice $\Lambda$. More generally of $\Lambda$ is an integral lattice, then $V_\Lambda$ is a vertex operator superalgebra. In Section \ref{sec:abelian} we considered the case $\Lambda = \sqrt{k} \mathbb Z$. 
 Recall that 
\begin{equation}
V_\Lambda = \bigoplus_{\nu \in \Lambda}  \pi_\nu
\end{equation}
with $\pi_\nu$ the Fock module of the Heisenberg vertex subalgebra of weight $\nu$.
The Heisenberg VOA fusion rules $\pi_\nu \boxtimes \pi_\mu = \pi_{\nu+\mu}$ correspond to the OPE \cite{FBZ}
\begin{equation}
\phi_\nu(z) \phi_\mu(w) = (z-w)^{\nu \mu} :\!\phi_\lambda(z) \phi_\mu(w)\!:
\end{equation}
of the fields $\phi_\nu(z), \phi_\mu(w)$ associated to the top-level vectors of the Fock modules. 
An example of a (potentially) twisted module is
\begin{equation}
V_{\Lambda+\mu} := \bigoplus_{\nu \in \Lambda+\mu}  \pi_\nu\,.
\end{equation}
Let $\alpha_1, \dots, \alpha_r$ be a $\mathbb Z$-basis of $\Lambda$; then the monodromy can be characterized by numbers
$\lambda_1 = e^{2\pi i \alpha_1 \mu}, \dots, \lambda_r = e^{2\pi i \alpha_r \mu}$, that is by an element $\vec{\lambda}$ in $(\mathbb C^*)^r$. 
The monodromies of $V_{\Lambda+\mu}$ and $V_{\Lambda+\mu'}$ with $V_\Lambda$ coincide if and only if $\mu \alpha_i = \mu' \alpha_i \mod \mathbb Z$ for all $i=1, \dots, n$; that is $\mu = \mu' \mod \Lambda'$ with $\Lambda'$ the lattice dual to $\Lambda$. 
Let us restrict to monodromy valued in the unit circles $(S^1)^r$ (this ensures that vertex tensor category theory applies, as this theory requires real conformal weights, which means unitary monodromy).
It follows then from vertex tensor category theory (see Example \ref{ex:freeboson} of Section \ref{sec:tensor}) that the $V_{\Lambda+\mu}$ are simple;  that every simple twisted module is of this form; and that
$V_{\Lambda+\mu} = V_{\Lambda+\mu'}$ if and only if $\mu = \mu' \mod \Lambda$, so that inequivalent $\vec{\lambda}$-twisted modules are parameterized by the set $\{ \mu' \in \mathbb C^r/\Lambda | \mu = \mu' \mod \Lambda' \} \cong \Lambda'/\Lambda$. 

Now take the lattice $\Lambda = \sqrt{k}Q$, so that $\FT_k(\g)\subset V_\Lambda$. We conjecture that the $\vec{\lambda}$-twisted modules of $\FT_k(\mathfrak{g})$ and $V_\Lambda$ coincide for generic $\vec{\lambda}$. In particular the category of $\vec{\lambda}$-twisted modules for generic $\vec{\lambda}$ is semisimple with 
\be
|\Lambda'/\Lambda| = \left| (\sqrt{k}Q)'/\sqrt{k}Q\right| = \left| \frac{1}{\sqrt{k}}Q' /\sqrt{k} Q \right| = k^{\text{rank}(Q)} |Q'/Q|
\ee
simple modules. For $\mathfrak{g} = \mathfrak{sl}_n$ this is 
\be \label{eq:twistedmodules}
|\Lambda'/\Lambda| =  k^{n-1} n\,.
\ee
We observe that this coincides with the counting of Bethe roots in $\CT_{n,k}^A$, as in \eqref{count-PkQ}, \eqref{eq:genericindextypeA}.

Note that the conjecture follows from work in progress \cite{CKMY} for the case $\mathfrak{g} = \mathfrak{sl}_2$ (see Example \ref{ex:singlet} of Section \ref{sec:tensor}). In this case all $\vec{\lambda} \in S^1\setminus \{1\}$ are generic. The higher-rank proof requires an understanding of representation theory of the higher-rank VOA's that needs to be developed.

\subsection{$\CN_{n,k}$ and the Langlands kernel}
\label{sec:A-N}

We now turn to the second boundary VOA for $\CT_{n,k}^A$, denoted $\CN_{n,k}$, and initially defined in Section \ref{sec:VOAnk} using a Neumann-like boundary condition in field theory. We would ultimately like to argue that (after a slight modification) $\CN_{n,k}$ and the Feigin-Tipunin algebra $\CD_{n,k}$ have equivalent braided tensor categories of modules, and that this equivalence is induced by a level-rank-like duality, wherein $\CN_{n,k}$ and $\CD_{n,k}$ appear as mutual commutants inside $nk$ copies of free fermions. We need to build up some technology to get to this statement.

The VOA $\CN_{n,k}$ was constructed by starting with the boundary VOA
\be A(\sln) := \CV_{T[SU(n)]}^A\,. \ee
for the A-twisted theory $T[SU(n)]^A$, tensoring with a number of free fermions, and taking derived $SL(n,\C[\![z]\!])$  invariants. We argued in Section \ref{sec:Tnk-VOA} that, as long as the Chevalley-Eilenberg (ghost) complex computing derived invariants has no higher cohomology, the process of taking derived invariants should be equivalent to taking a coset, resulting in
\be \CN_{n,k} \simeq \text{Com}\big( V^k(\sln), A(\sln) \otimes \FF(n(k-1))\big)\,. \label{N-Com} \ee

We will take \eqref{N-Com} as a working definition of $\CN_{n,k}$ in this section. Our plan is to use identities among deformable families of corner VOA's to produce a decomposition of $\CV_{T[SU(n)]}^A \otimes \FF(n(k-1))$ that makes the coset of $V^k(\sln)$ manifest --- and thus leads to a putative decomposition of $\CN_{n,k}$.
We will approach this  by generalizing $\CN_{n,k}$ to a deformable family $\CN_{n,k}^\psi$.

\subsubsection{The kernel $A^\psi(\g)$}
\label{sec:kernel}

The A-twisted boundary VOA $ A(\sln)=\CV_{T[SU(n)]}^A$ was first constructed and studied in \cite{CostelloGaiotto,CCG}. It is the classical Langlands duality kernel. The original construction via BRST reduction of beta-gamma systems, reviewed in Section \ref{sec:VOAnk}, is unfortunately not too enlightening. A more useful definition comes by relating $A(\sln)$ to corner VOA's.

The 3d theory $T[SU(n)]$ is engineered by sandwiching an S-duality interface between two Dirichlet boundary conditions $\wt B_{0,1}$ for 4d $\CN=4$ Yang-Mills theory. Colliding the interface with one of the two boundary conditions yields $SU(n)$ Yang-Mills sandwiched between one $\wt B_{0,1}$ boundary and one $\wt  B_{1,0}$ boundary (its S-dual). The VOA $A(\sln)$ should then be supported at a corner
\be  \raisebox{-.4in}{\includegraphics[width=3in]{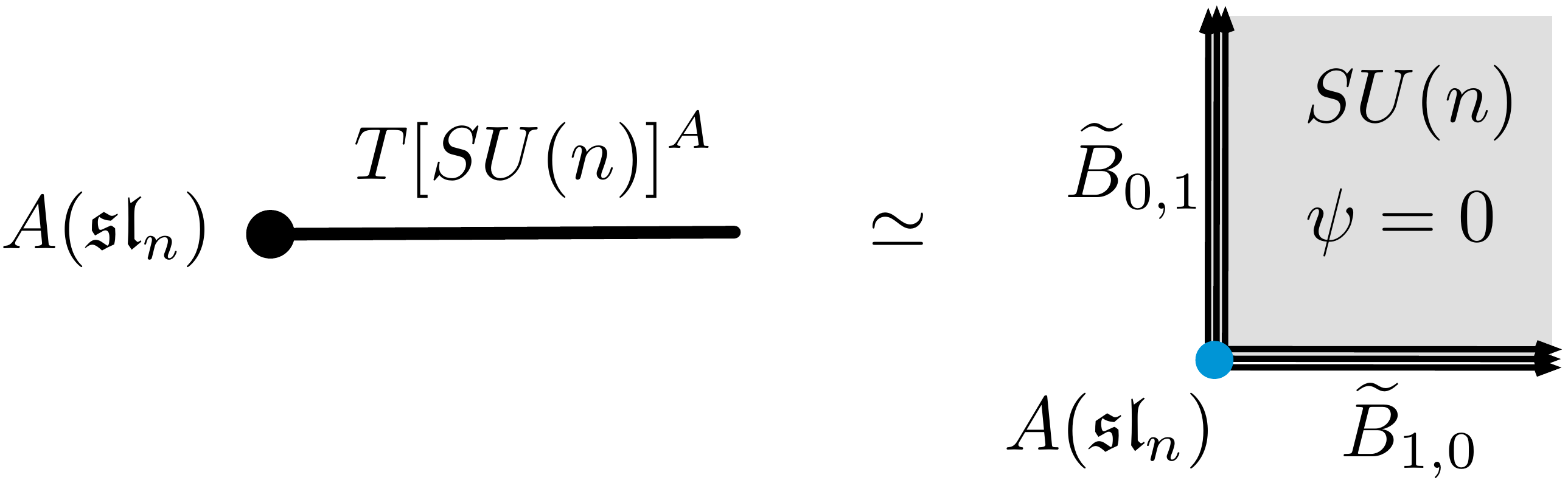}} \ee
with bulk twisting parameter $\psi=0$, as appropriate for the 3d A-twist of $T[SU(n)]$.

We expect $A(\sln)$ to arise as a limit of the deformable family of quantum Langlands kernel VOA's $A^{\psi+1}(\sln)$ defined by the corner on the RHS of \eqref{corner-fund}:
\be A(\sln) \otimes Z(\sln) = \lim_{\psi\to 0} A^{\psi+1}(\sln)\,. \ee
The limit has a large vertex-Poisson-algebra center $Z(\sln)$, which must be factored out, just like in the limit of Section \ref{sec:Dnk}. (Such a center appears whenever there is a $\wt B_{1,0}$ b.c. and the bulk parameter tends to zero; or, dually, when there is a Dirichlet $\wt B_{0,1}$ b.c. and the bulk parameter tends to infinity.)

The corner VOA, in turn, may be resolved as an extension of two affine algebras, by intersecting with a third boundary condition:
\be \raisebox{-.6in}{\includegraphics[width=3.7in]{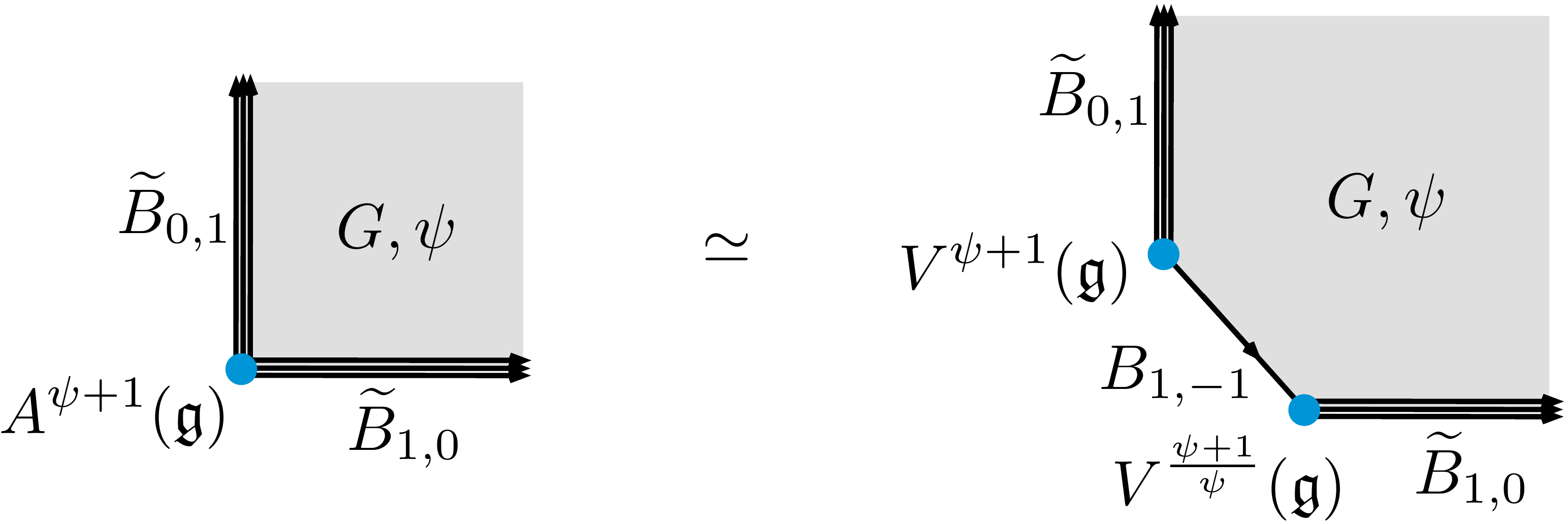}} \label{corner-Ares} \ee
The levels of the two affine algebras are determined to be $\psi+1$ and $\phi = \frac{\psi+1}{\psi}$ by applying suitable $SL(2,\Z)$ transformations to bring the new corners to canonical form:
\be \raisebox{-.6in}{\includegraphics[width=5.5in]{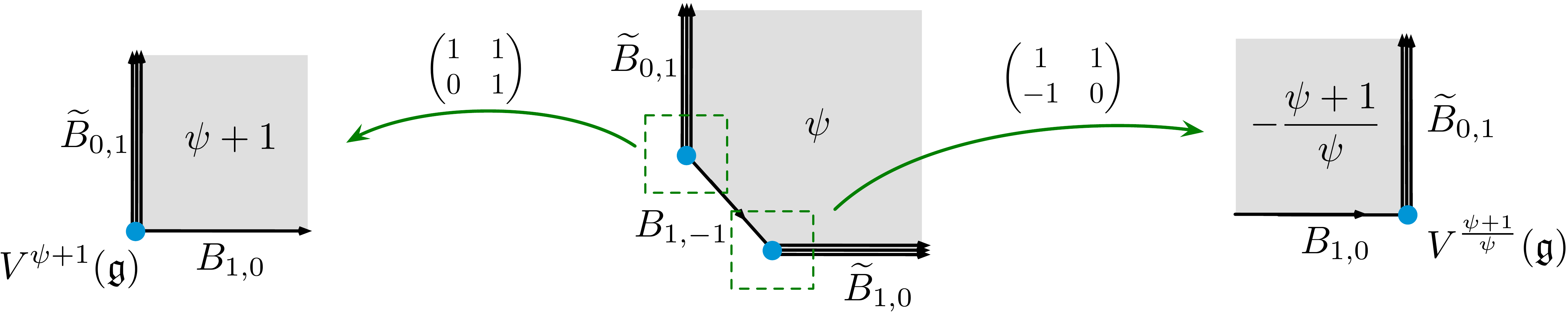}} \label{corner-Ag} \ee
The precise extension corresponding to the corner VOA takes the form \cite{CreutzigGaiotto-S}
\begin{equation}\label{eq:FTpsi}
	A^{\psi+1}(\sln) = \bigoplus_{\lambda \in P^+} V^{\psi+1}_\lambda \otimes V^\phi_\lambda \otimes L^{(n)}_{s(\lambda)}\,, \qquad \text{with} \ \ \frac{1}{\psi+1} + \frac{1}{\phi} =1\,.
\end{equation}
Here $L^{(n)}_{s(\lambda)}$ are modules for an additional lattice VOA $L^{(n)} = L_{\sqrt n \Z}$. Letting $\omega_1,\ldots,\omega_{n-1}$ be the usual fundamental weights, and setting $\omega_0=0$, the function $s:P^+\to \{0,\ldots n-1\}$ is defined by $s(\lambda) = r$ if $\lambda = \omega_r \mod Q$. Note that the decomposition \eqref{eq:FTpsi} makes manifest the symmetry
\be A^{\psi+1}(\sln) \simeq A^{\psi^{-1}+1}(\sln)\,. \label{A-dual} \ee

These VOA's --- taking \eqref{eq:FTpsi} as a definition --- were further studied in \cite{FrenkelGaiotto} and employed to understand trialities of $W$-algebras \cite{Creutzig:2020zaj,Creutzig:2021dda} and equivalences of module categories \cite{Creutzig:2021bmz}.
When $n=2$, the family $A^{\psi+1}(\mathfrak{sl}_2) = L_1(\mathfrak{d}(1, 2; -\psi))$ is the simple affine vertex superalgebra of the exceptional family $\mathfrak{d}(1, 2; -\psi)$ of simple Lie superalgebras at level one \cite{CreutzigGaiotto-S}. For higher rank it is an open and important conjecture (mathematically) that these objects can be given the structure of a deformable family of simple vertex superalgebras.

The limit relevant for the 3d A-twist of $T[SU(n)]$ is $\psi\to 0$, forcing $\phi\to \infty$.
The limit can be taken in such a way that $V^\phi(\mathfrak{sl}_n)$ becomes a Poisson vertex algebra $Z(\mathfrak{sl}_n)$, and it is then conjectured that
\begin{equation}
	\lim_{\psi \rightarrow 0} A^{\psi+1}(\sln) = Z(\mathfrak{sl}_n) \otimes  
	\bigoplus_{\lambda \in P^+} R_\lambda \otimes V^1_\lambda \otimes L^{(n)}_{s(\lambda)}\,,
\end{equation}
whence
\be A(\sln) = \bigoplus_{\lambda \in P^+} R_\lambda \otimes V^1_\lambda \otimes L^{(n)}_{s(\lambda)}\,. \label{An-decomp} \ee
Moreover, $A(\sln)$ is conjecturally a simple vertex superalgebra.
In the case of $\mathfrak{sl}_2$ both conjectures are true \cite{CreutzigGaiotto-S}.

The decomposition \eqref{An-decomp} was also conjectured to hold for the A-twisted VOA $\CV_{T[SU(n)]}^A$ defined by BRST reduction of beta-gamma systems in \cite{CostelloGaiotto,CCG}. It was supported there by computations of characters.

\subsubsection{Iterated slicing and the coset}
\label{sec:iterate}

\begin{figure}[htb]
\centering
\includegraphics[width=6in]{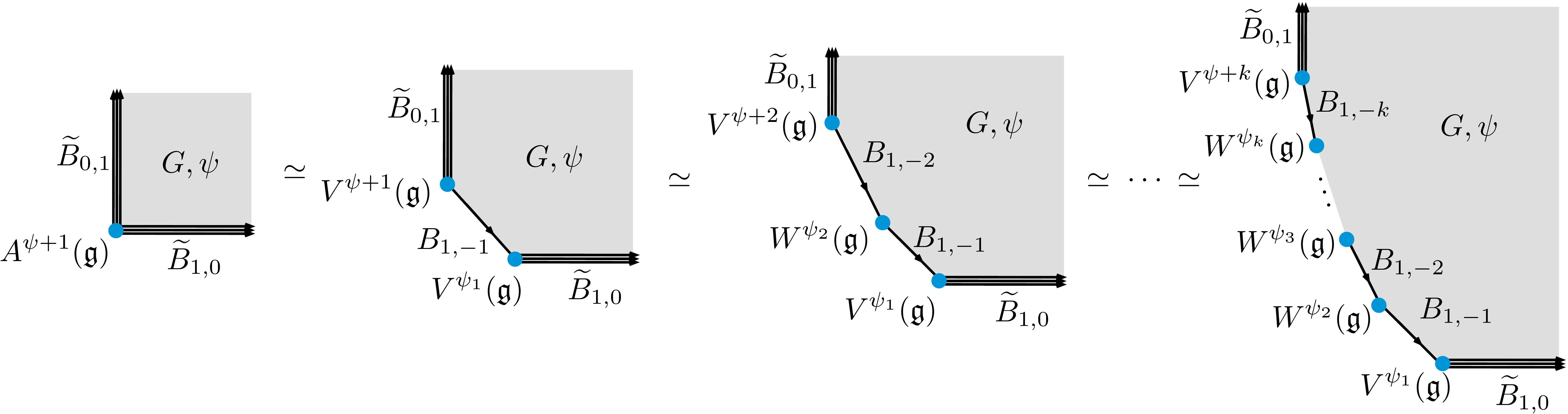} \vspace{-.3cm}
\caption{Iterated slices of the Langlands-kernel corner.}
\label{fig:corner-Ak}
\end{figure}

In order to make the coset \eqref{N-Com} explicit, we can keep slicing/regularizing the corner on the RHS of \eqref{corner-Ares} until a copy of $V^k(\g)$ appears in the limit $\psi\to 0$. We achieve this by introducing a sequence of boundary conditions $\wt B_{1,-1},\wt B_{1,-2},..., \wt B_{1,-k}$, as in Figure \ref{fig:corner-Ak}. Schematically, we expect
\be A^{\psi+1}(\g) \sim V^{\psi+k}\otimes W^{\psi_k}\otimes W^{\psi_{k-1}}\otimes \cdots \otimes W^{\psi_2} \otimes V^{\psi_1}\,, \ee
with parameters
\be \psi_r:= \frac{\psi+r}{\psi+r-1} \qquad\text{satisfying} \qquad  \frac{1}{\psi+r}+\frac{1}{\psi_r} = 1\,. \ee

To make this precise, we use the main Theorem of \cite{Arakawa:2018iyk} for $\mathfrak g = \sln$. The theorem says that, for non-rational $\psi$ and also over $\C(\psi)$,
for any $\lambda,\nu\in P^+$
\begin{equation}\label{eq:cosettheorem} 
	V^{\psi-1}_\lambda \otimes V_{Q + \nu} = \bigoplus_{\substack{\mu \in P^+ \\  \lambda + \mu + \nu \, \in\, Q}} V^{\psi}(\mu) \otimes \M^{\phi}_{\mu, \lambda}\,, \qquad \text{with} \quad \frac{1}{\psi} + \frac{1}{\phi}=1\,.
\end{equation}
This describes not just the decomposition of $V^{\psi-1}$ upon slicing, but of its modules. Note that the lattice VOA $V_Q$ coincides with the simple affine vertex algebra of $\sln$ at level one. The cosets of the root lattice $Q$ in the weight lattice $P$ are parameterized by zero and the fundamental weights $\omega_1, \dots, \omega_{n-1}$. We set $\omega_0 := 0$.

Let $\FF(m)$ be the vertex superalgebra of $m$ pairs of free fermions, and recall that by bosonization $\FF(m) \cong V_\Lambda$ for $\Lambda = \mathbb Z^m$. In particular,
\be
\FF(n) \cong  \bigoplus_{i=0}^{n-1} V_{Q + \omega_i} \otimes L^{(n)}_i
\ee
as a $V_Q \otimes L^{(n)}$ module (recall that $L^{(n)} = V_{\sqrt{n}\mathbb Z}$). 
Iterating the coset theorem \eqref{eq:cosettheorem} we get
\begin{equation}
	\begin{split}
		V^{\psi+1}_{\lambda_1} \otimes \FF(n) &= \bigoplus_{\lambda_2 \in P^+} V^{\psi+2}_{\lambda_2} \otimes \M^{\psi_2}_{\lambda_2, \lambda_1} \otimes L^{(n)}_{s(\lambda_1-\lambda_2)} \\
		V^{\psi+1}_{\lambda_1} \otimes \FF(2n) &= \bigoplus_{(\lambda_3, \lambda_2) \in (P^+)^2} V^{\psi+3}_{\lambda_3} \otimes \M^{\psi_3}_{\lambda_3, \lambda_2} \otimes \M^{\psi_2}_{\lambda_2, \lambda_1} \otimes L^{(n)}_{s(\lambda_2-\lambda_3)} \otimes L^{(n)}_{s(\lambda_1-\lambda_2)} \\
		& \vdots \\
		V^{\psi+1}_{\lambda_1} \otimes \FF((k-1)n) &= \bigoplus_{(\lambda_k,\lambda_{k-1},...,\lambda_2) \in (P^+)^{k-1}} V^{\psi+k}_{\lambda_{k}} \otimes \M^{\vec{\psi}}_{\vec{\lambda}}\otimes L_{\vec{\lambda}}\,,
	\end{split}
\end{equation} 
with $\vec{\psi}= (\psi_{k}, \psi_{k-1}, \dots, \psi_2)$, $\vec{\lambda}= (\lambda_{k}, \lambda_{k-1}, \dots, \lambda_2,\lambda_1)$, and
\be \begin{split}
		\M^{\vec{\psi}}_{\vec{\lambda}} &:= \M^{\psi_{k}}_{\lambda_k, \lambda_{k-1}} \otimes \M^{\psi_{k-1}}_{\lambda_{k-1}, \lambda_{k-2}} \otimes \dots \otimes \M^{\psi_2}_{\lambda_2, \lambda_1} \\
		L_{\vec{\lambda}} &:= L^{(n)}_{s(\lambda_{k-1}-\lambda_k)} \otimes L^{(n)}_{s(\lambda_{k-2}-\lambda_{k-1})} \otimes \dots \otimes L^{(n)}_{s(\lambda_1-\lambda_2)} \,.
	\end{split}
\end{equation} 
Combined with \eqref{eq:FTpsi}, it follows that
\begin{equation}
	\begin{split}
		A^{\psi+1}(\sln) \otimes \FF((k-1)n) &=  \bigoplus_{\lambda_1 \in P^+} V^{\psi+1}_{\lambda_1} \otimes \FF((k-1)n)\otimes V^{\psi_1}(\lambda_1) \otimes L^{(n)}_{s(\lambda_1)} \\
		&= \bigoplus_{\vec{\lambda} \in (P^+)^{k}} V^{\psi+k}_{\lambda_k} \otimes \M^{\vec{\psi}}_{\vec{\lambda}}
		\otimes V^{\psi_1}(\lambda_1) \ \otimes L_{(\vec{\lambda}, 0)}\,.
	\end{split}
\end{equation}

From here, we can explicitly identify the deformable family of cosets
\begin{align} \CN_{n,k}^\psi &:= \text{Com}\big( V^{\psi+k}(\sln),A^{\psi+1}(\sln)\otimes \FF((k-1)n)\big) \\
 &= \bigoplus_{\vec{\lambda} = (\lambda_{k-1}, \dots, \lambda_1) \in (P^+)^{k-1}}  \M^{\vec{\psi}}_{(0,\vec{\lambda})}
		\otimes V^{\psi_1}_{\lambda_1} \ \otimes L_{(0, \vec{\lambda}, 0)}\,. \label{eq:Cpsi}
\end{align}
By Theorem 8.1 of \cite{Creutzig:2014lsa} together with the proof of Theorem 4.3 of \cite{ACK21}, the simple quotient of the limit $\lim_{\psi\to 0}\CN_{n,k}^\psi$ (in particular, with a large $Z(\sln)$ center removed) will satisfy
\be \big[\lim_{\psi\to 0} \CN_{n,k}^\psi\big]_{\text{simp}} = \text{Com}\big( V^{k}(\sln),A(\sln)\otimes \FF((k-1)n)\big) = \CN_{n,k}\,. \ee
Naively, one would expect that
\be \label{N-naive}
\CN_{n,k} \overset{\text{naive}}{=} \bigoplus_{\vec{\lambda} = (\lambda_{k-1}, \dots, \lambda_1) \in (P^+)^{k-1}}  \M^{\vec{\psi}}_{(0, \vec{\lambda})}
\otimes R_{\lambda_1} \ \otimes L_{(0, \vec{\lambda}, 0)}\,,
\ee
where in this final formula $\vec\psi = (\psi_k,...,\psi_2) = \big(\frac{k}{k-1},\frac{k-1}{k-2},...,\frac{2}{1})$\,.
However, 
since the levels $\psi_r$ are non-generic there might be non-trivial extensions beyond modules, \ie\ the right-hand side might actually not be a direct sum decomposition. 
Nevertheless, our naive expectation \eqref{N-naive} is at least an identity in the Grothendieck ring of the Deligne product of the categories of the underlying principal W-algebras and lattice VOA's.

\subsection{``Level-rank'' duality}
\label{sec:levelrank}

The free fermion VOA $\FF(kn)$ has the simple vertex algebras of $\mathfrak{sl}_n$ at level $k$ as subalgebra and also the simple vertex algebras of $\mathfrak{sl}_k$ at level $n$. There is also an additional lattice VOA $V_{\sqrt{kn}\mathbb Z}$. This implies an braid-reversed equivalence between representation categories of $\mathfrak{sl}_n$ at level $k$ and $\mathfrak{sl}_k$ at level $n$ \cite{MR3162483}. Due to the additional lattice VOA this relation
is a bit subtle to formulate and one needs cyclic orbifolds and simple currents for that.
 We seem to encounter a similar subtlety when seeking a level-rank duality involving $\FT_{k}(\mathfrak{sl}_n)$ and $\CN_{n,k}$.
In fact, whenever one has a mutually commuting pair of VOAs inside a VOA with trivial representation category (like the free fermions) and certain technical vertex tensor category assumptions are satisfied then one has a braid-reversed equivalence of categories of modules of the two VOAs \cite{Creutzig:2019psu}.

We conjecture that there exists an embedding of $\FT_{k}(\mathfrak{sl}_n)$ in $\FF(nk)$, as well as an embedding of a slight modification $\wt{\CN}_{n,k}$ of $\CN_{n,k}$ in $\FF(nk)$. The modification $\wt{\CN}_{n,k}$ is obtained 
 as a $\Z_n$ orbifold of a simple current extension of $\CN_{n,k}$, and is such that $\widetilde \CN_{n,k}$ and $\CN_{n,k}$ have equivalent linear categories, though their associators might differ.  Our main conjecture is then the following:

\vspace{5mm}
\noindent{\bf Conjecture}
\begin{enumerate}
	\item $\FT_{k}(\mathfrak{sl}_n)$ and $\widetilde \CN_{n,k}$ form a mutually commuting pair inside $\FF(nk)$.
	\item $\FF(nk)$ is projective as a  $\FT_{k}(\mathfrak{sl}_n)$-module as well as an $\widetilde \CN_{n,k}$-module.
	\item There is a braid-reversed equivalence $\tau$ between $\FT_{k}(\mathfrak{sl}_n)$-mod and $\widetilde \CN_{n,k}$-mod.
	\item For a simple $\FT_{k}(\mathfrak{sl}_n)$-module $S$ denote by $P_S$ its projective cover, then
	\begin{equation}\label{eq:decleft}
		\FF(nk) = \bigoplus_S P_S \otimes \tau(S)^* 
	\end{equation}
	as a $\FT_{k}(\mathfrak{sl}_n)$-module
	and  
	\begin{equation}\label{eq:decright}
		\FF(nk) = \bigoplus_S S \otimes \tau(P_S)^* 
	\end{equation}
	as a $\widetilde \CN_{n,k}$-module. Here the sums are over all inequivalent simple objects in $\FT_{k}(\mathfrak{sl}_n)$-mod.
\end{enumerate}

\begin{rmk}
The central charges of Feigin-Tipunin algebra and $\CN_{n,k}$ (or $\wt \CN_{n,k}$) are
\begin{equation}
\begin{split}
c(\FT_{k}(\mathfrak{sl}_n)) &= -\frac{n^3-n}{k} + 2(n^3-n) +n-1 - k(n^3-n) \\
 c(\CN_{n,k}) &= nk -n^3+1 +\frac{n^3-n}{k}
\end{split}
\end{equation}
so that 
\begin{equation}
c(\FT_{k}(\mathfrak{sl}_n)) +  c(\CN_{n,k}) =  nk - (n^3-n)(k-1)
\end{equation}
which is exactly the central charge of $\FF(nk)$ with twisted conformal vector. This twist is due to a coset realization of $W$-algebras (that we will use), where the conformal vector of the lattice vertex algebra involved is twisted \cite{Arakawa:2020oqo}. 
\end{rmk}

We now explain our understanding that leads to the conjecture:
\begin{itemize}
	\item In subsection \ref{sec:dual-def} we derive a duality of deformable families of VOAs. Taking a suitable limit then gives us information about decompositions of $\FF(kn)$. 
	\item In subsection \ref{sec:tensor} we explain how cosets, vertex algebra extensions and braid-reversed equivalences of vertex tensor categories interplay.
	\item In subsection \ref{sec:SF} we discuss the example of two pairs of symplectic fermions embedding into two pairs of free fermions. We see that there is a family of embeddings and for the generic embedding we get exactly a decomposition of type \eqref{eq:decleft} and  \eqref{eq:decright}.
	\item In subsections \ref{sec:vir}, \ref{sec:genericcoset} and \ref{sec:coset} we consider the case of $\sltwo$. In particular we make branching rule observations, see \eqref{eq:branching},  that indicate a decomposition of the form of \eqref{eq:decright}.
\end{itemize}

\subsubsection{Duality of deformable families}
\label{sec:dual-def}

\begin{figure}[htb]
\centering
\includegraphics[width=5.5in]{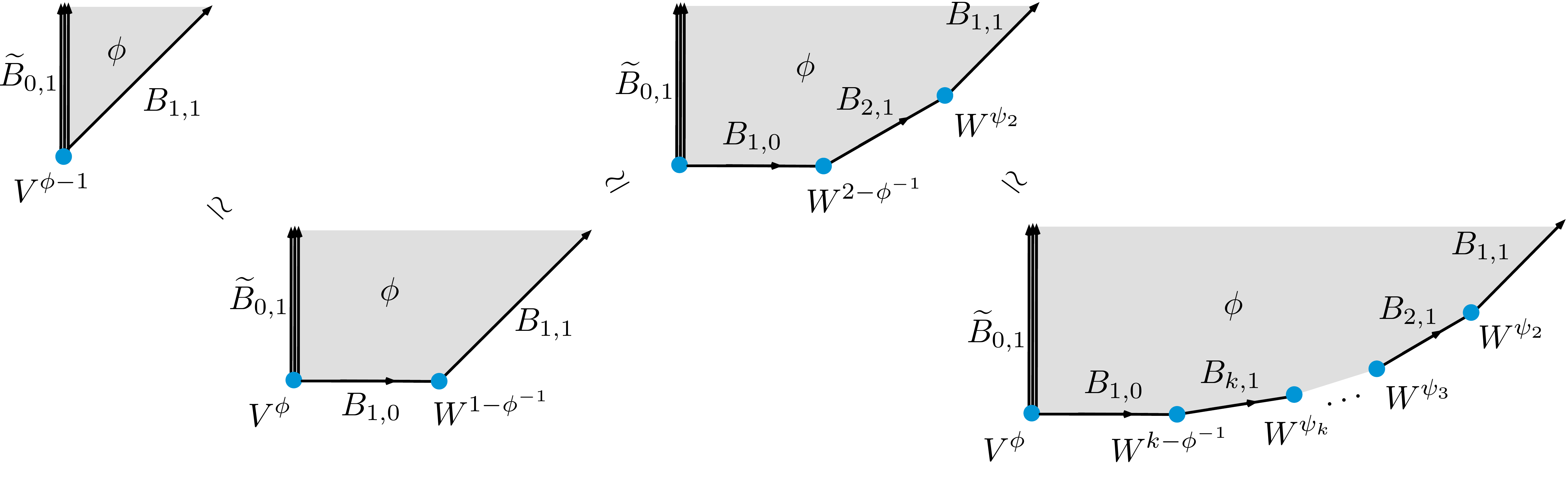}
\caption{Iterated slicing leading to a decomposition of $V^{\phi-1}(\sln)\otimes \FF(kn)$.}
\label{fig:corner-FF}
\end{figure}

We begin by deriving a duality of deformable families that is very close to the level-rank-like duality we are aiming for. We want to show that
\be \text{Com}\big(\CD^{\phi}_{n,k}, V^{\phi-1}(\sln)\otimes \FF(nk)\big) = \text{Com}\big(V^{1-\phi}(\sln), \CN^\psi_{n,k}\big)\,,\qquad \phi = -\frac{1}{\psi}\,. \label{voas2ways} \ee
This follows schematically from the iterated slicing of a $V^{\phi-1}$ corner shown in Figure \ref{fig:corner-FF}, which  leads us to expect that
\be V^{\phi-1} \sim V^\phi \otimes W^{\phi'} \otimes W^{\psi_{k}}\otimes W^{\psi_{k-1}}\otimes \cdots \otimes W^{\psi_2}\,, \ee
where $\phi',\psi_r$ are defined by
\be \frac{1}{\phi}+\frac{1}{\phi'} = k\,,\qquad \frac{1}{\psi+r}+\frac{1}{\psi_r}=1\,,\quad i.e.\quad \psi_r = \frac{r\phi-1}{(r-1)\phi-1}\,.\ee
Since, schematically, $\CD^\phi_{n,k} \sim V^\phi\otimes W^{\phi'}$ and $\CN^\psi_{n,k}\sim W^{\psi_{k}}\otimes W^{\psi_{k-1}}\otimes \cdots \otimes W^{\psi_2}\otimes V^{\psi_1}$ (where $\psi_1=1-\phi$), it seems plausible that taking cosets would produce \eqref{voas2ways}, with both sides being an extension of the form $W^{\psi_{k}}\otimes W^{\psi_{k-1}}\otimes \cdots \otimes W^{\psi_2}$.

To do this carefully, we begin with $V^{\phi-1}(\sln) \otimes \FF(n)$, which decomposes as
\begin{equation}
	\begin{split}
		V^{\phi-1}(\sln) \otimes \FF(n) &=  \bigoplus_{\lambda_0 \in P^+} V^\phi_{\lambda_0} \otimes \M^{\frac{\phi}{\phi-1}}_{\lambda_0, 0} \otimes L^{(n)}_{s(\lambda_0)} \\
		&=  \bigoplus_{\lambda_0 \in P^+} V^\phi_{\lambda_0} \otimes \M^{1-\phi^{-1}}_{0, \lambda_0} \otimes L^{(n)}_{s(\lambda_0)} 
	\end{split}
\end{equation}
where we used Feigin-Frenkel duality of modules $\M^\xi_{\lambda, \mu} = \M^{\xi^{-1}}_{\mu, \lambda}$, \emph{cf.} \cite{Arakawa:2018tzn}.
The quantum-Hamiltonian-reduction functor commutes with tensoring with integrable representations; in particular, Theorem 8.2 of \cite{Arakawa:2020oqo}   says  
\be
\M^{\xi-1}_{\mu, \mu'} \otimes V_{Q+\nu} = \bigoplus_{\substack{\lambda \in P^+ \\ \lambda = \mu + \mu' + \nu \mod Q }} \M^{\xi}_{\lambda, \mu'} \otimes \M^\zeta_{\lambda, \mu}\,, \qquad \text{where} \ \frac{1}{\xi} + \frac{1}{\zeta} =1\,. 
\ee
This allows us to decompose $V^{\phi-1}(\sln) \otimes \FF(nm)$ iteratively as:
\begin{align}	
		V^{\phi-1}(\sln) \otimes \FF(n)  &=  \bigoplus_{\lambda_0 \in P^+} V^\phi_{\lambda_0} \otimes \M^{1-\phi^{-1}}_{0, \lambda_0} \otimes L^{(n)}_{s(\lambda_0)}  \\
		V^{\phi-1}(\sln) \otimes \FF(2n)  &=  \bigoplus_{\lambda_0,\lambda_1 \in P^+} V^\phi_{\lambda_0} \otimes \M^{2-\phi^{-1}}_{\lambda_1, \lambda_0} \otimes \M^{\psi_2}_{\lambda_1, 0} \otimes  L^{(n)}_{s(\lambda_1-\lambda_0)} \otimes  L^{(n)}_{s(\lambda_0)}  \notag\\
		& \vdots \notag \\
		V^{\phi-1}(\sln) \otimes \FF(kn)  &=  \bigoplus_{\substack{ \lambda_0 \in P^+ \\ \vec{\lambda}  \in (P^+)^{m}}} V^\phi_{\lambda_0} \otimes \M^{k-\phi^{-1}}_{\lambda_m, \lambda_0} \otimes \M^{\vec{\psi}}_{ (\vec{\lambda}, 0)} \otimes L_{-(\vec{\lambda}, \lambda_0, 0)}\,, \notag
\end{align}
where
\be	\vec{\psi} := (\psi_k \dots, \psi_2)\,,\qquad \vec{\lambda} := (\lambda_{k-1}, \dots, \lambda_1)\,. \notag \ee

Note that rescaling the Heisenberg subalgebra generators of the lattice vertex algebra by minus one has the effect that the weight of  the module $L_{\vec{\lambda}}$ becomes $L_{-\vec{\lambda}}$; \ie\ we can (and will) replace $L_{-(\vec{\lambda}, \lambda_0, 0)}$ by $L_{(\vec{\lambda}, \lambda_0, 0)}$ by simply rescaling our generating Heisenberg fields. 
Comparing now with \eqref{eq:Cpsi} and \eqref{FeiginFrenkel} we recover \eqref{voas2ways}.

We now take the limit $\phi \rightarrow \infty $ so that $\psi \rightarrow 0$.
As before this is a naive limit, i.e. should be viewed as an identity in the Grothendieck ring of the categories. 
We take it in such a way that the affine vertex algebra becomes the commutative algebra $Z(\sln)$, and as before this limit should be 
\begin{equation}
	\begin{split}
		Z(\sln) \otimes &\FF(nk)  = \lim _{\phi \rightarrow \infty}  V^{\phi-1}(\sln) \otimes \FF(nk) \\
		&= \lim _{\phi \rightarrow \infty}  \bigoplus_{\substack{ \lambda_0 \in P^+ \\ \vec{\lambda} = (\lambda_m, \dots, \lambda_1) \in (P^+)^{m}}} V^\phi_{\lambda_0} \otimes \M^{k+\psi}_{\lambda_m, \lambda_0} \otimes \M^{\vec{\psi}}_{(\vec{\lambda}, 0)} \otimes L_{(\vec{\lambda}, \lambda_0, 0)}\\
		&= Z(\sln) \otimes  \bigoplus_{\substack{ \lambda_0 \in P^+ \\ \vec{\lambda} = (\lambda_m, \dots, \lambda_1) \in (P^+)^{m}}} R_{\lambda_0} \otimes \M^{k}_{\lambda_m, \lambda_0} \otimes \M^{\vec{\psi}}_{(\vec{\lambda}, 0)} \otimes L_{(\vec{\lambda}, \lambda_0, 0)}\,
	\end{split}
\end{equation}
with specialized levels $\vec\psi=\big(\frac{k}{k-1},\frac{k-1}{k-2},...,\frac{3}{2},2\big)$.
Decoupling the center $Z(\sln)$ gives
\begin{equation}
	\begin{split}
		\FF(nk)   &= \bigoplus_{\substack{ \lambda_0 \in P^+ \\ \vec{\lambda} = (\lambda_m, \dots, \lambda_1) \in (P^+)^{m}}} R_{\lambda_0} \otimes \M^{k}_{\lambda_m, \lambda_0} \otimes \M^{\vec{\psi}}_{(\vec{\lambda}, 0)} \otimes L_{(\vec{\lambda}, \lambda_0, 0)}\,.
	\end{split}
\end{equation}
$\FF(nk)$ is isomorphic to the lattice VOA of the lattice $\mathbb Z^{nk}$, but the latter contains the lattice $\sqrt{k}A_{n-1}$ as a sublattice. The Feigin-Tipunin algebra in turn is a subalgebra of the lattice vertex operator algebra of the lattice $\sqrt{k}A_{n-1}$. It follows that $\FT_{k}(\mathfrak{sl}_n)$ is a subalgebra of $\FF(nk)$.

\subsubsection{Vertex tensor category and cosets}
\label{sec:tensor}

Let $V$ be a vertex operator algebra and $A$ an object in a vertex tensor category $\CC$ of $V$-modules. Assume that $\text{Hom}_\CC(V, A) = \mathbb C$. Then $A$ can be given the structure of a vertex (super)algebra that extends the algebra structure on $V$ if and only of $A$ can be given the structure of a commutative (super)algebra in $\CC$ \cite{Huang:2014ixa,Creutzig:2019inr}. Moreover the vertex tensor category of $A$-modules that lie in $\CC$ is braided equivalent to the category of local modules $\rep(A)^{\text{loc}}$ for $A$ in $\CC$ \cite{Creutzig:2017anl}. This means that understanding the relation between representation categories of vertex algebras and their extension is a purely categorical question. We now state important properties obtained in \cite{Creutzig:2017anl}. In particular there is an induction functor $\CF : \CC \rightarrow \rep(A)$ that is monoidal, \ie\
\be
\CF(X \otimes Y) \cong \CF(X) \otimes \CF(Y)\,.
\ee
Moreover, the induction of an object $X$ is local if and only if the monodromy with $A$ is trivial. The right adjoint to $\CF$ is the restriction functor $\CG: \rep(A) \rightarrow \CC$ that forgets the action of $A$ on modules. In particular
\be
\CG(\CF(X)) = A \otimes X.
\ee

A useful concept to compare modules in $\CC$ and $\rep(A)$ is Frobenius reciprocity
\be
\text{Hom}_\CC(X, \CG(Y)) \cong \text{Hom}_{\rep(A)}(\CF(X), Y)
\ee
for any two objects $X$ in $\CC$ and $Y$ in $\rep(A)$.

\begin{ex}\label{ex:freeboson}
Let $V$ be the rank $n$ Heisenberg vertex algebra and $A = V_\Lambda$ be the lattice VOA of a positive even lattice. Let $\pi_\lambda$ be the Fock module of top level $\lambda \in \mathbb C^n$ and let us restrict to $\lambda \in \mathbb R^n$, since in that case it is known that the category of Fock modules is a vertex tensor category \cite{Creutzig:2016ehb} and the fusion rules are $\pi_\lambda \otimes \pi_\mu = \pi_{\lambda+\mu}$. The braiding is $e^{\pi i \lambda \mu}$ times the identity on $\pi_{\lambda+\mu}$ and the monodromy, that is, the double braiding, is its square. 
We have 
\begin{equation}
V_\Lambda = \bigoplus_{\nu \in \Lambda} \pi_\nu
\end{equation}
and  so $V_\Lambda$ is a vertex algebra in a completion of the category of Fock modules. The categorical framework also works in the completions \cite{Creutzig:2020smh}. 
We have
\be
\CG(\CF(\pi_\mu)) = V_\Lambda \otimes \pi_\mu  = \bigoplus_{\nu \in \Lambda} \pi_\nu \otimes \pi_\mu = \bigoplus_{\nu \in \Lambda} \pi_{\nu+\mu}.
\ee
We denote $\CF(\pi_\mu)$ by $V_{\Lambda+\mu}$.  It is a simple $V_\Lambda$-module and all simple $V_\Lambda$-modules are of this form  by Proposition 3.4 of \cite{Creutzig:2020jxj}.
By Frobenius reciprocity
\be
\text{Hom}_\CC( \pi_\mu, \bigoplus_{\nu \in \Lambda} \pi_{\nu+\mu'}) \cong \text{Hom}_{\rep(A)}(V_{\Lambda+\mu}, V_{\Lambda+\mu'})
\ee
and so $V_{\Lambda+\mu} \cong V_{\Lambda+\mu'}$ if and only if $\mu = \mu' \mod \Lambda$. 
\end{ex}
\begin{ex}\label{ex:singlet}
Now let $\Lambda =\sqrt{2k}\mathbb Z$, so that the triplet algebra $\FT_k(\mathfrak{sl}_2)$ is a subalgebra. 
The singlet algebra $\FT^0_k(\mathfrak{sl}_2)$ is the restriction of the triplet to the Heisenberg weight zero subspace, that is $\FT^0_k(\mathfrak{sl}_2)  = \FT_k(\mathfrak{sl}_2) \cap \pi_0$. Define $J_\lambda := \FT_k(\mathfrak{sl}_2) \cap \pi_\lambda$  so that 
\be 
\FT_k(\mathfrak{sl}_2) = \bigoplus_{\lambda \in \Lambda} J_\lambda
\ee
as a singlet module. The vertex tensor category of an interesting subcategory of singlet modules is obtained in \cite{Creutzig:2020qvs},  with the property that the $J_\lambda$ are simple currents $J_\lambda \otimes J_\mu = J_{\lambda+\mu}$. For every $\mu \in \mathbb R$ (or $\mathbb C$ if desired) $\pi_\mu$ is a singlet module. The fusion rules are conjectured via Verlinde's formula $J_\lambda \otimes \pi_\mu = \pi_{\lambda+\mu}$ \cite{Creutzig:2013zza}. This conjecture and complete vertex tensor category structure (for real weights) of the singlet will appear in \cite{CKMY}. In particular, it will be proven that $\pi_\mu$ is simple and projective if $\mu \notin \Lambda'$. 
Let $\mu \notin \Lambda'$.
As in the previous example the $\CF(\pi_\mu)$ are simple and 
\be
\CG(\CF(\pi_\mu)) =\FT_k(\mathfrak{sl}_2)  \otimes \pi_\mu  = \bigoplus_{\nu \in \Lambda} J_\nu \otimes \pi_\mu = \bigoplus_{\nu \in \Lambda} \pi_{\nu+\mu}, 
\ee
i.e. $V_{\Lambda+\mu}$ is also simple as a twisted triplet module.
\end{ex}

Let us consider the special case that $A$ is simple and an extension of $V \otimes W$ for two vertex algebras $V, W$. Assume that $V$ and $W$ are a mutually commuting pair inside $A$, meaning  $\text{Com}(V, A) =W$ and $\text{Com}(W, A) = V$. Assume that 
\be
A = \bigoplus_{i\in I} V_i \otimes W_i\,,
\ee
with $I$ an index set, such that the $V_i$ are inequivalent simple $V$-modules in a vertex tensor category $\CC_V$ of $V$. $I$ need not be finite and if it is infinite one replaces $\CC$ by the direct limit completion \cite{Creutzig:2020smh}. Assume that the $W_i$ are objects in a vertex tensor category $\CC_W$ of $W$. Assume that both $\CC_V$ and $\CC_W$ are rigid categories and let $\CD_V$ be the subcategory of $\CC_V$ whose simple objects are the $V_i$ for $i \in I$. Let $\CD_W$ be the subcategory of $\CC_W$ whose objects are direct sums of the $W_i$ for $i \in I$.
The main result of \cite{Creutzig:2019psu} says that $\CD_V$ and $\CD_W$ are semisimple tensor subcategories of $\CC_V$ and $\CC_W$ and they are braid-reversed equivalent, where the equivalence $\tau$ maps $V_i$ to $W_i^*$, the dual of $W_i$. 

Assume now that we are given two VOA's $V$ and $W$ with  vertex tensor subcategories $\CC$ and $\CD$ that are braid reversed equivalent, then one can view $\CC$ as a module category for $\CC \boxtimes \CD$, via the action $X \boxtimes Y$ mapping $Z \in \CC$ to $X \otimes Z \otimes Y$. There is then an internal End, call it $A$. It is an object in $\CC \boxtimes \CD$ (or the direct limit completion if $\CC$ is not finite) satisfying the property that
\begin{equation}
\text{Hom}_{\CC}(X \otimes Y,  \mathbf 1) \cong \text{Hom}_{\CC \boxtimes \CD}(X \boxtimes Y, A)
\end{equation}
This internal End $A$, also called the canonical algebra, is a vertex algebra extension of $V \otimes W$ since Theorem 3.3 of \cite{Creutzig:2019psu} also holds in the non-semisimple setting.

Let us take the example that $V$ is the triplet VOA and $W$ is a VOA that has a representation category $\CD$ that is braid-reversed equivalent to the category $\CC$ of modules of $V$. Denote this equivalence by $\tau$.%
\footnote{This should not be confused with the nilpotent endomorphism denoted $\tau$ in Section \ref{sec:CGP-0}.} %
Then consider objects (note that simple and projective triplet modules are all their own dual) characterized by the Loewy diagram

\begin{center}
	\begin{tikzpicture}[thick,>=latex]
		\node (t11) at (12,4) [] {$ S_{k-i}^- \otimes \tau(S_{k-i}^-)$};
		\node (t33) at (4,4) [] {$ S_i^+ \otimes \tau(S_{i}^+)$};
		\node (b11) at (12,0) [] {$ S_{k-i}^- \otimes \tau(S_{k-i}^-)$};
		\node (b33) at (4,0) [] {$ S_i^+ \otimes \tau(S_i^+)$};
		\node (01') at (14,2) [] {$ S_{k-i}^- \otimes \tau(S_i^+)$};
		\node (31') at (10,2) [] {$ S_{k-i}^- \otimes \tau(S_i^+)$};
		\node (13') at (6,2) [] {$ S_i^+ \otimes \tau(S_{k-i}^-)$};
		\node (63') at (2,2) [] {$ S_i^+ \otimes \tau(S_{k-i}^-)$};
		\draw [->] (t11) -- (01');
		\draw [->] (01') -- (b11);
		\draw [->] (t11) -- (31');
		\draw [->] (31') -- (b11);
		\draw [->] (t33) -- (13');
		\draw [->] (13') -- (b33);
		\draw [->] (t33) -- (63');
		\draw [->] (63') -- (b33);
		\draw [->,dotted] (t11) -- (63');
		\draw [->,dotted] (13') -- (b11);
		\draw [->,dotted] (t33) -- (31');
		\draw [->,dotted] (31') -- (b33);
		\draw [->,dotted] (t11) -- (13');
		\draw [->,dotted] (t33) -- (01');
		\draw [->,dotted] (01') -- (b33);
		\draw [->,dotted] (63') -- (b11);
	\end{tikzpicture}
\end{center}

Here the thick arrows denote arrows in $\CC$ and the dotted ones in $\CD$.
Let us denote these modules by $R_i$ for $i=1, \dots, k-1$. Also set $R_k = S_k^+ \otimes \tau(S_k^+) \oplus S_k^- \otimes \tau(S_k^-)$. 
Then each $R_s$ for $s=1, \dots, k-1$ is as an object in $\CC$ of the form
\begin{equation}
R_s = P^+_s \otimes \tau(S^+_s) \oplus P^-_{k-s} \otimes \tau(S^-_{k-s})
\end{equation}
and as an object in $\CD$ it is of the form
\begin{equation}
R_s = S^+_s \otimes \tau(P^+_s) \oplus S^-_{k-s} \otimes \tau(P^-_{k-s}).
\end{equation}
The algebra object $A$ is then 
\begin{equation}
A = \bigoplus_{s=1}^k R_s\,.
\end{equation}
In particular taking Frobenius-Perron (FP) dimensions of the objects in $\CD$ provides a functor from $\CC \boxtimes \CD$ mapping $A$ to 
\begin{equation}
\bigoplus_{s=1}^k P^+_s \text{FP}(S^+_s) \oplus P^-_s \text{FP}(S^-_s) 
\end{equation}
which is exactly the regular representation of $\CC$ \cite{tensorcategories}. Compare this with \eqref{eq:coend}.
The Frobenius-Perron dimension of the regular representation of a category is also called its dimension, in this case we have
\begin{equation}
\text{dim}(\CC) = \text{dim}(\CD) =  \sum_{s=1}^k \text{FP}(P^+_s) \text{FP}(S^+_s) +  \text{FP}(P^-_s) \text{FP}(S^-_s) = \text{FP}(A).
\end{equation} 
Let $\text{Rep}(A)$ be the category of $A$-modules that lie in $\CC \boxtimes \CD$. This is a tensor category, but locality might fail and so it is not braided, only the subcategory $\text{Rep}^{\text{loc}}(A)$ of local modules is. $\text{Rep}^{\text{loc}}(A)$ is precisely the category of modules for the VOA $A$ \cite{Creutzig:2017anl}.
Lemma 6.2.4 of \cite{tensorcategories} applied to $S=A$ yields the identity
\begin{equation}
\text{FP}(A) = \frac{\text{dim}(\CC \boxtimes \CD) }{\text{dim}(\text{Rep}(A))}
\end{equation}
which in our case becomes
\begin{equation}
\text{dim}(\text{Rep}(A)) = \frac{\text{dim}(\CC \boxtimes \CD) }{\text{FP}(A)} = \frac{\text{dim}(\CC)^2}{\text{dim}(\CC)} =\text{dim}(\CC).
\end{equation}
We have a tensor functor from $\CC$ to $\text{Rep}(A)$ mapping an object $S$ to $\CF(S \otimes W)$, i.e. first embedding $\CC$ in $\CC \boxtimes \CD$ and then applying the induction functor $\CF$. Frobenius reciprocity immediately implies that this functor is fully faithful and so $\text{Rep}(A)$ is the image of this functor. It is easy to check that $\CC$ is non-degenerate, this also  follows from  \cite{McRae21}. But only transparent modules can induce to local $A$-modules and so $A$ is the only indecomposable object in $\text{Rep}^{\text{loc}}(A)$, i.e. $A$ is a self-dual VOA, in particular it is rational. 

Conversely, assume that we have a rational and $C_2$-cofinite VOA $A$ that is self-dual in the sense that the only simple module is the VOA $A$ itself. Assume that the triplet algebra $V$ is a subalgebra of $A$ and assume that its commutant, call it $W$, is $C_2$-cofinite and its representation category is rigid. Moreover assume that $A$ is of the form
\begin{equation}
A = \bigoplus_{s=1}^k R_s 
\end{equation}
where the $R_s$ are indecomposable and projective as $V$ modules as well as $W$-modules. Moreover assume that the multiplicity of each projective triplet module is a simple $W$-module and vice versa, i.e. the $R_s$ are of the form as in  above Loewy diagram (except for $s=k$) for some map $\tau$ that maps triplet modules to $W$-modules. It is reasonable to conjecture that this is possible if and only if $\tau$ comes from a braid-reversed equivalence. One direction, that is braid-reversed equivalence implying such extensions holds by \cite{Creutzig:2019psu}. For the converse direction one needs to lift the proof of the key Lemma of that work beyond semisimplicity.
This conjecture is the main motivation for our conjecture and in particular for the decompositions \eqref{eq:decleft} and \eqref{eq:decright}.
Next we will demonstrate an explicit example of this behavior.

\subsubsection{The symplectic fermion example}\label{sec:SF}

The simplest example that illustrates our idea are symplectic fermions. Consider two pairs of symplectic fermions $\text{SF}(2)$ generated by $\chi_1^\pm, \chi_2^\pm$ with non-vanishing operator products
\begin{equation}
	\chi_1^+(z)\chi_1^-(w) = (z-w)^{-2} = \chi_2^+(z)\chi_2^-(w).
\end{equation}
Consider two pairs of free fermions $\FF(2)$ generated by $b_1, c_1, b_2, c_2$ with non-vanishing operator products
\begin{equation}
	b_1(z)c_1(w) = (z-w)^{-1} = b_2(z)c_2(w).
\end{equation}
The map 
\begin{equation}
	\chi_1^+ \mapsto b_1, \quad \chi_1^- \mapsto \partial c_1, \qquad \chi_2^+ \mapsto b_2, \quad \chi_2^- \mapsto \partial c_2
\end{equation}
embeds the symplectic fermions in the free fermions. Via this embedding we clearly have that each pair of free fermions $\FF(1)$ generated by $b_a, c_a$ (for fixed $a=1$ or $a=2$) satisfies the non-split exact sequence
\begin{equation}
	0 \rightarrow \text{SF}(1) \rightarrow \FF(1) \rightarrow \text{SF}(1) \rightarrow 0
\end{equation}
as a module for the symplectic fermion algebra $\text{SF}(1) $ generated by $\chi^\pm_a$ for $a \in \{1, 2\}$. In terms of Loewy diagrams
\begin{center}
	\begin{tikzpicture}[thick,>=latex,nom/.style={circle,draw=black!20,fill=black!20,inner sep=1pt}]
		\node (top1) at (5,1.5) [] {$\text{SF}(1)$};
		\node (bot1) at (5,0) [] {$\text{SF}(1)$};
		\draw [->] (top1) -- (bot1);

	\end{tikzpicture}
\end{center}
It follows that $\FF(2)$ is just the tensor product of these self-extensions of the two copies of symplectic fermions, i.e the Loewy diagram is 
\begin{center}
	\begin{tikzpicture}[thick,>=latex,nom/.style={circle,draw=black!20,fill=black!20,inner sep=1pt}]
		\node (top1) at (5,1.5) [] {$\text{SF}(1)\otimes \text{SF}(1)$};
		\node (left1) at (3.5,0) [] {$\text{SF}(1)\otimes \text{SF}(1)$};
		\node (right1) at (6.5,0) [] {$\text{SF}(1)\otimes \text{SF}(1)$};
		\node (bot1) at (5,-1.5) [] {$\text{SF}(1)\otimes \text{SF}(1)$};
		\node (top1LD) at (4.5,1.4)[]{};
		\node (top1RD) at (5.5,1.4)[]{};
		\node (left1LU) at (3,0.1)[]{};
		\node (left1LD) at (3,-0.1)[]{};
		\node (left1RU) at (4,0.1)[]{};
		\node (left1RD) at (4,-0.1)[]{};
		\node (right1LU) at (6,0.1)[]{};
		\node (right1LD) at (6,-0.1)[]{};
		\node (right1RU) at (7,0.1)[]{};
		\node (right1RD) at (7,-0.1)[]{};
		\node (bot1LU) at (4.5,-1.4)[]{};
		\node (bot1RU) at (5.5,-1.4)[]{};
		\draw [->] (top1LD) -- (left1LU);
		\draw [->,dotted] (top1RD) -- (right1RU);
		\draw [->,dotted] (left1RD) -- (bot1RU);
		\draw [->] (right1LD) -- (bot1LU);

	\end{tikzpicture}
\end{center}
Here and as before thick arrows denote left action and dotted one right action.
Clearly $\FF(2)$ is neither projective as a module for the extension of the first symplectic fermion algebra nor the second one. This is due to a singular choice of embedding of the symplectic fermions. To rectify it, let $A_1 = \begin{pmatrix} \alpha_1 & \beta_1 \\ \gamma_1 & \delta_1 \end{pmatrix},  A_2 = \begin{pmatrix} \alpha_2 & \beta_2 \\ \gamma_2 & \delta_2 \end{pmatrix}  \in \text{SL}(2, \mathbb C)$ and consider the more general maps
\begin{equation}
	\begin{split}
		\chi_1^+ \mapsto \alpha_1 b_1 + \beta_1 \partial c_2, \qquad  \chi_1^- \mapsto \gamma_1 b_2 + \delta_1 \partial c_1,  \\
		\chi_2^+ \mapsto \alpha_2 b_1 + \beta_2 \partial c_2, \qquad  \chi_2^- \mapsto \gamma_2 b_2 + \delta_2 \partial c_1.
	\end{split}
\end{equation}
This gives an embedding of $\text{SF}(2)$ in $\FF(2)$ provided that
\begin{equation}\label{eq;cond}
	\alpha_1\delta_2-\beta_1\gamma_2 = 0 = \alpha_2\delta_1 - \beta_2\gamma_1.
\end{equation}
A generic choice of $A_1, A_2 \in \text{SL}(2, \mathbb C)$ satisfying \eqref{eq;cond} has the property that all coefficients are non-zero. In particular if $\alpha_1, \alpha_2, \gamma_1, \gamma_2$ are all non-zero, then the Loewy diagram of $\FF(2)$ as an $\text{SF}(2) = \text{SF}(1) \otimes  \text{SF}(1)$-module via this embedding is
\begin{center}
	\begin{tikzpicture}[thick,>=latex,nom/.style={circle,draw=black!20,fill=black!20,inner sep=1pt}]
		\node (top1) at (5,1.5) [] {$\text{SF}(1)\otimes \text{SF}(1)$};
		\node (left1) at (3.5,0) [] {$\text{SF}(1)\otimes \text{SF}(1)$};
		\node (right1) at (6.5,0) [] {$\text{SF}(1)\otimes \text{SF}(1)$};
		\node (bot1) at (5,-1.5) [] {$\text{SF}(1)\otimes \text{SF}(1)$};
		\node (top1LD) at (4.5,1.4)[]{};
		\node (top1RD) at (5.5,1.4)[]{};
		\node (left1LU) at (3,0.1)[]{};
		\node (left1LD) at (3,-0.1)[]{};
		\node (left1RU) at (4,0.1)[]{};
		\node (left1RD) at (4,-0.1)[]{};
		\node (right1LU) at (6,0.1)[]{};
		\node (right1LD) at (6,-0.1)[]{};
		\node (right1RU) at (7,0.1)[]{};
		\node (right1RD) at (7,-0.1)[]{};
		\node (bot1LU) at (4.5,-1.4)[]{};
		\node (bot1RU) at (5.5,-1.4)[]{};
		\draw [->] (top1LD) -- (left1LU);
		\draw [->, dotted] (top1RD) -- (left1RU);
		\draw [->] (top1LD) -- (right1LU);
		\draw [->,dotted] (top1RD) -- (right1RU);
		\draw [->] (left1LD) -- (bot1LU);
		\draw [->,dotted] (left1RD) -- (bot1RU);
		\draw [->] (right1LD) -- (bot1LU);
		\draw [->, dotted] (right1RD) -- (bot1RU);

	\end{tikzpicture}
\end{center}
Denote by $P_{ \text{SF}(1)}$ the projective cover of  $\text{SF}(1)$.
Then we have 
\begin{equation}
	\FF(2) \cong P_{ \text{SF}(1)} \otimes  \text{SF}(1) \cong \text{SF}(1) \otimes P_{ \text{SF}(1)} 
\end{equation}
as a module for the action of the first (respectively, second) copy of the symplectic fermions.

\subsubsection{The Virasoro algebra and $\widehat{\mathfrak{sl}}_2$}\label{sec:vir}

Two of the best known families of VOA's are surely the Virasoro vertex algebra $\text{Vir}^{c}$ at central charge $c$ and the affine vertex algebra of $\widehat{\mathfrak{sl}}_2$ at level $k$, denoted by $V^{k+2}(\mathfrak{sl}_2)$. The two families are related via quantum Hamiltonian reduction. Denoting the reduction functor by $H$, then $H(V^{k+2}(\mathfrak{sl}_2)) = \text{Vir}^{c(k)}$ with $c(k) = 13 - 6(k+2) -6(k+2)^{-1}$. 
The representation categories of interest are the categories of ordinary modules. 

In the affine case, this category is usually denoted by KL$^k$ and for generic level it is a semisimple vertex tensor category \cite{KL-I, KL-III, KL-IV}. For admissible levels this result is established in \cite{Creutzig:2020cvf} and for $k=-1$, which is neither generic nor admissible but most important to us, it is due to \cite{Creutzig:2020cvf, Adamovic:2020lvj}.  For a review of all this, see \cite{Huang:2018qpr}.

For generic level and also $k=-1$ the set of inequivalent simple ordinary modules is $\{ V^{k+2}_\lambda | \lambda \in P^+\}$, the set of Weyl modules whose top level is an integrable $\mathfrak{sl}_2$-module. The set of dominant positive weights  are non-negative integer multiples of the fundamental weight. Let us write $V^{k+2}_n$ for $V^{k+2}_{(n-1)\omega}$, so that the label $n$ denotes the dimension of the top level subspace. 

In the Virasoro case and for generic $c_\psi = 13- 6(\psi+\psi^{-1})$ the category of ordinary modules has inequivalent simple modules $\M^\psi_{r, s}$ where both $r, s$ are positive integers and $H(V^{k+2}_n) = M_{n, 1}$, while $H(V^{\ell+2}_m) = \M^\psi_{1, m}$ for $\ell$ the Feigin-Frenkel dual level, that is $(k+2)(\ell+2) =1$. 
Moreover there are also Arakawa-Frenkel twists of the functor \cite{Arakawa:2018tzn}, where one twists the character of the reduction. These twists are also parameterized by positive dominant weights and so we denote them by $H^m$ with $H= H^1$ the untwisted case. The resulting modules are $H^m(V^{k+2}_n) = M_{n, m} = H^n(V^{\ell+2}_m) = M_{1, m}$.

The category of ordinary modules for generic central charge is actually also a rigid vertex tensor category \cite{Creutzig:2020zvv}. We don't need the complete fusion rules, but only 
\[
M_{n, 1} \boxtimes M_{1, m} = M_{n, m}.
\]
We denote by $\CC^\psi$ the category of ordinary modules of the Virasoro algebra at level $\psi-2$ and we denote by $\CO^\psi$ the category of ordinary modules of the universal affine vertex algebra of $\sltwo$ at level $\psi-2$; and by $\CO_\psi$ the category of ordinary modules of its simple quotient.

\subsubsection{The generic coset $\CN^{\psi-1}_{2,k}$}\label{sec:genericcoset}

Fix a positive integer $k$ and let $\psi$ be generic (not a rational number). 
Let $\psi_r = \frac{\psi+r}{\psi+r-1}$ and $\vec{\psi} = (\psi_{k-1}, \dots, \psi_1)$. Let $\CC^{\vec{\psi}} = \CC^{\psi_{k-1}} \boxtimes \dots \boxtimes \CC^{\psi_1} \boxtimes \CC_3^{\boxtimes k}$
and $\CD^{\vec{\psi}}  = \CC^{\vec{\psi}} \boxtimes \CO^\phi$ with $\frac{1}{\psi} +\frac{1}{\phi} =1$. 
Then 
\begin{equation} \label{C-sl2}
	\begin{split}
		\CN^{\psi-1}_{2,k} &= \bigoplus_{\vec{n} = (n_{k-2}, \dots, n_0) \in \mathbb Z_{>0}^{k-1}}  \M^{\vec{\psi}}_{(1, \vec{n})}
		\otimes V^\phi_{n_0} \ \otimes L_{(1, \vec{n}, 1)}\,,
	\end{split}
\end{equation}
with 
\[
\M^{\vec{\psi}}_{(r, \vec{n})} = \M^{\psi_{k-1}}_{r, n_{k-2}} \otimes \M^{\psi_{k-2}}_{n_{k-2}, n_{k-3}} \otimes \dots \otimes \M^{\psi_{1}}_{n_1, n_{0}}, \qquad 
\]
and 
\[
L_{(r, \vec{n}, s)} = L^{(2)}_{\overline{r+n_{k-2}}} \otimes L^{(2)}_{\overline{n_{k-2}+n_{k-3}}} \otimes \dots \otimes L^{(2)}_{\overline{n_{0}+s}}
\]
where $\overline{t}$ is zero if $t$ is even and one if $t$ is odd. 
Let $\CF : \CD^{\vec{\psi}} \rightarrow \rep \ \CN^{\psi-1}_{2,k}$ be the induction functor and $\CG$ its right adjoint. Let $\vec{1} = (1, \dots, 1) \in \mathbb Z^{k-1}_{>0}$ and $\vec{s} \in \{ 0, 1\}^{k+1}$, then 
\begin{equation}\nonumber
	\begin{split}
		X^\psi_{r, 1, \vec{s}} :&= \CF\left(\M^{\vec{\psi}}_{(r, \vec{1})} \otimes V^\phi_{1} \otimes  L_{\vec{s}} \right) \\
		\CG\left(X^\psi_{r, 1, \vec{s}}\right)  &=  \bigoplus_{\vec{n} = (n_{k-2}, \dots, n_0) \in \mathbb Z_{>0}^{k-1}}  \M^{\vec{\psi}}_{(r, \vec{n})}
		\otimes V^\phi_{n_0} \ \otimes L_{(1, \vec{n}, 1)+\vec{s}}\,.
	\end{split}
\end{equation}
and by Frobenius reciprocity
\begin{equation} 
	\begin{split}
		\text{Hom}_{\rep \ \CN^{\psi-1}_{2,k}}\left( X^\psi_{r, 1, \vec{s}}, X^\psi_{r', 1, \vec{s'}} \right) &= 
		\text{Hom}_{\CD^{\vec{\psi}}}\left( \M^{\vec{\psi}}_{(r, \vec{1})} \otimes V^\phi_{1} \otimes L_{\vec{s}} ,\CG\left(X^\psi_{r', 1, \vec{s'}}\right)  \right)\\  &=  \delta_{r, r' }\delta_{\vec{s}, \vec{s'}}\mathbb C\,.
	\end{split}
\end{equation}
It follows that the subcategory of  $\rep \ \CN^{\psi-1}_{2,k}$ whose objects are direct sums of those $\CF(\M^{\vec{\psi}}_{r, \vec{1}} \boxtimes L_{\vec{s}}  )$ that are integer graded by conformal weight is a semisimple category of modules for the vertex algebra $\CN^{\psi-1}_{2,k}$ with simple objects precisely the $\CF(\M^{\vec{\psi}}_{(r, \vec{1})} \boxtimes L_{\vec{s}}  )$ that are integer graded by conformal weight. 
Consider
\[
\text{Com}(V^\phi(\sltwo), \CN^{\psi-1}_{2,k}) =  \bigoplus_{\vec{n} = (n_{m-1}, \dots, n_1) \in \mathbb Z_{>0}^{k-2}}  \M^{\vec{\psi}}_{(1, \vec{n}, 1)}
\ \otimes L_{(1, \vec{n}, 1, 1)}
\] 
and let $\CF' : \CC^{\vec{\psi}} \rightarrow \rep \ B^\psi(\sltwo, m)$ be the induction functor and $\CG'$ its right adjoint
\begin{equation}
	\begin{split}
		Y^\psi_{r, t, \vec{s}} :&= \CF'\left(\M^{\vec{\psi}}_{(r, \vec{1}, t)} \otimes   L_{\vec{s}} \right) \\
		\CG'\left(Y^\psi_{r, t, \vec{s}}\right)  &=  \bigoplus_{\vec{n} = (n_{k-2}, \dots, n_1) \in \mathbb Z_{>0}^{k-2}}  \M^{\vec{\psi}}_{(r, \vec{n}, t)}
		\otimes L_{(1, \vec{n}, 1, 1)+\vec{s}}\,.
	\end{split}
\end{equation}
By Frobenius reciprocity
\begin{equation}\label{eq:Frob}
	\begin{split}
		\text{Hom}_{\rep \ \text{Com}(V^\phi(\sltwo), \CN^{\psi-1}_{2,k}) }\left( Y^\psi_{r, t, \vec{s}}, Y^\psi_{r', t', \vec{s'}} \right) &= 
		\text{Hom}_{\CC^{\vec{\psi}}}\left( \M^{\vec{\psi}}_{(r, \vec{1}, t)} \otimes L_{\vec{s}} ,\CG\left(Y^\psi_{r', t', \vec{s'}}\right)  \right)\\  &=  \delta_{r, r' }\delta_{t, t' }\delta_{\vec{s}, \vec{s'}}\mathbb C\,.
	\end{split}
\end{equation}
Here we note that all Virasoro modules and lattice VOA modules are self-dual and categories are rigid so that we actually can interchange the order of the objects, i.e. we can replace $\text{Hom}(X, Y)$ by $\text{Hom}(X^*, Y^*) = \text{Hom}(X, Y)$.  
This gives us the decompositions
\begin{equation}
	\begin{split}
		\CN^{\psi-1}_{2,k}  &= \bigoplus_{ n_0 \in \mathbb Z_{>0}}  Y^\psi_{1, n_0, \vec{1}+(n_0+1)e}  \otimes V^\phi_{n_0}, \qquad e=(0, \dots, 0, 1)\,,
	\end{split}
\end{equation}
and
\begin{equation}\nonumber
	\begin{split}
		V^{\phi-1}(\sln) \otimes F(nk)  &=  \bigoplus_{\substack{ \lambda_0 \in P^+ \\ \vec{\lambda} = (\lambda_{k-1}, \dots, \lambda_1) \in (P^+)^{k-1}}} V^\phi_{\lambda_0} \otimes \M^{\psi^{-1}+k-1}_{\lambda_{k-1}, \lambda_0} \otimes \M^{\vec{\phi}}_{(\vec{\lambda}, 0)} \otimes L_{(\vec{\lambda}, \lambda_0, 0)}\\
		V^{\phi-1}(\sltwo) \otimes F(2k)  &=  \bigoplus_{\substack{ r_0 \in \mathbb Z_{>0} \\ \vec{r} = (r_{k-1}, \dots, r_1) \in \mathbb Z_{>0}^{k-1}}} V^\phi_{r_0} \otimes \M^{\psi^{-1}+k-1}_{r_{k-1}, r_0} \otimes \M^{\vec{\phi}}_{(\vec{r}, 1)} \otimes L_{(\vec{r}, r_0, 1)}\\
		&=  \bigoplus_{r_0, r_{k-1} \in \mathbb Z_{>0}} V^\phi_{r_0} \otimes \M^{\psi^{-1}+k-1}_{r_{k-1}, r_0} \otimes Y^\phi_{r_{k-1}, 1, (\vec{1}, r_0)}\,.
	\end{split}
\end{equation}

\subsubsection{The coset $\CN_{2,k}$}\label{sec:coset}

We now specialize to $\psi \rightarrow 1$. In this case $\CO^\phi$ as $\phi \rightarrow \infty$ becomes just $\rep(SU(2))$ and $SU(2)$ acts via automorphisms on $\CN_{2,k}$. We prefer to consider $\CN_{2,k}$ as an object in $\CC:=\CC^{\vec{\psi}}$ for $\psi =1$. We are interested in the multiplicity of the triplet module $X^+_s$. Since the Virasoro module $M_{1, s}^{1, k}$ appears with multiplicity one in $X^+_s$ and is not a submodule of any other simple triplet module it is enough to study the multiplicity of this Virasoro module. 

For $\psi \rightarrow 1$ we have $\psi_r \rightarrow \frac{r+1}{r}$ and let us write $\M^{r+1, r}_{n_{r-1}, n_{r-2}}$ for $\M^{\frac{r+1}{r}}_{n_{r-1}, n_{r-2}}$.
In this case the lattices $\psi_r \mathbb Z$ and $\psi_r^{-1} \mathbb Z$ intersect non-trivially and this translates into the isomorphisms $\M^{r+1, r}_{a, b} \cong \M^{r+1, r}_{t(r+1) + a, tr +b}$ for any positive integer $t$. Combining with the invarince under Weyl reflection, that is $\M^{r+1, r}_{a, b} \cong \M^{r+1, r}_{r+1- a, r-b},$ this yields $\M^{r+1, r}_{a, b} \cong \M^{r+1, r}_{t(r+1) \pm a, tr \pm b}$.
Similarly one also has   $\M^{p, 1}_{a, b} \cong \M^{p, 1}_{tp \pm a, t \pm b}$.
Set $\M_{(r, \vec{n}, t)} = \lim_{\psi \rightarrow 1} \M^{\vec{\psi}}_{(r, \vec{1}, t)}$ and ${\vec{x}} = (k-1, \dots, 3, 2)$ and $\vec{1} = (1, \dots, 1)$ both in $\mathbb Z_{>0}^{k-2}$.
Then we get the following identities:
\begin{equation} 
	\begin{split}
		\M_{(s, \vec{1}, t)} &= \M^{k, k-1}_{s, 1} \otimes \M^{k-1, k-2}_{1, 1} \otimes \dots \otimes \M^{2, 1}_{1, t} \\
		&= \M^{k, k-1}_{s-kt, 1 -(k-1)t} \otimes \M^{k-1, k-2}_{1 -(k-1)t, 1-(k-2)t} \otimes \dots \otimes \M^{2, 1}_{1-2t, 0} \\
		&= \M^{k, k-1}_{k-s+kt, k-2 +(k-1)t} \otimes \M^{k-1, k-2}_{k-2 +(k-1)t, (k-2)-1+(k-2)t} \otimes \dots \otimes \M^{2, 1}_{1+2t, 1} \\
		&= \M_{(k-s+kt, (t+1)\vec{x}-\vec{1}, 1)}
	\end{split}
\end{equation} 
and 
\begin{equation} 
	\begin{split}
		\M_{(s, \vec{1}, t)} &= \M^{k, k-1}_{s, 1} \otimes \M^{k-1, k-2}_{1, 1} \otimes \dots \otimes \M^{2, 1}_{1, t} \\
		&= \M^{k, k-1}_{s, 1} \otimes \M^{k-1, k-2}_{1, 1} \otimes \dots \otimes \M^{2, 1}_{1, 1-t} \\
		&= \M^{k, k-1}_{s+kt, 1 +(k-1)t} \otimes \M^{k-1, k-2}_{1 +(k-1)t, 1+(k-2)t} \otimes \dots \otimes \M^{2, 1}_{1+2t, 1} \\
		&= \M_{(s+kt, t\vec{x}+\vec{1}, 1)}\,.
	\end{split}
\end{equation}
Set $Y_{s, t, \vec{r}} = \lim_{\psi \rightarrow 1} Y^\psi_{s, t, \vec{r}}$
and denote by $[Y_{s, t, \vec{r}}]$ the image of $Y_{s, t, \vec{r}}$ in the Grothendieck ring. 
We 
assume that 
\eqref{eq:Frob} also holds in the Grothendieck ring of the limit $\psi \rightarrow 1$. It follows that there are non-zero homomorphisms
\begin{equation}\label{eq:mor1}
	[Y_{s, t, \vec{r}}] \rightarrow [Y_{k-s+tk, 1, \vec{r} + (t+1)\vec{y}}]\qquad \text{and}\qquad 
	[Y_{s, t, \vec{r}}] \rightarrow [Y_{s+tk, 1, \vec{r} + t\vec{y}}]\,.
\end{equation}
With $\vec{y} = ( 0 , 1, 1,  0, 0, 1, 1, 0, 0, 1, 1, \dots)$ if $k-1$ is even and $\vec{y} = ( 1, 1,  0, 0, 1, 1, 0, 0, 1, 1, \dots)$ if $k-1$ is odd.
We have $\M^{1, k}_{1, s} = \M_{r_0, r_{k-1}}^{1, k}$ if and only if either $r_0 = 1+t$ and $r_{k-1}= s+tk$ or $r_0 =t$ and $r_{k-1} = k-s+kt$.  
It follows that the multiplicity of $\M^{1, k}_{1, s}$ in $\FF(2k)$ is
\begin{equation}
	\begin{split}
		\text{mult} \ \M^{1, k}_{1, s} &= \bigoplus_{ t\geq 1} R_{t} \otimes \left(Y_{s - k +tk, 1, \vec{r}_t}\ \oplus \
		Y_{k-s+tk, 1, \vec{r}_t}\right) \,,
	\end{split}
\end{equation}  
with $\vec{r}_t := \vec{1} + (t+1)e+(s-1+(t+1)k)\vec{1}$\,.

Next we compute
\begin{equation} 
	\begin{split}
		\M_{(k-s+kt, \vec{1}, 1)} &= \M^{k, k-1}_{k-s+kt, 1} \otimes \M^{k-1, k-2}_{1, 1} \otimes \dots \otimes \M^{2, 1}_{1, 1} \\
		&= \M^{k, k-1}_{s-kt, k-2} \otimes \M^{k-1, k-2}_{k-2, (k-2)-1} \otimes \dots \otimes \M^{2, 1}_{1, 1} \\
		&= \M^{k, k-1}_{s, k-2+(k-1)t} \otimes \M^{k-1, k-2}_{k-2+(k-1)t, (k-2)-1+(k-2)t} \otimes \dots \otimes \M^{2, 1}_{1+2t, 1+t} \\
		&= \M_{(s,  (t+1)\vec{x}-\vec{1}, 1+t)}
	\end{split}
\end{equation} 
and we again assume that  \eqref{eq:Frob} holds in the $\psi\rightarrow 1$ limit in the Grothendieck ring
\begin{equation}\label{eq:mor2}
	[Y_{k-s+tk, 1, \vec{r}}]  \rightarrow  [Y_{s, t+1, \vec{r} + (t+1)\vec{y}}]\, .
\end{equation}
We now assume that there exists one more type of morphism 
\begin{equation}
	\begin{split}
		[Y_{s+tk, 1, \vec{r}}] \rightarrow [Y_{s, t+1, \vec{r} + t\vec{y}}]
	\end{split}
\end{equation}
and in fact that we have the embeddings
\begin{equation}
	\begin{split}
		[Y_{s+tk, 1, \vec{r}}] &\hookrightarrow [Y_{s, t, \vec{r} + t \vec{y}}]\ \oplus  \
		 [Y_{s, t+1, \vec{r}  +  t\vec{y}}]\ \oplus \ 
	         [Y_{k - s, t, \vec{r} + (t+1)\vec{y}}]\ \oplus  \
		[Y_{k-s, t+1, \vec{r} + (t+1) \vec{y}}]  
		 \\
		[Y_{k-s+tk, 1, \vec{r} } ] &\hookrightarrow [Y_{s, t+1, \vec{r} +  (t+1) \vec{y}}]\ \oplus  \
		 [Y_{s, t, \vec{r} + (t+1)\vec{y}}]\ \oplus  \
		 [Y_{k - s, t+1, \vec{r} + t\vec{y}}]\ \oplus  \
		 [Y_{k-s, t, \vec{r} + t \vec{y}}] \,.
	\end{split}
\end{equation}  
This assumption yields the embedding  (we use that $R_t = \mathbb C^t$)
\begin{equation}\label{eq:branching}
	\begin{split}
		\text{mult} \ \M^{1, k}_{1, s} &\hookrightarrow
		\bigoplus_{t \geq 1} \mathbb C^2 \otimes  \mathbb C^t \otimes  Y_{s, t, \vec{r}_{t+1} + t\vec{y}}  \ \oplus \\
		&\qquad \bigoplus_{t \geq 1} \mathbb C^2 \otimes  \mathbb C^t \otimes Y_{s, t, \vec{r}_t + (t+1)\vec{y}}  \ \oplus \\
		&\qquad \bigoplus_{t \geq 1}  \mathbb C^2 \otimes  \mathbb C^t \otimes Y_{k-s, t, \vec{r}_{t+1} + (t+1)\vec{y}}  \ \oplus \\
		&\qquad \bigoplus_{t \geq 1} \mathbb C^2 \otimes  \mathbb C^t \otimes Y_{k-s, t, \vec{r}_t + t\vec{y}} \,. \\
	\end{split}
\end{equation}  
Set $s=1$ so that we get the multiplicity of the triplet algebra itself.
The first two lines coincide with the image in the Grothendieck ring of two copies of a  $\mathbb Z_2 \times \mathbb Z_2$   simple current extension of $\CN_{2, k}^{\mathbb Z_2}$ if $k$ is odd and for $k$ even it is a $\mathbb Z_2$ simple current extension of  $\CN_{2, k}$.
Moreover 
 the last two 
lines can be identified with two copies of a module for this simple current extension.

\subsection{$\CN_{n, k}$ and rectangular $W$-algebras}\label{sec:rect}

The Feigin-Tipunin algebra $\FT_{k}(\mathfrak{sl}_n)$ is an extension of the principal $W$-algebra of $\mathfrak{sl}_n$ at shifted level $1/k$. Here we ask the question whether $\CN_{n, k}$ is possibly also related to some interesting $W$-algebra.

Let $\mathfrak{g} = \mathfrak{sl}_{nm}$ and let $f_{\text{rect}}$ be the nilpotent element corresponding to the partition $(n, n, ... , n)$ of $nm$. Then the $W$-algebra obtained from the affine vertex algebra of $\mathfrak{sl}_{nm}$ at level $\ell$ via Quantum-Hamiltonian reduction associated to the nilpotent element $f_{\text{rect}}$ is called rectangular. It has an affine subalgebra of type $\mathfrak{sl}_m$ and has $m^2$ fields at conformal weights $2, 3, \dots, n$ transforming in the adjoint plus trivial representation. For more information on these algebras, see \cite{Creutzig:2018pts}. We use the data from \cite[Sec. 3.1]{Creutzig:2018pts}.

Let $m=k-1$ and the critically shifted level $\psi = \ell + h^\vee = k$. Denote this algebra by $W_k(\mathfrak{sl}_{n(k-1)}, f_{\text{rect}})$. Its affine vertex subalgebra is of type $\mathfrak{sl}_{k-1}$ and has level $n$. This is exactly as for $\CN_{n, k}$. Moreover it turns out that the central charge of $W_k(\mathfrak{sl}_{n(k-1)}, f_{\text{rect}})$ plus two coincides with the central charge of $\CN_{n, k}$. This suggests that $W_k(\mathfrak{sl}_{n(k-1)}, f_{\text{rect}})$ is related to a coset by two free bosons, call them $\pi^{\otimes 2}$, of $\CN_{n, k}$. 

The Feigin-Tipunin algebra $\FT_{k}(\mathfrak{sl}_n)$ is an extension of the principal $W$-algebra of $\mathfrak{sl}_n$ at critically shifted level $1/k$. By Theorem 10.2 of \cite{Linshaw:2017tvv} this algebra is isomorphic to the principal $W$-algebra of $\mathfrak{sl}_{n(k-1)}$ at critically shifted level $1- 1/k$. 
By \cite{Arakawa:2020oqo},  the principal $W$-algebra of $\mathfrak{sl}_{n(k-1)}$ at critically shifted level $1- 1/k$ and $W_k(\mathfrak{sl}_{n(k-1)}, f_{\text{rect}})$ extend to $W_{k-1}(\mathfrak{sl}_{n(k-1)}, f_{\text{rect}})$ times the lattice VOA of the root lattice of $\mathfrak{sl}_{n(k-1)}$. Here the conformal vector of the root lattice is twisted. Moreover the central charge of $W_{k-1}(\mathfrak{sl}_{n(k-1)}, f_{\text{rect}})$ is $n-1$ and the affine subalgebra vanishes at this level (since the affine subalgebra has level zero in this instance). 
It is thus natural to conjecture that $W_{k-1}(\mathfrak{sl}_{n(k-1)}, f_{\text{rect}})$ is the $SU(n)$ orbifold of the lattice VOA of $\mathfrak{sl}_n$. At least for $k=2$ this is true \cite[Example 7.13]{Creutzig:2014lsa}. 
In summary, there are the following embeddings
\begin{equation}\label{variousembeddings}
\begin{split}
W_{1- 1/k}(\mathfrak{sl}_{n(k-1)}) &\hookrightarrow  \FT_{k}(\mathfrak{sl}_n) \\
W_k(\mathfrak{sl}_{n(k-1)}, f_{\text{rect}}) &\overset{Conj.}{\hookrightarrow} \text{Com}(\pi^{\otimes 2},  \CN_{n, k}) \\
W_{1- 1/k}(\mathfrak{sl}_{n(k-1)}) \otimes W_k(\mathfrak{sl}_{n(k-1)}, f_{\text{rect}}) & \hookrightarrow W_{k-1}(\mathfrak{sl}_{n(k-1)}, f_{\text{rect}})  \otimes V_{A_{n(k-1)-1}}\\
W_{k-1}(\mathfrak{sl}_{n(k-1)}, f_{\text{rect}}) &\overset{Conj.}{\cong} V_{A_{n-1}}^{SU(n)} \\
V_{A_{n-1}}^{SU(n)}  \otimes V_{A_{n(k-1)-1}} &\hookrightarrow  \text{Com}(\pi^{\otimes 2}, \FF(nk)).
\end{split}
\end{equation}

\appendix

\section{Hochschild homology of $U_q(\mathfrak{sl}_2)$-mod}
\label{app:U}

In this appendix we illustrate several direct computations of Hochschild homology of (stalks of) the category $\CC=D^b(U_q(\mathfrak{sl}_2)\text{-mod})$, supplementing more abstract discussions in Section~\ref{sec:Uqsl2}. This should match the torus state spaces of the QFT $\CT_{2,k}^A$.

We first focus on the identity stalk $\CC_1 = D^b(u_q(\mathfrak{sl}_2)\text{-mod})$, which is the most nontrivial.
This part is well studied in the literature; see \emph{e.g.} \cite{LQ-derived} and \cite[Sec. 7]{CGP2} for similar computations. We then consider deformations by generic flat connections, and verify that the two descriptions from Section \ref{sec:Hoch-flavor} --- ordinary Hochschild homology of $\CC_g$ and $g$-twisted Hochschild homology of $\CC_1$ --- are indeed equivalent.

As in Section \ref{sec:Uqsl2}, we assume throughout that $k\geq 2$ and the corresponding root of unity $q=e^{i\pi/k}$ are fixed. We write $\mb u:= u_q(\mathfrak{sl}_2)$.

\subsection{From the quantum group itself}
\label{app:algebra}

One should in principle be able to compute Hochschild homology of $\CC_1 = D^b(\mb u\text{-mod})$ by constructing a Hochschild complex for the associative algebra $\mb u$ itself. This computation is only feasible for small $k$ and small cohomological degree.

Recall that $\mb u$ is generated by $E,F,K^{\pm1}$, with relations
\be \begin{array}{c} \ds KE = q^2 EK\,,\qquad KF=q^{-2}FK\,,\qquad [E,F] = \frac{K-K^{-1}}{q-q^{-1}}\,, \\[.2cm]
  E^k=F^k=0\,,\qquad K^{2k} = 1\,, \end{array}
\ee
This is an associative algebra of dimension $2k^3$. (The cubic increase in dimension is what makes direct computations difficult for large $k$.) 

The zeroth Hochschild homology is just the co-center 
\be HH_0(\CC_1) \simeq \mb u/[\mb u,\mb u]\,. \ee
For $k=2$ we find 
\be  HH_0(\CC_1) \simeq  \C\langle 1,K,K^2,EF+FE,(EF+FE)K^2\rangle\,. \ee
For $k=3$ we find $\text{dim}\, HH_0(\CC_1) = 8$ with assistance from \emph{Mathematica}. This is consistent with the general result $\text{dim} \mb u/[\mb u,\mb u] = \text{dim}\,Z(\mb u) = 3k-1$ \cite{Kerler, Lachowska-center}.

At $k=2$ it is feasible to code the first few degrees of the Hochschild complex \eqref{HH-complex} for $\mb u$. (Note that $\text{dim}(\mb u)= 16$, $\text{dim}(\mb u^{\otimes 2}) = 256$, $\text{dim}(\mb u^{\otimes 3}) =4096$. One ends up dealing with very large matrices!) In degree $-1$, we find
\be \text{dim}\, HH_{-1}(\CC_1)  = 4\,. \ee
After inverting the cohomological degree, as discussed in Section \ref{sec:grading}, this matches $\CH_1=\rho_1^{\oplus k-1}\oplus\rho_3^{\oplus k-1} \simeq \C^4$ from \eqref{Hilb-T2}.

\subsection{Koszul-dual quiver algebras}
\label{app:quivers}

A smarter way to proceed is to represent the category $\CC_1$ as modules for the endomorphism algebra of a generating object. There are two natural choices, either taking the derived endomorphism algebra of the (direct sum of) simple objects in the abelian category $\CC_1^{\rm ab} = \mb u\text{-mod}$, or taking the endomorphism algebra of the (direct sum of) projective objects in $\CC_1^{\rm ab}$. These ultimately yields two Koszul-dual algebras --- studied in \cite{ABG,BL} and also very similar to those studied in \cite{BGS} and much more recently \cite{Gale,BLPW} in the context of categories $\CO$ and Symplectic Duality.  The two choices are direct analogues of our two ways to represent the category of line operators in $\CT_{\rm hyper}$ in Section \ref{sec:hyper-lines}:
\begin{itemize}
\item The derived endomorphism algebra $A$ of the simple modules is analogous to the symmetric algebra $\C[X,Y]$ from Section \ref{sec:hyper-lines}. It is infinite dimensional but finitely generated, with generators in cohomological degree 1. The Hochschild complex \eqref{HH-complex} with negative grading convention yields \emph{non-negatively} graded homology, due to the internal degree in the algebra.

Geometrically, $A$ is related to the endomorphism algebra of the structure sheaf and a twisted structure sheaf on the moduli space $T^*[2]\mathbb{P}^1$, just as $\C[X,Y]$ is the endomorphism algebra on the moduli space $T^*[2](\C[1])$. A particular quotient of $A$ (described below) is isomorphic to $\C[X,Y]$.

 \item The endomorphism algebra $A'$ of the projectives is analogous to the exterior algebra $\C[\xi,\psi]$ from Section \ref{sec:hyper-lines}. It is finite dimensional and its generators are all in degree zero. Its Hochschild complex should use an inverted, positive grading convention to produce a non-negatively graded homology groups (as discussed in Section \ref{sec:grading}). A particular quotient of $A'$ is isomorphic to $\C[\xi,\psi]$.
\end{itemize}

Let's describe $A$ and $A'$. Recall from \eqref{block-C1} that, if we ignore braided tensor structure, the abelian category $\CC_1^{\rm ab}=\mb u\text{-mod}$ and its derived category $\CC_1$ decompose as direct sums of blocks,
\be \label{block-app}  \CCab_1 \simeq \CB_k^{\text{ab}\,+}\oplus \CB_k^{\text{ab}\,-} \oplus \bigoplus_{j=1}^{k-1} \CB_j^{\rm ab}\,,\qquad\quad  \CC_1 \simeq \CB_k^+\oplus \CB_k^- \oplus \bigoplus_{j=1}^{k-1} \CB_j\,, \qquad \ee
with each block on the right the derived category of the block on the left.
The first two blocks $\CB_{k}^{\text{ab}\,\pm}$ are semisimple with a single simple object each, so they will just contribute a $\C^2$ summand in degree zero to $HH_\bullet(\CC_1)$. The remaining blocks $\CB_j^{\rm ab}$ ($j=1,...,k-1$) are all isomorphic. Thus it suffices to analyze $\CB_1^{\rm ab}$, known as the principal block.

\begin{figure}[htb]
\centering
\includegraphics[width=5in]{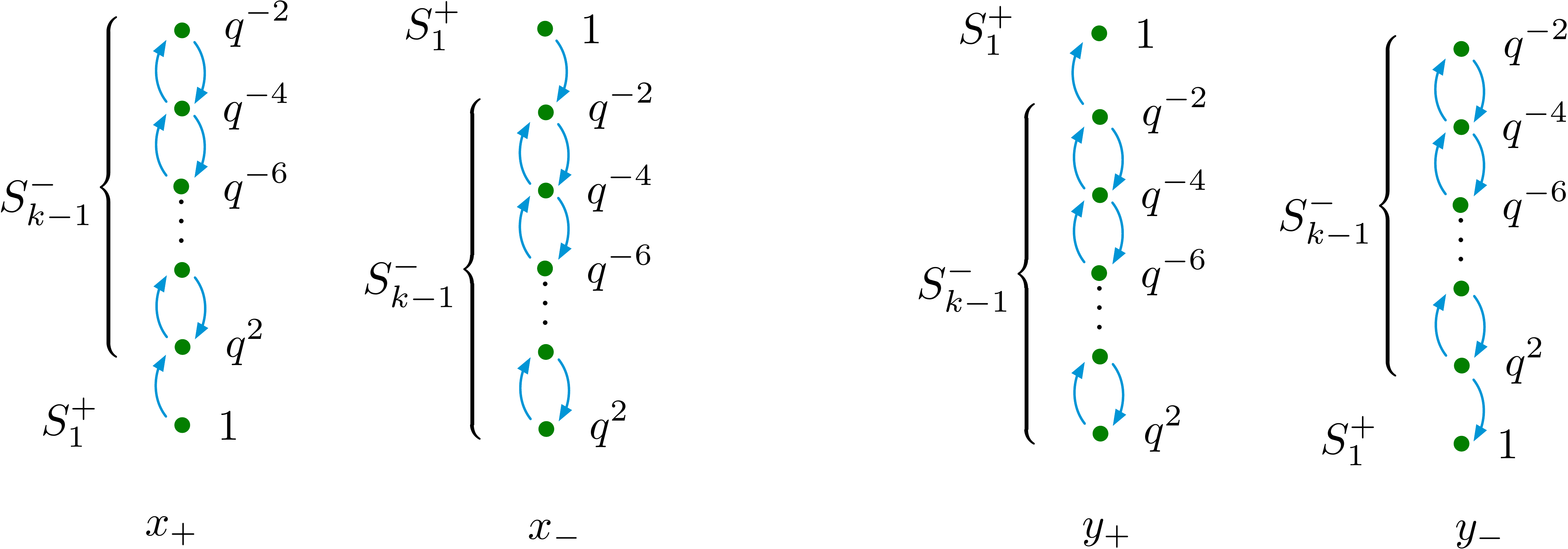}
\caption{Left: two extensions of $S_1^+$ by $S_{k-1}^-$. Right: two extensions of $S_{k-1}^-$ by $S_1^+$.}
\label{fig:extS}
\end{figure}

$\CB_1$ contains the simple modules $S_1^+$ and $S_{k-1}^-$, in the notation of \eqref{reps-small}. There are two extensions $x_\pm$ of $S_1^+$ by $S_{k-1}^-$, producing a Verma module and a dual Verma module. There are also two extensions $y_\pm$ of $S_{k-1}^-$ by $S_1^+$, also producing a Verma module and a dual Verma module. See Figure \ref{fig:extS}. Together with the idempotent projections $e_1$ (projection onto $S_1^+$) and $e_{k-1}$ (projection onto $S_{k-1}^-1$), $x_\pm$ and $y_\pm$ generate the derived endomorphism algebra
\be A_1 :=  \text{Ext}^\bullet(S_1^+\oplus S_{k-1}^-)\,, \ee
subject to the slightly nontrivial relations $x_+y_-=x_-y_+$ and $y_+x_-=y_-x_+$. Note that $e_1,e_{k-1}$ are in cohomological degree 0 and $x_\pm,y_\pm$ are in cohomological degree 1.

We may further organize $x_\pm$ and $y_\pm$ as two doublets for (a double cover of) the global symmetry group $PGL(2,\C)$. Then the relations are more suggestively denoted $\epsilon^{ab}x_ay_b = \epsilon^{ab}y_ax_b=0$, with $\epsilon=\bsp 0&1\\-1&0\esp$ the antisymmetric tensor. We may also encode $A_1$ as a quiver path algebra as in \eqref{simple-quiver},
\be \hspace{.2in} A_1\,:\quad  \raisebox{-.35in}{$\includegraphics[width=1.5in]{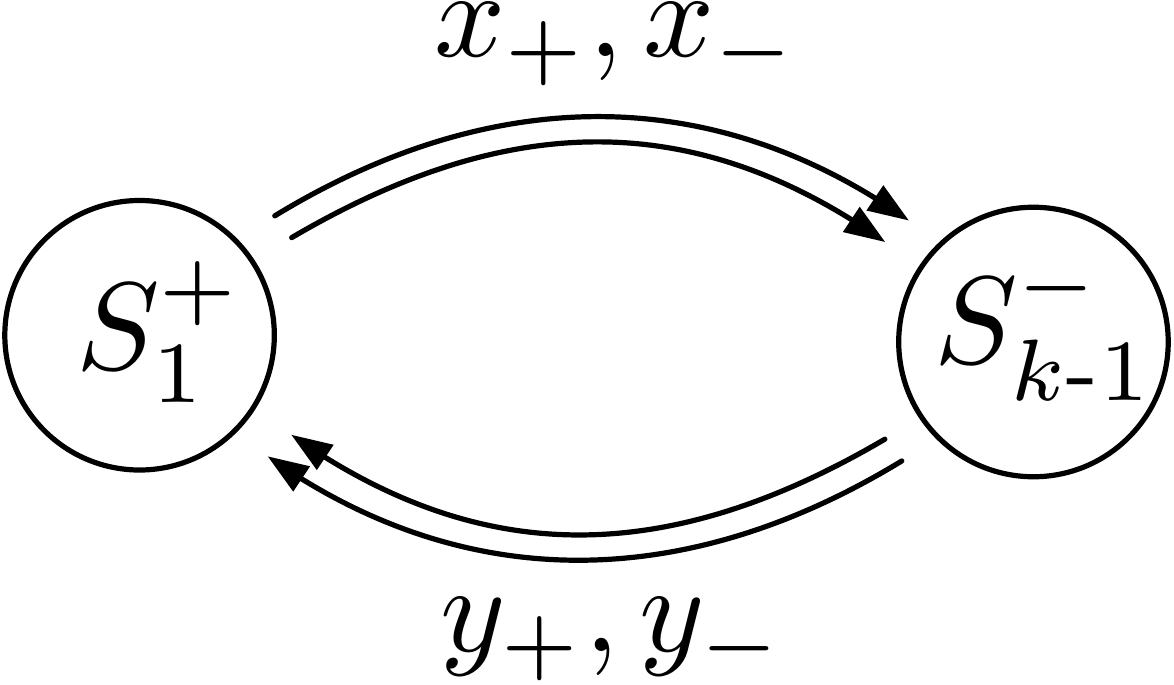}$}  \qquad \text{w/ relations $\epsilon^{ab}x_ay_b=\epsilon^{ab}y_ax_b=0$}\,. \label{simple-A} \ee
(The idempotents $e_1,e_{k-1}$ are usually not drawn in the quiver.)

It is now easy to see which quotient of $A_1$ gives the symmetric algebra $\C[X,Y]$. If we identify the objects $S_1^+,S_{k-1}^-$, and correspondingly identify $x_+=y_+$ and $x_-=y_-$, the relations just say that $x_+,x_-$ commute. Thus we obtain the symmetric algebra $\C[X,Y]$ generated by $X=x_+$ and $Y=x_-$.

We may similarly consider the two projective modules $P_1^+,P_{k-1}^-$ in the block $\CB_1$. They are depicted in \eqref{reps-proj}.  The projective modules do not admit nontrivial extensions, so their derived endomorphism algebra is equivalent to their ordinary endomorphism algebra. It is easy to see that the only maps among the two projectives are those depicted in \eqref{P-maps} 
\be \raisebox{-.5in}{$\includegraphics[width=2.6in]{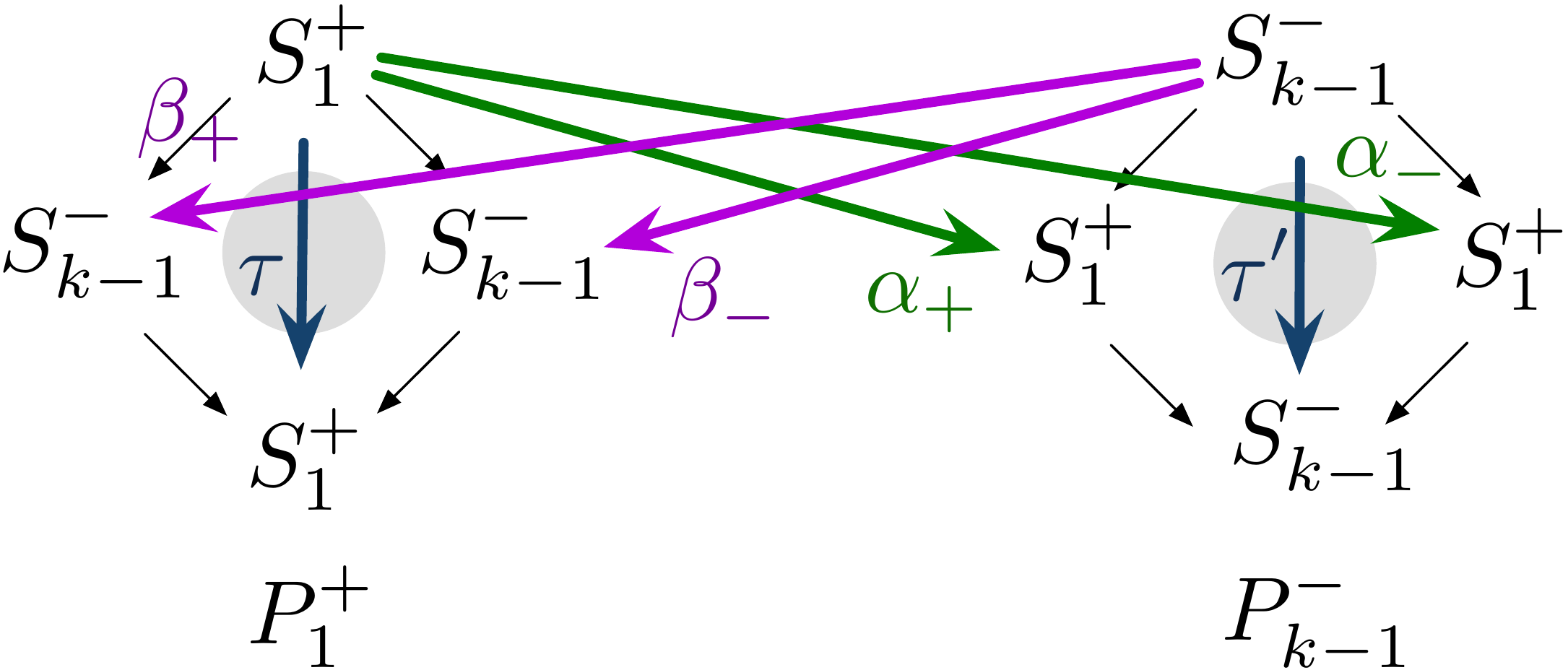}$} \label{P-maps} \ee
Namely, we have the two projections of heads (maximal quotients) onto socles (minimal submodules) $\tau$ and $\tau'$; maps $\alpha_\pm$ and $\beta_\pm$ induced by projecting the heads $S_1^+,S_{k-1}^-$ of one projective onto the intermediate composition factors of the other projective; and (as always) the two idempotent projections $e$ and $e'$, onto $P_1^+$ and $P_{k-1}^-$, respectively. Thus the algebra
\be A_1' := \text{End}(P_1^+\oplus P_{k-1}^-) \ee
is just eight-dimensional! (In contrast, the Ext algebra of the simples is clearly infinite-dimensional, since there is a quotient $A_1\to \C[X,Y]$.)

The relations in the algebra are again easiest to write down after organizing its elements in representations of a double cover of $PGL(2,\C)$. The maps $\alpha_a$ and $\beta_b$ are doublets, while $\tau,\tau',e,e'$ are invariant. We find
\be \epsilon^{ab}\beta_a\alpha_b=\tau\,,\quad \epsilon^{ab}\alpha_a\beta_b=\tau'\,,\qquad \sigma_\mu^{ab}\alpha_a\beta_b=\sigma_\mu^{ab}\beta_a\alpha_b=0 \quad (\mu=1,2,3)\,, \label{P-rels}\ee
where  $\sigma_1=\bsp 1&0\\0&1\esp$, $\sigma_2=\bsp 1&0\\0&-1\esp$ , $\sigma_3=\bsp 0&1\\1&0\esp$ are the  Pauli matrices. Notice that $\tau,\tau'$ are not independent. The algebra $A_1'$ is generated by $\alpha_\pm,\beta_\pm$ and the idempotents.

We may similarly encode $A_1'$ as a quiver algebra:
\be \label{Pquiver} A_1': \quad \raisebox{-.45in}{$\includegraphics[width=2.2in]{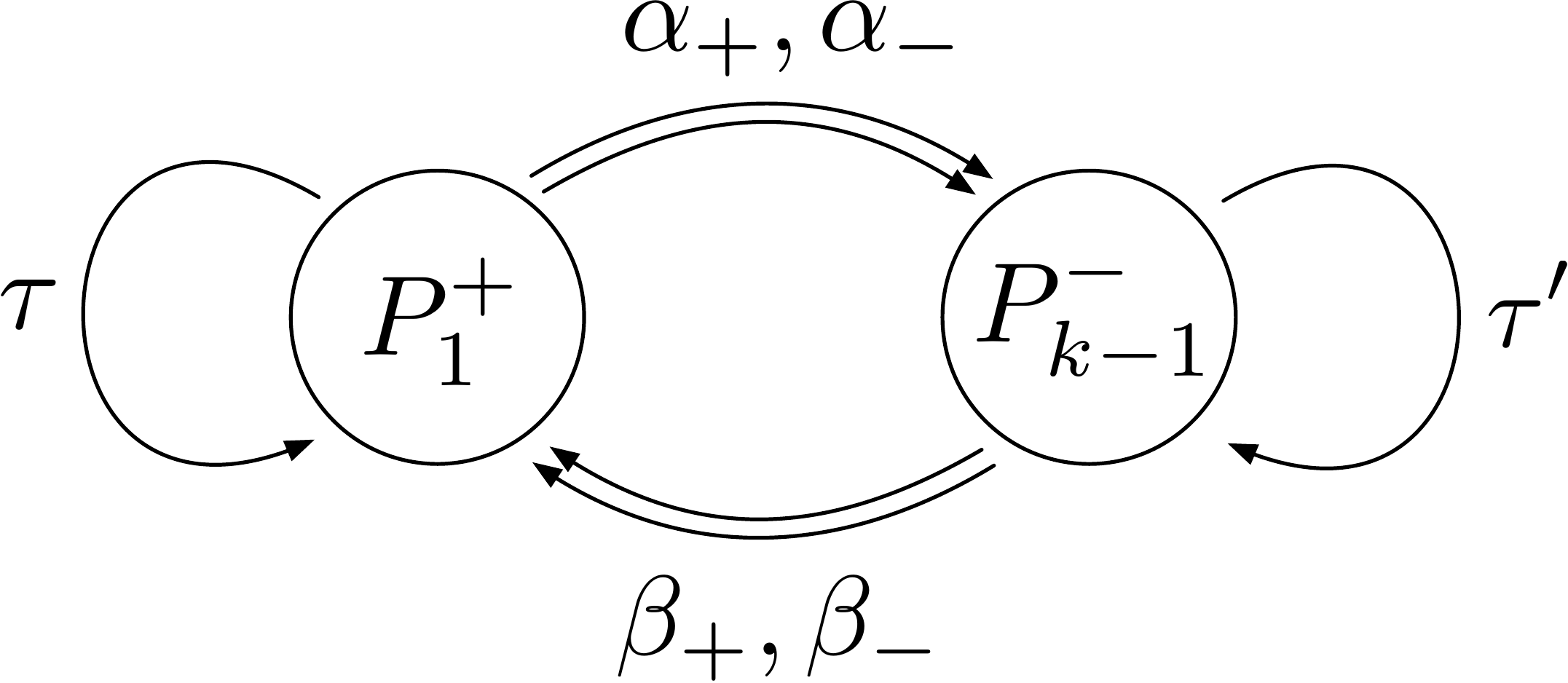}$} \quad\text{w/ relations $\sigma_\mu^{ab}\alpha_a\beta_b=\sigma_\mu^{ab}\beta_a\alpha_b=0\,.$}\ee
(Strictly speaking, one should not include $\tau,\tau'$ in the quiver, since they are not independent of $\alpha,\beta$. We include them to emphasize their existence, as the only other nontrivial maps.)

We now observe that quotienting the category $\CB_1$ in such a way that $S_1^+,S_{k-1}^-$ become identified also identifies the two projectives, and quotients the algebra $A_1'$ by setting $\alpha_+=\beta_+$ and $\alpha_-=\beta_-$. This quotient of $A_1'$ is isomorphic to the \emph{exterior} algebra $\C[\xi,\psi]$, where $\xi=\alpha_+$, $\psi=\alpha_-$, and $\xi\psi=\tau/2$.

We also observe that $A_1$ and $A_1'$ are Koszul-dual, as quadratic algebras (see \cite{Gale} and references therein for a review of this concept).  Both algebras can be given an additional non-cohomological grading, such that the generators $x_\pm,y_\pm$ and $\alpha_\pm,\beta_\pm$ all lie in degree one, and relations lie in degree two. Koszul duality amounts to the statement that, if we identify the spaces $V=\C\langle x_\pm,y_\pm\rangle$ and $W=\langle\alpha_\pm,\beta_\pm\rangle$ as linear duals of each other, the relations $\{\epsilon^{ab}x_ay_b,\epsilon^{ab}y_ax_b\}$ and $\{\sigma_\mu^{ab}\alpha_a\beta_b,\sigma_\mu^{ab}\beta_a\alpha_b\}$ span orthogonal complements of each other in $V^{\otimes 2}$ and $W^{\otimes 2}$.

\subsection{Hochschild homology from quivers}

Hochschild homology of the category $\CC_1$ decomposes by blocks, so
\be HH_\bullet(\CC) = \C^2\oplus HH_\bullet(\CB_1)^{\oplus k-1}\,. \label{blocks-app} \ee
In turn, one may compute Hochschild homology of the block $\CB_1$ by representing it as $\CB_1\simeq D^b(A_1\text{-mod})$ or $\CB_1\simeq D^b(A_1'\text{-mod})$.

The computation using $A_1$, the derived endomorphism algebra of the simples, is somewhat difficult. It ultimately leads to the geometric formulation in terms of Dolbeault cohomology of $T^*[2]\P^1$ from \cite{ABG,BL}, described in Section \ref{sec:U-1}.

The computation using $A_1'$, the derived endomorphism algebra of the projectives, can be done by hand (with some help from \emph{Mathematica}). Let $\rho_d$ denote the $d$-dimensional representation of $PGL(2,\C)$. Then we find in the first few degrees that%
\footnote{Here we use a Hochschild complex with inverted (positive) cohomological grading, in order to produce non-negatively graded spaces. See Section \ref{sec:grading}.}
\be \label{HH-P-list} \begin{array}{lcl@{\qquad}c}
&&& \text{$PGL(2,\C)$ highest weight vectors } \\
 HH_0(A_1')  &\simeq&  \rho_1^{\oplus 3} & e,\,e',\,\tau\\
HH_{1}(A_1') &\simeq& \rho_1\oplus \rho_3  & \id\otimes \tau\,,\; \beta_+\otimes \alpha_+ \quad (\id := e+e')\\
HH_{2}(A_1') &\simeq& \rho_3^{\oplus 2} & \id\otimes (\beta_+\otimes \alpha_+-\alpha_+\otimes\beta_+)\,,\; \tau\otimes \beta_+\otimes \alpha_+ \\
HH_{3}(A_1')&\simeq& \rho_3\oplus \rho_5  & \id\otimes(\tau\otimes \beta_+\otimes \alpha_++\beta_+\otimes \alpha_+\otimes\tau+\alpha_+\otimes\tau\otimes\beta_+)\,, \\
 &&& \beta_+\otimes\alpha_+\otimes\beta_+\otimes\alpha_+ \end{array}\ee
On the right, we have listed explicit Hochschild cycles that play the role of highest-weight vectors for the $PGL(2,\C)$ representations appearing on the left. The list \eqref{HH-P-list} is clearly compatible with the geometric result \eqref{dol-P1}. The generators $e,e',\tau$ of $HH_0(A_1')$ are also the ones found in the CGP approach of Section \ref{sec:CGP-0}.

The general pattern of highest-weight vectors appears to be
\be HH_{2i}(A') \simeq \rho_{2i+1}^{\oplus 2}\quad \text{w/ h.w. vecs} \quad  \begin{array}{c}(\beta_+\otimes\alpha_+)^{\otimes i}\,,\\ \id\otimes\big[\tau\otimes (\beta_+\otimes\alpha_+)^{\otimes i-1}+\text{cyclic perms.}\big]  \end{array}\,, \label{HH-guess} \ee
\be HH_{2i-1}(A') \simeq \rho_{2i-1}\oplus\rho_{2i+1} \quad \text{w/ h.w. vecs} \quad  \begin{array}{c} \id\otimes\big[(\beta_+\otimes\alpha_+)^{\otimes i-1}-(\alpha_+\otimes\beta_+)^{\otimes i-1}\big] \\  \tau\otimes(\beta_+\otimes\alpha_+)^{\otimes i-1} \end{array}\,, \notag \ee
for higher even and odd homology classes, respectively. It is easy to check that these proposed highest-weight vectors are indeed closed.

\subsection{Flat connections and twisting}

Suppose we are interested in computing the genus-one state space $\CH(T^2,\CA)$ in the presence of a  flat connection $\CA$ with generic diagonal holonomy $g$. As discussed in Section \ref{sec:Hoch-flavor}, we may then compute the torus state in two different ways, which should produce equivalent results:
\begin{itemize}
\item[1)] by deforming the category $\CC_1\leadsto \CC_g$ and computing ordinary Hochschild homology
\item[2)] by deforming the Hochschild differential as in \eqref{dHA} and computing twisted Hochschild homology  $HH_\bullet^g(\CC_1)$
\end{itemize}

We already  saw in Section \ref{sec:U-1} that method (1) leads to a semisimple category $\CC^{\rm ab}_g$ with $2k$ simple objects, so
\be HH_0(\CC_g) \simeq [1]^{\oplus 2k} = \C^{2k}\,,\qquad HH_{i<0}(\CC_g) \equiv 0\,. \label{HH-k-twist0} \ee
We now describe how to apply method (2).

Let $g = \text{diag}(\gamma,1)$ be the $PGL(2,\C)$ holonomy. We saw above that $PGL(2,\C)$ acts on the endomorphism algebra $A_1'$ of the projectives in block $\CB_1$, so we compute the corresponding twisted Hochschild homology of $A_1'$. In degree zero, we find that the new co-center of $A_1'$ is two-dimensional, generated by $e,e'$; in particular, $\tau$ has been removed, since
\be \begin{array}{c}  d_H^g(\beta_-\otimes \alpha_+) = \beta_-\alpha_+-\gamma\alpha_+\beta_- = -\tau-\gamma \tau'\,,\\
d_H^g(\alpha_-\otimes\beta_+) = \alpha_-\beta_+-\gamma\beta_+\alpha_- = -\tau'-\gamma\tau\,,
 \end{array} \ee
whence both $\tau$ and $\tau'$ are in the image of $d_H^g$ when $\gamma$ is generic. In higher degrees, we find that homology vanishes entirely. Therefore,
\be HH_0^g(\CB_1) \simeq \C^2\,,\qquad HH_{i>0}^g(\CB_1) \equiv 0\,. \ee
From the block decomposition \eqref{blocks-app} we then obtain $HH_0^g(\CC_1)\simeq \C^{2k}$ and $HH_{i>0}^g(\CC_1)\equiv 0$, in perfect agreement with  \eqref{HH-k-twist0}.

The deformation of the category $\CC_1$ to $\CC_g$ may also be understood somewhat intuitively, from the perspective of the algebra $A_1'$. Notice that $A_1'$ is the algebra of local operators on the line operator $P_1^+\oplus P_{k-1}^-$.
A deformation by a flavor holonomy along a loop linking the line operator $P_1^+\oplus P_{k-1}^-$ ought to remove all charged operators from $A_1'$, making them massive. In addition, it should remove any uncharged operators that can be created as products of charged ones. We would expect the deformed block $(\CB_1)_g$ to be the category of modules for the resulting invariant algebra.

If we decompose the algebra $A_1'$ into subspaces
\be A_1' = A_1'^{<0}\oplus A_1'^0 \oplus A_1'^{>0} \ee
according to weights of a maximal torus of $SL(2,\C)$, this procedure tells us that the flavor deformation should effectively leave us with the quotient
\be A_1'^g := \frac{A_1'^0}{A_1'^0 \cap A_1'^{<0}A_1'^{>0}}\,, \ee
where $ A_1'^{<0}A_1'^{>0}$ denotes elements of $A_1'^{<0}$ and $A_1'^{>0}$ combined in any order.

The charged operators of $A_1'$ are  $\alpha_\pm$ and $\beta_\pm$, and their products contain $\tau = \beta_+\alpha_-$ and $\tau'=\alpha_+\beta_-$. Thus the flavor deformation reduces $A_1'$ to the two-dimensional algebra
\be A_1'{}^g = \frac{\C\langle e,e',\tau,\tau'\rangle}{\C\langle \tau,\tau'\rangle} \simeq \C\langle e,e'\rangle \ee
with the usual idempotent relations $e^2=e$, $e'{}^2=e'$, $ee'=e'e=0$. The category $(\CB^{\rm ab}_1)_g=A_1^g{}'$-mod thus becomes semisimple, with two simple blocks.

\section{Computations for $T[SU(2)]/SU(2)_k$}
\label{app:TSU2}
In this Appendix we discuss the specific example of the 3d $\CN=4$ theory obtained by gauging the $SU(2)$ flavor symmetry of $T[SU(2)]$ with Chern-Simons gauge fields. We start with a description of the $A$-twist of $T[SU(2)]$ in the twisted formalism, as well as the boundary VOA of \cite{CostelloGaiotto}, in Section \ref{sec:TSU2}. We then discuss the effects of the Chern-Simons gauging in Section \ref{sec:TSU2CS} and show that the boundary VOA is concentrated in cohomological degree 0, at least for conformal dimension $\Delta \leq 2$.

\subsection{$T[SU(2)]$ in the twisted formalism}
\label{sec:TSU2}
In this subsection we discuss the example of $U(1)$ gauge theory with two charge 1 hypermultiplets. This theory is known to flow to $T[SU(2)]$ in the IR, and the boundary chiral algebra is computed in \cite{CostelloGaiotto}; here we review this result using the twisted formalism and to simplify Section \ref{sec:TSU2CS}.

Denote the charge $-1$ chiral multiplets within the hypermultiplets by $\bX^n := (\bX_1)^n$ and the charge $1$ chiral multiplets by $\bY_m := (\bY_1)_m$, $n,m = 1,2$. We additionally have a $U(1)$ vector multiplet $\bA = \bA_1$ and adjoint (charge 0) chiral multiplet $\bPhi = \bPhi_1$, which make up the $\CN=4$ vector multiplet. As discussed in \cite{twistedN=4}, the $A$-twisted action for this theory is given by
\be
\label{eq:AtwistedTSU2}
	S_{T[SU(2)]} = \int \bB \diff'\bA + \bLambda \diff' \bPhi + \bPsi_{\bX}\diff'_{\bA}\bX + \bPsi_{\bY}\diff'_{\bA}\bY + \Tr(\bX \Phi \bY) + \bB \bPhi - \bPsi_{\bX} \bPsi_{\bY}.
\ee
The resulting action of the $A$-twist supercharge $Q_A$ is given by
\be
\begin{aligned}
	\label{eq:QTSU2}
	Q_A \bA & = \diff'\bA + \bPhi \qquad & Q_A \bB & = \diff' \bB - \bX \bPsi_{\bX} + \bPsi_{\bY} \bY\\
	Q_A \bPhi & = \diff' \bPhi & Q_A \bLambda & = \diff' \bLambda + \bX \bY + \bB\\
	Q_A \bX & = \diff'_{\bA} \bX - \bPsi_{\bY} \qquad & Q_A \bPsi_{\bX} & = \diff'_{\bA} \bPsi_{\bX} + \bPhi \bY\\
	Q_A \bY & = \diff'_{\bA} \bY + \bPsi_{\bX} \qquad & Q_A \bPsi_{\bY} & = \diff'_{\bA} \bPsi_{\bY} + \bX \bPhi\\
\end{aligned},
\ee
where we have suppressed the $SU(2)$ flavor indices $n,m$.

We introduce the following boundary conditions for the fields:
\begin{itemize}
	\item Neumann boundary conditions for the vector multiplet. ($\bB|_{\pd} = 0$)
	\item Dirichlet boundary conditions for the adjoint chiral multiplet. ($\bPhi|_{\pd} = 0$)
	\item Neumann boundary conditions for the charge $\pm1$ chiral multiplets. ($\bPsi_{\bX}|_{\pd}, \bPsi_{\bY}|_{\pd} = 0$)
\end{itemize}
As discussed in \cite{CDG}, the bulk superpotential terms $\Tr(\bX \Phi \bY) + \bB \bPhi - \bPsi_{\bX} \bPsi_{\bY}$ can introduce boundary OPE's of bulk fields that are not $Q_A$-exact. In particular, if we denote the lowest component (0-form) of the twisted superfields $\bA, \bLambda, \bX, \bY$ by $c, \lambda, X, Y$ then the superpotential $\bB \bPhi - \bPsi_{\bX} \bPsi_{\bY}$ introduces the following OPE's
\be
	c(z) \lambda(w) \sim \frac{1}{z-w} \qquad X^n(z) Y_m(w) \sim \frac{\delta^n{}_m}{z-w}.
\ee
The other superpotential term $\Tr(\bX \Phi \bY)$ does not introduce an OPE.

The 3d bulk fields contribute $-2$ to the anomaly coefficient for the boundary $U(1)$ gauge symmetry, which can be compensated for by introducing two boundary Fermi multiplets denoted $\bGamma^\alpha$ of gauge charge $1$, with their conjugates $\wt{\bGamma}_\alpha$ of charge $-1$, for $\alpha = 1,2$; we will denote the lowest components of these boundary fermions by $\gamma$ and $\widetilde{\gamma}$. There is additionally a mixed anomaly involving the $U(1)_{\rm top}$ topological flavor symmetry, which we can remedy by giving $\bGamma$ charge $-\tfrac{1}{2}$ and $\wt{\bGamma}$ charge $\tfrac{1}{2}$ under this symmetry. The presence of the boundary degrees of freedom modify the boundary condition $\bB|_\pd = 0 \rightsquigarrow \bB|_\pd = \wt{\bGamma}_\alpha \bGamma^\alpha$. 
Putting this together, the boundary VOA is generated by the fields $c, \lambda, X^n, Y_m, \gamma^\alpha, \widetilde{\gamma}_\beta$ with OPE's given by
\be
c(z) \lambda(w) \sim \frac{1}{z-w} \qquad X^n(z) Y_m(w) \sim \frac{\delta^n{}_m}{z-w} \qquad \gamma^\alpha(z) \widetilde{\gamma}_\beta(w) \sim \frac{\delta^\alpha{}_\beta}{z-w}
\ee
subject to the differential
\be
\begin{aligned}
	Q_A c & = 0 \qquad & Q_A \lambda & = \norm{Y_n X^n} + \norm{\widetilde{\gamma}_\alpha \gamma^\alpha}\\
	Q_A X^n & = c X^n \qquad & Q_A Y_m & = - c Y_m\\
	Q_A \gamma^\alpha & = c \gamma^\alpha \qquad & Q_A \widetilde{\gamma}_\beta & = - c \widetilde{\gamma}_\beta\\
\end{aligned}.
\ee
From the form of $Q_A$ and the OPEs of these fields, it is clear that the boundary VOA agrees with \cite{CostelloGaiotto}. Namely, it is the $U(1)$ BRST reduction of the algebra generated by the symplectic bosons $X,Y$ and the complex fermions $\gamma, \wt{\gamma}$.

We compute the cohomology by considering the gauge invariant combinations of the fundamental fields subject to the differential induced from the above, only keeping $\pd^\ell c$ for $\ell > 0$. The basic gauge invariant operators are the bilinears $\norm{Y_m X^n}$, $\widetilde{\gamma}_\beta X^n $, $Y_m \gamma^\alpha$, $\norm{\widetilde{\gamma}_\beta \gamma^\alpha}$; the antighost $\lambda$; and the derivative of the ghost $\pd c$. A straightforward computation shows that the derivative of the ghost is exact in two ways, \cf\, Section 6.4 of \cite{CDG}:
\be
\begin{aligned}
	Q_A \norm{Y_m X^n}(z) &= \lim\limits_{w\to z} (-c(w) Y_m(w))X^n(z) + (c(z) X^n(z)) Y_m(w)\\
	& = \lim\limits_{w\to z} (c(z) - c(w))\bigg(:X^n(z) Y_m(w): + \frac{\delta^n{}_m}{z-w}\bigg)\\
	& = \delta^n{}_m \pd c(z)
\end{aligned}
\ee
and, similarly,
\be
\begin{aligned}
	Q_A \norm{\widetilde{\gamma}_\beta \gamma^\alpha}(z) & = \lim\limits_{w\to z}(-c(w) \widetilde{\gamma}_\beta(w))\gamma^\alpha(z) - \widetilde{\gamma}_\beta(w) (c(z) \gamma^\alpha(z))\\
	& = -\delta^\alpha{}_\beta \pd c(z)
\end{aligned}.
\ee
In particular, we can remove the $Q_A$-exact, gauge-invariant operator $\pd c$ and its primitive $\tfrac{1}{4}(\norm{Y_n X^n} - \norm{\widetilde{\gamma}_\alpha \gamma^\alpha})$. Similarly, we can remove $\norm{Y_n X^n} + \norm{\widetilde{\gamma}_\alpha \gamma^\alpha}$ and its primitive $\lambda$. The same should be true for derivatives of these fields.

The remaining bosonic bilinears can be organized into the currents
\be
	J_a = \norm{Y \sigma_a X} = \norm{Y_m (\sigma_a)^m{}_n X^n} \qquad L_a = \norm{\widetilde{\gamma} \sigma_a \gamma} = \norm{\widetilde{\gamma}_\beta (\sigma_a)^\beta{}_\alpha \gamma^\alpha}\,,
\ee
which generate a copy of $\mathfrak{su}(2)_{-1}$ and a simple quotient ${\rm su}(2)_1$ of $\mathfrak{su}(2)_{1}$, respectively. Similarly, the fermionic bilinears $N^n{}_{\beta} = \widetilde{\gamma}_\beta X^n$ and $M^\alpha{}_m = Y_m \gamma^\alpha$ represent non-trivial cohomology classes and together with the bosonic currents generate a ${\rm psu}(2|2)_{1}$ current algebra, in agreement with \cite{CostelloGaiotto}.

It is straightforward to deform this theory with a background flat connection $\CA$ associated to the $U(1)_{\rm top}$ flavor symmetry, \cf\, Section \ref{sec:bdy-VOA} and \ref{sec:Tnk-BV}. We work in a holomorphic gauge, so that $\CA = \CA_z(z) \diff z,$ and deform the bulk superpotential by $\W_{\rm flavor} = -\CA_z \diff z \bPhi$. This superpotential does not change any of the OPE's prior to taking $Q_A$ cohomology, but changes the action of $Q_A$:%
\be
\begin{aligned}
	Q_A c & = 0 \qquad & Q_A \lambda & = \norm{Y_n X^n} + \norm{\widetilde{\gamma}_\alpha \gamma^\alpha} - \CA_z\\
	Q_A X^n & = c X^n \qquad & Q_A Y_m & = - c Y_m\\
	Q_A \gamma^\alpha & = c \gamma^\alpha \qquad & Q_A \widetilde{\gamma}_\beta & = - c \widetilde{\gamma}_\beta\\
\end{aligned},
\ee
and hence changes the OPE's of cohomology classes. As an example of this phenomenon, consider the operators $\eta = \Tr(N)$ and $\overline{\eta} = \Tr(M)$. Prior to taking $Q_A$ cohomology, their OPE is given by
\be
	\eta(z) \overline{\eta}(w) \sim \frac{2}{(z-w)^2} + \frac{1}{z-w}\big(\norm{Y_n X^n}(w) + \norm{\wt{\gamma}_\alpha \gamma^\alpha}(w)\big) = \frac{2}{(z-w)^2} + \frac{\CA_z(w)}{z-w} + Q_A\lambda\,.
\ee
Thus, the OPE at the level of cohomology gets deformed
\be
	\eta(z) \overline{\eta}(w) \sim \frac{2}{(z-w)^2} \quad \rightsquigarrow \quad \eta(z) \overline{\eta}(w) \sim \frac{2}{(z-w)^2} + \frac{\CA_z(w)}{z-w}\,,
\ee
as expected.

\subsection{Gauging $SU(2)_k$}
\label{sec:TSU2CS}
We now discuss how the boundary chiral algebra changes after gauging the $SU(2)$ flavor symmetry of $T[SU(2)]$ with Chern-Simons fields at level $k$. Just as with the Yang-Mills gauge fields, we give the Chern-Simons gauge fields Neumann boundary conditions. As discussed in Section \ref{sec:Atwist}, the $A$-twist of an $\CN=4$ Chern-Simons matter theory doesn't introduce any superpotential involving the Chern-Simons fields (unlike the $\bB \bPhi$ term used for gauging with Yang-Mills fields), the corresponding boundary chiral algebra (in the $A$-twist) should be obtained by taking derived $SL(2,\C[\![z]\!])$ invariants of the boundary algebra discussed in Section \ref{sec:TSU2}, possibly dressed by boundary degrees of freedom to cancel for any gauge anomalies. Importantly, this is {\em not} the same chiral algebra that would result from gauging this symmetry with an $\CN=4$ vectormultiplet; there is no gaugino to implement the vanishing of the moment map.

Since the Chern-Simons gauge fields (at level $k$) and the hypermultiplets $X^n, Y_m$ are given Neumann boundary conditions, the bulk fields contribute $-k+2-1 = -k+1$ to the boundary anomaly coefficient. We shall assume that $k \geq 1$ so that we can remedy this boundary anomaly with $k-1$ $SU(2)$ doublets of boundary complex fermions, which we denote $\rho^n_i, \widetilde{\rho}^j_m$. We then take derived invariants with respect to the $SL(2,\C[\![z]\!])$ acting on the doublet indices $n,m$. Just as above, we compute this cohomology by first restricting to $SU(2)$ invariants of the algebra generated by the $\textrm{psu}(2|2)_1$ currents $J_a, N^n{}_\beta, M^\alpha{}_m, L_a$ together with the fermions $\rho^n_i, \widetilde{\rho}^j_m$. We then compute the cohomology of the resulting operators with respect to $Q_A$. Since the scaling dimension of $Q_A$ is zero, we can perform this computation at fixed scaling dimension. In what follows, we will show that the cohomology of $Q_A$ is concentrated in cohomological degree $0$ for scaling dimension $\Delta \leq 2$ and expect this feature to persist to higher $\Delta$.

As above, the action of $Q_A$ on normal-ordered products can be subtle. For example, one finds
\be
\begin{aligned}
	Q_A J_a(z) & = \lim\limits_{w\to z} \norm{Y(w) \sigma_a X(z)} = \lim\limits_{w\to z} Y(w) \sigma_a X(z)\\
	& = \lim\limits_{w\to z} (-c^b(w) Y(w)) \sigma_b \sigma_a X(z) + c^b(z) Y(w) \sigma_a \sigma_b X(z)\\
	& = f^c{}_{ab} c^b(z) \norm{Y \sigma_c X}(z) + \pd c^b(z) \Tr[\sigma_a \sigma_b]\\
	& =  f^c{}_{ab} c^b J_c(z) + K_{ab} \pd c^b(z)
\end{aligned},
\ee
where $K_{ab} = \Tr[\sigma_a \sigma_b] = 2 \delta_{ab}$ and $f^c{}_{ab} = 2 i \epsilon_{abc'}\delta^{c' c}$ are the $\mathfrak{su}(2)$ structure constants; the first term of the last line is the usual term associated to a gauge transformation with parameter $c^a(z)$ but the second term is new. In situations where the insertion points of operators is obvious, we will leave it implicit. For example, the above would read
\be
	Q_A J_a = f^c{}_{ab} c^b J_c + K_{ab} \pd c^b.
\ee

We denote $(R_a)^i{}_j = \norm{\widetilde{\rho}^i \sigma_a \rho_j}$ and $R_a = (R_a)^i{}_i$; note that $R_a$ is an $\mathfrak{su}(2)_{k-1}$ current and thus the current $J^{\textrm{tot}}_a = J_a + R_a$ is an $\mathfrak{su}(2)_{k-2}$ current, as expected. Indeed, for operators $\CO$ not involving $\pd c^a$, the action of $Q_A$ can be expressed as in terms of the OPE of $\CO$ and $J^{\textrm{tot}}_a$:
\be
Q_A \CO(z) = -\sum \limits_{n \geq 0} \frac{1}{n!} \pd^n c^a(z) \oint (w-z)^n J^{\textrm{tot}}_a(w) \CO(z) \diff w.
\ee
The above expression for $Q_A J_a$ is equally straightforward to using this formula and the OPE
\be
J^{\rm tot}_a(z) J_b(w) \sim J_a(z) J_b(w) \sim \frac{f^c{}_{ab}}{z-w} J_c(w) - \frac{K_{ab}}{(z-w)^2}.
\ee
For later convenience we also denote $(Z_a)_{ij} = (\epsilon \rho_i)\sigma_a \rho_j$ and $(\widetilde{Z}_a)^{ij} = \widetilde{\rho}^i \sigma_a (\widetilde{\rho}^j \epsilon)$, where $(\epsilon \rho_i)_m = \epsilon_{mn} \rho^n_i$ and $(\widetilde{\rho}^i \epsilon)^n = \widetilde{\rho}^j_m \epsilon^{mn}$.

Note that there are no $SU(2)$-invariant operators with $\Delta < 1$. At $\Delta = 1$, the $SU(2)$-invariant operators are concentrated in degree 0, hence all survive in cohomology. They are the $\textrm{su}(2)_{1}$ currents $L_a$ together with the currents $U^i{}_j(z) = \norm{\widetilde{\rho}^i \cdot \rho_j }$, $Z_{ij} = (\epsilon \rho_i)\cdot \rho_j$, and $\widetilde{Z}^{ij} = \widetilde{\rho}^i \cdot (\widetilde{\rho}^j\epsilon)$, where $\cdot$ denotes contracting the $SU(2)$ indices. These operators have non-vanishing OPE's given by
\be
\begin{aligned}
	U^i{}_j(z) U^k{}_l(w) & \sim  \frac{\delta^k{}_j}{z-w} U^i{}_l(w) - \frac{\delta^i{}_l}{z-w} U^k{}_j(w) + \frac{2 \delta^i{}_l \delta^k{}_j}{(z-w)^2}\\
	U^i{}_j(z) Z_{kl}(w) & \sim \frac{\delta^i{}_k}{z-w} Z_{jl}(w) + \frac{\delta^i{}_l}{z-w} Z_{kj}(w)\\
	U^i{}_j(z) \widetilde{Z}^{kl}(w) & \sim - \frac{\delta^k{}_j}{z-w} \widetilde{Z}^{il}(w) - \frac{\delta^l{}_j}{z-w} \widetilde{Z}^{ki}(w)\\
	Z_{ij}(z) \widetilde{Z}^{kl}(w) & \sim -\frac{\delta^{k}{}_i}{z-w}U^l{}_j(w) - \frac{\delta^{l}{}_i}{z-w}U^k{}_j(w) - \frac{\delta^{k}{}_j}{z-w}U^l{}_i(w) - \frac{\delta^{l}{}_j}{z-w}U^k{}_i(w) - 2\frac{\delta^{k}{}_i \delta^{l}{}_j + \delta^{l}{}_i \delta^{k}{}_j}{(z-w)^2}\\
\end{aligned}
\ee
Together, these OPE's imply that the currents $U, Z, \widetilde{Z}$ form a copy of the simple quotient ${\rm usp}(k-1)_{1}$ of $\mathfrak{usp}(k-1)_{1}$.

The $SU(2)$-invariant operators at $\Delta = \tfrac{3}{2}$ are again concentrated in degree 0. We will denote these operators $\widetilde{S}^i{}_\beta = \widetilde{\rho}^i \cdot N_\beta$, $S_{j \beta} = (\epsilon \rho_j) \cdot N_\beta,$ which together transform in the fundamental representation of ${\rm usp}(k-1)_{1}$ and separately in anti-fundamental representations of $\rm{su}(2)_1$, and $\widetilde{T}^{i \alpha} = M^\alpha \cdot (\widetilde{\rho}^i \epsilon)$, $T^{\beta}{}_i = M^\alpha \cdot \rho_i,$ which together transform in the fundamental representation of ${\rm usp}(k-1)_{1}$ and separately in fundamental representations of $\rm{su}(2)_1$.

Finally, consider scaling dimension $\Delta = 2$. The operators of interest are $J_a \pd c^a$, $(R_a)^i{}_j \pd c^a$, $(Z_a)_{ij} \pd c^a$, and $(\widetilde{Z}_a)^{ij} \pd c^a$.\footnote{The gauge-invariant bilinear $K_{ab} \pd c^a \pd c^b$ vanishes because $K_{ab}$ is symmetric and $\pd c^a$ is Grassmann odd.} We can see that the last three operators as $Q_A$-exact fairly easily. In particular,
\be
\begin{aligned}
	Q_A \big[\norm{\widetilde{\rho}^i \cdot \pd \rho_j}\big] = - Q_A \big[\norm{\pd \widetilde{\rho}^i \cdot \rho_j}\big] & = (R_a)^i{}_j \pd c^a\\
	Q_A \big[(\epsilon \rho_i) \cdot \pd \rho_j\big] = - Q_A \big[(\epsilon \pd \rho_i) \cdot \rho_j\big]& =  (Z_a)_{ij} \pd c^a\\
	Q_A \big[\widetilde{\rho}^i \cdot (\pd \widetilde{\rho}^j \epsilon)\big] = - Q_A \big[\pd \widetilde{\rho}^i \cdot ( \widetilde{\rho}^j \epsilon)\big] & = -(\widetilde{Z}_a)^{ij} \pd c^a\\
\end{aligned}.
\ee
For the first operator, we find that
\be
Q_A \big[K^{ab}\norm{J_a J_b}\big] = 2 J_a \pd c^a
\ee
and similarly
\be
\begin{aligned}
	Q_A \big[K^{ab} \norm{(R_a)^i{}_j (R_b)^k{}_l}\big] & = -\big(\delta^i{}_j (R_a)^k{}_l + \delta^k{}_l (R_a)^i{}_j\big) \pd c^a\\
	Q_A \big[K^{ab} J_a (R_b)^i{}_j\big] & = -\big(\delta^i{}_j J_a - (R_a)^i{}_j\big) \pd c^a\\
	Q_A \big[K^{ab} \norm{(R_a)^i{}_j (Z_b)_{kl}}\big] & = -\delta^i{}_j (Z_a)_{kl} \pd c^a\\
	Q_A \big[K^{ab} J_a (Z_b)_{kl}\big] & = (Z_a)_{kl} \pd c^a\\
	Q_A \big[K^{ab} \norm{(R_a)^i{}_j (\widetilde{Z}_b)^{kl}}\big] & = -\delta^i{}_j (\widetilde{Z}_a)^{kl} \pd c^a\\
	Q_A \big[K^{ab} J_a (\widetilde{Z}_b)^{kl}\big] & = (\widetilde{Z}_a)^{kl} \pd c^a\\
\end{aligned}
\ee
All of these expressions can be obtained from an OPE of these operators with $J^{\textrm{tot}}$.

The above yield some interesting cohomology classes besides derivatives or bilinears of the currents $L, U, Z, \widetilde{Z}$:
\be
\begin{aligned}
	Q_A \big[K^{ab} \norm{(\delta^i{}_j J_a + (R_a)^i{}_j)(\delta^k{}_l J_b + (R_b)^k{}_l)}\big] = 0\,\,\\
	Q_A \big[K^{ab} \norm{(\delta^i{}_j J_a + (R_a)^i{}_j)(Z_b)_{kl}}\big] = 0\,\,\\
	Q_A \big[K^{ab} \norm{(\delta^i{}_j J_a + (R_a)^i{}_j)(\widetilde{Z}_b)^{kl}}\big] = 0\,\,\\
	Q_A \big[\delta^i{}_j \norm{\widetilde{\rho}^k \cdot \rho_l} + \delta^k{}_l \norm{\widetilde{\rho}^i \cdot \rho_j} + K^{ab}\norm{(R_a)^i{}_j (R_b)^k{}_l} \big] = 0\,\,\\
	Q_A \big[(\epsilon \rho_i) \cdot \pd \rho_j - K^{ab} J_a (Z_b)_{ij}\big] = 0\,\,\\
	Q_A \big[(\widetilde{\rho}^i) \cdot (\pd \widetilde{\rho}^j) + K^{ab} J_a (\widetilde{Z}_b)^{ij}\big] = 0\,.\\
\end{aligned}
\ee


\newpage

\bibliographystyle{JHEP}

\providecommand{\href}[2]{#2}\begingroup\raggedright\endgroup

\end{document}